\documentclass[11pt]{report}
\usepackage[
  dissertation
 ,final
 ,raggedbottom
 ,tocbold
]{USCthesis}

\usepackage[export]{adjustbox} 
\usepackage{amsmath}
\usepackage{amssymb}
\usepackage{array}
\usepackage[utf8]{inputenc} 
\usepackage[english]{babel}
\usepackage[
  backend     = biber,
  doi         = true,
  hyperref    = true,
  maxbibnames = 99,
  style       = numeric,
]{biblatex}
\usepackage{booktabs}
\usepackage{color, colortbl}
\usepackage{csquotes}
\usepackage{efbox}
\usepackage{enumitem}
\usepackage[shortcuts]{extdash} 
\usepackage[tt=false]{libertine} 
\usepackage[T1]{fontenc} 
\usepackage[
  showframe = false,
  pass      = true, 
]{geometry}
\usepackage{graphicx}
\usepackage{hyphenat}
\usepackage{ifthen}
\usepackage{lipsum}
\usepackage{multirow}
\usepackage{parnotes}
\usepackage{pdflscape} 
\usepackage{pifont}
\usepackage{ragged2e}
\usepackage{seqsplit}
\usepackage{siunitx}
\usepackage{subcaption}
\usepackage{tabularx}
\usepackage{xcolor}
\usepackage{xspace}
\usepackage{url}
\usepackage{mathptmx}
\usepackage{diagbox}
\usepackage{rotating}

\DeclareMathAlphabet{\mathcal}{OMS}{cmsy}{m}{n}
\usepackage{algorithmic}
\usepackage[linesnumbered,vlined,ruled]{algorithm2e}
\SetKwInOut{Parameters}{Parameters}

\DeclareMathAlphabet{\mathbf}{OT1}{cmr}{bx}{n}
\newcommand{\Mod}[1]{\ (\mathrm{mod}\ #1)}

\usepackage{mathtools}
\DeclarePairedDelimiter{\ceil}{\lceil}{\rceil}
\DeclareMathOperator*{\argmin}{arg\,min}
\DeclareMathOperator*{\subjectto}{subject\,to}

\usepackage{appendix}
\AtBeginEnvironment{subappendices}{%
\chapter*{Appendix}
\counterwithin{table}{section}
}

\AtEndEnvironment{subappendices}{%
\counterwithin{table}{chapter}
}

\usepackage[
  breaklinks    = true,
  colorlinks    = true,
  hypertexnames = false,
  pdfpagelabels = false,
  citecolor     = {blue!80!black},
  linkcolor     = {blue!80!black},
  urlcolor      = {blue!80!black},
]{hyperref} 

\AtEveryBibitem{\iffieldundef{doi}{}{\clearfield{url}}} 
\addto\extrasenglish{%

}

\begin{document}

\thispagestyle{empty}

\noindent
\begin{minipage}{\textwidth}
\centering
\textbf{{\fontsize{28pt}{30pt}\selectfont Revisiting FastMap: New Applications}}\\
{\fontsize{16pt}{18pt}\selectfont Ang Li}\\
{\fontsize{16pt}{18pt}\selectfont T.~K.~Satish Kumar}\\
\textbf{{\fontsize{18pt}{20pt}\selectfont University of Southern California}}
\end{minipage}

\vspace{3cm}

This is a copy of Ang Li's PhD dissertation submitted to the University of Southern California (Department of Computer Science) in August 2024. It is presented here as an article coauthored by him and his PhD adviser T.~K.~Satish Kumar.

\newpage

\title{REVISITING FASTMAP: NEW APPLICATIONS}

\author{Ang Li}

\majorfield{COMPUTER SCIENCE}

\submitdate{August 2024}

\begin{preface}

\prefacesection{Acknowledgements}
First, I would like to thank my adviser, T.~K.~Satish Kumar, for his valuable guidance on my research and his unwavering support for me at every stage of my doctoral education at the University of Southern California. I would also like to thank him for helping me revise this document.

Second, I would like to thank the rest of my PhD dissertation committee, including John Carlsson, Emilio Ferrara, Sven Koenig, and Aiichiro Nakano, for their valuable time, guidance, support, and feedback. I would especially like to thank Sven for his feedback on many of my published research papers. I would also like to thank Peter Stuckey from Monash University, Australia, for his valuable discussions with me on my research projects.

Third, I would like to thank my other collaborators and other members of my adviser's research group, including Nori Nakata, Kushal Sharma, Omkar Thakoor, Malcolm White, Han Zhang, and Kexin Zheng, for productive and friendly discussions.

Finally, I would like to thank my family members and friends for their understanding and constant support during the intense period of my doctoral education.

My doctoral education at the University of Southern California has shaped me as a researcher and as a person in many ways. I will cherish the memories of it for a long time to come, thanks to all the people mentioned above for creating and engaging me in a prolific research environment.

The research presented in this dissertation is the culmination of multiple published research papers. My research was partially supported by DARPA under grant number HR001120C0157 and by NSF under grant numbers 1409987, 1724392, 1817189, 1837779, 1935712, and 2112533.

{
\hypersetup{hidelinks}
\tableofcontents
\listoftables
\listoffigures
}

\prefacesection{Abstract}
Many recent breakthroughs in Artificial Intelligence (AI) have stemmed from our ability to create~\emph{embeddings} of complex objects. A ``good'' embedding of the objects in a given domain assigns numerical coordinates to every object so as to capture the properties of individual objects as well as the relationships between them. Hence, embeddings open up the possibility of using powerful analytical and geometric techniques to reason about complex objects. For example, recent breakthroughs in Natural Language Processing (NLP) utilize word and sentence embeddings. This dissertation is also based on the same paradigm but with a strong emphasis on the efficiency of the embedding procedure. Towards this end, we build on an algorithm called FastMap that was originally introduced in the Data Mining community.

FastMap is a clever embedding algorithm that circumvents the complexity of representing~\emph{individual} objects by leveraging a distance (similarity) function on~\emph{pairs} of them. On the one hand, objects in many real-world domains often require complex representations and cannot always be visualized as points in geometric space. Examples of such complex objects include long deoxyribonucleic acid (DNA) strings, multi-media datasets like voice excerpts or images, and medical datasets like electrocardiograms (ECGs) or magnetic resonance images (MRIs). On the other hand, many of the same real-world domains naturally lend themselves to well-defined distance functions on pairs of objects. Examples include the~\emph{edit distance} between two DNA strings, the~\emph{Minkowski distance} between two images, and the~\emph{cosine similarity} between two text documents. FastMap leverages such a distance function efficiently: It embeds a collection of $N$ complex objects in an artificially created Euclidean space in only $O(\kappa N)$ time, where $\kappa$ is the user-specified number of dimensions. While FastMap attempts to preserve all of the quadratic number of pairwise distances between the objects in the Euclidean embedding, it remarkably does so in only linear time. Its efficiency has already lead to many applications in Data Mining, particularly with regard to fast indexing, searching, and enabling clustering algorithms on complex objects that otherwise require a collection of points in geometric space as input.

In the first part of this dissertation, we present a generalization of FastMap to graphs: This graph version of FastMap, also called FastMap for convenience, embeds the vertices of a undirected edge-weighted graph as points in a Euclidean space in near-linear time (linear time after ignoring logarithmic factors). The pairwise Euclidean distances between the points approximate a desired graph-based distance function on the corresponding vertices. With a proper choice of the distance function, FastMap allows us to efficiently interpret a graph-theoretic problem in geometric space. This leads to an important upshot: In the modern era, graphs are used to represent social networks, communication networks, and transportation networks, among similar structures in many other domains with entities and relationships between them. These graphs can be very large with millions of vertices and hundreds of millions of edges. Therefore, algorithms with a running time that is quadratic or more in the size of the input are undesirable. In fact, algorithms with any super-linear running times, discounting logarithmic factors, are also largely undesirable. Hence, a desired algorithm should have a near-linear running time close to that of merely reading the input. FastMap tries to address these requirements by first creating a geometric interpretation of a given graph-theoretic problem in only near-linear time and consequently enabling analytical and geometric techniques that are better at absorbing large input sizes compared to discrete algorithms that work directly on the input graph. We apply FastMap to solve various new graph-theoretic problems of significant interest in AI: including facility location, top-$K$ centrality computations, community detection and block modeling, and graph convex hull computations. Through comprehensive experimental results, we show that our FastMap-based approaches outperform many state-of-the-art competing methods for these problems, in terms of both the efficiency and the effectiveness.

In the second part of this dissertation, we propose a novel Machine Learning (ML) framework, called FastMapSVM, which combines FastMap and Support Vector Machines (SVMs) to classify complex objects. While Neural Networks (NNs) and related Deep Learning (DL) methods are popularly used for classifying complex objects, they are generally based on the paradigm of~\emph{characterizing individual} objects. In contrast, FastMapSVM is generally based on the paradigm of~\emph{comparing pairs} of objects via a distance function. One benefit of FastMapSVM is that the distance function can encapsulate and invoke the intelligence of other powerful algorithms such as the $A^*$ search procedure and maxflow computations, among many other optimization methods. The distance function can also incorporate domain-specific knowledge that otherwise may be hard for ML algorithms to automatically extract from the data. FastMapSVM serves as a lightweight alternative to NNs for classifying complex objects, particularly when training data or time is limited. It also extends the applicability of SVMs to domains with complex objects by combining the complementary strengths of FastMap and SVMs. Furthermore, FastMapSVM provides a perspicuous visualization of the objects and the classification boundaries between them. This aids human interpretation of the data and results. It also enables a human-in-the-loop framework for refining the processes of learning and decision making. We apply FastMapSVM to solve various classification problems of significant interest in AI: including the problem of identifying and classifying seismograms in Earthquake Science and the problem of predicting the satisfiability of Constraint Satisfaction Problems (CSPs). Through comprehensive experimental results, we show that FastMapSVM outperforms many state-of-the-art competing methods for these problems, particularly in terms of the training data and time required for producing high-quality results.

In summary, FastMap plays a key role in representing the vertices of a graph, or complex objects in other domains, as points in Euclidean spaces. The ability of FastMap to efficiently generate these ``simplified'' representations of the vertices, or the complex objects, enables many powerful downstream algorithms developed in diverse research communities such as AI, ML, Computational Geometry, Mathematics, Operations Research, and Theoretical Computer Science. Hence, we envision that FastMap can facilitate and harness the confluence of these algorithms and find future applications in many other problem domains that are not necessarily discussed here.

\end{preface}

\chapter{Introduction to FastMap}
\label{ch:fastmap}
In this chapter, we first review the FastMap algorithm as it was originally presented in the Data Mining community. We then review the graph version of the FastMap algorithm. We finally review some of the previous applications of the FastMap algorithm and its graph version. Hence, this chapter is intended to provide the necessary background material for the remaining chapters in this dissertation. It is also intended to provide some related work, although each following chapter provides more context and related work on the subject matter it broaches.

\section{The Original FastMap Algorithm in Data Mining}

Many algorithms developed in Machine Learning (ML) and Computational Geometry require the input to be a collection of points in Euclidean space. For example, routinely used clustering algorithms, such as the $K$-means algorithm, assume that the input is presented as a collection of points in Euclidean space. Other ML algorithms that make the same assumption include Gaussian Mixture Model (GMM) clustering, Principal Component Analysis (PCA), and Support Vector Machines (SVMs), among many others. Many important algorithms developed in Computational Geometry also work on a collection of points in Euclidean space. For example, algorithms that construct Voronoi diagrams~\cite{a91} to answer nearest-neighbor queries efficiently require such an assumption. Similarly, many analytical techniques developed in Mathematics generally require the conceptualization of the objects in the domain as points in Euclidean space.

\begin{figure}[t!]
\centering
\includegraphics[width=0.8\columnwidth]{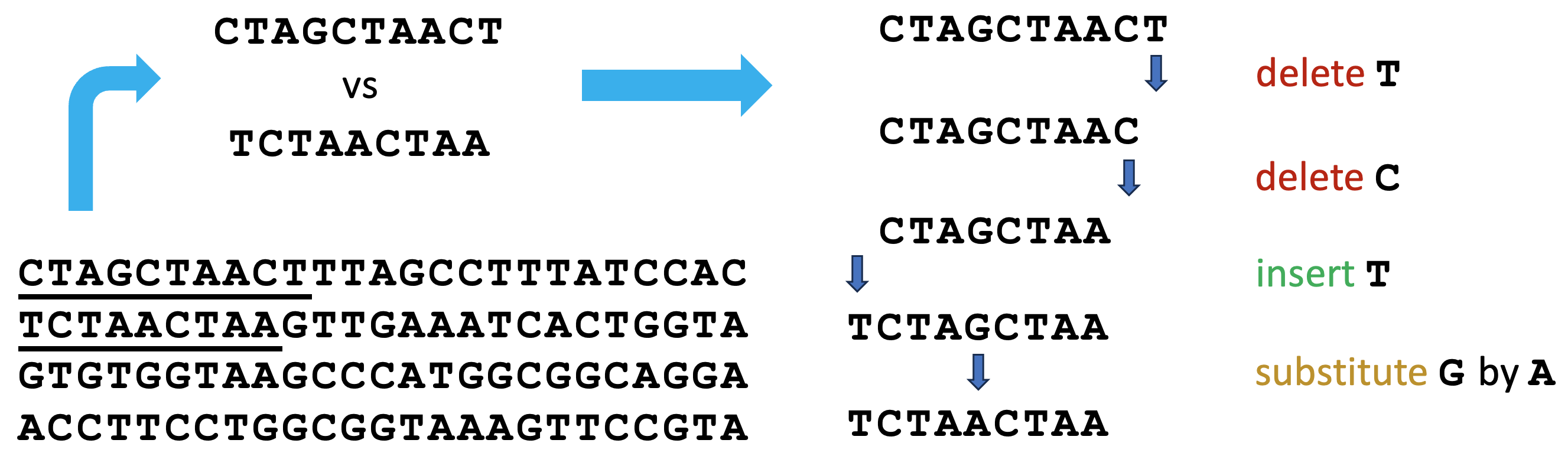}
\caption[An illustration of the edit distance between two DNA strings.]{Illustrates the edit distance between two DNA strings. The left half shows two snippets of DNA strings extracted from a collection of them. The right half shows the minimum number of edit operations required to convert one to the other.}
\label{fig:DNAs}
\end{figure}

\begin{figure}[t!]
\centering
\includegraphics[width=0.6\columnwidth]{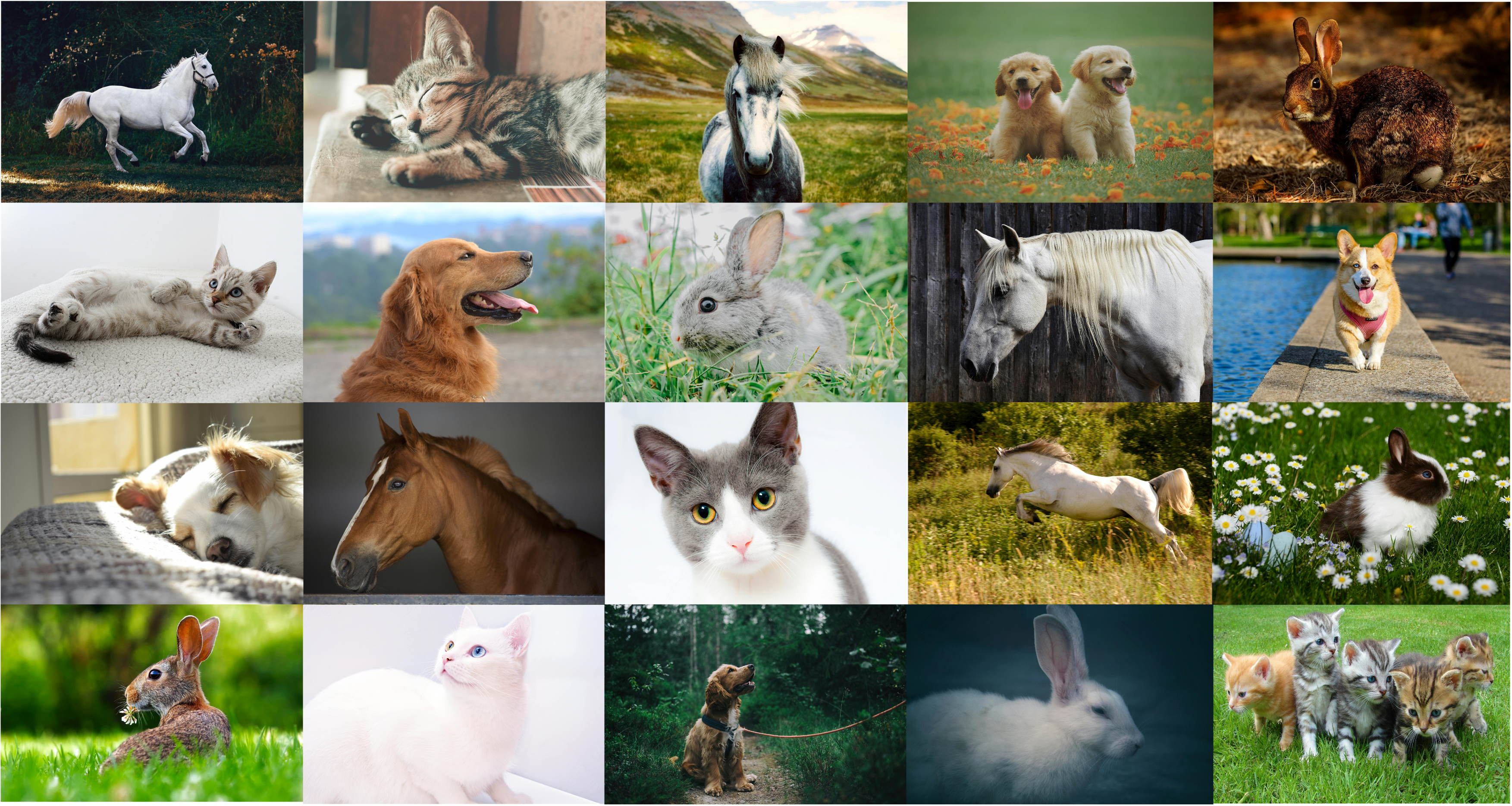}
\caption[A clustering task on a domain with images of animals.]{Shows a domain where the complex objects are images of animals. A well-defined clustering task is to group the images that portray the same animal species.}
\label{fig:animals}
\end{figure}

\begin{figure}[t!]
\centering
\includegraphics[width=0.6\columnwidth]{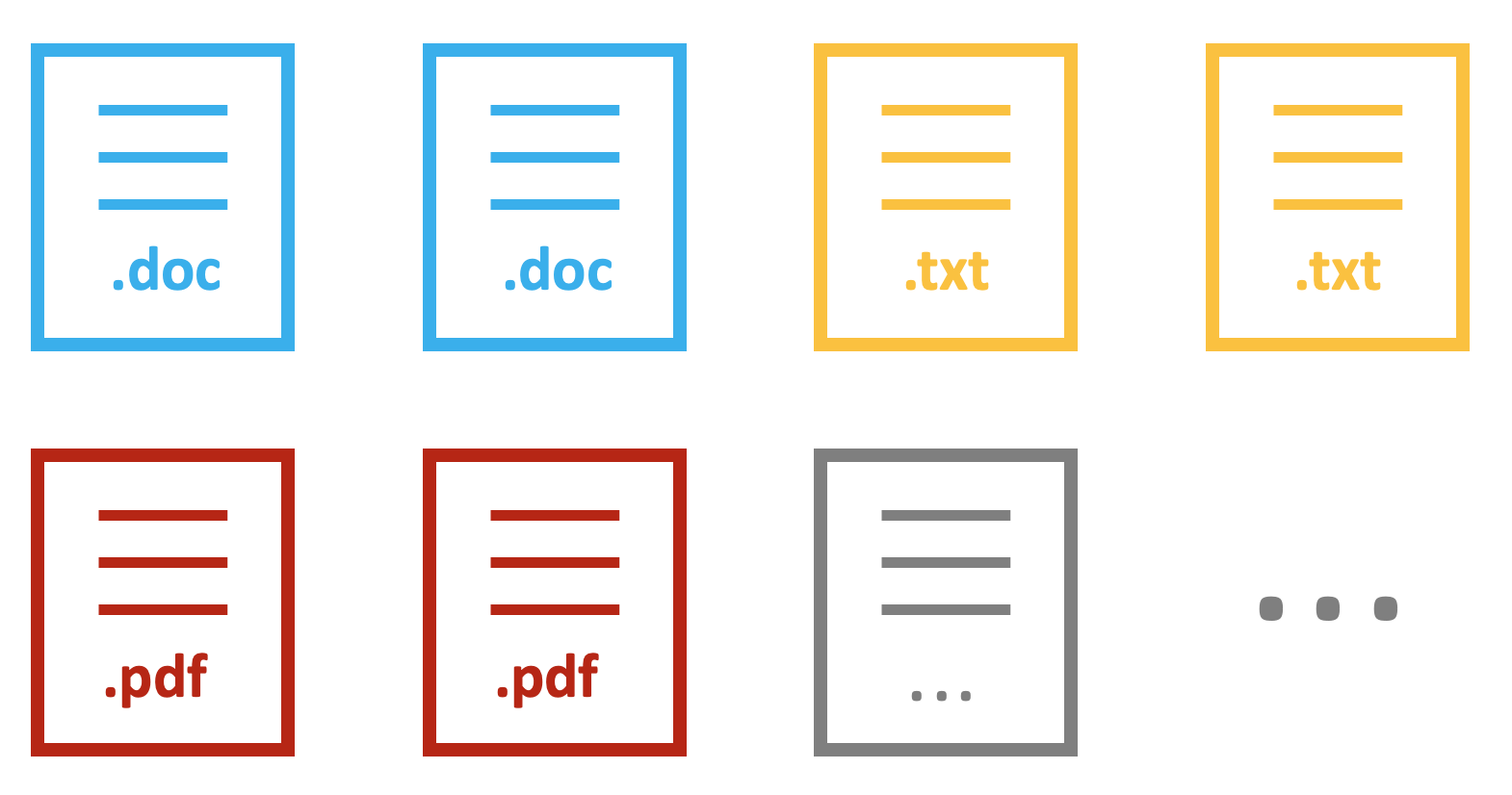}
\caption[A clustering task on a domain with text documents.]{Shows a domain where the complex objects are text documents (albeit in different formats). A well-defined clustering task is to group the text documents that have similar content.}
\label{fig:docs}
\end{figure}

Unfortunately, a plethora of such useful algorithms are rendered useless in domains with complex objects that cannot be easily conceptualized as points in Euclidean space. Figures~\ref{fig:DNAs},~\ref{fig:animals}, and~\ref{fig:docs} show three such domains in which the complex objects are deoxyribonucleic acid (DNA) strings, images, and text documents, respectively. In these domains, although it is unwieldy to conceptualize the objects as points in Euclidean space, it is easy to observe that the clustering tasks are well defined. Hence, the quest for a universal way of extending the applicability of algorithms that work on a collection of points in Euclidean space to domains with complex objects is of immense significance.

Apropos this quest, FastMap~\cite{fl95} was originally introduced in the Data Mining community to circumvent the complexity of representing~\emph{individual} objects by leveraging a distance (similarity) function on~\emph{pairs} of them. On the one hand, many real-world domains have objects such as DNA strings, multi-media datasets like voice excerpts or images, or medical datasets like electrocardiograms (ECGs) or magnetic resonance images (MRIs) that require complex representations and cannot always be visualized as points in geometric space. On the other hand, many of the same real-world domains naturally lend themselves to well-defined distance functions on pairs of objects. Examples include the~\emph{edit distance} between two DNA strings~\cite{ry98}, the~\emph{Minkowski distance} between two images~\cite{lkd07}, and the~\emph{cosine similarity} between two text documents~\cite{rka12}. Figure~\ref{fig:DNAs} illustrates the edit distance between two snippets of DNA strings: It is the minimum number of insertions, deletions, or substitutions that are needed to transform one to the other.

FastMap embeds a collection of $N$ complex objects in an artificially created Euclidean space in only $O(\kappa N)$ time, where $\kappa$ is the user-specified number of dimensions. While FastMap attempts to preserve all of the $O(N^2)$ number of pairwise distances between the objects in the Euclidean embedding, it remarkably does so in only linear time. Hence, it efficiently enables geometric interpretations, algebraic manipulations, and downstream ML algorithms.

\subsection{An Algorithmic Description of FastMap}

\begin{figure}[t!]
\centering
\subfloat[the ``cosine law'' projection in a triangle]{
\begin{minipage}[b]{0.45\textwidth}
\centering
\includegraphics[width=0.9\textwidth]{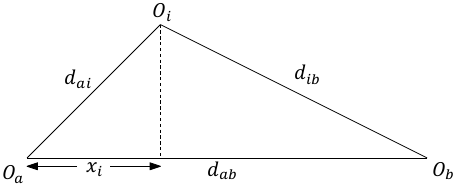}
\label{fig:cosine_law}
\end{minipage}
}
\subfloat[projection onto a hyperplane that is perpendicular to $\overline{O_aO_b}$]{
\begin{minipage}[b]{0.45\textwidth}
\centering
\includegraphics[width=0.9\textwidth]{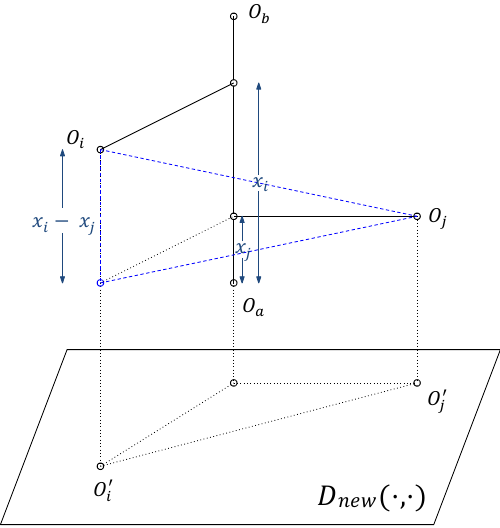}
\label{fig:hyperplane}
\end{minipage}
}
\caption[Iterative computation of coordinates in FastMap.]{Illustrates how coordinates are computed and recursion is carried out in FastMap, borrowed from~\cite{cujakk18}.}
\label{fig:fastmap}
\end{figure}

\begin{algorithm}[t!]
\caption{{\sc FastMap-DataMining}: A linear-time algorithm for embedding complex objects in a Euclidean space.}
\label{alg:fastmap_dm}
\textbf{Input}: ${\mathcal O}$, $\kappa$, and $\epsilon$\\
\textbf{Output}: $p_i \in \mathbb{R}^r$ for all $O_i \in {\mathcal O}$
\begin{algorithmic}[1]
\FOR{$r = 1, 2 \ldots \kappa$}
\STATE Choose $O_a \in {\mathcal O}$ randomly and let $O_b = O_a$.\label{line:pivot_changing_start}
\FOR{$t = 1, 2 \ldots Q$ (a small constant)}
\STATE $\{d_{ai}: O_i \in {\mathcal O}\} \leftarrow \{D(O_a, O_i): O_i \in {\mathcal O}\}$.
\STATE $O_c \leftarrow \mbox{argmax}_{O_i} \{d_{ai}^2 - \sum_{j=1}^{r-1} ([p_i]_j - [p_a]_j)^2\}$.\label{line:dist_update1}
\IF{$O_c == O_b$}
\STATE Break.
\ELSE
\STATE $O_b \leftarrow O_a$; $O_a \leftarrow O_c$.
\ENDIF
\ENDFOR\label{line:pivot_changing_end}
\STATE $d'_{ab} \leftarrow  D(O_a, O_b)^2 - \sum_{j=1}^{r-1} ([p_b]_j - [p_a]_j)^2$.\label{line:dist_update2}
\IF{$d'_{ab} < \epsilon$}
\STATE $r \leftarrow r - 1$; Break.
\ENDIF
\FOR{each $O_i \in {\mathcal O}$}
\STATE $d'_{ai} \leftarrow  D(O_a, O_i)^2 - \sum_{j=1}^{r-1} ([p_i]_j - [p_a]_j)^2$.\label{line:dist_update3}
\STATE $d'_{ib} \leftarrow  D(O_i, O_b)^2 - \sum_{j=1}^{r-1} ([p_b]_j - [p_i]_j)^2$.\label{line:dist_update4}
\STATE $[p_i]_r \leftarrow (d'_{ai} + d'_{ab} - d'_{ib})/(2\sqrt{d'_{ab}})$.
\ENDFOR\label{line:compute_coord_end}
\ENDFOR
\STATE \textbf{return} $p_i \in \mathbb{R}^r$ for all $O_i \in {\mathcal O}$.
\end{algorithmic}
\end{algorithm}

FastMap gets as input a collection of $N$ complex objects ${\mathcal O}$, a domain-specific distance function $D(\cdot, \cdot)$ defined on all pairs of objects, and a user-specified value of $\kappa$. It outputs a $\kappa$-dimensional Euclidean embedding of the complex objects. A Euclidean embedding assigns a $\kappa$-dimensional point $p_i \in \mathbb{R}^{\kappa}$ to each object $O_i$. A good Euclidean embedding is one in which the Euclidean distance $\chi_{ij}$ between any two points $p_i$ and $p_j$ closely approximates $D(O_i, O_j)$. Here, for $p_i = ([p_i]_1, [p_i]_2 \dots [p_i]_{\kappa})$ and $p_j = ([p_j]_1, [p_j]_2 \dots [p_j]_{\kappa})$, $\chi_{ij} = \sqrt{\sum_{r = 1}^{\kappa} ([p_j]_r - [p_i]_r)^2}$; and $D(O_i, O_j)$ is the domain-specific distance between the objects $O_i, O_j \in {\mathcal O}$.

In the very first iteration, FastMap heuristically identifies the farthest pair of objects $O_a$ and $O_b$ in linear time. Once $O_a$ and $O_b$ are determined, every other object $O_i$ defines a triangle with sides of lengths $d_{ai} = D(O_a, O_i)$, $d_{ab} = D(O_a, O_b)$, and $d_{ib} = D(O_i, O_b)$, as shown in Figure~\ref{fig:cosine_law}. The sides of the triangle define its entire geometry, and the projection of $O_i$ onto the line $\overline{O_aO_b}$ is given by:
\begin{equation}
x_i = (d_{ai}^2 + d_{ab}^2 - d_{ib}^2) / (2d_{ab}).
\label{eqn:x_i}
\end{equation}

FastMap sets the first coordinate of $p_i$, the embedding of $O_i$, to $x_i$. In the subsequent $\kappa - 1$ iterations, the same procedure is followed for computing the remaining $\kappa - 1$ coordinates of each object. However, the distance function is adapted for different iterations. For example, for the first iteration, the coordinates of $O_a$ and $O_b$ are $0$ and $d_{ab}$, respectively. Because these coordinates fully explain the true domain-specific distance between these two objects, from the second iteration onward, the rest of $p_a$ and $p_b$'s coordinates should be identical. Intuitively, this means that the second iteration should mimic the first one on a hyperplane that is perpendicular to the line $\overline{O_aO_b}$, as shown in Figure~\ref{fig:hyperplane}. Although the hyperplane is never constructed explicitly, its conceptualization implies that the distance function for the second iteration should be changed for all $i$ and $j$ in the following way:
\begin{equation}
D_{new}(O'_i, O'_j)^2 = D(O_i, O_j)^2 - (x_j - x_i)^2.
\label{eqn:D}
\end{equation}
Here, $O'_i$ and $O'_j$ are the projections of $O_i$ and $O_j$, respectively, onto this hyperplane, and $D_{new}(\cdot, \cdot)$ is the new distance function. The same reasoning is used to derive the distance function for the third iteration from the distance function for the second iteration, and so on.

In each of the $\kappa$ iterations, FastMap heuristically finds the farthest pair of objects according to the distance function defined for that iteration. These objects are called~\emph{pivots} and can be stored as reference objects for future use. There are very few, that is, $\leq 2\kappa$, reference objects. Technically, finding the farthest pair of objects in any iteration takes $O(N^2)$ time. However, FastMap uses a linear-time ``pivot changing'' heuristic~\cite{fl95} to efficiently and effectively identify a pair of objects $O_a$ and $O_b$ that is very often the farthest pair. It does this by initially choosing a random object $O_b$ and then choosing $O_a$ to be the farthest object away from $O_b$. It then reassigns $O_b$ to be the farthest object away from $O_a$, reassigns $O_a$ to be the farthest object away from $O_b$, and so on, until convergence or a maximum of $Q$ iterations, for a small constant $Q \leq 10$.

Algorithm~\ref{alg:fastmap_dm} presents the pseudocode for the FastMap algorithm described above. It uses a threshold parameter $\epsilon$ to detect large values of $\kappa$ that have diminishing returns on the accuracy of approximating the pairwise distances between the objects. Hence, it returns an embedding in $\mathbb{R}^r$ for a certain $r \leq \kappa$. On Lines~\ref{line:dist_update1},~\ref{line:dist_update2},~\ref{line:dist_update3}, and~\ref{line:dist_update4}, the algorithm invokes the arguments explained in Figure~\ref{fig:hyperplane} and updates the distance function as required by the current iteration. On Lines~\ref{line:pivot_changing_start} to~\ref{line:pivot_changing_end}, it finds the farthest pair of objects via the pivot changing heuristic. On Lines~\ref{line:dist_update2} to~\ref{line:compute_coord_end}, Algorithm~\ref{alg:fastmap_dm} invokes the arguments explained in Figure~\ref{fig:cosine_law} and computes the next coordinate of each object.

\section{FastMap for Embedding Graphs}

FastMap can also be used to embed the vertices of a undirected edge-weighted graph in a Euclidean space. The idea is to view the vertices of a given undirected edge-weighted graph $G = (V, E, w)$ as the objects to be embedded. However, the distance function on the vertices can be defined in many ways. For illustration and without much loss of generality, we assume that the distance function of interest is the one that returns the shortest-path distances between vertices. Hence, the objective is to embed the vertices of $G$ in a Euclidean space so as to preserve the pairwise shortest-path distances between them as Euclidean distances.

As such, the original FastMap algorithm described in Algorithm~\ref{alg:fastmap_dm} cannot be directly used for generating the required Euclidean embedding of the vertices in linear time. This is because it assumes that the distance $d_{ij}$ between any two objects $O_i$ and $O_j$ can be computed in constant time, independent of the number of objects in the problem domain. However, computing the shortest-path distance between two vertices depends on the size of the graph.

\begin{figure}[t!]
\centering
\includegraphics[width=0.6\columnwidth]{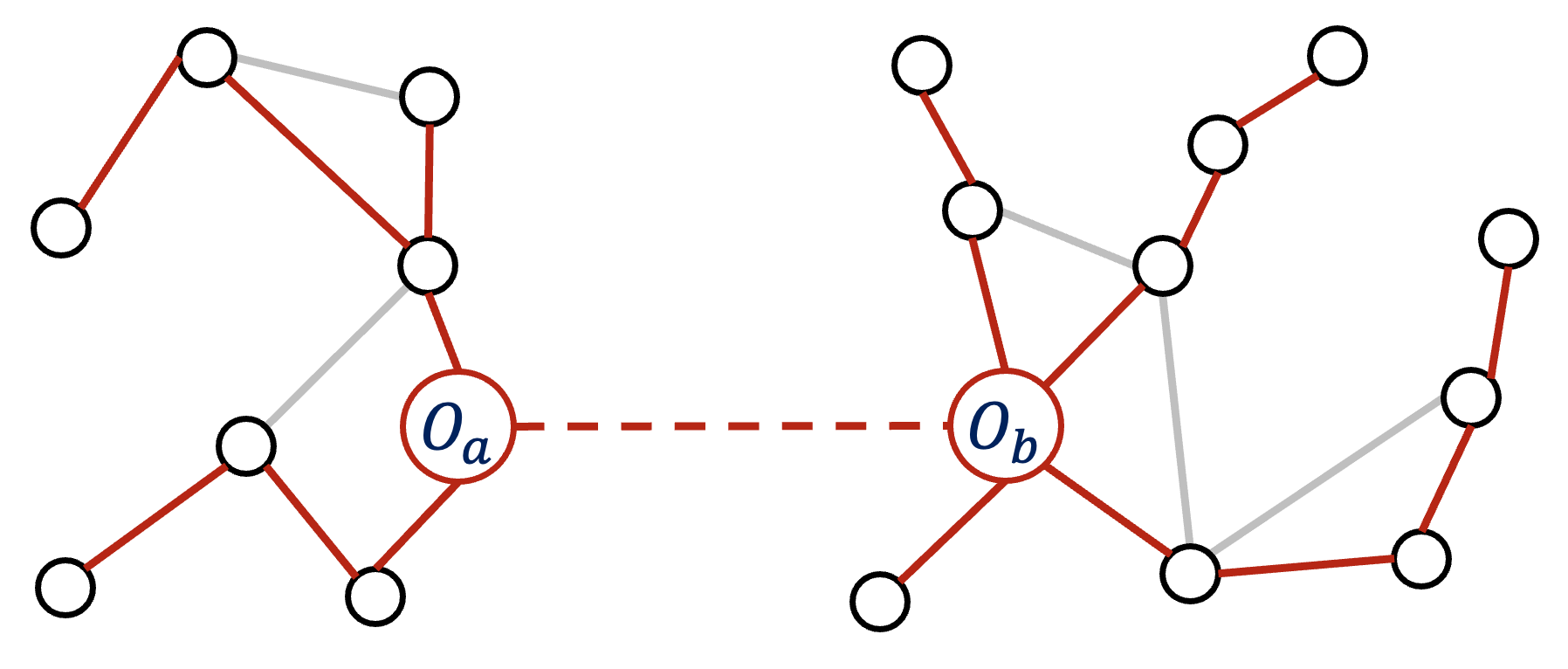}
\caption[An illustration of the two shortest-path trees rooted at the pivots in each iteration of FastMap on graphs.]{Illustrates the two shortest-path trees rooted at the pivots in each iteration of FastMap on graphs.}
\label{fig:shortest_path_trees}
\end{figure}

The issue of having to retain (near-)linear time complexity can be addressed as follows: In each iteration, after we heuristically identify the farthest pair of vertices $O_a$ and $O_b$, the distances $d_{ai}$ and $d_{ib}$ need to be computed for~\emph{all} other vertices $O_i$. Computing $d_{ai}$ and $d_{ib}$ for any single vertex $O_i$ can no longer be done in constant time but requires $O(|E| + |V|\log |V|)$ time instead~\cite{ft87}. However, since we need to compute these distances for all vertices, computing two shortest-path trees rooted at each of the vertices $O_a$ and $O_b$ yields all necessary shortest-path distances in one shot. Figure~\ref{fig:shortest_path_trees} illustrates the construction of thees two shortest-path trees. The complexity of doing so is also $O(|E| + |V|\log |V|)$, which is only linear in the size of the graph\footnote{unless $|E| = O(|V|)$, in which case the complexity is near-linear in the size of the input because of the $\log |V|$ factor}. This simple yet critical observation revives the applicability of FastMap on graphs.

\begin{algorithm}[t!]
\caption{{\sc FastMap}: A near-linear-time algorithm for embedding the vertices of a given undirected edge-weighted graph in a Euclidean space.}
\label{alg:fastmap}
\textbf{Input}: $G = (V, E, w)$, $\kappa$, and $\epsilon$\\
\textbf{Output}: $p_i \in \mathbb{R}^r$ for all $v_i \in V$
\begin{algorithmic}[1]
\FOR{$r = 1, 2 \ldots \kappa$}
\STATE Choose $v_a \in V$ randomly and let $v_b = v_a$.\label{line:g_pivot_changing_start}
\FOR{$t = 1, 2 \ldots Q$ (a small constant)}
\STATE $\{d_{ai}: v_i \in V\} \leftarrow \mbox{ShortestPathTree}(G, v_a)$.\label{line:spt1}
\STATE $v_c \leftarrow \mbox{argmax}_{v_i} \{d_{ai}^2 - \sum_{j=1}^{r-1} ([p_i]_j - [p_a]_j)^2\}$.\label{line:g_dist_update1}
\IF{$v_c == v_b$}
\STATE Break.
\ELSE
\STATE $v_b \leftarrow v_a$; $v_a \leftarrow v_c$.
\ENDIF
\ENDFOR\label{line:g_pivot_changing_end}
\STATE $\{d_{ai}: v_i \in V\} \leftarrow \mbox{ShortestPathTree}(G, v_a)$.\label{line:spt2}
\STATE $\{d_{ib}: v_i \in V\} \leftarrow \mbox{ShortestPathTree}(G, v_b)$.\label{line:spt3}
\STATE $d'_{ab} \leftarrow d^2_{ab} - \sum_{j=1}^{r-1} ([p_b]_j - [p_a]_j)^2$.\label{line:g_dist_update2}
\IF{$d'_{ab} < \epsilon$}
\STATE $r \leftarrow r - 1$; Break.
\ENDIF
\FOR{each $v_i \in V$}
\STATE $d'_{ai} \leftarrow d^2_{ai} - \sum_{j=1}^{r-1} ([p_i]_j - [p_a]_j)^2$.\label{line:g_dist_update3}
\STATE $d'_{ib} \leftarrow d^2_{ib} - \sum_{j=1}^{r-1} ([p_b]_j - [p_i]_j)^2$.\label{line:g_dist_update4}
\STATE $[p_i]_r \leftarrow (d'_{ai} + d'_{ab} - d'_{ib})/(2\sqrt{d'_{ab}})$.
\ENDFOR\label{line:g_compute_coord_end}
\ENDFOR
\STATE \textbf{return} $p_i \in \mathbb{R}^r$ for all $v_i \in V$.
\end{algorithmic}
\end{algorithm}

The foregoing observations are used in~\cite{lfkk19} to build a graph version of FastMap that embeds the vertices of a given undirected edge-weighted graph in a Euclidean space in near-linear time: The Euclidean distances approximate the pairwise shortest-path distances between vertices.

Algorithm~\ref{alg:fastmap} presents the pseudocode for FastMap on graphs. It gets as input the undirected edge-weighted graph $G = (V, E, w)$, where $w(e)$ is the non-negative weight on edge $e \in E$, and the two parameters $\kappa$ and $\epsilon$. It outputs an embedding of the vertices in $\mathbb{R}^r$ for a certain $r \leq \kappa$. As in Algorithm~\ref{alg:fastmap_dm}, $r$ detects the point of diminishing returns on the accuracy of the embedding. Moreover, the overall control structure of Algorithm~\ref{alg:fastmap} mirrors that of Algorithm~\ref{alg:fastmap_dm} with the difference that the distances from the pivots to all other objects in each iteration of Algorithm~\ref{alg:fastmap} are computed in one shot via the function $\mbox{ShortestPathTree}(\cdot, \cdot)$ on Lines~\ref{line:spt1},~\ref{line:spt2}, and~\ref{line:spt3}. On Lines~\ref{line:g_dist_update1},~\ref{line:g_dist_update2},~\ref{line:g_dist_update3}, and~\ref{line:g_dist_update4}, Algorithm~\ref{alg:fastmap} invokes the arguments explained in Figure~\ref{fig:hyperplane} and updates the distances as required by the current iteration. On Lines~\ref{line:g_pivot_changing_start} to~\ref{line:g_pivot_changing_end}, it finds the farthest pair of vertices via the pivot changing heuristic. On Lines~\ref{line:g_dist_update2} to~\ref{line:g_compute_coord_end}, it invokes the arguments explained in Figure~\ref{fig:cosine_law} and computes the next coordinate of each vertex.

\section{Applications}

\begin{figure}[t!]
\centering
\includegraphics[width=0.4\columnwidth]{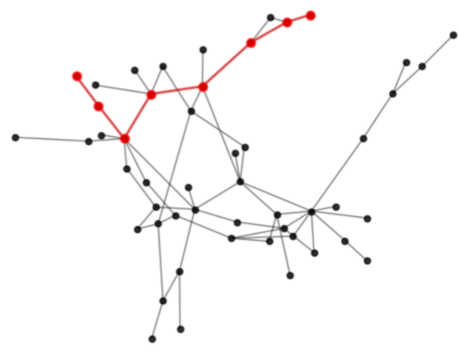}
\hspace{0.05\columnwidth}
\includegraphics[width=0.4\columnwidth]{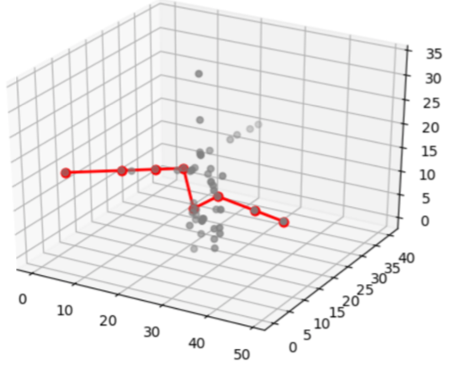}
\caption[An illustration of the Euclidean embedding produced by the graph version of FastMap for shortest-path computations.]{Illustrates the graph version of FastMap for shortest-path computations. The left panel shows an undirected edge-weighted graph. The right panel shows the $3$-dimensional Euclidean embedding of it produced by FastMap. The Euclidean distances can be used for heuristic guidance while computing shortest paths (shown in red) via $A^*$ search.}
\label{fig:shortest_path_computations}
\end{figure}

\begin{figure}[t!]
\centering
\includegraphics[width=0.8\columnwidth]{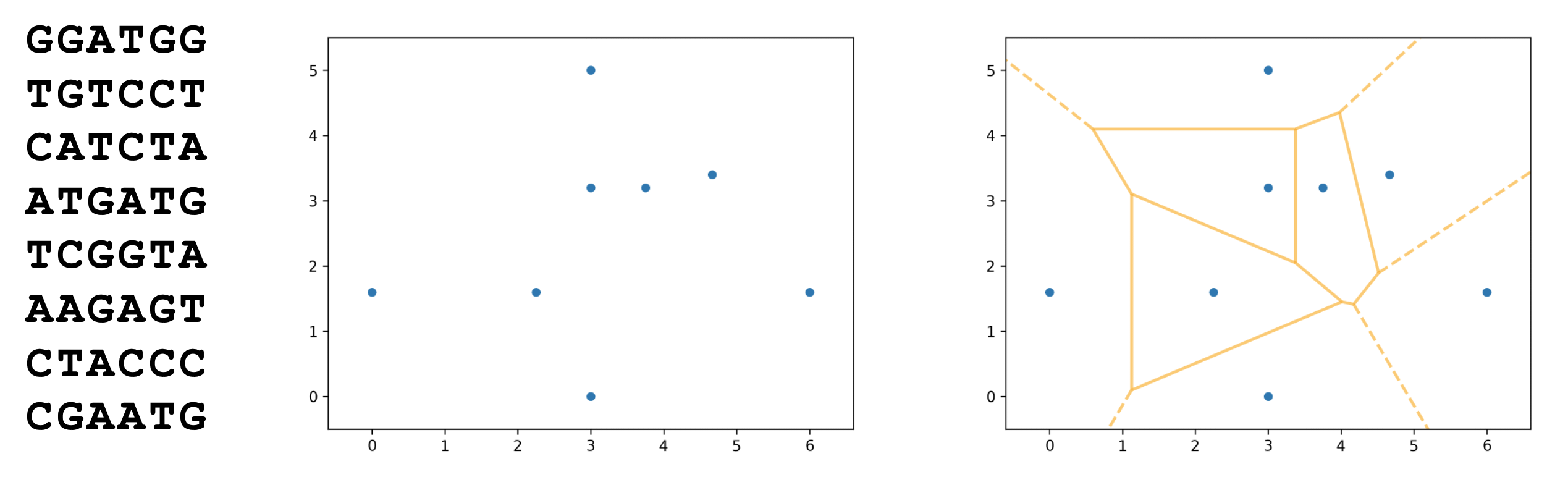}
\caption[An illustration of how the FastMap embedding can be used to answer nearest-neighbor queries efficiently.]{Illustrates how the FastMap embedding can be used to answer nearest-neighbor queries efficiently. The left panel shows snippets of DNA strings, on pairs of which the edit distance is well defined. The middle panel shows a $2$-dimensional Euclidean embedding of these snippets of DNA strings produced by FastMap. The right panel shows a Voronoi diagram that is constructed on these point representations to answer nearest-neighbor queries efficiently at runtime.}
\label{fig:DNA_fm_voronoi}
\end{figure}

In this section, we briefly mention some of the applications of the original FastMap algorithm and the graph version of it. The efficiency of the original FastMap algorithm has already lead to many applications in Data Mining, particularly with regard to fast indexing, searching, and enabling clustering algorithms on complex objects that otherwise require a collection of points in geometric space as input~\cite{fl95}. In the same contexts, the embeddings produced by FastMap allow us to visualize the complex objects, their spread, and the outliers, thereby facilitating human-in-the-loop reasoning methods. As for the graph version of FastMap, a slight modification of it, presented in~\cite{cujakk18}, preserves the~\emph{admissibility} and the~\emph{consistency} of the Euclidean distance approximation, required in an $A^*$ search framework, if used as a heuristic. This version of FastMap, used in the $A^*$ search framework for heuristic guidance, leads to one of the state-of-the-art algorithms for shortest-path computations. Figure~\ref{fig:shortest_path_computations} provides an illustration.

With a runtime complexity close to that of merely reading the input, FastMap produces a Euclidean embedding and a geometric interpretation of combinatorial problems on complex objects and graphs. In doing so, it delegates the combinatorial heavy-lifting to analytical and geometric techniques that are often better equipped for absorbing large input sizes. The properties of the Euclidean space can be leveraged in many ways. First, it empowers analytical methods to set up and solve equations or establish other conditions of optimality. Second, it empowers geometric methods to conceive of structures such as straight lines, angles, and bisectors, which facilitate visual intuition and techniques from Computational Geometry.

\subsection{Complementing FastMap with Locality Sensitive Hashing}

\begin{figure}[t!]
\centering
\includegraphics[width=0.9\columnwidth]{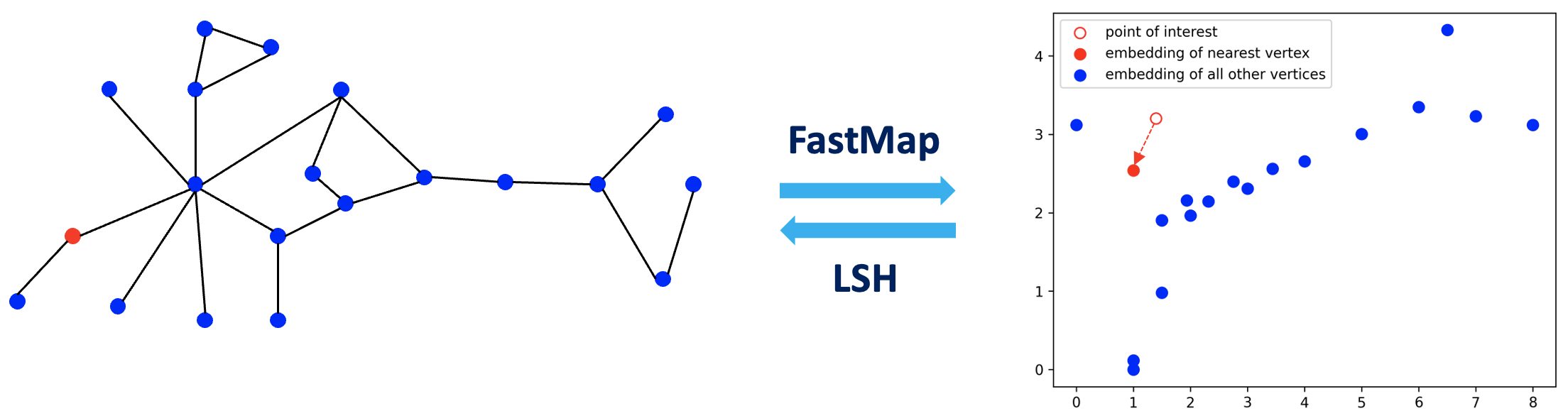}
\caption[An illustration of the coupling of FastMap and Locality Sensitive Hashing.]{Illustrates how FastMap and LSH are coupled with each other. FastMap allows us to efficiently interpret a graph (a collection of complex objects) in geometric space. LSH allows us to efficiently map a point of interest in the geometric space to the nearest vertex (object).}
\label{fig:fm_lsh}
\end{figure}

While the input of FastMap and its graph version consists only of a finite number of objects, the Euclidean space it generates is a continuous space and has an infinite number of points. Therefore, while every object maps to a point in the Euclidean space, not every point in the Euclidean space maps to an object. In fact, a point in the Euclidean space deemed as belonging to a solution by a downstream algorithm\textemdash or any other point of interest in the Euclidean space\textemdash may not map to an object in the original problem domain.

To address the foregoing concern, we assign any point of interest in the Euclidean space to its nearest neighbor that maps to an object. This requires us to answer nearest-neighbor queries very efficiently. Fortunately, this problem is well studied in Computational Geometry. For example, in a $2$-dimensional Euclidean space with straight-line distances, nearest-neighbor queries can be answered in logarithmic time using Voronoi diagrams~\cite{a91}. Figure~\ref{fig:DNA_fm_voronoi} illustrates this.

Although Voronoi diagrams are well defined in higher dimensions as well, constructing and using them in higher-dimensional spaces may become computationally expensive. Therefore, in higher dimensions, we use the idea of Locality Sensitive Hashing (LSH)~\cite{diim04}. LSH is a hashing technique that maps similar input items to the same hash buckets with high probability. It answers nearest-neighbor queries very efficiently, practically matching a near-logarithmic time complexity.

The efficiency and effectiveness of both FastMap and LSH allow us to combine them and rapidly switch between the original problem domain and its geometric interpretation. Hence, solutions produced in one space can be quickly interpreted in the other. Figure~\ref{fig:fm_lsh} illustrates this coupling, henceforth referred to as the FastMap+LSH framework.

\section{Overview and Contributions}

In this dissertation, we present many new applications of FastMap, its graph version, and the FastMap+LSH framework described above.

In the first part of the dissertation, we apply the graph version of FastMap to solve various new graph-theoretic problems of significant interest in AI. Chapters~\ref{ch:facility_location},~\ref{ch:centrality},~\ref{ch:block_modeling}, and~\ref{ch:convex_hull} describe such applications for facility location, top-$K$ centrality computations, community detection and block modeling, and graph convex hull computations, respectively. In each case, we choose a proper distance function to interpret the graph-theoretic problem in geometric space, after which the appropriate analytical and geometric techniques are invoked to better absorb large input sizes compared to discrete algorithms that work directly on the input graph. In each case, we also conduct comprehensive experiments and show that our FastMap-based approach outperforms state-of-the-art competing methods, in terms of both the efficiency and the effectiveness. Overall, our approach leads to an important upshot: In the modern era, graphs are used to represent social networks, communication networks, and transportation networks, among similar structures in many other domains with entities and relationships between them. These graphs can be very large with millions of vertices and hundreds of millions of edges. Therefore, algorithms with a running time that is quadratic or more in the size of the input are undesirable. In fact, algorithms with any super-linear running times, discounting logarithmic factors, are also largely undesirable. Hence, a desired algorithm should have a near-linear running time close to that of merely reading the input. Our approach addresses these requirements by first creating a geometric interpretation of a given graph-theoretic problem in only near-linear time and consequently enabling powerful downstream algorithms.

In the second part of the dissertation, we propose a novel ML framework, called FastMapSVM, which combines FastMap and SVMs to classify complex objects. Chapter~\ref{ch:fastmapsvm} presents FastMapSVM and its underlying concepts. It also presents an application of FastMapSVM for identifying and classifying seismograms in Earthquake Science. Chapter~\ref{ch:constraint} presents another important application of FastMapSVM for predicting the satisfiability of Constraint Satisfaction Problems (CSPs). In both cases, we show that FastMapSVM outperforms many state-of-the-art competing methods, particularly in terms of the training data and time required for producing high-quality results. Overall, our approach leads to an important upshot: While Neural Networks (NNs) and related Deep Learning (DL) methods are popularly used for classifying complex objects, they are generally based on the paradigm of~\emph{characterizing individual} objects. In contrast, FastMapSVM is generally based on the paradigm of~\emph{comparing pairs} of objects via a distance function. One benefit of FastMapSVM is that the distance function can encapsulate and invoke the intelligence of other powerful algorithms such as the $A^*$ search procedure and maxflow computations, among many other optimization methods. The distance function can also incorporate domain-specific knowledge that otherwise may be hard for ML algorithms to automatically extract from the data. FastMapSVM serves as a lightweight alternative to NNs for classifying complex objects, particularly when training data or time is limited. It also extends the applicability of SVMs to domains with complex objects by combining the complementary strengths of FastMap and SVMs. Furthermore, FastMapSVM provides a perspicuous visualization of the objects and the classification boundaries between them. This aids human interpretation of the data and results. It also enables a human-in-the-loop framework for refining the processes of learning and decision making.

In summary, our contributions show that FastMap plays a key role in representing the vertices of a graph, or complex objects in other domains, as points in Euclidean spaces. The ability of FastMap to efficiently generate these ``simplified'' representations of the vertices, or the complex objects, enables many powerful downstream algorithms developed in diverse research communities such as AI, ML, Computational Geometry, Mathematics, Operations Research, and Theoretical Computer Science. Hence, we envision that FastMap can facilitate and harness the confluence of these algorithms and find future applications in many other problem domains that are not necessarily discussed here.

\chapter{FastMap for Facility Location Problems}
\label{ch:facility_location}
Facility Location Problems (FLPs) arise while serving multiple customers in a shared environment, minimizing transportation and other costs. Hence, they involve the optimal placement of facilities. They are defined on graphs as well as in Euclidean spaces with or without obstacles; and they are typically NP-hard to solve optimally. There are many heuristic algorithms tailored to different kinds of FLPs. However, FLPs defined in Euclidean spaces without obstacles are the most amenable to efficient and effective heuristic algorithms. This motivates the idea of quickly reformulating FLPs on graphs and in Euclidean spaces with obstacles to FLPs in Euclidean spaces without obstacles. In this chapter, we propose a new approach towards this end based on FastMap and LSH. Through extensive experiments, we show that our approach significantly outperforms other state-of-the-art competing algorithms on a variety of FLPs: the Multi-Agent Meeting (MAM) problem, Vertex $K$-Median (VKM) problem, Weighted VKM (WVKM) problem, and the Capacitated VKM (CVKM) problem.

\section{Introduction}

\begin{figure}[t!]
\centering
\includegraphics[width=0.8\columnwidth]{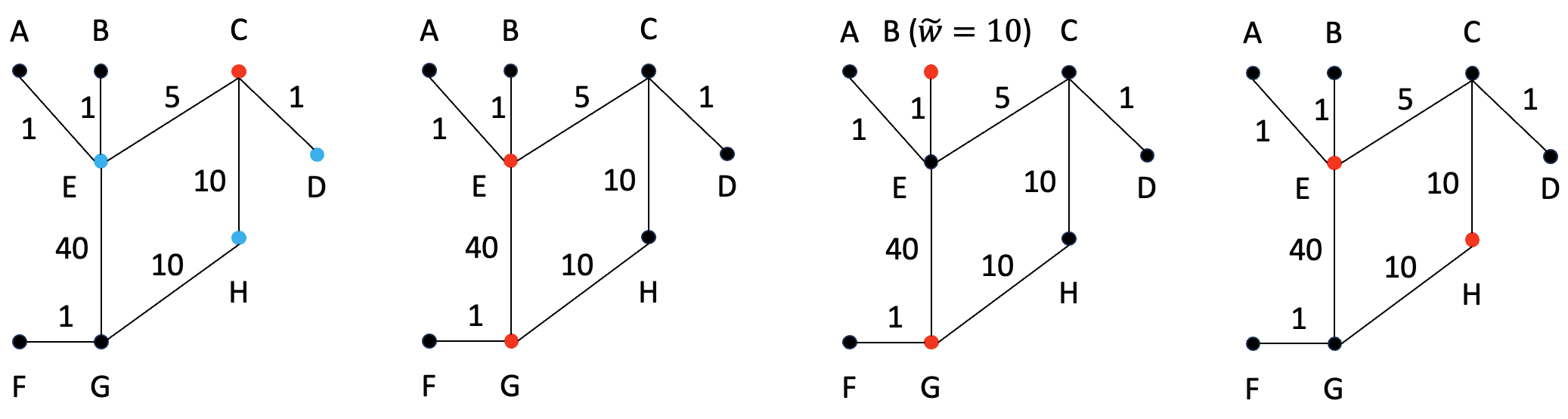}
\caption[An illustration of the Multi-Agent Meeting problem, Vertex $K$-Median problem, Weighted Vertex $K$-Median problem, and the Capacitated Vertex $K$-Median problem on the same input graph.]{Illustrates the MAM, VKM, WVKM, and the CVKM problems. The first, second, third, and the fourth panels show the optimal solution in red for the MAM, VKM, WVKM, and the CVKM problems, respectively, on the same input graph. In the MAM problem, there are $3$ agents, each initially on a different vertex shown in blue. The VKM, WVKM, and the CVKM problems have different optimal solutions for the same value of $K = 2$. In the WVKM problem, all vertices have weight $1$; except `B' has weight $10$. In the CVKM problem, $\tau = 4$.}
\label{fig:flp_demo}
\end{figure}

FLPs are constrained optimization problems that seek the optimal placement of facilities for providing resources and services to multiple customers in a shared environment. That is, FLPs serve the purpose of orchestrating shared resources between multiple customers. They are used to model decision problems related to transportation, warehousing, polling, and healthcare, among many other tasks, for maximizing efficiency, impact, and/or profit. FLPs can be defined on graphs or in geometric spaces, in continuous or discrete environments, and with a variety of distance metrics and objectives. A compendium of FLPs along with various algorithms and case studies can be found in~\cite{fh09}.

FLPs defined on graphs as well as in Euclidean spaces with or without obstacles are NP-hard to solve optimally~\cite{gx98,od98}. Nonetheless, there are many heuristic algorithms tailored to different kinds of FLPs. FLPs defined on graphs are broadly applicable, since most environments can be represented as a graph (even if discretization is required). Modulo discretization, they are more general compared to FLPs defined in Euclidean spaces with obstacles, which, in turn, are more general compared to FLPs defined in Euclidean spaces without obstacles. However, FLPs defined in Euclidean spaces without obstacles are the most amenable to efficient and effective heuristic algorithms. In fact, FLPs in Euclidean spaces without obstacles are definitionally very close to clustering problems, which, in turn, are amenable to popular clustering algorithms such as the $K$-means and GMM clustering~\cite{m12}.

The foregoing summary motivates the idea of quickly reformulating FLPs on graphs and in Euclidean spaces with obstacles to FLPs in Euclidean spaces without obstacles. In this chapter, we propose a new approach towards this end based on FastMap and LSH. We use FastMap to efficiently reformulate an FLP on a graph to an FLP in a Euclidean space without obstacles.\footnote{As explained later, an FLP in Euclidean space with obstacles is also amenable to a similar reformulation.} We use LSH to efficiently interpret a solution found in the FastMap embedding as a solution on the original graph.

We address four well-known FLPs in this chapter: the MAM problem, VKM problem, WVKM problem, and the CVKM problem. Below, we briefly describe each of these problems on graphs. Their counterparts in Euclidean spaces with or without obstacles have analogous definitions. Moreover, we assume that the graphs are undirected for two reasons: for the ease of exposition and to preserve the analogy in Euclidean spaces where distances are inherently symmetric.

In the MAM problem~\cite{aflssk23}, the input is a graph and a set of agents, each initially on a different start vertex. The task is to find a common vertex where all the agents should meet so as to minimize the sum of the agents' shortest-path distances to it.\footnote{There are a few other variants of the MAM problem described in~\cite{aflssk23}. These differ in being conflict-tolerant or conflict-free and having different objective functions.} The VKM problem seeks $K$ vertices on the input graph for the placement of facilities so as to minimize the sum of the shortest-path distances over each vertex to its nearest facility. The WVKM problem is similar to the VKM problem, except that the objective is to minimize the sum of the weighted shortest-path distances over each vertex to its nearest facility. Here, each vertex is given a weight that measures its importance. The CVKM problem is also similar to the VKM problem, except that no facility is allowed to serve more than $\tau$ vertices.\footnote{In a more general version of the CVKM problem, there is a supply and a demand associated with each facility and vertex, respectively. No facility is allowed to serve a total demand that exceeds its supply.}

The MAM, VKM, WVKM, and the CVKM problems have many real-world applications. For example, in multi-agent coordination tasks~\cite{aflssk23}, they can be used to choose a gathering point; in urban development~\cite{ffrha19}, they can be used to optimally place various public service centers within a city; and in communication networks~\cite{mc03}, they can be used to determine the optimal placement of computation sites for critical multiplexing. Figure~\ref{fig:flp_demo} shows examples of these four problems posed on the same input graph.

On each of the FLPs described above, including their Euclidean variants, we demonstrate the efficiency and effectiveness of our approach through extensive experimentation: We show that our approach significantly outperforms other state-of-the-art competing algorithms. In discretized Euclidean spaces, we also show that it is possible to combine FastMap with an any-angle path planner, such as Anya~\cite{hgoa16}.

\section{Preliminaries}

In this section, we define the MAM, VKM, WVKM, and the CVKM problems. We first define the graph variants of these problems. We then briefly describe their Euclidean variants.

The MAM problem is as follows: Given an undirected edge-weighted graph $G = (V, E, w)$, where $w(e)$ is the non-negative weight on edge $e \in E$, and the start vertices $s_1, s_2 \ldots s_k \in V$ of $k$ agents, the task is to find a vertex $v^* \in V$ such that $v^* = \argmin_{v \in V} \sum_{i = 1}^k d_G(s_i, v)$. Here, $d_G(v_i, v_j)$, for $v_i, v_j \in V$, is the shortest-path distance between $v_i$ and $v_j$ in $G$ with respect to the edge weights.

The VKM problem is as follows: Given an undirected edge-weighted graph $G = (V, E, w)$, where $w(e)$ is the non-negative weight on edge $e \in E$, and a positive integer $K$, the task is to find a subset of vertices $U^* \subseteq V$ of cardinality $K$ such that $U^* = \argmin_{U} \sum_{v \in V} \min_{u \in U} d_G(v, u)$.

The WVKM problem is as follows: Given an undirected vertex-weighted and edge-weighted graph $G = (V, E, \tilde{w}, w)$, where $\tilde{w}(v)$ is the non-negative weight on vertex $v \in V$ and $w(e)$ is the non-negative weight on edge $e \in E$, and a positive integer $K$, the task is to find a subset of vertices $U^* \subseteq V$ of cardinality $K$ such that $U^* = \argmin_{U} \sum_{v \in V} \min_{u \in U} \tilde{w}(v) d_G(v, u)$.

The CVKM problem is as follows: Given an undirected edge-weighted graph $G = (V, E, w)$, where $w(e)$ is the non-negative weight on edge $e \in E$, and positive integers $K$ and $\tau$, the task is to find a subset of vertices $U^* \subseteq V$ of cardinality $K$ and an assignment function $f^*: V \rightarrow U^*$ such that $(U^*, f^*) =$
\begin{equation}
\begin{array}{ll}
\argmin_{U, f} &\sum_{v \in V} d_G(v, f(v))\\
\subjectto &\forall u \in U: |\{v \in V: f(v) = u\}| \leq \tau.
\end{array}
\end{equation}

The Euclidean variants of the MAM, VKM, WVKM, and the CVKM problems are defined in Euclidean spaces, which are continuous. In a Euclidean space without obstacles, a given set of $N$ points corresponds to $V$; and the straight-line distances between pairs of these points correspond to the shortest-path distances between the pairs of vertices. However, the solution may be allowed to contain points in the Euclidean space outside of the given $N$ points. In a Euclidean space with obstacles, shortest-path distances via free space, that is, avoiding obstacle regions, replace straight-line distances; and the solution can only include points in free space.

\section{Solving Facility Location Problems via FastMap and Locality Sensitive Hashing}

In this section, we present our approach for solving FLPs via FastMap and LSH. We illustrate it on the MAM, VKM, WVKM, and the CVKM problems. The FastMap component of our approach allows us to quickly render the FLP in Euclidean space without obstacles: This enables efficient and effective geometric and analytical techniques for solving the problem. The LSH component of our approach allows us to quickly interpret the solution obtained in Euclidean space as a viable solution on the original graph.

\subsection{Solving Facility Location Problems in Euclidean Space without Obstacles}

Most FLPs defined on graphs are also defined in Euclidean spaces without obstacles. As described before, the MAM, VKM, WVKM, and the CVKM problems are defined in such a space using Euclidean distances instead of graph-based distances. There are two benefits of using FastMap to convert an FLP specified on a graph to an FLP specified in the Euclidean embedding of that graph. First, FLPs defined in Euclidean spaces without obstacles are the most amenable to efficient and effective heuristic algorithms. Second, invoking FastMap with an intelligently designed distance function can simplify the problem even more.

For illustration of the above arguments, we consider the MAM problem on an input graph with $k$ agents. Solving it optimally requires the computation of $k$ shortest-path trees rooted at the individual start vertices of the agents. The same problem in Euclidean space without obstacles is referred to as the Fermat-Weber problem~\cite{dm85}. This problem is also NP-hard to solve optimally but is amenable to very effective heuristics~\cite{fmb05}. Moreover, if FastMap is invoked on the input graph to preserve the square-roots of the shortest-path distances\textemdash instead of the shortest-path distances\textemdash the problem in the resulting Euclidean space becomes one of finding a point that minimizes the sum of the squared distances to $k$ given points. This is a significantly easier problem since the required point is the centroid of the $k$ given points.

Algorithm~\ref{alg:fastmap} (from Chapter~\ref{ch:fastmap}) can be easily modified to incorporate the square-root of the shortest-path distance function $\sqrt{d_G(\cdot, \cdot)}$ between vertices. This is done by returning the square-roots of the shortest-path distances found by the procedure $\mbox{ShortestPathTree}(\cdot, \cdot)$ on Lines~\ref{line:spt1},~\ref{line:spt2}, and~\ref{line:spt3}.

The VKM, WVKM, and the CVKM problems can also utilize the FastMap embedding with the square-root of the shortest-path distance function. Doing so makes them very similar to clustering problems popularly studied in ML. For example, in a Euclidean space without obstacles, clustering algorithms such as the $K$-means algorithm intend to minimize the sum of the squared Euclidean distances over each data point to its nearest centroid. With the Euclidean distances representing the square-roots of the shortest-path distances in the input graph, this is equivalent to solving the VKM problem of minimizing the sum of the shortest-path distances over each vertex to its nearest facility. Similarly, the WVKM problem can also be solved by invoking the $K$-means algorithm in the FastMap embedding that preserves the square-roots of the shortest-path distances: The $K$-means algorithm is also given a weight associated with each data point that measures its importance. Finally, the CVKM problem can also be solved by invoking the constrained $K$-means algorithm~\cite{bbd00} in the FastMap embedding that preserves the square-roots of the shortest-path distances: The constrained $K$-means algorithm restricts the size of each cluster to be no more than a user-specified parameter $\tau$.

Although we have described how to solve FLPs in the Euclidean space generated by FastMap on an input graph, the solutions produced reside in the Euclidean space and are not yet interpretable on the original graph. As described in Chapter~\ref{ch:fastmap}, we use LSH towards this end.

\subsection{FastMap with Anya}

Compared to FLPs in Euclidean spaces without obstacles, FLPs in Euclidean spaces with obstacles are much harder to solve. Primarily, this is because the shortest-path distances in the latter are no more straight-line distances. In fact, even by itself, computing the shortest path between two points in a Euclidean space with obstacles may be very hard. Even shortest-path algorithms that operate in a Euclidean space with obstacles have to make various kinds of assumptions on the nature of the obstacles and the acceptable paths that maneuver through them. For the same reason, we define FLPs in Euclidean spaces with obstacles only when the environment also supports Anya~\cite{hgoa16}, a popular any-angle path planner. We note that this is not a restriction on the kinds of FLPs that can be discussed but is a standardization of the input environment that is also applicable to the state-of-the-art shortest-path algorithms.

Anya~\cite{hgoa16} is a recent any-angle shortest-path algorithm for grid-worlds. Given any two discrete points on a $2$-dimensional grid-world, Anya finds a shortest any-angle path between them, if one exists. It uses a variant of $A^*$ search over sets of states represented as intervals. Anya is very efficient since it does not require preprocessing or the introduction of additional memory overheads.

In our FastMap-based approach for solving FLPs, the FastMap component always generates a Euclidean space without obstacles. It can be used to transform an input Euclidean space with obstacles to an output Euclidean space without obstacles if the straight-line distances in the output space preserve the desired distances in the input space. Towards this end, we can use the any-angle shortest-path distance function on the discrete points of a $2$-dimensional grid-world, as generated by Anya. However, using the any-angle shortest-path distance function with FastMap has the same fundamental challenge as using the regular shortest-path distance function with FastMap: To retain the near-linear time complexity of FastMap, the distances should not be computed from a root vertex to all other vertices independently. The computations have to be amortized to yield all of them simultaneously, as shown on Lines~\ref{line:spt1},~\ref{line:spt2}, and~\ref{line:spt3} of Algorithm~\ref{alg:fastmap} (from Chapter~\ref{ch:fastmap}). Thus, we modify Anya to compute the entire tree of any-angle shortest-path distances from a root vertex to all other vertices. We call this version as Anya-Dijkstra. Hence, FastMap with Anya is similar to Algorithm~\ref{alg:fastmap}, except that it replaces the procedure $\mbox{ShortestPathTree}(\cdot, \cdot)$ by Anya-Dijkstra and returns the square-roots of the any-angle shortest-path distances found by Anya-Dijkstra on Lines~\ref{line:spt1},~\ref{line:spt2}, and~\ref{line:spt3}.

\section{Competing Algorithms}

The MAM, VKM, WVKM, and the CVKM problems can be formulated using Integer Linear Programming (ILP). We use the template for the CVKM problem presented below. In this template, $d_{ij}$ is a shorthand for $d_G(v_i, v_j)$; $c_j$ is a Boolean variable that is `$1$' iff $v_j$ is a facility; and $b_{ij}$ is a Boolean variable that is `$1$' iff $v_j$ is the facility assigned to $v_i$. For the VKM problem, $\tau = |V|$. For the WVKM problem, $d_{ij} = \tilde{w}(v_i)d_G(v_i, v_j)$ and $\tau = |V|$. For the MAM problem, the outer summation of the objective function spans only the start vertices of the agents, $\tau = |V|$, and $K = 1$.
\begin{equation}
\begin{array}{ll}
\min &\sum_{v_i \in V}\sum_{v_j \in V} b_{ij}d_{ij}\\
\subjectto &\forall v_i \in V: \sum_{v_j \in V} b_{ij} = 1\\
&\forall v_j \in V: \sum_{v_i \in V} b_{ij} \leq \tau\\
&\forall v_i, v_j \in V: b_{ij} \leq c_j\\
&\sum_{v_j \in V} c_j = K.
\end{array}
\end{equation}

The MAM problem can be solved optimally in polynomial time and does not require the ILP solver: The algorithm computes a shortest-path tree rooted at each of the $k$ start vertices of the agents.\footnote{FastMap, as in Algorithm~\ref{alg:fastmap} (from Chapter~\ref{ch:fastmap}), has been used for the MAM problem~\cite{lfkk19}; but there, it does not use the square-roots of the shortest-path distances and, consequently, uses a heuristically computed solution\textemdash instead of the centroid\textemdash in the Euclidean space.} When heuristic guidance is available, $MM^*$~\cite{aflssk23} can also be used to solve the MAM problem optimally. For the VKM problem, the state-of-the-art algorithm is FasterPAM~\cite{sr21}, the successor of the Partition Around Medoids (PAM) algorithm~\cite{kr87}. FasterPAM conducts local search by repeatedly swapping a vertex from its current solution $S$ with a vertex in $V \setminus S$. It runs in $O(K|V|^2)$ time. For the WVKM problem, the state-of-the-art implementation of the PAM algorithm is available in the procedure `wcKMedoids'~\cite{m18} in R 4.3. The CVKM problem is significantly harder: To the best of our knowledge, there are no good solvers for this problem that scale to the problem sizes discussed in this chapter.

The ILP solver and the PAM algorithms require the precomputation of the all-pairs shortest-path distances, which can be done via the Floyd-Warshall algorithm.

\section{Experimental Results}

In this section, we provide experimental results that compare our approach to competing algorithms on the MAM, VKM, WVKM, and the CVKM problems.

We implemented our approach in Python3. For the LSH module, we used the `FALCONN' library~\cite{ailrs15} that has many code-level optimizations. For the $K$-means procedure without weights, required for solving the VKM problem, and for the $K$-means procedure with weights, required for solving the WVKM problem, we used the `scikit-learn' library~\cite{pvgmtgbpwdvpcbpd11}. For the constrained $K$-means procedure, required for solving the CVKM problem, we used the `k-means-constrained' library. For the ILP solver, we used the Gurobi Optimizer 10.0~\cite{g23}. We used the Anya procedure via a Python interface to its implementation in Java. For FasterPAM~\cite{sr21}, we used the `kmedoids' library. However, for the PAM clustering of weighted data, we used the procedure `wcKMedoids'~\cite{m18} implemented in R 4.3. For the Floyd-Warshall algorithm, we used the `NetworkX' library~\cite{hss08}. We conducted all experiments on a laptop with an Apple M2 Max chip and 96 GB memory. For evaluation purposes, we chose two categories of problem instances, both of which contain realistic FLPs.

In the first category, we chose problem instances representative of FLPs that arise in warehousing, urban planning, and transportation domains, among others. In such cases, the environment is essentially a $2$-dimensional map. Moreover, in such cases, both the regular FastMap (FM), that is, FastMap that implements the procedure $\mbox{ShortestPathTree}(\cdot, \cdot)$ using Dijkstra's algorithm (DJK), and FastMap with Anya (FMA), that is, FastMap that implements the procedure $\mbox{ShortestPathTree}(\cdot, \cdot)$ using Anya-Dijkstra (ADJK), are defined. This enables a more direct comparison of the various algorithms. Such instances are available in the movingAI dataset~\cite{s12}: Each instance serves as both a graph instance and a $2$-dimensional grid-world instance. In it, each discrete point on a traversable cell\footnote{as defined in~\cite{hgoa16} for the application of Anya} is represented as a vertex. Adjacent vertices, corresponding to discrete points on the same traversable cell, are connected by an edge. A horizontal or vertical edge has unit weight but a diagonal edge has weight $\sqrt{2}$. If the graph constructed this way has multiple connected components, only the largest one is used to represent the instance. We used five representative benchmark suites in this category: `Dragon Age: Origins', `Warcraft III', `Baldurs Gate II', `City/Street Maps', and `Mazes'. The first three are from commercial game environments; the fourth is from the real world; and the fifth is artificial. Both FastMap and FastMap with Anya use $\kappa = 10$ for these instances.

In the second category, we chose problem instances representative of FLPs that arise in communication networks. In the field of Computer and Communication Networks, Waxman graphs~\cite{w88} are used as realistic communication networks. Hence, we generated Waxman graph instances using NetworkX~\cite{hss08} with commonly used parameter values $\alpha = 0.3$ and $\beta = 0.1$, within a rectangular domain of $100 \times 100$, and with the weight on each edge set to the Euclidean distance between its endpoints. FastMap uses $\kappa = 100$ for these instances.\footnote{The Normalized Root Mean Square Deviation, as used in~\cite{lfkk19} to measure the accuracy of the FastMap embedding, is much higher for Waxman graphs even with $\kappa = 100$ compared to movingAI instances with $\kappa = 10$.}

\begin{table}[!t]
\footnotesize
\centering
\scalebox{0.76}{
\begin{tabular}{|l|r|r|r|r|r|r|r|r|}
\cline{1-9}
\multirow{3}{*}{Instance} &\multirow{3}{*}{Size ($|V|$, $|E|$)} &\multicolumn{1}{c|}{Preprocessing} &\multicolumn{3}{c|}{$k = 50$} &\multicolumn{3}{c|}{$k = 100$}\\
\cline{4-9}
&&\multicolumn{1}{c|}{for FM:} &\multicolumn{2}{c|}{Time (s)} &SO (\%) &\multicolumn{2}{c|}{Time (s)} &SO (\%)\\
\cline{4-9}
&&\multicolumn{1}{c|}{FM\_pre (s)} &DJK &FM &FM &DJK &FM &FM\\
\cline{1-9}
orz102d &(738, 2632) &0.05 &0.09 &0.00 &1.23 &0.19 &0.00 &0.93\\
den407d &(852, 3054) &0.07 &0.09 &0.00 &0.28 &0.18 &0.00 &0.98\\
lak526d &(954, 3329) &0.07 &0.09 &0.00 &0.56 &0.19 &0.00 &1.17\\
den009d &(1003, 3620) &0.07 &0.10 &0.00 &1.49 &0.21 &0.00 &1.55\\
\cline{1-9}
AR0512SR &(896, 3275) &0.06 &0.10 &0.00 &6.35 &0.19 &0.00 &2.31\\
AR0402SR &(1075, 3796) &0.10 &0.13 &0.00 &6.49 &0.26 &0.00 &1.70\\
AR0517SR &(1083, 4078) &0.09 &0.12 &0.00 &4.47 &0.24 &0.00 &1.17\\
AR0530SR &(1092, 3885) &0.08 &0.11 &0.00 &6.45 &0.22 &0.00 &0.57\\
\cline{1-9}
Shanghai\_0\_256 &(48696, 190303) &4.81 &6.67 &0.01 &1.15 &13.26 &0.00 &2.26\\
blastedlands &(131342, 505974) &12.95 &18.58 &0.02 &5.87 &37.18 &0.02 &0.90\\
maze512-32-5 &(253856, 990715) &21.82 &35.25 &0.04 &1.09 &70.54 &0.04 &2.81\\
\cline{1-9}
wm00800 &(800, 9498) &1.06 &0.25 &0.00 &7.08 &0.50 &0.00 &5.81\\
wm01000 &(1000, 14923) &1.78 &0.38 &0.00 &9.37 &0.76 &0.00 &5.80\\
wm05000 &(5000, 374925) &88.64 &13.43 &0.00 &3.80 &26.76 &0.00 &3.56\\
wm10000 &(10000, 1499713) &360.29 &54.00 &0.00 &1.65 &107.38 &0.00 &1.13\\
\cline{1-9}
\end{tabular}
}
\caption[Results for the Multi-Agent Meeting problem on various graph instances.]{Shows the results for the MAM problem on various graph instances. `FM', `FM\_pre', `DJK', and `SO' stand for `FastMap', `FastMap preprocessing', `Dijkstra', and `suboptimality', respectively.}
\label{tab:mam_graphs}
\end{table}

\begin{table}[!t]
\footnotesize
\centering
\scalebox{0.76}{
\begin{tabular}{|l|r|r|r|r|r|r|r|r|r|r|}
\cline{1-11}
\multirow{3}{*}{Instance} &\multirow{3}{*}{Size ($|V|$, $|E|$)} &\multicolumn{1}{c|}{Preprocessing} &\multicolumn{4}{c|}{$k = 50$} &\multicolumn{4}{c|}{$k = 100$}\\
\cline{4-11}
&&\multicolumn{1}{c|}{for FMA:} &\multicolumn{2}{c|}{Time (s)} &\multicolumn{2}{c|}{SO (\%)} &\multicolumn{2}{c|}{Time (s)} &\multicolumn{2}{c|}{SO (\%)}\\
\cline{4-11}
&&\multicolumn{1}{c|}{FMA\_pre (s)} &ADJK &FMA &FM &FMA &ADJK &FMA &FM &FMA\\
\cline{1-11}
orz102d &(738, 2632) &3.29 &5.42 &0.00 &0.96 &0.30 &10.86 &0.00 &1.45 &0.46\\
den407d &(852, 3054) &2.66 &3.92 &0.00 &0.91 &0.58 &8.07 &0.00 &0.09 &0.13\\
lak526d &(954, 3329) &3.77 &5.12 &0.00 &2.52 &2.52 &10.01 &0.00 &2.90 &5.15\\
den009d &(1003, 3620) &2.00 &3.46 &0.00 &3.65 &1.08 &6.96 &0.00 &0.05 &0.29\\
\cline{1-11}
AR0512SR &(896, 3275) &10.81 &13.84 &0.00 &4.91 &2.71 &27.62 &0.00 &2.76 &3.47\\
AR0402SR &(1075, 3796) &8.54 &10.54 &0.00 &1.07 &2.53 &21.33 &0.00 &0.03 &3.53\\
AR0517SR &(1083, 4078) &9.84 &13.89 &0.00 &6.33 &4.28 &27.95 &0.00 &1.82 &2.72\\
AR0530SR &(1092, 3885) &12.09 &15.37 &0.00 &0.64 &0.21 &30.87 &0.00 &1.33 &0.43\\
\cline{1-11}
Shanghai\_0\_256 &(48696, 190303) &16.77 &16.09 &0.01 &1.56 &1.27 &32.12 &0.01 &1.06 &3.07\\
blastedlands &(131342, 505974) &44.81 &35.70 &0.02 &6.87 &7.52 &71.59 &0.02 &3.70 &5.50\\
maze512-32-5 &(253856, 990715) &34.69 &55.79 &0.04 &2.68 &3.03 & 111.51 &0.04 &1.43 &1.48\\
\cline{1-11}
\end{tabular}
}
\caption[Results for the Multi-Agent Meeting problem on various grid-world instances.]{Shows the results for the MAM problem on various grid-world instances. `FMA', `FMA\_pre', and `ADJK' stand for `FastMap with Anya', `FastMap with Anya preprocessing', and `Anya-Dijkstra', respectively.}
\label{tab:mam_grids}
\end{table}

\begin{table}[!t]
\footnotesize
\centering
\scalebox{0.76}{
\begin{tabular}{|l|r|r|r|r|r|r|r|r|r|r|r|r|r|}
\cline{1-14}
\multirow{3}{*}{Instance} &\multirow{3}{*}{Size ($|V|$, $|E|$)} &\multicolumn{1}{c|}{Preprocessing} &\multicolumn{1}{c|}{Preprocessing} &\multicolumn{5}{c|}{$K = 10$} &\multicolumn{5}{c|}{$K = 20$}\\
\cline{5-14}
&&\multicolumn{1}{c|}{for ILP, PAM:} &\multicolumn{1}{c|}{for FM:} &\multicolumn{3}{c|}{Time (s)} &\multicolumn{2}{c|}{SO (\%)} &\multicolumn{3}{c|}{Time (s)} &\multicolumn{2}{c|}{SO (\%)}\\
\cline{5-14}
&&\multicolumn{1}{c|}{FW (s)} &\multicolumn{1}{c|}{FM\_pre (s)} &ILP &PAM &FM &PAM &FM &ILP &PAM &FM &PAM &FM\\
\cline{1-14}
orz102d &(738, 2632) &27.10 &0.05 &79.34 &0.00 &0.00 &1.47 &3.70 &175.76 &0.00 &0.00 &1.84 &1.01\\
den407d &(852, 3054) &41.24 &0.07 &48.70 &0.00 &0.00 &1.34 &3.44 &132.63 &0.00 &0.01 &1.95 &1.78\\
lak526d &(954, 3329) &56.36 &0.07 &170.70 &0.01 &0.02 &0.23 &1.09 &71.53 &0.00 &0.01 &7.43 &2.72\\
den009d &(1003, 3620) &64.38 &0.07 &83.20 &0.12 &0.06 &0.21 &1.23 &96.65 &0.08 &0.06 &4.23 &1.60\\
\cline{1-14}
AR0512SR &(896, 3275) &48.20 &0.06 &52.95 &0.00 &0.00 &54.19 &2.52 &485.62 &0.00 &0.01 &0.99 &2.98\\
AR0402SR &(1075, 3796) &82.27 &0.10 &73.03 &0.07 &0.05 &1.85 &2.68 &74.51 &0.11 &0.16 &0.11 &4.06\\
AR0517SR &(1083, 4078) &85.25 &0.09 &1556.52 &0.15 &0.12 &0.22 &0.80 &4467.56 &0.19 &0.06 &0.45 &2.66\\
AR0530SR &(1092, 3885) &84.71 &0.08 &121.16 &0.08 &0.03 &1.43 &0.41 &209.84 &0.13 &0.10 &0.12 &5.69\\
\cline{1-14}
Shanghai\_0\_256 &(48696, 190303) &- &4.40 &- &- &0.15 &- &- &- &- &0.25 &- &-\\
blastedlands &(131342, 505974) &- &12.88 &- &- &0.49 &- &- &- &- &0.72 &- &-\\
maze512-32-5 &(253856, 990715) &- &22.61 &- &- &0.65 &- &- &- &- &1.00 &- &-\\
\cline{1-14}
wm00800 &(800, 9498) &35.94 &1.06 &1650.12 &0.00 &0.02 &0.03 &16.50 &3302.97 &0.00 &0.02 &0.04 &13.16\\
wm01000 &(1000, 14923) &70.30 &1.78 &9188.83 &0.06 &0.04 &0.07 &15.04 &9970.25 &0.09 &0.07 &0.19 &16.66\\
wm05000 &(5000, 374925) &- &88.64 &- &- &0.12 &- &- &- &- &0.16 &- &-\\
wm10000 &(10000, 1499713) &- &360.29 &- &- &0.33 &- &- &- &- &0.49 &- &-\\
\cline{1-14}
\end{tabular}
}
\caption[Results for the Vertex $K$-Median problem on various graph instances.]{Shows the results for the VKM problem on various graph instances. `FW', `ILP', and `PAM' stand for `Floyd-Warshall', `ILP solver', and `PAM algorithm', respectively. The ILP solver and the PAM algorithm require the Floyd-Warshall algorithm in a preprocessing phase for the computation of the all-pairs shortest-path distances.}
\label{tab:vkm_graphs}
\end{table}

\begin{table}[!t]
\footnotesize
\centering
\scalebox{0.76}{
\begin{tabular}{|l|r|r|r|r|r|r|r|r|r|r|r|r|}
\cline{1-13}
\multirow{3}{*}{Instance} &\multirow{3}{*}{Size ($|V|$, $|E|$)} &\multicolumn{1}{c|}{Preprocessing} &\multicolumn{5}{c|}{$K = 10$} &\multicolumn{5}{c|}{$K = 20$}\\
\cline{4-13}
&&\multicolumn{1}{c|}{for FMA:} &\multicolumn{2}{c|}{Time (s)} &\multicolumn{3}{c|}{SO (\%)} &\multicolumn{2}{c|}{Time (s)} &\multicolumn{3}{c|}{SO (\%)}\\
\cline{4-13}
&&\multicolumn{1}{c|}{FMA\_pre (s)} &ILP &FMA &PAM &FM &FMA &ILP &FMA &PAM &FM &FMA\\
\cline{1-13}
orz102d &(738, 2632) &3.29 &77.38 &0.28 &6.44 &1.69 &1.35 &214.98 &0.28 &38.30 &1.57 &2.87\\
den407d &(852, 3054) &2.66 &49.04 &0.28 &0.51 &3.65 &1.32 &144.29 &0.23 &31.33 &2.42 &4.77\\
lak526d &(954, 3329) &3.77 &170.97 &0.28 &0.04 &1.80 &0.12 &72.49 &0.28 &0.31 &1.98 &6.70\\
den009d &(1003, 3620) &2.00 &83.16 &0.28 &-0.02 & 2.41 &0.89 &95.91 &0.28 &0.28 &3.83 &4.14\\
\cline{1-13}
AR0512SR &(896, 3275) &10.81 &55.76 &0.27 &0.00 &1.29 &3.26 &519.66 &0.28 &0.99 &2.20 &4.25\\
AR0402SR &(1075, 3796) &8.54 &74.70 &0.29 &2.01 &1.21 &5.10 &75.09 &0.29 &0.51 &4.07 &4.27\\
AR0517SR &(1083, 4078) &9.84 &1581.05 &0.28 &3.68 &0.78 &0.73 &5905.02 &0.28 &0.56 &1.50 &0.72\\
AR0530SR &(1092, 3885) &12.09 &122.93 &0.28 &1.90 &3.64 &3.33 &211.75 &0.30 &1.58 &3.91 &1.26\\
\cline{1-13}
Shanghai\_0\_256 &(253856, 990715) &14.07 &- &0.10 &- &- &- &- &0.30 &- &- &-\\
blastedlands &(131342, 505974) &37.21 &- &0.47 &- &- &- &- &0.67 &- &- &-\\
maze512-32-5 &(253856, 990715) &39.19 &- &0.68 &- &- &- &- &1.11 &- &- &-\\
\cline{1-13}
\end{tabular}
}
\caption[Results for the Vertex $K$-Median problem on various grid-world instances.]{Shows the results for the VKM problem on various grid-world instances.}
\label{tab:vkm_grids}
\end{table}

\begin{table}[!t]
\footnotesize
\centering
\scalebox{0.76}{
\begin{tabular}{|l|r|r|r|r|r|r|r|r|r|r|r|r|r|}
\cline{1-14}
\multirow{3}{*}{Instance} &\multirow{3}{*}{Size ($|V|$, $|E|$)} &\multicolumn{1}{c|}{Preprocessing} &\multicolumn{1}{c|}{Preprocessing} &\multicolumn{5}{c|}{$K = 10$} &\multicolumn{5}{c|}{$K = 20$}\\
\cline{5-14}
&&\multicolumn{1}{c|}{for ILP, PAM:} &\multicolumn{1}{c|}{for FM:} &\multicolumn{3}{c|}{Time (s)} &\multicolumn{2}{c|}{SO (\%)} &\multicolumn{3}{c|}{Time (s)} &\multicolumn{2}{c|}{SO (\%)}\\
\cline{5-14}
&&\multicolumn{1}{c|}{FW (s)} &\multicolumn{1}{c|}{FM\_pre (s)} &ILP &PAM &FM &PAM &FM &ILP &PAM &FM &PAM &FM\\
\cline{1-14}
orz102d &(738, 2632) &27.10 &0.05 &44.49 &0.06 &0.00 &2.61 &4.26 &95.91 &0.14 &0.00 &4.12 &5.50\\
den407d &(852, 3054) &41.24 &0.07 & 52.29 & 0.15 & 0.00 &1.69 &0.72 &74.92 &0.18 &0.01 &4.91 &4.90\\
lak526d &(954, 3329) &56.36 &0.07 &101.26 &0.09 &0.02 &1.19 &2.78 &60.25 &0.21 &0.01 &3.35 &2.49\\
den009d &(1003, 3620) &64.38 &0.07 &79.18 &0.11 &0.01 &1.12 &1.63 &137.58 &0.35 &0.01 &2.39 & 4.09\\
\cline{1-14}
AR0512SR &(896, 3275) &48.20 &0.06 &146.60 &0.08 &0.00 &1.57 &0.72 &119.31 &0.33 &0.01 &1.68 &3.78\\
AR0402SR &(1075, 3796) &82.27 &0.10 &80.94 &0.11 &0.02 &3.92 &3.18 &92.47 &0.25 &0.03 &4.92 &6.26\\
AR0517SR &(1083, 4078) &85.25 &0.09 &2907.62 &0.19 &0.27 &0.93 &1.09 &571.17 &0.45 &0.02 &2.85 &1.93\\
AR0530SR &(1092, 3885) &84.71 &0.08 &318.33 &0.14 &0.02 &2.15 &0.90 &127.48 &0.28 &0.02 &3.67 &5.11\\
\cline{1-14}
Shanghai\_0\_256 &(48696, 190303) &- &4.77 &- &- &0.17 &- &- &- &- &0.25 &- &-\\
blastedlands &(131342, 505974) &- &13.10 &- &- &0.38 &- &- &- &- &0.70 &- &-\\
maze512-32-5 &(253856, 990715) &- &21.89 &- &- &0.68 &- &- &- &- &1.07 &- &-\\
\cline{1-14}
wm00800 &(800, 9498) &35.94 &1.06 &1946.18 &0.12 &0.24 &0.00 &15.84 &1388.69 &0.21 &0.01 &0.00 &17.19\\
wm01000 &(1000, 14923) &70.30 &1.78 &5135.40 &0.20 &0.02 &0.00 &11.62 &15112.05 &0.21 &0.02 &0.01 &12.68\\
wm05000 &(5000, 374925) &- &88.64 &- &- &0.07 &- &- &- &- &0.07 &- &-\\
wm10000 &(10000, 1499713) &- &360.29 &- &- &0.32 &- &- &- &- &0.37 &- &-\\
\cline{1-14}
\end{tabular}
}
\caption[Results for the Weighted Vertex $K$-Median problem on various graph instances.]{Shows the results for the WVKM problem on various graph instances.}
\label{tab:wvkm_graphs}
\end{table}

\begin{table}[!t]
\footnotesize
\centering
\scalebox{0.76}{
\begin{tabular}{|l|r|r|r|r|r|r|r|r|r|r|r|r|}
\cline{1-13}
\multirow{3}{*}{Instance} &\multirow{3}{*}{Size ($|V|$, $|E|$)} &\multicolumn{1}{c|}{Preprocessing} &\multicolumn{5}{c|}{$K = 10$} &\multicolumn{5}{c|}{$K = 20$}\\
\cline{4-13}
&&\multicolumn{1}{c|}{for FMA:} &\multicolumn{2}{c|}{Time (s)} &\multicolumn{3}{c|}{SO (\%)} &\multicolumn{2}{c|}{Time (s)} &\multicolumn{3}{c|}{SO (\%)}\\
\cline{4-13}
&&\multicolumn{1}{c|}{FMA\_pre (s)} &ILP &FMA &PAM &FM &FMA &ILP &FMA &PAM &FM &FMA\\
\cline{1-13}
orz102d &(738, 2632) &3.29 &37.61 &0.28 &2.63 &1.49 &2.60 &37.60 &0.22 &3.97 &3.51 &2.95\\
den407d &(852, 3054) &2.66 &50.66 &0.28 &1.69 &5.39 &0.14 &61.92 &0.22 &3.97 &2.41 &2.18\\
lak526d &(954, 3329) &3.77 &105.26 &0.28 &1.66 &1.61 &7.51 &72.85 &0.28 &3.90 &3.63 &2.45\\
den009d &(1003, 3620) &2.00 &128.13 &0.28 &2.18 &0.81 &0.54 &91.62 & 0.28 &2.60 &4.08 &3.88\\
\cline{1-13}
AR0512SR &(896, 3275) &10.81 &53.07 &0.28 &1.21 &0.13 &2.82 &129.15 &0.28 &2.70 &2.98 &5.86\\
AR0402SR &(1075, 3796) &8.54 &84.35 &0.29 &2.31 &4.10 &4.63 &100.25 &0.29 &1.90 &2.52 &5.81\\
AR0517SR &(1083, 4078) &9.84 &92.75 &0.28 &0.97 &0.51 &1.11 &1358.58 &0.28 &2.11 &1.98 &1.67\\
AR0530SR &(1092, 3885) &12.09 &204.61 &0.27 &1.29 &4.39 &2.22 &139.88 &0.28 &4.71 &3.88 &2.83\\
\cline{1-13}
Shanghai\_0\_256 &(253856, 990715) &14.98 &- &0.21 &- &- &- &- &0.40 &- &- &-\\
blastedlands &(131342, 505974) &24.34 &- &0.22 &- &- &- &- &0.45 &- &- &-\\
maze512-32-5 &(253856, 990715) &37.99 &- &0.67 &- &- &- &- &1.03 &- &- &-\\
\cline{1-13}
\end{tabular}
}
\caption[Results for the Weighted Vertex $K$-Median problem on various grid-world instances.]{Shows the results for the WVKM problem on various grid-world instances.}
\label{tab:wvkm_grids}
\end{table}

\begin{table}[!t]
\footnotesize
\centering
\scalebox{0.76}{
\begin{tabular}{|l|r|r|r|r|r|r|r|r|r|}
\cline{1-10}
\multirow{3}{*}{Instance} &\multirow{3}{*}{Size ($|V|$, $|E|$)} &\multicolumn{1}{c|}{Preprocessing} &\multicolumn{1}{c|}{Preprocessing} &\multicolumn{3}{c|}{$K = 10$} &\multicolumn{3}{c|}{$K = 20$}\\
\cline{5-10}
&&\multicolumn{1}{c|}{for ILP:} &\multicolumn{1}{c|}{for FM:} &\multicolumn{2}{c|}{Time (s)} &\multicolumn{1}{c|}{SO (\%)} &\multicolumn{2}{c|}{Time (s)} &\multicolumn{1}{c|}{SO (\%)}\\
\cline{5-10}
&&\multicolumn{1}{c|}{FW (s)} &\multicolumn{1}{c|}{FM\_pre (s)} &ILP &FM &FM &ILP &FM &FM\\
\cline{1-10}
orz102d &(738, 2632) &27.10 &0.05 &101.07 &0.09 &1.80 &162.59 &0.13 &3.57\\
den407d &(852, 3054) &41.24 &0.07 &49.35 &0.04 &4.02 &166.29 &0.09 &2.88\\
lak526d &(954, 3329) &56.36 &0.07 &184.90 &0.04 &7.82 &84.68 &0.28 &3.34\\
den009d &(1003, 3620) &64.38 &0.07 &86.95 &0.17 &1.45 &114.34 &0.26 &4.10\\
\cline{1-10}
AR0512SR &(896, 3275) &48.20 &0.06 &66.96 &0.06 &2.89 &495.82 &0.10 &3.19\\
AR0402SR &(1075, 3796) &82.27 &0.10 &90.72 &0.06 &3.99 &85.61 &0.28 &5.63\\
AR0517SR &(1083, 4078) &85.25 &0.09 &1942.77 &0.24 &3.17 &3615.14 &0.29 &2.85\\
AR0530SR &(1092, 3885) &84.71 &0.08 &175.05 &0.30 &0.54 &242.01 &0.23 &3.94\\
\cline{1-10}
Shanghai\_0\_256 & (48696, 190303) &- &4.85 &- &6.19 &- &- &33.48 &-\\
blastedlands &(131342, 505974) &- &15.18 &- &59.19 &- &- &62.15 &-\\
maze512-32-5 &(253856, 990715) &- &22.50 &- &13.18 &- &- &20.08 &-\\
\cline{1-10}
wm00800 &(800, 9498) &35.94 &1.06 &2721.67 &0.13 &48.18 &4590.93 &0.27 &62.34\\
wm01000 &(1000, 14923) &70.30 &1.78 &10531.79 &0.13 &35.18 &23911.17 &0.24 &58.05\\
wm05000 &(5000, 374925) &- &88.64 &- &1.08 &- &- &2.56 &-\\
wm10000 &(10000, 1499713) &- &360.29 &- &3.07 &- &- &6.04 &-\\
\cline{1-10}
\end{tabular}
}
\caption[Results for the Capacitated Vertex $K$-Median problem on various graph instances.]{Shows the results for the CVKM problem on various graph instances.}
\label{tab:cvkm_graphs}
\end{table}

\begin{table}[!t]
\footnotesize
\centering
\scalebox{0.76}{
\begin{tabular}{|l|r|r|r|r|r|r|r|r|r|r|}
\cline{1-11}
\multirow{3}{*}{Instance} &\multirow{3}{*}{Size ($|V|$, $|E|$)} &\multicolumn{1}{c|}{Preprocessing} &\multicolumn{4}{c|}{$K = 10$} &\multicolumn{4}{c|}{$K = 20$}\\
\cline{4-11}
&&\multicolumn{1}{c|}{for FMA:} &\multicolumn{2}{c|}{Time (s)} &\multicolumn{2}{c|}{SO (\%)} &\multicolumn{2}{c|}{Time (s)} &\multicolumn{2}{c|}{SO (\%)}\\
\cline{4-11}
&&\multicolumn{1}{c|}{FMA\_pre (s)} &ILP &FMA &FM &FMA &ILP &FMA &FM &FMA\\
\cline{1-11}
orz102d &(738, 2632) &3.29 &90.55 &0.05 &1.52 &2.44 &192.79 &0.19 &2.40 &3.64\\
den407d &(852, 3054) &2.66 &66.04 &0.09 &2.12 &4.45 &153.89 &0.10 &2.43 &2.32\\
lak526d &(954, 3329) &3.77 &187.95 &0.09 &2.01 &2.08 &78.51 &0.27 &3.91 &1.65\\
den009d &(1003, 3620) &2.00 &116.48 &0.08 &1.00 &1.05 &106.17 &0.25 &1.59 &1.60\\
\cline{1-11}
AR0512SR &(896, 3275) &10.81 &62.47 &0.26 &5.51 &3.47 &521.99 &0.13 &4.24 &8.09\\
AR0402SR &(1075, 3796) &8.54 &84.41 &0.21 &2.02 &4.80 &90.01 &0.14 &5.27 &7.96\\
AR0517SR &(1083, 4078) &9.84 &1237.97 &0.42 &3.20 &3.22 &4324.49 &0.35 &3.35 &2.45\\
AR0530SR &(1092, 3885) &12.09 &138.18 &0.11 &1.66 &3.64 &254.33 &0.21 &5.71 &1.68\\
\cline{1-11}
Shanghai\_0\_256 &(48696, 190303) &17.02 &- &12.27 &- &- &- &21.37 &- &-\\
blastedlands &(131342, 505974) &42.48 &- &33.20 &- &- &- &34.67 &- &-\\
maze512-32-5 &(253856, 990715) &39.58 &- &14.39 &- &- &- &24.48 &- &-\\
\cline{1-11}
\end{tabular}
}
\caption[Results for the Capacitated Vertex $K$-Median problem on various grid-world instances.]{Shows the results for the CVKM problem on various grid-world instances.}
\label{tab:cvkm_grids}
\end{table}

Tables~\ref{tab:mam_graphs}-\ref{tab:cvkm_grids} show the performance results of various algorithms on the MAM, VKM, WVKM, and the CVKM problems. Although our experiments are extensive and conclusive, we present only representative results in these tables. In each table, representative results are shown in sets of rows: the first set on instances from `Dragon Age: Origins', the second set on instances from `Baldurs Gate II', and the third set on the largest instances from `City/Street Maps', `Warcraft III', and `Mazes', in that order. These three sets are from the first category and serve as both graph and grid-world instances. The odd-numbered tables also have a fourth set of rows from the second category that serve only as graph instances. A `-' is associated with any instance whose preprocessing time exceeds~\SI{60}{\minute}. The suboptimality `SO' columns report (cost - optimal cost)/(optimal cost) as a percentage. In general, our approach is the only one that can scale to large input sizes for the VKM, WVKM, and the CVKM problems.

Table~\ref{tab:mam_graphs} compares FastMap (FM) and the brute-force algorithm (DJK) on the MAM problem. DJK computes the ground truth by rooting a shortest-path tree at each of the $k$ start vertices of the agents. Here, each graph instance is designed by picking the $k$ start vertices at random. The FastMap preprocessing time (FM\_pre) refers to the time taken by FastMap to generate the Euclidean embedding of the graph plus the time taken by LSH for the initial indexing. This preprocessing time is required only once per graph, independent of $k$ and the start vertices. We observe that FastMap is significantly faster than the brute-force algorithm on larger instances, up to $4$-$5$ orders of magnitude.\footnote{Based on the results reported in~\cite{aflssk23}, FastMap also seems significantly faster\textemdash and more scalable to large graphs with large values of $k$\textemdash compared to $MM^*$.} In fact, FastMap is very often more efficient even with the preprocessing time included. It also produces solutions that are within just $7\%$ suboptimality on instances from the first category and within $10\%$ suboptimality on instances from the second category. Table~\ref{tab:mam_grids} shows a similar dominance of FastMap and FastMap with Anya (FMA) over the brute-force algorithm (ADJK) that uses Anya to compute an any-angle shortest-path tree rooted at each of the $k$ start vertices for generating the ground truth on grid-world instances. The FastMap times are excluded from Table~\ref{tab:mam_grids} since they appear in Table~\ref{tab:mam_graphs}. The suboptimality of FastMap is different in Tables~\ref{tab:mam_graphs} and~\ref{tab:mam_grids}, since the quality of a solution is measured using any-angle shortest-path distances in Table~\ref{tab:mam_grids}.

Table~\ref{tab:vkm_graphs} compares FastMap, the ILP solver, and FasterPAM for solving graph instances of the VKM problem. Both the ILP solver and FasterPAM use the Floyd-Warshall algorithm (FW) in a preprocessing step to compute the all-pairs shortest-path distances. The preprocessing time of FastMap is significantly smaller than that of the Floyd-Warshall algorithm: The latter is about $3$ orders of magnitude slower and not even viable for large graphs. Moreover, at query time, FastMap and FasterPAM are both significantly faster than the ILP solver. FastMap produces solutions within just $6\%$ suboptimality on instances from the first category and within $17\%$ suboptimality on instances from the second category. FasterPAM also produces high-quality solutions but it does so with occasional outliers and does not scale to large instances. Table~\ref{tab:vkm_grids} on grid-world instances shows a similar dominance of FastMap and FastMap with Anya\textemdash within similar suboptimality ranges\textemdash over the ILP solver and FasterPAM with respect to the preprocessing time and over the ILP solver with respect to the query time. The FastMap times, the Floyd-Warshall preprocessing times, and the FasterPAM query times are excluded from Table~\ref{tab:vkm_grids} since they appear in Table~\ref{tab:vkm_graphs}.\footnote{The computation of the all-pairs any-angle shortest-path distances is very expensive. Therefore, the Floyd-Warshall algorithm is invoked by treating the grid-world as a graph. This also creates the remote possibility of the ILP solver producing a suboptimal solution, although it is practically still treated as the ground truth.} However, the query times of the ILP solver vary across different runs on the same instance and, thus, are included in Table~\ref{tab:vkm_grids}. Tables~\ref{tab:wvkm_graphs} and~\ref{tab:wvkm_grids} show the same trends for the WVKM problem,\footnote{The ILP solver can also be an anytime solver. However, its performance is nowhere comparable to that of FastMap. For example, on the representative instance `wm01000' with $K = 20$, compared to FM, ILP takes about $400 \times$ time to produce a mere $34\%$ suboptimal solution for both the VKM and the WVKM problems.} instances of which are generated by assigning to each vertex a weight chosen uniformly at random from the interval $[0, 1)$. Here, the PAM algorithm is the `wcKMedoids' procedure in R 4.3.

Table~\ref{tab:cvkm_graphs} compares FastMap and the ILP solver for solving graph instances of the CVKM problem. For representative results, these instances use $\tau = \ceil{2|V|/K}$. Although the CVKM problem renders many approaches ineffective due to its comparative hardness over the VKM and the WVKM problems, FastMap can still solve all the instances, even those with nearly a quarter-million vertices and a million edges, in mere seconds. On smaller instances, FastMap is also significantly faster than the ILP solver, which generates the ground truth. Moreover, FastMap produces solutions within just $8\%$ suboptimality on instances from the first category and within $63\%$ suboptimality\footnote{$63\%$ suboptimality is still impressive, since the CVKM problem is combinatorially very hard and, till date, there is no known polynomial-time constant-factor approximation algorithm for it.} on instances from the second category. Table~\ref{tab:cvkm_grids} shows the same trends for solving the grid-world instances of the CVKM problem with the same $\tau$, within just $9\%$ suboptimality using FastMap and FastMap with Anya.

\section{Discussion}

\begin{figure}[t!]
\centering
\includegraphics[width=0.9\columnwidth]{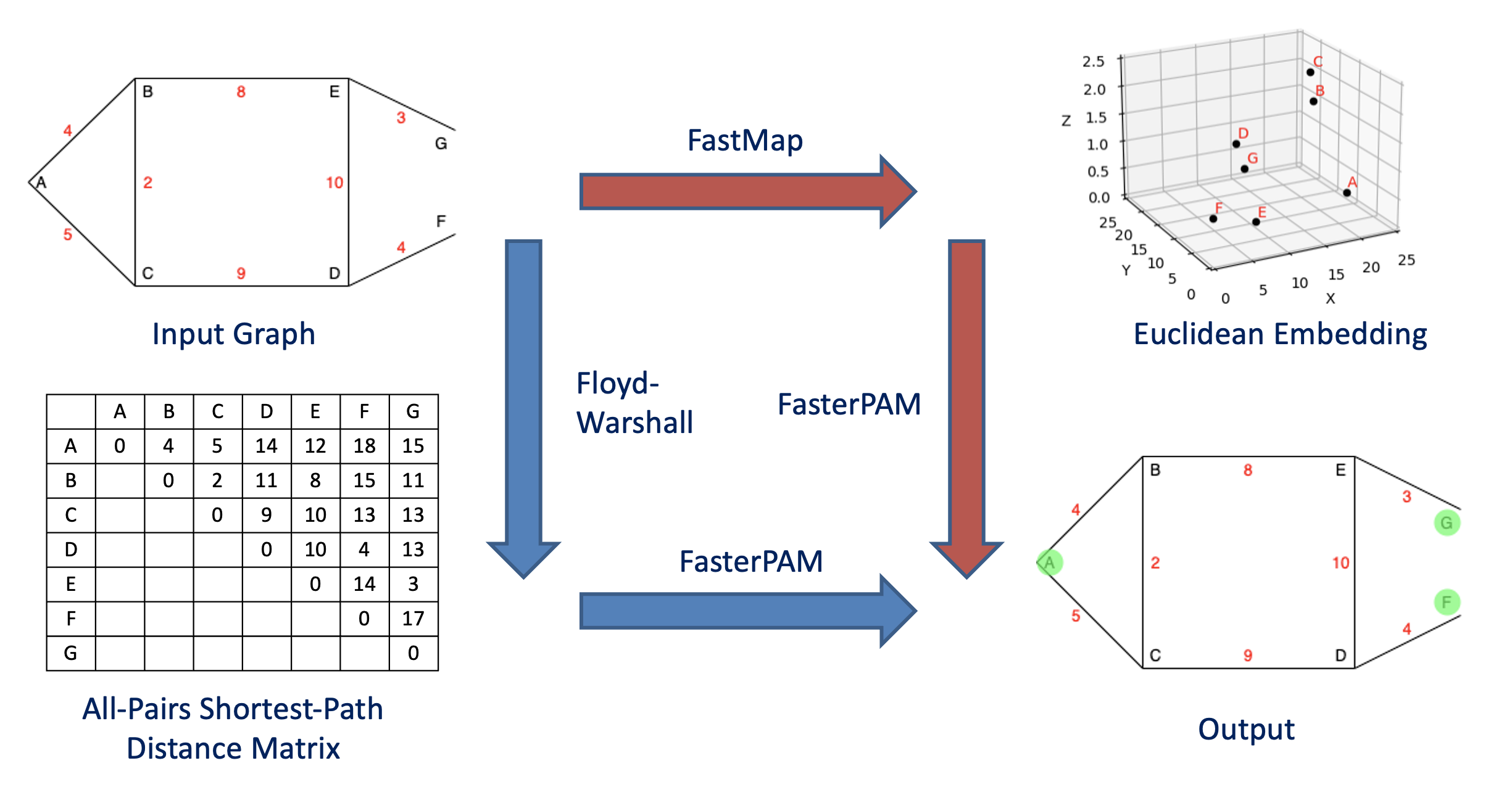}
\caption[The FastMap pipeline for the fast approximation of the all-pairs shortest-path distances required by existing graph algorithms for facility location.]{Illustrates the FastMap pipeline\textemdash as an alternative to the Floyd-Warshall pipeline\textemdash for the fast approximation of the all-pairs shortest-path distances. Here, the input graph is for the VKM problem with $K = 3$. The distortion in the pairwise shortest-path distances produced by FastMap is largely absorbed by FasterPAM and does not affect the quality of the final solution (shown in green).}
\label{fig:pipelines}
\end{figure}

\begin{figure}[t!]
\centering
\includegraphics[width=\columnwidth]{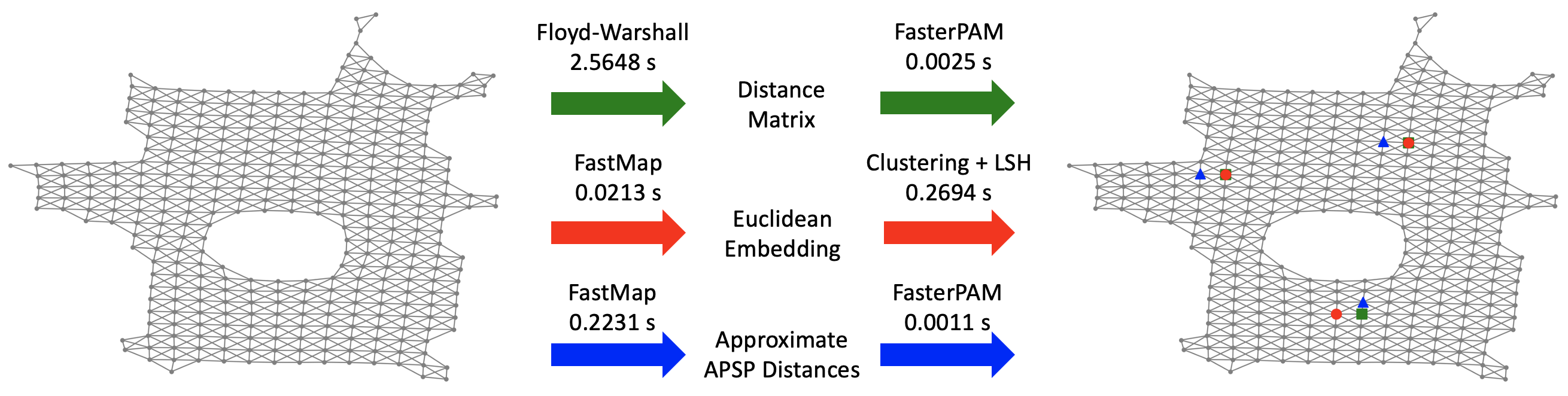}
\caption[A comparison of three methods for solving the Vertex $K$-Median problem on a representative movingAI instance.]{Compares three methods for solving the VKM problem on a representative movingAI instance. These three methods are generally applicable for solving FLPs on graphs, as mentioned in this chapter. On this instance, all three methods produce solutions within $1\%$ suboptimality for $K = 3$. However, they do so with different preprocessing and query times. `APSP Distances' refer to the all-pairs shortest-path distances. The outputs of the three methods are shown in the same color used for each of them.}
\label{fig:pipelines_comparison}
\end{figure}

While we have shown how FastMap can be used to solve FLPs on graphs by quickly reformulating them to FLPs in Euclidean spaces without obstacles, there is another way in which FastMap can be used for solving FLPs on graphs. In this method, the idea is to merely empower the existing state-of-the-art heuristic algorithms that work directly on the input graph $G$. Many such existing algorithms first compute the metric closure of $G$ in a preprocessing phase, which, in turn, requires the computation of the all-pairs shortest-path distances. As observed in the foregoing sections, this preprocessing phase can easily become a bottleneck. FastMap can alleviate this issue by computing the all-pairs shortest-path distances approximately but very efficiently: They are approximated by the Euclidean distances in the FastMap embedding. Figure~\ref{fig:pipelines} shows the suggested pipeline of this method.

Figure~\ref{fig:pipelines_comparison} compares the three methods discussed so far. The first method refers to the existing method of computing the metric closure of the input graph in a preprocessing phase followed by the application of a state-of-the-art algorithm that works directly on the graph. The second method refers to our method discussed in the previous sections of this chapter: It uses FastMap to create a Euclidean embedding, solves the FLP in this Euclidean space without obstacles by relating it to a clustering problem, and finally uses LSH to interpret the solution back on the original graph. The third method refers to the suggested method from the previous paragraph: It resembles the first method, except that it uses FastMap to merely approximate the all-pairs shortest-path distances required for computing the metric closure of the input graph.

In general, the third method is not as efficient and effective as the second method. However, it can still be useful when the FLP in the Euclidean space without obstacles is not readily relatable to a well-studied clustering problem. The Vertex $K$-Center (VKC) problem~\cite{mlh03} may be one such problem: It is like the VKM problem, except that the objective is to minimize the maximum of the shortest-path distances over each vertex to its nearest facility. A comprehensive study of this third method can be found in~\cite{tlkrkk22}.
 
\section{Conclusions}

In this chapter, we studied four representative FLPs: the MAM, VKM, WVKM, and the CVKM problems. Like most FLPs, these problems are well defined on graphs as well as in Euclidean spaces with or without obstacles. While they are generally difficult to solve optimally, the ones defined in a Euclidean space without obstacles are akin to clustering problems. We used the idea of FastMap to reformulate FLPs defined on a graph to FLPs defined in a Euclidean space without obstacles. Subsequently, we used standard clustering algorithms to solve the problems in the resulting Euclidean space and LSH to interpret the solutions back on the original graph. End to end, our approach produces high-quality solutions with orders-of-magnitude speedup over state-of-the-art competing algorithms.

\begin{subappendices}

\section{Table of Notations}

\begin{table}[h]
\centering
\begin{tabular}{|l|p{0.75\linewidth}|}
\cline{1-2}
Notation &Description\\
\cline{1-2}
$K$ &The number of facilities in the VKM, WVKM, and the CVKM problems.\\
\cline{1-2}
$\tau$ &The upper bound on the number of vertices that any facility can serve.\\
\cline{1-2}
$G = (V, E, w)$ &An undirected edge-weighted graph, where $w(e)$ is the non-negative weight on edge $e \in E$.\\
\cline{1-2}
$k$ &The number of agents in the MAM problem.\\
\cline{1-2}
$d_G(v_i, v_j)$ &The shortest-path distance between $v_i$ and $v_j$ in $G$ with respect to the edge weights.\\
\cline{1-2}
$G = (V, E, \tilde{w}, w)$ &An undirected vertex-weighted and edge-weighted graph, where $\tilde{w}(v)$ is the non-negative weight on vertex $v \in V$ and $w(e)$ is the non-negative weight on edge $e \in E$.\\
\cline{1-2}
$\kappa$ &The user-specified number of dimensions of the FastMap embedding.\\
\cline{1-2}
\end{tabular}
\caption[Notations used in Chapter~\ref{ch:facility_location}.]{Describes the notations used in Chapter~\ref{ch:facility_location}.}
\label{tab:facility_location_notations}
\end{table}

\end{subappendices}

\chapter{FastMap for Efficiently Computing Top-$K$ Projected Centrality}
\label{ch:centrality}
In this chapter, we describe how to use FastMap for efficient top-$K$ centrality computations. In graph theory and network analysis, various measures of centrality are used to characterize the importance of vertices in a graph. Although different measures of centrality have been invented to suit the nature and requirements of different underlying problem domains, their application is restricted to explicit graphs. Here, we first define implicit graphs that involve auxiliary vertices in addition to the pertinent vertices. We then generalize the various measures of centrality on explicit graphs to corresponding measures of projected centrality on implicit graphs. Finally, we propose a FastMap-based unifying framework for approximately, but very efficiently computing the top-$K$ pertinent vertices in explicit graphs for various measures of centrality and in implicit graphs for the generalizations of these measures to projected centrality.

\section{Introduction}

\begin{figure}[t!]
\centering
\includegraphics[width=0.5\textwidth]{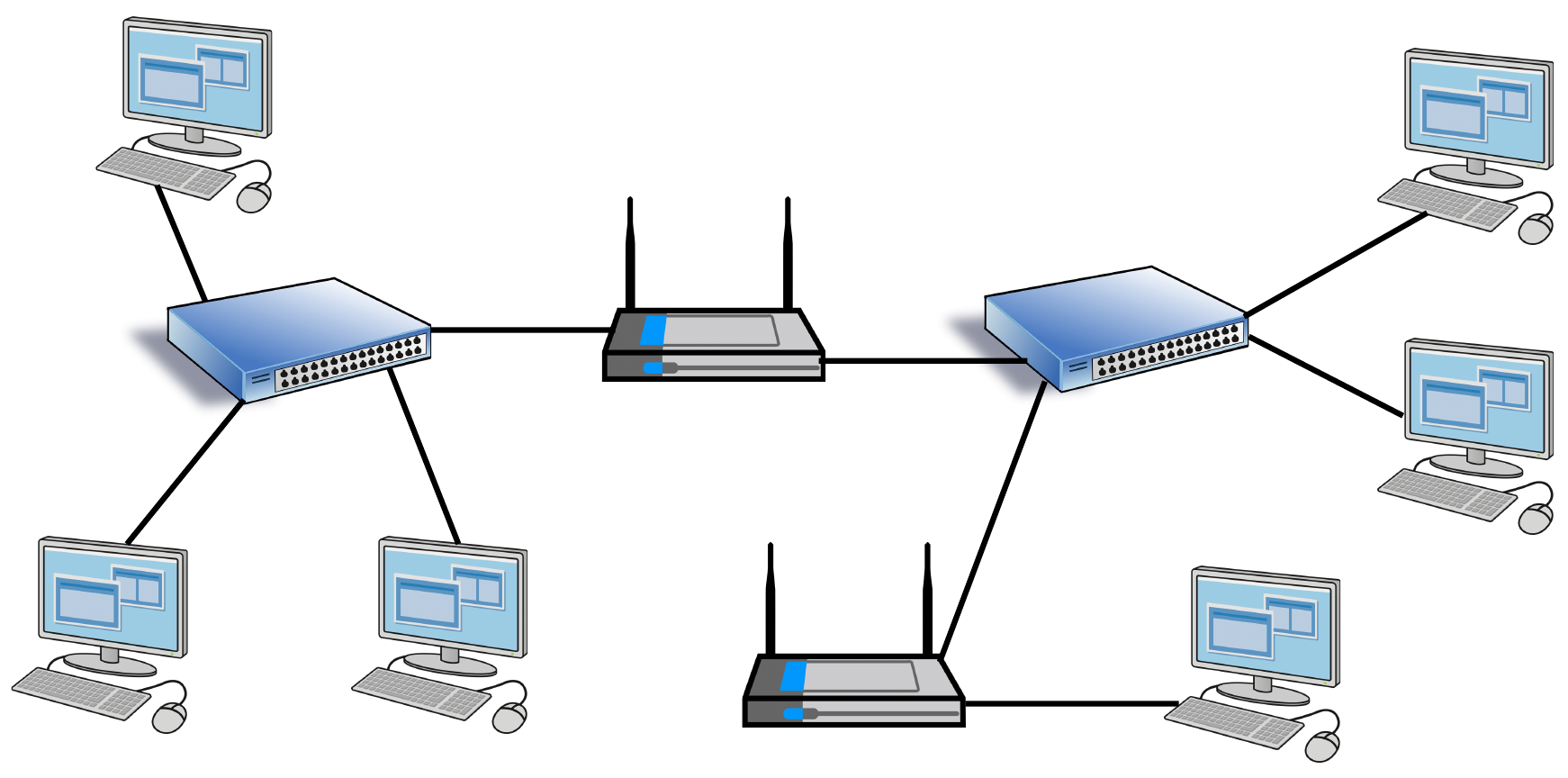}
\caption[A communication network in which user terminals represent the pertinent vertices and routers and switches represent the auxiliary vertices.]{Shows a communication network with user terminals, routers, and switches. The solid lines indicate direct communication links. Depending on the application, the user terminals may be considered as the pertinent vertices while the routers and the switches may be considered as the auxiliary vertices.}
\label{fig:network}
\end{figure}

Graphs are used to represent entities in a domain and important relationships between them: Often, vertices represent the entities and edges represent the relationships. However, graphs can also be defined implicitly by using two kinds of vertices and edges between the vertices: The~\emph{pertinent} vertices represent the main entities, that is, the entities of interest; the~\emph{auxiliary} vertices represent the hidden entities; and the edges represent relationships between the vertices. For example, in an air transportation domain, the pertinent vertices could represent the international airports, the auxiliary vertices could represent the domestic airports, and the edges could represent the flight connections between the airports. In a social network, the pertinent vertices could represent individuals, the auxiliary vertices could represent communities, and the edges could represent friendships or memberships. Figure~\ref{fig:network} shows another example in the domain of communication networks. Here, depending on the application, the pertinent vertices could represent the user terminals, the auxiliary vertices could represent the routers and the switches, and the edges could represent the direct communication links between them.

Explicit and implicit graphs can be used to model transportation networks, social networks, communication networks, and biological networks, among many others. In most of these domains, the ability to identify the ``important'' pertinent vertices has many applications. For example, the important pertinent vertices in an air transportation network could represent transportation hubs, such as Amsterdam and Los Angeles for Delta Airlines. The important pertinent vertices in a social network could represent highly influential individuals. Similarly, the important pertinent vertices in a communication network could represent the admin-users, and the important pertinent vertices in a properly modeled biological network could represent biochemicals critical for cellular operations.

The important pertinent vertices in a graph (network) as well as the task of identifying them depend on the definition of ``importance''. Such a definition is typically domain-specific. It has been studied in explicit graphs and is referred to as a~\emph{measure of centrality}. For example, the~\emph{page rank} is a popular measure of centrality used in Internet search engines~\cite{pbmw99}. In general, there are several other measures of centrality defined on explicit graphs, such as the~\emph{degree centrality}, the~\emph{closeness centrality}~\cite{f79}, the~\emph{harmonic centrality}~\cite{bv14}, the~\emph{current-flow closeness centrality}~\cite{sz89,bf05}, the~\emph{eigenvector centrality}~\cite{b87}, the~\emph{Katz centrality}~\cite{k53}, and the~\emph{betweenness centrality}~\cite{f77,b01,b08}.

The degree centrality of a vertex measures the immediate connectivity of it, that is, the number of its neighbors. The closeness centrality of a vertex is the average shortest-path distance between that vertex and all other vertices. The harmonic centrality resembles the closeness centrality but reverses the sum and reciprocal operations in its mathematical definition to be able to handle disconnected vertices and infinite distances. The current-flow closeness centrality also resembles the closeness centrality but uses an ``effective resistance'' between two vertices instead of the shortest-path distance between them. The eigenvector centrality scores the vertices based on the eigenvector corresponding to the largest eigenvalue of the adjacency matrix. The Katz centrality generalizes the degree centrality by incorporating a vertex's $k$-hop neighbors with a weight $\alpha^k$, where $\alpha \in (0, 1)$ is an attenuation factor. The betweenness centrality of a vertex measures the number of shortest paths between any two vertices that utilize it.

While many measures of centrality are frequently used on explicit graphs, they are not frequently used on implicit graphs. However, for any measure of centrality, a measure of ``projected'' centrality can be defined on implicit graphs. The measure of projected centrality is equivalent to the regular measure of centrality applied on a graph that ``factors out'' the auxiliary vertices from the implicit graph. Auxiliary vertices can be factored out by conceptualizing a clique on the pertinent vertices, in which an edge connecting two pertinent vertices is annotated with a graph-based distance between them that, in turn, is derived from the implicit graph.\footnote{The clique on the pertinent vertices is a mere conceptualization. Constructing it explicitly may be expensive and/or practically prohibitive for large graphs since it requires the computation of the graph-based distance between every pair of the pertinent vertices.} The graph-based distance can be the shortest-path distance or any other domain-specific distance. If there are no auxiliary vertices, the graph-based distance function is expected to be such that the measure of projected centrality reduces to the regular measure of centrality.

For a given measure of projected centrality, the projected centrality of a pertinent vertex is referred to as its~\emph{projected centrality value}. Identifying the important pertinent vertices in a network is equivalent to identifying the top-$K$ pertinent vertices with the highest projected centrality values. Graph theoretically, the projected centrality values can be computed for all pertinent vertices of a network in polynomial time, for most measures of projected centrality. However, in practical domains, the real challenge is to achieve scalability to very large networks with millions of vertices and hundreds of millions of edges. Therefore, algorithms with a running time that is quadratic or more in the size of the input are undesirable. In fact, algorithms with any super-linear running times, discounting logarithmic factors, are also largely undesirable. In other words, modulo logarithmic factors, a desired algorithm should have a near-linear running time close to that of merely reading the input.

Although attempts to achieve such near-linear running times exist, they are applicable only for certain measures of centrality on explicit graphs. For example,~\cite{ew06,cdpw14} approximate the closeness centrality using sampling-based procedures. A number of approaches~\cite{gss08,y14} approximate the betweenness centrality using refined estimators and a near-linear-time hypergraph sketching procedure.~\cite{bm15,bms15} approximate the betweenness centrality on dynamic graphs.~\cite{bbcmm19} maintains and updates a lower bound for each vertex, utilizing the bound to skip the analysis of a vertex when appropriate. It supports fairly efficient approximation algorithms for computing the top-$K$ vertices for the closeness and the harmonic centrality measures.~\cite{bdr16} provides a survey on many of the algorithms mentioned above. However, algorithms of the aforementioned kind are known only for a few measures of centrality on explicit graphs. Moreover, such algorithms do not provide a general framework since they are tied to specific measures of centrality.

In this chapter, we generalize the various measures of centrality on explicit graphs to corresponding measures of projected centrality on implicit graphs. Importantly, we also propose a framework for computing the top-$K$ pertinent vertices approximately, but very efficiently, using FastMap, for various measures of centrality and projected centrality. The FastMap framework allows us to conceptualize the various measures of centrality and projected centrality in Euclidean space, thereby facilitating a variety of geometric and analytical techniques for efficiently computing the top-$K$ pertinent vertices. It is extremely valuable because it implements this reformulation for different measures of centrality and projected centrality in only near-linear time and delegates the combinatorial heavy-lifting to geometric and analytical techniques that are better equipped for efficiently absorbing large input sizes.

Computing the top-$K$ pertinent vertices in the FastMap framework for different measures of projected centrality often requires interpreting analytical solutions found in the FastMap embedding back in the graphical space. We achieve this via nearest-neighbor queries and LSH. Through experimental results on a comprehensive set of benchmark and synthetic instances, we show that the FastMap+LSH framework is both efficient and effective for many popular measures of centrality and their generalizations to projected centrality. For our experiments, we also implement generalizations of some competing algorithms on implicit graphs. Overall, our approach demonstrates the benefits of drawing power from analytical techniques via FastMap for efficiently computing the top-$K$ projected centrality.

\section{Measures of Projected Centrality}

In this section, we generalize measures of centrality to corresponding measures of projected centrality. Consider an implicit graph $G = (V, E, w)$, where $V^P \subseteq V$ and $V^A \subseteq V$, for $V^P \cup V^A = V$ and $V^P \cap V^A = \emptyset$, are the pertinent vertices and the auxiliary vertices, respectively. We define a graph $G^P = (V^P, E^P, w^P)$, where, for any two distinct vertices $v_i^P, v_j^P \in V^P$, the edge $(v_i^P, v_j^P) \in E^P$ is annotated with the weight $w((v_i^P, v_j^P)) = \mathcal{D}_G(v_i^P, v_j^P)$. Here, $\mathcal{D}_G(\cdot, \cdot)$ is a distance function defined on pairs of vertices in $G$. For any measure of centrality $\mathcal{M}$ defined on explicit graphs, an equivalent measure of projected centrality $\mathcal{M}^P$ can be defined on implicit graphs as follows: $\mathcal{M}^P$ on $G$ is equivalent to $\mathcal{M}$ on $G^P$.

The distance function $\mathcal{D}_G(\cdot, \cdot)$ can be the shortest-path distance function or any other domain-specific distance function. If it is a graph-based distance function, computing it would typically require the consideration of the entire graph $G$, including the auxiliary vertices $V^A$. For example, computing the shortest-path distance between $v_i^P$ and $v_j^P$ in $V^P$ requires us to utilize the entire graph $G$. Other graph-based distance functions are the probabilistically-amplified distance function, introduced in Chapter~\ref{ch:block_modeling}, and the effective resistance between two vertices when interpreting the non-negative weights on edges as electrical resistance values.

\section{FastMap for Top-$K$ Centrality and Projected Centrality}

In this section, we first show how to use the FastMap framework, coupled with LSH, for efficiently computing the top-$K$ pertinent vertices with the highest centrality values in a given explicit graph (network), for various measures of centrality. We then generalize our methodology to the corresponding measures of projected centrality on implicit graphs. We note that the FastMap framework is applicable as a general paradigm, independent of the measure of centrality or projected centrality: The measure of centrality or projected centrality that is specific to the problem domain affects only the distance function used in the FastMap embedding and the analytical techniques that work on it. In other words, the FastMap framework allows us to interpret and reason about the various measures of centrality or projected centrality by invoking the power of analytical techniques. This is in stark contrast to other approaches that are tailored to a specific measure of centrality or its corresponding measure of projected centrality.

We recollect that, in the FastMap framework, any point of interest computed analytically in the Euclidean embedding may not map to a vertex in the original graph. Therefore, we use LSH to find the point closest to the point of interest that corresponds to any of the vertices. In fact, LSH not only answers nearest-neighbor queries very efficiently but also finds the top-$K$ nearest neighbors of a query point efficiently.

We assume that the input is an undirected edge-weighted graph $G = (V, E, w)$, where $V$ is the set of vertices, $E$ is the set of edges, and for any edge $e \in E$, $w(e)$ is the non-negative weight on it. We also assume that $G$ is connected since several measures of centrality and projected centrality are not very meaningful for disconnected graphs.\footnote{For disconnected graphs, we usually consider the measures of centrality and projected centrality on each connected component separately.} For simplicity, we further assume that there are no self-loops or multiple edges between any two vertices.

In the rest of this section, we first show how to use the FastMap framework for computing the top-$K$ vertices in explicit graphs, for some popular measures of centrality. We then show how to use the FastMap framework more generally for computing the top-$K$ pertinent vertices in implicit graphs, for the corresponding measures of projected centrality.

\subsection{FastMap for Closeness Centrality on Explicit Graphs}
\label{subsec:closeness}

Let $d_G(u, v)$ denote the shortest-path distance between two distinct vertices $u, v \in V$. The closeness centrality~\cite{f79} of $v$ is the reciprocal of the average shortest-path distance between $v$ and all other vertices. It is defined as follows:
\begin{equation}
C_{clo}(v) = \frac{|V| - 1}{\sum_{u \in V, u \neq v} d_G(u, v)}.
\label{eqn:clo_centrality}
\end{equation}

Computing the closeness centrality values of all vertices and identifying the top-$K$ vertices with the highest such values require calculating the shortest-path distances between all pairs of vertices. All-pair shortest-path computations generally require $O(|V||E| + |V|^2 \log |V|)$ time via the Floyd–Warshall algorithm~\cite{f62}.

The FastMap framework allows us to avoid the above complexity and compute the top-$K$ vertices using a geometric interpretation. We know that given $N$ points $q_1, q_2 \ldots q_N$ in Euclidean space $\mathbb{R}^\kappa$, finding the point $q$ that minimizes $\sum_{i = 1}^{N}(q - q_i)^2$ is easy. In fact, it is the centroid given by $q = (\sum_{i = 1}^{N}q_i)/N$. Therefore, we can use the distance function $\sqrt{d_G(\cdot, \cdot)}$ in Algorithm~\ref{alg:fastmap} (from Chapter~\ref{ch:fastmap}) to embed the square-root of the shortest-path distances between vertices. This is done by returning the square-roots of the shortest-path distances found by ShortestPathTree() on Lines~\ref{line:spt1},~\ref{line:spt2}, and~\ref{line:spt3}. Computing the centroid in the resulting embedding minimizes the sum of the shortest-path distances to all vertices. This centroid is mapped back to the original graphical space via LSH.

Overall, we use the following steps to find the top-$K$ vertices: (1) Use FastMap with the square-root of the shortest-path distance function between vertices to create a Euclidean embedding; (2) Compute the centroid of all points corresponding to vertices in this embedding; and (3) Use LSH to return the top-$K$ nearest neighbors of the centroid.

\subsection{FastMap for Harmonic Centrality on Explicit Graphs}

The harmonic centrality~\cite{bv14} of a vertex $v$ is the sum of the reciprocal of the shortest-path distances between $v$ and all other vertices. It is defined as follows:
\begin{equation}
C_{har}(v) = \sum_{u \in V, u \neq v}\frac{1}{d_G(u, v)}.
\label{eqn:har_centrality}
\end{equation}

As in the case of closeness centrality, the time complexity of computing the top-$K$ vertices, based on shortest-path algorithms, is $O(|V||E| + |V|^2 \log |V|)$. However, the FastMap framework once again allows us to avoid this complexity and compute the top-$K$ vertices using analytical techniques. Given $N$ points $q_1, q_2 \ldots q_N$ in Euclidean space $\mathbb{R}^\kappa$, finding the point $q$ that maximizes $\sum_{i = 1}^{N}\frac{1}{\|q - q_i\|}$ is not easy. However, the Euclidean space enables gradient ascent and the standard ingredients of local search to avoid local maxima and efficiently arrive at good solutions. In fact, the centroid obtained after running Algorithm~\ref{alg:fastmap} (from Chapter~\ref{ch:fastmap}) is a good starting point for the local search.

Overall, we use the following steps to find the top-$K$ vertices: (1) Use Algorithm~\ref{alg:fastmap} to create a Euclidean embedding; (2) Compute the centroid of all points corresponding to vertices in this embedding; (3) Perform gradient ascent starting from the centroid to maximize $\sum_{i = 1}^{N}\frac{1}{\|q - q_i\|}$; and (4) Use LSH to return the top-$K$ nearest neighbors of the result of the previous step.

\subsection{FastMap for Current-Flow Closeness Centrality on Explicit Graphs}

The current-flow closeness centrality~\cite{sz89,bf05} is a variant of the closeness centrality based on ``effective resistance'', instead of the shortest-path distance, between vertices. It is also known as the~\emph{information centrality}, under the assumption that information spreads like electrical current. The current-flow closeness centrality of a vertex $v$ is the reciprocal of the average effective resistance between $v$ and all 
other vertices. It is defined as follows:
\begin{equation}
C_{cfc}(v) = \frac{|V| - 1}{\sum_{u \in V, u \neq v}R_G(u, v)}.
\label{eqn:cfc_centrality}
\end{equation}
The term $R_G(u, v)$ represents the effective resistance between $u$ and $v$. A precise mathematical definition for it can be found in~\cite{bf05}.

Computing the current-flow closeness centrality values of all vertices and identifying the top-$K$ vertices with the highest such values are slightly more expensive than calculating the shortest-path distances between all pairs of vertices. The best known time complexity is $O(|V||E| \log |V|)$~\cite{bf05}.

Once again, the FastMap framework allows us to avoid the above complexity and compute the top-$K$ vertices by merely changing the distance function used in Algorithm~\ref{alg:fastmap} (from Chapter~\ref{ch:fastmap}). We use the probabilistically-amplified shortest-path distance (PASPD) function presented in Algorithm~\ref{alg:PASPD} (from Chapter~\ref{ch:block_modeling}).\footnote{We ignore edge-complement graphs by deleting Line~\ref{line:repeat_for_G_bar} of this algorithm.} The PASPD function computes the sum of the shortest-path distances between two vertices in a set of graphs $G_{set}$. $G_{set}$ contains different lineages of graphs, each starting from the given graph. In each lineage, a fraction of probabilistically-chosen edges is progressively dropped to obtain nested subgraphs. The probabilistically-amplified shortest-path distance captures the effective resistance between two vertices for the following two reasons: (a) Larger the $d_G(u, v)$, larger the probabilistically-amplified shortest-path distance between $u$ and $v$, as larger the effective resistance between them should be; and (b) Larger the number of paths between $u$ and $v$ in $G$, smaller the probabilistically-amplified shortest-path distance between them, as smaller the effective resistance between them should be.

Overall, we use the same steps as in Section~\ref{subsec:closeness} to find the top-$K$ vertices, except for the distance function being modified to the PASPD function.

\subsection{FastMap for Normalized Eigenvector Centrality on Explicit Graphs}

Suppose $G$ is an undirected unweighted graph with the adjacency matrix $\textbf{A}$, where $\textbf{A}_{ij}$ is set to $1$ if $(v_i, v_j) \in E$ and to $0$ otherwise. The eigenvector centrality~\cite{b87} of $v_i$ is the corresponding component of the eigenvector for the largest eigenvalue of $\textbf{A}$. In essence, the eigenvector centrality of a vertex depends on those of its neighbors, imparting to it the semblance of the page rank. In fact, the~\emph{normalized eigenvector centrality} is a measure that is intimately related to the page rank~\cite{pbmw99}. It uses a matrix $\textbf{N}$ that is more generally defined for directed unweighted graphs as follows:
\begin{equation}
\textbf{N}_{ij} =
\begin{cases} 
\frac{1}{od(v_i)} &\text{if } (v_i, v_j) \in E\\
0 &\text{otherwise}
\end{cases}.
\label{eqn:row_stochastic}
\end{equation}
Here, $od(v_i)$ refers to the out-degree of $v_i$. If $G$ is undirected, $od(v_i)$ is just the degree of $v_i$, that is, the number of neighbors of $v_i$.

$\textbf{N}$ uses a row-wise normalization of the entries of $\textbf{A}$ and, therefore, is a row-stochastic matrix. When $G$ is edge-weighted, the weight of an edge from $v_i$ to $v_j$ can be interpreted as a ``resistance'' to the transition from $v_i$ to $v_j$. Therefore, $\textbf{N}$ can be generalized as follows:
\begin{equation}
\textbf{N}_{ij} =
\begin{cases} 
\frac{1 / w((v_i, v_j))}{\sum_{(v_i, u) \in E}{1 / w((v_i, u))}} &\text{if } (v_i, v_j) \in E\\
0 &\text{otherwise}
\end{cases}.
\label{eqn:row_stochastic_weighted}
\end{equation}

The normalized eigenvector centrality of $v_i$ is the corresponding component of the eigenvector $\textbf{e}$ for the largest eigenvalue $\lambda$ of $\textbf{N}^T$. Since $\textbf{N}$ is a row-stochastic matrix, $\lambda = 1$ and $\textbf{e}$ satisfies the equation $\textbf{e} = \textbf{N}^T \textbf{e}$. Moreover, by the Perron–Frobenius theorem~\cite{psc05}, $\textbf{e}$ has a unique solution with positive entries.

In general, for a directed graph on $|V|$ vertices, computing the eigenvector centrality values of all vertices and identifying the top-$K$ vertices with the highest such values are more expensive than calculating the largest eigenvalue of a $|V| \times |V|$ matrix. Although these tasks are easier on undirected graphs, producing accurate results for them via the FastMap framework is a significant step towards generalization of the framework to directed graphs.\footnote{FastMap has already been generalized to directed graphs~\cite{gckk20}. However, the subsequent analytical techniques have to be generalized.}

For undirected graphs, the FastMap framework allows us to compute the top-$K$ vertices using analytical techniques. It is well known that row-stochastic matrices relate to infinite-length random walks and stationary distributions.\footnote{In such random walks, we start from an initial vertex and, in each iteration, we hop from the current vertex $u$ to one of its neighbors $v$ chosen with a probability proportional to $1 / w((u, v))$.} In turn, random walks are related to Brownian motion, diffusion equations, and heat equations~\cite{l10}. Therefore, it is conceivable that setting up a proper heat equation in the FastMap embedding could efficiently generate high-quality solutions. However, this refined approach is left for future work. Here, we use a simpler approach for a proof of concept. Since Gaussian distributions solve certain kinds of heat equations, we fit a mixture of Gaussian distributions on the point representations of vertices in the FastMap embedding. The vertices close to the centers of the dominant Gaussian distributions are more likely to have higher normalized eigenvector centrality values.

Towards this end, we invoke a GMM clustering procedure\footnote{say, the one available in Python3}. GMM clustering generates $k$ clusters, for a specified value of $k$. Each cluster $c \in \{1, 2 \ldots k\}$ is characterized by a Gaussian distribution $\pi_c N(\vec{x}; \vec{\mu}_c, \mathbf{\Sigma}_c)$, where $\pi_c$ is the amplitude representing its total probability mass, $\vec{\mu}_c$ is its center, and $\mathbf{\Sigma}_c$ is its covariance matrix. While it is possible to choose $k$ automatically via the Silhouettes measure~\cite{r87}, doing so is computationally expensive. Instead, we choose $k = 3$ and hierarchically decompose each resulting cluster into $3$ smaller clusters. We set the total probability mass of each cluster $c$ in the resulting $9$ clusters to $\pi_c \pi_{pa(c)}$, where $pa(c)$ is the parent cluster of $c$. Using LSH, we return the top-$K$ nearest neighbors of $\vec{\mu}_{c^*}$, where $c^* = \mbox{argmax}_{c} \pi_c \pi_{pa(c)}$.

Overall, we use the following steps to find the top-$K$ vertices: (1) Use Algorithm~\ref{alg:fastmap} (from Chapter~\ref{ch:fastmap}) to create a Euclidean embedding; (2) Use hierarchical GMM clustering and find the dominant cluster; and (3) Use LSH to return the top-$K$ nearest neighbors of the center of the Gaussian distribution representing this cluster.

\subsection{Generalization to Projected Centrality}

We now generalize the FastMap framework to compute the top-$K$ pertinent vertices in implicit graphs for different measures of projected centrality. There are several methods to do this. The first method is to create an explicit graph by factoring out the auxiliary vertices, that is, the explicit graph $G^P = (V^P, E^P, w^P)$ has only the pertinent vertices $V^P$ and is a complete graph on them, where for any two distinct vertices $v_i^P, v_j^P \in V^P$, the edge $(v_i^P, v_j^P) \in E^P$ is annotated with the weight $\mathcal{D}_G(v_i^P, v_j^P)$. This is referred to as the All-Pairs Distance (APD) method. For the closeness, harmonic, and the normalized eigenvector centrality measures, $\mathcal{D}_G(\cdot, \cdot)$ is the shortest-path distance function. For the current-flow closeness centrality measure, $\mathcal{D}_G(\cdot, \cdot)$ is the PASPD function. The second method also constructs the explicit graph $G^P$ but computes the weight on each edge only approximately using differential heuristics~\cite{sfbsb09}. This is referred to as the Differential Heuristic Distance (DHD) method. The third method is similar to the second, except that it uses the FastMap heuristics~\cite{lfkk19} instead of the differential heuristics. This is referred to as the FastMap Distance (FMD) method.

The foregoing three methods are inefficient because they construct $G^P$ explicitly by computing the distances between all pairs of pertinent vertices. To avoid this inefficiency, we propose the fourth and the fifth methods. The fourth method is to directly create the FastMap embedding for all vertices of $G$ but apply the analytical techniques only to the points corresponding to the pertinent vertices. This is referred to as the FastMap All-Vertices (FMAV) method. The fifth method is to create the FastMap embedding only for the pertinent vertices of $G$ and apply the analytical techniques to their corresponding points. This is referred to as the FastMap Pertinent-Vertices (FMPV) method.

\section{Experimental Results}

\begin{table}[!t]
\footnotesize
\centering
\scalebox{0.85}{
\begin{tabular}{|l|r|r|r|r|r|r|r|r|r|r|r|r|r|}
\cline{1-14}
\multirow{2}{*}{Instance} &\multirow{2}{*}{Size ($|V|$, $|E|$)} &\multicolumn{3}{c|}{Closeness} &\multicolumn{3}{c|}{Harmonic} &\multicolumn{3}{c|}{Current-Flow} &\multicolumn{3}{c|}{Eigenvector}\\
\cline{3-14}
&&GT &FM &nDCG &GT &FM &nDCG &GT &FM &nDCG &GT &FM &nDCG\\
\cline{1-14}
myciel5 &(47, 236) &0.01 &0.01 &0.8810 &0.01 &0.08 &0.8660 &0.00 &0.06 &0.7108 &0.00 &0.02 &0.4507\\
games120 &(120, 638) &0.06 &0.03 &0.9619 &0.06 &0.21 &0.9664 &0.02 &0.12 &0.9276 &0.01 &0.05 &0.8335\\
miles1500 &(128, 5198) &0.41 &0.09 &0.9453 &0.42 &0.29 &0.8888 &0.05 &0.79 &0.9818 &0.01 &0.12 &0.9379\\
queen16\_16 &(256, 6320) &1.06 &0.12 &0.9871 &1.07 &0.49 &0.9581 &0.11 &0.84 &0.9381 &0.04 &0.14 &0.8783\\
le450\_5d &(450, 9757) &3.13 &0.23 &0.9560 &3.11 &0.91 &0.9603 &0.30 &1.59 &0.8648 &0.13 &0.31 &0.7077\\
\cline{1-14}
myciel4 &(23, 71) &0.00 &0.01 &0.9327 &0.00 &0.04 &0.8299 &0.00 &0.02 &0.7697 &0.00 &0.02 &0.7180\\
games120 &(120, 638) &0.07 &0.03 &0.8442 &0.07 &0.21 &0.8004 &0.02 &0.14 &0.9032 &0.01 &0.06 &0.6868\\
miles1000 &(128, 3216) &0.28 &0.07 &0.9427 &0.28 &0.25 &0.8510 &0.04 &0.47 &0.7983 &0.03 &0.15 &0.5904\\
queen14\_14 &(196, 4186) &0.56 &0.09 &0.8866 &0.55 &0.38 &0.8897 &0.06 &0.73 &0.9188 &0.03 &0.15 &0.7399\\
le450\_5c &(450, 9803) &3.32 &0.24 &0.8843 &3.27 &0.88 &0.9203 &0.29 &2.09 &0.8196 &0.13 &0.29 &0.7285\\
\cline{1-14}
kroA200 &(200, 19900) &2.59 &0.52 &0.9625 &2.54 &0.50 &0.7275 &0.14 &2.95 &0.7589 &0.04 &0.30 &0.7336\\
pr226 &(226, 25425) &4.01 &0.51 &0.9996 &4.06 &0.63 &0.6803 &0.18 &3.52 &0.7978 &0.06 &0.36 &0.5372\\
pr264 &(264, 34716) &6.87 &0.62 &0.9911 &6.95 &0.84 &0.6506 &0.30 &6.30 &0.8440 &0.07 &0.61 &0.9816\\
lin318 &(318, 50403) &13.62 &1.00 &0.9909 &13.16 &1.19 &0.9537 &0.43 &8.07 &0.7243 &0.10 &0.89 &0.7981\\
pcb442 &(442, 97461) &39.97 &1.91 &0.9984 &39.25 &2.01 &0.9757 &0.84 &17.77 &0.7283 &0.21 &2.10 &0.9445\\
\cline{1-14}
orz203d &(244, 442) &0.11 &0.06 &0.9975 &0.11 &0.41 &0.9943 &0.05 &0.13 &0.8482 &0.04 &0.06 &1.0000\\
den404d &(358, 632) &0.23 &0.08 &0.9969 &0.23 &0.58 &0.8879 &0.10 &0.14 &0.9471 &0.12 &0.06 &1.0000\\
isound1 &(2976, 5763) &18.19 &0.63 &0.9987 &18.55 &4.86 &0.9815 &6.18 &1.95 &0.9701 &13.77 &0.28 &1.0000\\
lak307d &(4706, 9172) &46.74 &1.02 &0.9996 &48.56 &7.66 &0.9866 &15.82 &2.90 &0.9845 &50.07 &0.50 &1.0000\\
ht\_chantry\_n &(7408, 13865) &131.30 &1.51 &0.9969 &134.42 &12.29 &0.9144 &37.92 &3.64 &0.9189 &183.27 &0.69 &0.8694\\
\cline{1-14}
n0100 &(100, 99) &0.01 &0.02 &0.9171 &0.01 &0.17 &0.8102 &0.01 &0.05 &0.9124 &0.01 &0.04 &0.4799\\
n0500 &(500, 499) &0.34 &0.10 &0.8861 &0.33 &0.79 &0.6478 &0.18 &0.16 &0.9466 &0.21 &0.06 &0.1748\\
n1000 &(1000, 999) &1.33 &0.19 &0.9125 &1.38 &1.63 &0.9477 &0.70 &0.34 &0.7292 &0.86 &0.09 &0.2075\\
n1500 &(1500, 1499) &3.00 &0.28 &0.8856 &3.15 &2.39 &0.6360 &1.61 &0.41 &0.8118 &2.25 &0.11 &0.2140\\
n2000 &(2000, 1999) &5.25 &0.37 &0.9078 &5.56 &3.21 &0.8925 &2.71 &0.75 &0.9516 &4.97 &0.15 &0.1330\\
\cline{1-14}
n0100k4p0.3 &(100, 262) &0.02 &0.03 &0.9523 &0.03 &0.17 &0.9308 &0.01 &0.08 &0.9326 &0.03 &0.05 &0.7818\\
n0500k6p0.3 &(500, 1913) &0.83 &0.11 &0.9340 &0.85 &0.83 &0.8951 &0.23 &0.51 &0.8411 &0.16 &0.11 &0.6807\\
n1000k4p0.6 &(1000, 3192) &3.00 &0.24 &0.9119 &3.07 &1.68 &0.8975 &1.13 &0.83 &0.8349 &0.70 &0.22 &0.6386\\
n4000k6p0.6 &(4000, 19121) &86.37 &1.09 &0.9095 &85.18 &6.70 &0.9160 &214.99 &8.13 &0.8054 &32.05 &0.86 &0.6185\\
n8000k6p0.6 &(8000, 38517) &387.25 &2.86 &0.9240 &392.76 &14.65 &0.9368 &474.30 &11.08 &0.7466 &235.38 &1.83 &0.6376\\
\cline{1-14}
\end{tabular}
}
\caption[Results of competing algorithms for top-$K$ centrality computations.]{Shows results for various measures of centrality. Entries show running times in seconds and nDCG values.}
\label{tab:regular_centrality}
\end{table}

\begin{table}[!t]
\footnotesize
\centering
\scalebox{0.85}{
\begin{tabular}{|c|l|r|r|r|r|r|r|r|r|r|r|r|}
\cline{1-13}
\multirow{2}{*}{\diagbox{~}{~}} &\multirow{2}{*}{Instance} &\multicolumn{6}{c|}{Running Time (s)} &\multicolumn{5}{c|}{nDCG}\\
\cline{3-13}
&&APD &DHD &FMD &ADT &FMAV &FMPV &DHD &FMD &ADT &FMAV &FMPV\\
\cline{1-13}
\multirow{12}{*}{\rotatebox[origin=c]{90}{Closeness}} &queen16\_16 &0.55 &0.32 &0.12 &0.05 &0.12 &0.10 &0.9827 &0.9674 &0.9839 &0.9707 &0.9789\\
&le450\_5d &1.58 &0.98 &0.29 &0.09 &0.27 &0.26 &0.9576 &0.9621 &0.9872 &0.9577 &0.9521\\
&queen14\_14  &0.29 &0.19 &0.08 &0.03 &0.08 &0.09 &0.9274 &0.8996 &0.9639 &0.9377 &0.8898\\
&le450\_5c &1.71 &1.00 &0.32 &0.09 &0.28 &0.23 &0.9169 &0.9231 &0.9700 &0.8959 &0.8798\\
&lin318 &7.01 &0.68 &0.50 &0.45 &0.97 &1.05 &0.9285 &1.0000 &0.9645 &0.9867 &1.0000\\
&pcb442 &18.03 &1.31 &0.93 &0.84 &2.26 &1.97 &0.9455 &1.0000 &0.9663 &0.9969 &0.9950\\
&lak307d  &24.78 &116.69 &25.85 &1.98 &0.36 &0.35 &0.8991 &0.9928 &0.9387 &0.9994 &0.9960\\
&ht\_chantry\_n &67.59 &266.07 &58.18 &4.88 &0.47 &0.45 &0.8956 &0.9952 &0.9522 &0.9879 &0.9879\\
&n1500 &1.71 &10.87 &2.35 &0.19 &0.06 &0.06 &0.7990 &0.9485 &0.9830 &0.7759 &0.7758\\
&n2000 &2.96 &19.45 &4.32 &0.35 &0.08 &0.09 &0.8187 &0.9691 &0.9720 &0.9284 &0.9229\\
&n4000k6p0.6 &44.54 &78.68 &16.96 &1.37 &0.84 &0.68 &0.9403 &0.9172 &0.9582 &0.9146 &0.9299\\
&n8000k6p0.6 &207.24 &319.56 &67.23 &5.39 &1.37 &1.58 &0.9432 &0.9376 &0.9522 &0.9274 &0.9296\\
\cline{1-13}
\multirow{12}{*}{\rotatebox[origin=c]{90}{Harmonic}} &queen16\_16 &0.54 &0.31 &0.12 &0.00 &0.40 &0.32 &0.9730 &0.9602 &0.9729 &0.9466 &0.9730\\
&le450\_5d &1.59 &0.95 &0.31 &0.00 &0.65 &0.52 &0.9467 &0.9499 &0.9901 &0.9447 &0.9578\\
&queen14\_14 &0.31 &0.18 &0.08 &0.00 &0.29 &0.22 &0.9235 &0.8912 &0.9903 &0.9381 &0.8767\\
&le450\_5c &1.74 &0.98 &0.30 &0.00 &0.65 &0.51 &0.8905 &0.8694 &0.9962 &0.8485 &0.8866\\
&lin318 &6.93 &0.72 &0.52 &0.00 &1.22 &1.06 &0.9922 &1.0000 &0.8974 &0.8128 &0.8289\\
&pcb442 &20.79 &1.39 &1.04 &0.01 &2.07 &1.85 &0.9924 &1.0000 &0.9859 &0.9274 &0.9755\\
&lak307d &24.76 &105.98 &23.11 &0.16 &4.09 &3.94 &0.9265 &0.9915 &0.9908 &0.9819 &0.9762\\
&ht\_chantry\_n &65.83 &262.83 &58.63 &0.37 &6.34 &6.37 &0.7549 &0.6484 &0.9947 &0.8909 &0.9925\\
&n1500 &1.74 &10.83 &2.41 &0.01 &1.33 &1.22 &0.6280 &0.7079 &0.9858 &0.6575 &0.7469\\
&n2000 &2.95 &19.09 &4.15 &0.01 &1.71 &1.61 &0.6695 &0.6647 &0.9354 &0.9227 &0.9252\\
&n4000k6p0.6 &44.53 &78.08 &17.17 &0.10 &3.74 &3.76 &0.9252 &0.9083 &0.9971 &0.9163 &0.9072\\
&n8000k6p0.6 &215.14 &314.85 &67.36 &0.43 &7.94 &7.39 &0.9351 &0.9111 &0.9999 &0.9074 &0.9439\\
\cline{1-13}
\multirow{12}{*}{\rotatebox[origin=c]{90}{Current-Flow}} &queen16\_16 &5.92 &0.63 &0.72 &- &1.21 &1.26 &0.9638 &0.9542 &- &0.9690 &0.9683\\
&le450\_5d &17.50 &1.47 &1.45 &- &2.02 &2.05 &0.9463 &0.9678 &- &0.9529 &0.9632\\
&queen14\_14 &3.25 &0.38 &0.59 &- &1.29 &1.16 &0.9243 &0.8916 &- &0.8867 &0.8946\\
&le450\_5c &30.90 &2.03 &2.58 &- &5.55 &5.03 &0.9275 &0.9372 &- &0.8832 &0.8778\\
&lin318 &61.02 &3.29 &5.22 &- &11.69 &10.99 &0.8995 &1.0000 &- &0.9993 &0.9990\\
&pcb442 &180.43 &6.96 &12.75 &- &24.14 &23.77 &0.9781 &1.0000 &- &0.9997 &0.9997\\
&lak307d &265.38 &105.91 &24.77 &- &3.28 &2.71 &0.6864 &0.6158 &- &0.6800 &0.6316\\
&ht\_chantry\_n &499.27 &260.40 &58.07 &- &1.88 &3.54 &0.4634 &0.4568 &- &0.4314 &0.4997\\
&n1500 &5.70 &10.62 &2.45 &- &0.55 &0.30 &0.6731 &0.6820 &- &0.6436 &0.6487\\
&n2000 &10.01 &19.05 &4.21 &- &0.58 &0.63 &0.6689 &0.6847 &- &0.6487 &0.6468\\
&n4000k6p0.6 &448.20 &76.73 &19.42 &- &7.74 &7.65 &0.9218 &0.9130 &- &0.9320 &0.9126\\
&n8000k6p0.6 &2014.92 &306.53 &73.18 &- &16.04 &14.40 &0.9325 &0.9174 &- &0.9189 &0.9229\\
\cline{1-13}
\multirow{12}{*}{\rotatebox[origin=c]{90}{Eigenvector}} &queen16\_16 &0.56 &0.33 &0.13 &- &0.12 &0.13 &0.9536 &0.9541 &- &0.9576 &0.9383\\
&le450\_5d &1.61 &1.01 &0.32 &- &0.29 &0.21 &0.9569 &0.9469 &- &0.9410 &0.9474\\
&queen14\_14 &0.30 &0.19 &0.08 &- &0.13 &0.14 &0.9270 &0.8715 &- &0.8630 &0.9077\\
&le450\_5c &1.71 &1.03 &0.35 &- &0.27 &0.25 &0.8920 &0.8864 &- &0.9322 &0.8841\\
&lin318 &6.94 &0.71 &0.50 &- &0.78 &0.88 &0.9276 &1.0000 &- &0.8120 &0.7834\\
&pcb442 &17.90 &1.36 &1.05 &- &1.85 &1.65 &0.9823 &1.0000 &- &0.9331 &0.9626\\
&lak307d &31.23 &112.31 &30.54 &- &0.39 &0.57 &0.8147 &0.9783 &- &0.9045 &0.9728\\
&ht\_chantry\_n &87.72 &287.02 &79.17 &- &0.61 &0.73 &0.5739 &0.6980 &- &0.7092 &0.7838\\
&n1500  &1.89 &11.47 &2.67 &- &0.10 &0.11 &0.4143 &0.5353 &- &0.5712 &0.5383\\
&n2000 &3.35 &20.17 &4.81 &- &0.14 &0.12 &0.5533 &0.6402 &- &0.5640 &0.5308\\
&n4000k6p0.6 &45.17 &82.48 &20.50 &- &0.81 &0.69 &0.9272 &0.8970 &- &0.9129 &0.9114\\
&n8000k6p0.6 &234.65 &344.21 &94.71 &- &2.51 &1.45 &0.9473 &0.9209 &- &0.9294 &0.9286\\
\cline{1-13}
\end{tabular}
}
\caption[Results of competing algorithms for top-$K$ projected centrality computations.]{Shows results for various measures of projected centrality. Entries show running times in seconds and nDCG values.}
\label{tab:projected_cetrality}
\end{table}

We used six datasets in our experiments: DIMACS, wDIMACS, TSP, movingAI, Tree, and SmallWorld. The DIMACS dataset\footnote{\url{https://mat.tepper.cmu.edu/COLOR/instances.html}} is a standard benchmark dataset of unweighted graphs. We obtained edge-weighted versions of these graphs, constituting our wDIMACS dataset, by assigning an integer weight chosen uniformly at random from the interval $[1, 10]$ to each edge. We also obtained edge-weighted graphs from the TSP (Traveling Salesman Problem) dataset~\cite{r91} and large unweighted graphs from the movingAI dataset~\cite{s12}.\footnote{The movingAI dataset contains grid-world maps with free cells and obstacle cells. Each such map is converted to a graph with vertices representing the free cells and unweighted edges representing adjacent free cells with a common side. Only those resulting graphs that are connected are retained for our experiments.} In addition to these benchmark datasets, we synthesized Tree and SmallWorld graphs using the Python library NetworkX~\cite{hss08}. For the trees, we assigned an integer weight chosen uniformly at random from the interval $[1, 10]$ to each edge. We generated the small-world graphs using the Newman-Watts-Strogatz model~\cite{nw99}. For the regular measures of centrality, the graphs in the six datasets were used as such. For the projected measures of centrality, $50\%$ of the vertices in each graph were randomly chosen to be the pertinent vertices. (The choice of the percentage of pertinent vertices need not be $50\%$. This value is chosen merely for presenting illustrative results.)

We note that the largest graphs chosen in our experiments have about $18,500$ vertices and $215,500$ edges.\footnote{Tables~\ref{tab:regular_centrality} and~\ref{tab:projected_cetrality} show only representative instances that may not match these numbers.} Although FastMap itself runs in near-linear time and scales to much larger graphs, some of the baseline methods used for comparison in Tables~\ref{tab:regular_centrality} and~\ref{tab:projected_cetrality} are impeded by such large graphs. Nonetheless, our choice of problem instances and the experimental results on them illustrate the important trends in the effectiveness of our approach.

We used two metrics for evaluation: the normalized Discounted Cumulative Gain (nDCG), and the running time. The nDCG~\cite{wwlhl13} is a standard measure of the effectiveness of a ranking system. Here, it is used to compare the (projected) centrality values of the top-$K$ vertices returned by an algorithm against the (projected) centrality values of the top-$K$ vertices in the ground truth (GT). The nDCG value is in the interval $[0, 1]$, with higher values representing better results, that is, closer to the GT. We set $K = 10$. All experiments were done on a laptop with a 3.1 GHz Quad-Core Intel Core i7 processor and 16 GB LPDDR3 memory. We implemented FastMap in Python3 and set $\kappa = 4$.

Table~\ref{tab:regular_centrality} shows the performance of our FastMap (FM) framework against standard baseline algorithms that produce the GT for various measures of centrality. For the closeness, harmonic, and the current-flow closeness measures, the standard baseline algorithms are available in NetworkX.\footnote{\url{https://networkx.org/documentation/stable/reference/algorithms/centrality.html}} For the normalized eigenvector measure, a standard baseline algorithm can be implemented using matrix computations. The rows of the table are divided into six blocks corresponding to the six datasets in the order: DIMACS, wDIMACS, TSP, movingAI, Tree, and SmallWorld. For illustration, only five representative instances are shown in each block. For the closeness, harmonic, and the current-flow closeness measures, we observe that FM produces high-quality solutions. For the normalized eigenvector measure, FM generally produces good-quality solutions, with only occasional poor results. For all measures of centrality, FM is significantly faster than the standard baseline algorithms on large instances.

Table~\ref{tab:projected_cetrality} shows the performances of APD, DHD, FMD, FMAV, and FMPV for various measures of projected centrality. An additional column, called ``Adapted'' (ADT), is introduced for the closeness and the harmonic measures of projected centrality. For the closeness and the harmonic measures, ADT refers to our intelligent adaptations of state-of-the-art algorithms, presented in~\cite{cdpw14} and~\cite{bbcmm19}, respectively, to the projected case. The rows of the table are divided into four blocks corresponding to the four measures of projected centrality. For illustration, only twelve representative instances are shown in each block: the largest two from each block of Table~\ref{tab:regular_centrality}. The nDCG values for DHD, FMD, ADT, FMAV, and FMPV are computed against the GT produced by APD. We observe that all our algorithms produce high-quality solutions for the various measures of projected centrality. While the success of ADT is attributed to the intelligent adaptations of two separate algorithms, the success of FMAV and FMPV is attributed to the power of appropriate analytical techniques used in the same FastMap framework. The success of DHD and FMD is attributed to their ability to closely approximate the all-pairs distances. We also observe that FMAV and FMPV are significantly more efficient than APD, DHD, and FMD since they avoid the construction of explicit graphs on the pertinent vertices. For the same reason, ADT is also efficient when applicable.

For all measures of centrality and projected centrality considered in this chapter, Tables~\ref{tab:regular_centrality} and~\ref{tab:projected_cetrality} demonstrate that our FastMap approach is viable as a unified framework for leveraging the power of analytical techniques. This is in contrast to the nature of other existing algorithms that are tied to certain measures of centrality and have to be generalized to the projected case separately.

\section{Conclusions}

In this chapter, we generalized various measures of centrality on explicit graphs to corresponding measures of projected centrality on implicit graphs. Computing the top-$K$ pertinent vertices with the highest projected centrality values is not always easy for large graphs. To address this challenge, we proposed a unifying framework based on FastMap, exploiting its ability to embed a given undirected graph into a Euclidean space in near-linear time such that the pairwise Euclidean distances between vertices approximate a desired graph-based distance function between them. We designed different distance functions for different measures of projected centrality and invoked various procedures for computing analytical solutions in the resulting FastMap embedding. We also coupled FastMap with LSH to interpret analytical solutions found in the FastMap embedding back in the graphical space. Overall, we experimentally demonstrated that the FastMap+LSH framework is both efficient and effective for many popular measures of centrality and their generalizations to projected centrality.

Unlike other methods, our FastMap framework is not tied to a specific measure of projected centrality. This is because its power stems from its ability to transform a graph-theoretic problem into Euclidean space in only near-linear time close to that of merely reading the input. Consequently, it delegates the combinatorics tied to any given measure of projected centrality to various kinds of analytical techniques that are better equipped for efficiently absorbing large input sizes.

\begin{subappendices}

\section{Table of Notations}

\begin{table}[h]
\centering
\begin{tabular}{|l|p{0.75\linewidth}|}
\cline{1-2}
Notation &Description\\
\cline{1-2}
$K$ &Top-$K$ vertices with the highest centrality values.\\
\cline{1-2}
$G = (V, E, w)$ &An undirected edge-weighted graph, where $V$ is the set of vertices, $E$ is the set of edges, and for any edge $e \in E$, $w(e)$ is the non-negative weight on it.\\
\cline{1-2}
$G = (V, E, w)$ &An implicit graph, where $V^P \subseteq V$ and $V^A \subseteq V$, for $V^P \cup V^A = V$ and $V^P \cap V^A = \emptyset$, are the pertinent vertices and the auxiliary vertices, respectively.\\
\cline{1-2}
$G^P = (V^P, E^P, w^P)$ &An explicit graph constructed from $G = (V, E, w)$, where, for any two distinct vertices $v_i^P, v_j^P \in V^P$, the edge $(v_i^P, v_j^P) \in E^P$ is annotated with the weight $w((v_i^P, v_j^P)) = \mathcal{D}_G(v_i^P, v_j^P)$.\\
\cline{1-2}
$\mathcal{D}_G(\cdot, \cdot)$ &A distance function that can be the shortest-path distance function or any other domain-specific distance function.\\
\cline{1-2}
$\mathcal{M}$ &Any measure of centrality defined on explicit graphs.\\
\cline{1-2}
$\mathcal{M}^P$ &A measure of projected centrality defined on implicit graphs equivalent to $\mathcal{M}$ defined on explicit graphs.\\
\cline{1-2}
$d_G(u, v)$ &The shortest-path distance between two distinct vertices $u, v \in V$.\\
\cline{1-2}
$C_{clo}(v)$ &The closeness centrality value of $v \in V$.\\
\cline{1-2}
$\kappa$ &The user-specified number of dimensions of the FastMap embedding.\\
\cline{1-2}
$C_{har}(v)$ &The harmonic centrality value of $v \in V$.\\
\cline{1-2}
$C_{cfc}(v)$ &The current-flow closeness centrality value of $v \in V$.\\
\cline{1-2}
$R_G(u, v)$ &The effective resistance between two vertices $u, v \in V$.\\
\cline{1-2}
$\textbf{A}$ &The adjacency matrix of an undirected unweighted graph $G(V, E)$, where $\textbf{A}_{ij}$ is set to $1$ if $(v_i, v_j) \in E$ and to $0$ otherwise.\\
\cline{1-2}
$\textbf{N}$ &A row-stochastic matrix that uses a row-wise normalization of the entries of $\textbf{A}$.\\
\cline{1-2}
\end{tabular}
\caption[Notations used in Chapter~\ref{ch:centrality}.]{Describes the notations used in Chapter~\ref{ch:centrality}.}
\label{tab:centrality_notations}
\end{table}

\end{subappendices}

\chapter{FastMap for Community Detection and Block Modeling}
\label{ch:block_modeling}
Community detection and the more general block modeling algorithms are used to discover important latent structures in graphs. They are the graph equivalent of clustering algorithms. However, existing community detection and block modeling algorithms work directly on the given graphs, making them computationally expensive and less effective on large complex graphs. In this chapter, we propose a FastMap-based block modeling algorithm, FMBM, on single-view undirected unweighted graphs. In the first phase, FMBM uses FastMap with the PASPD function between vertices: Our novel PASPD function is explained in this chapter. In the second phase, it uses GMMs for identifying clusters (blocks) in the resulting Euclidean space. We show that FMBM outperforms other state-of-the-art methods on many benchmark and synthetic instances, in terms of both efficiency and solution quality. It also enables a perspicuous visualization of the clusters (blocks) in the graphs, not provided by other methods.

\section{Introduction}

Finding inherent groups in graphs, that is, the ``graph'' clustering problem, has important applications in many real-world domains, such as identifying communities in social networks~\cite{gn02}, analyzing the diffusion of ideas in them~\cite{lhwy15}, identifying functional modules in protein-protein interactions~\cite{lgl13}, and understanding the modular design of brain networks~\cite{a16}. Identifying the groups involves mapping each vertex in the graph to a group (cluster), where vertices in the same group share important properties in the underlying graph.

The conditions under which two vertices are deemed to be similar and therefore belonging to the same group are popularly studied in community detection and block modeling~\cite{a17}. In community detection, a group (community) implicitly requires its vertices to be more connected to each other than to vertices of other groups. Although this is justified in many real-world domains, such as social networks, it is not always justified in general. 

\begin{figure}[t!]
\centering
\subfloat[a core-periphery graph in the airport domain]{
\begin{minipage}[b]{0.4\textwidth}
\centering
\includegraphics[width=0.8\textwidth]{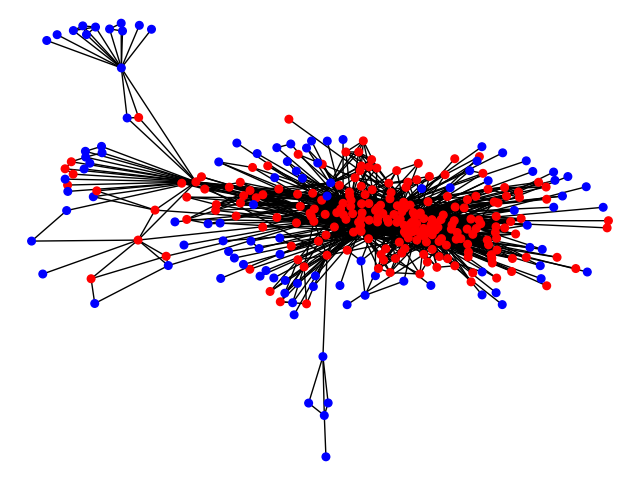}
\label{fig:airport_graph}
\end{minipage}
}
\hspace{0.05\textwidth}
\subfloat[a FastMap embedding of the graph on the left]{
\begin{minipage}[b]{0.4\textwidth}
\centering
\includegraphics[width=0.8\textwidth]{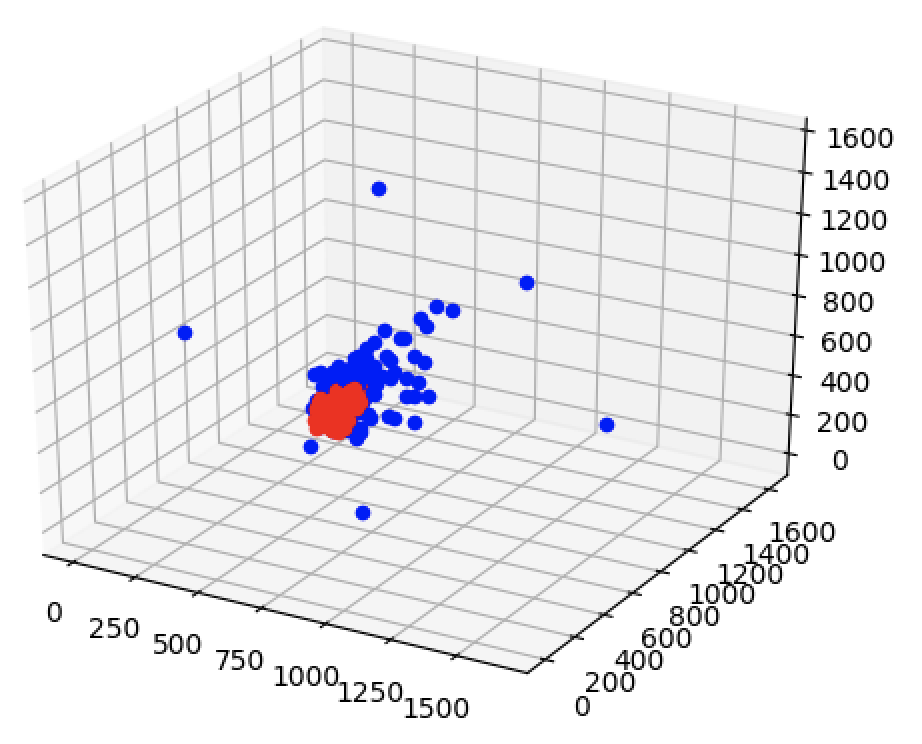}
\label{fig:airport_3d}
\end{minipage}
}
\caption[A core-periphery graph in an airport domain with its FastMap embedding.]{Shows a core-periphery graph in an airport domain with its FastMap embedding. (a) shows a core-periphery graph in the airport domain with edges representing flight connections, red vertices representing ``hub'' airports at the core, and blue vertices representing ``local'' airports at the periphery. (b) shows a FastMap embedding of the graph in Euclidean space, in which the red and blue vertices correctly appear in the core and periphery, respectively.}
\label{fig:airport}
\end{figure}

Block modeling uses more general criteria for identifying groups (blocks) where community detection fails. For example, block modeling can be used to correctly identify groups in~\emph{core-periphery} graphs characterized by a core of vertices tightly connected to each other and a peripheral set of vertices loosely connected to each other but well connected to the core.\footnote{The conditions used in community detection prevent the proper identification of peripheral groups.} Core-periphery graphs are common in many real-world domains, such as financial networks and flight networks~\cite{a17,rscrbld20}. Figure~\ref{fig:airport_graph} shows a core-periphery graph in an air transportation domain~\cite{d14}.

Existing community detection and block modeling algorithms work directly on the given graphs and are mostly inefficient. Block modeling algorithms typically use matrix operations that incur cubic time complexities even within their inner loops. For example, FactorBlock~\cite{clklbr13}, a state-of-the-art block modeling algorithm, uses matrix multiplications in its inner loop and an expectation-maximization-style outer loop. Due to their inefficiency, existing block modeling algorithms are not scalable and result in poor solution qualities on large complex graphs.

In this chapter, we propose a FastMap-based algorithm for block modeling on single-view\footnote{In a single-view graph, there is at most one edge between any two vertices.} undirected unweighted graphs. We also propose a novel distance function that probabilistically amplifies the shortest-path distances between vertices. We refer to this distance function as the PASPD function.

Our FastMap-based block modeling algorithm (FMBM) works in two phases. In the first phase, FMBM uses FastMap to efficiently embed the vertices of a given graph in a Euclidean space, preserving the probabilistically-amplified shortest-path distances between them. In the second phase, FMBM identifies clusters (blocks) in the resulting Euclidean space using standard methods from unsupervised learning. Therefore, the first phase of FMBM efficiently reformulates the block modeling problem from a graphical space to a Euclidean space, as illustrated in Figure~\ref{fig:airport_3d}; and the second phase of FMBM leverages any technique that is already known or can be developed for clustering in Euclidean space. In our current implementation of FMBM, we use GMMs for identifying clusters in the Euclidean space.

We empirically show that, in addition to the theoretical advantages of FMBM, it outperforms other state-of-the-art methods on many benchmark and synthetic test cases. We report on the superior performance of FMBM in terms of both efficiency and solution quality. We also show that it enables a perspicuous visualization of clusters in the graphs, beyond the capabilities of other methods.

\section{Preliminaries of Block Modeling}

In this section, we review some preliminaries of block modeling. Let $G = (V, E)$ be an undirected unweighted graph with vertices $V = \{v_1, v_2 \ldots v_n\}$ and edges $E = \{e_1, e_2 \ldots e_m\} \subseteq V \times V$. Let $\textbf{A} \in \{0, 1\}^{n \times n}$ be the adjacency matrix representation of $G$, where $\textbf{A}_{ij} = 1$ iff $(v_i, v_j) \in E$. 

A~\emph{block model} decomposes $G$ into a set of $k$ vertex partitions representing the blocks (groups), for a given value of $k$. The partitions are represented by the~\emph{membership matrix} $\textbf{C} \in \{0,1\}^{n \times k}$, where $\textbf{C}_{ij} = 0$ and $\textbf{C}_{ij} = 1$ represent vertex $v_i$ being absent from and being present in partition $j$, respectively. Therefore, $\sum_{j = 1}^k \textbf{C}_{ij} = 1$ for all $1 \leq i \leq n$. An~\emph{image matrix} is a matrix $\textbf{M} \in [0, 1]^{k \times k}$, where $\textbf{M}_{ij}$ represents the likelihood of an edge between a vertex in partition $i$ and a vertex in partition $j$. The block model decomposition of $G$, as discussed in~\cite{clklbr13}, tries to approximate $\textbf{A}$ by $\textbf{C}\textbf{M}\textbf{C}^\top$ with the best choice for $\textbf{C}$ and $\textbf{M}$. In other words, the objective is:
\begin{equation}
\min_{\textbf{C}, \textbf{M}}~\|\textbf{A} - \textbf{C}\textbf{M}\textbf{C}^\top\|_F^2,
\label{eqn:objective}
\end{equation}
where $\| \cdot \|_F$ is the Frobenius norm. An improved objective function is also considered in~\cite{clklbr13} to account for the imbalance of edges to non-edges\footnote{that is, a pair of vertices not connected by an edge} in $\textbf{A}$, since real-world graphs are typically sparse with significantly more non-edges than edges. The revised objective is:
\begin{equation}
\min_{\textbf{C}, \textbf{M}}~\|(\textbf{A} - \textbf{C}\textbf{M}\textbf{C}^\top) \circ (\textbf{A} - \textbf{R})\|_F^2,
\label{eqn:nullmodel-bm}
\end{equation}
where $\textbf{R} \in [0, 1]^{n \times n}$, $\textbf{R}_{ij} = \frac{m}{n^2}$, and $\circ$ represents element-wise multiplication.

The above formalization can be generalized to directed graphs and multi-view graphs~\cite{rscrbld20}. It can also be generalized to soft partitioning, where each vertex partially belongs to each partition, that is, $\textbf{C} \in [0, 1]^{n \times k}$ with $\sum_{j = 1}^k \textbf{C}_{ij} = 1$ for all $1 \leq i \leq n$.

\section{The FastMap-Based Block Modeling Algorithm (FMBM)}

\begin{figure}[t!]
\centering
\subfloat[]{
\centering
\includegraphics[width=0.3\textwidth]{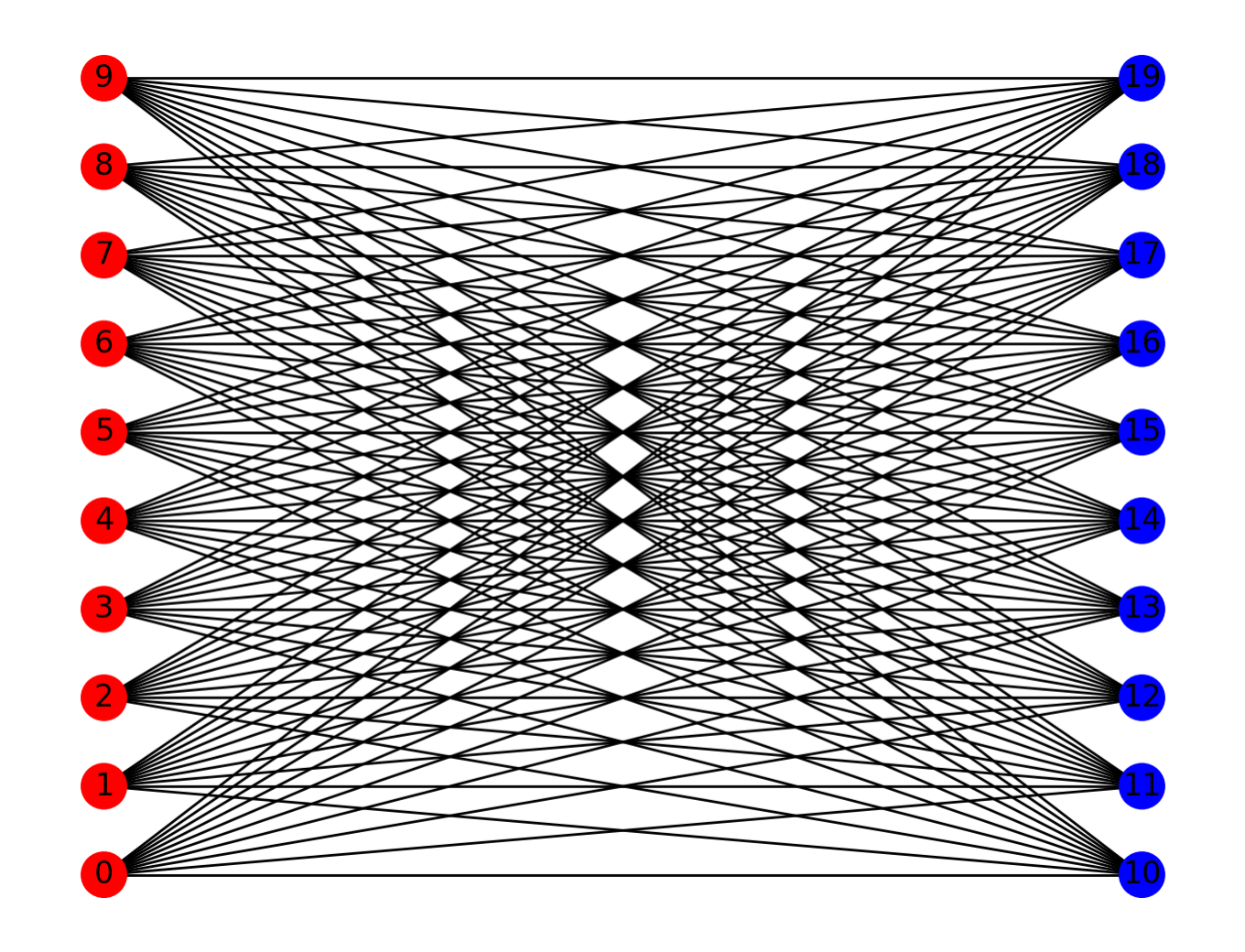}
\label{fig:bipartite}
}
\subfloat[]{
\centering
\includegraphics[width=0.3\textwidth]{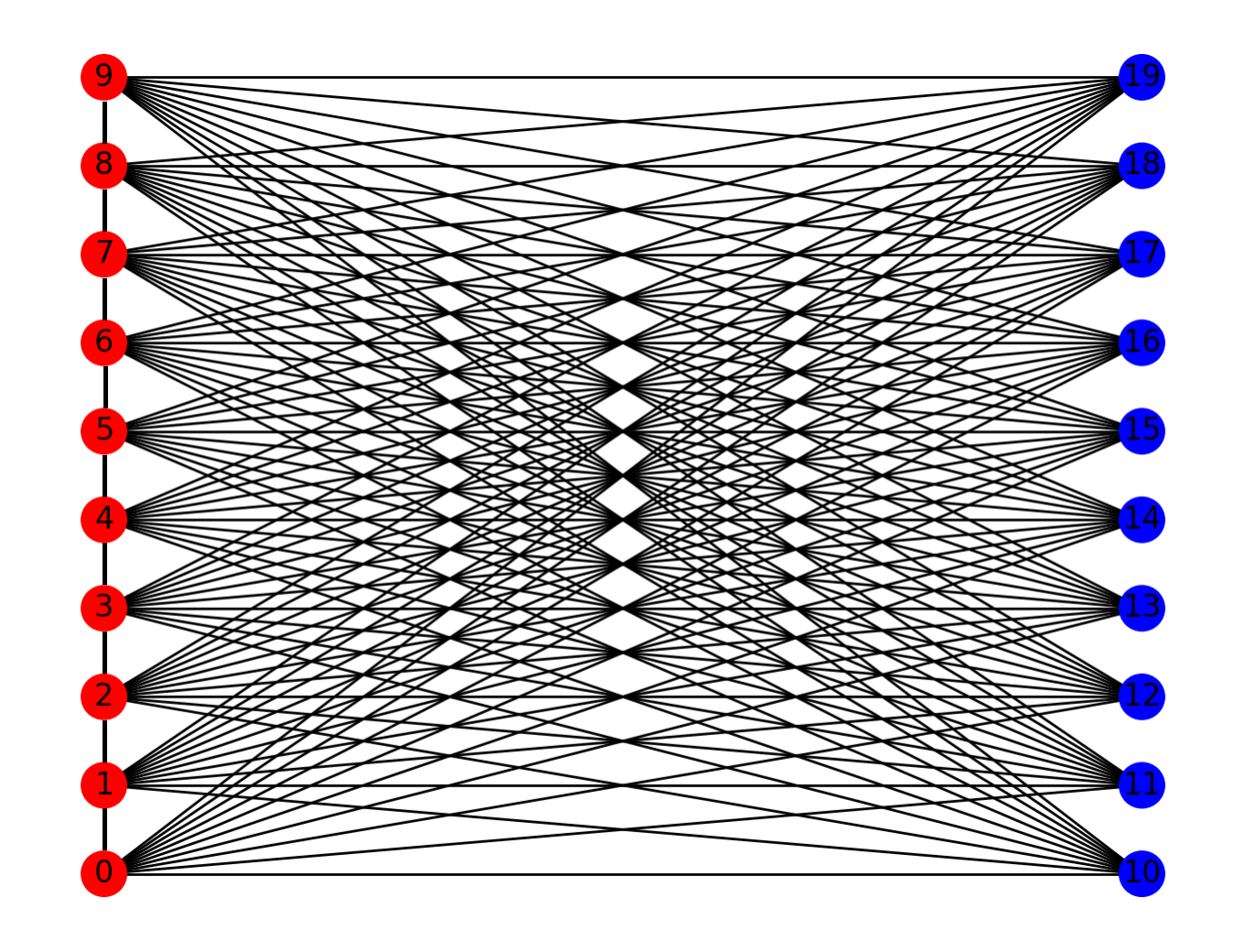}
\label{fig:core_periphery}
}
\subfloat[]{
\centering
\includegraphics[width=0.3\textwidth]{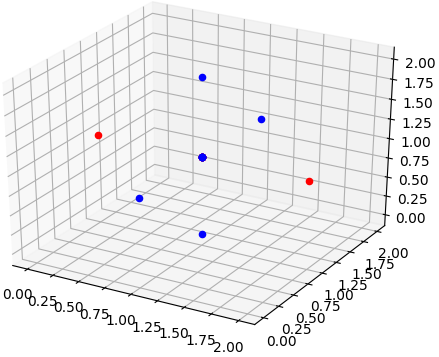}
\label{fig:naive_bp3d}
}

\subfloat[]{
\centering
\includegraphics[width=0.3\textwidth]{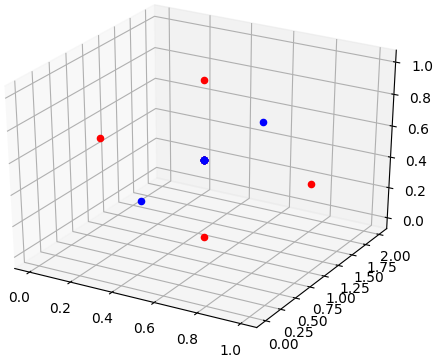}
\label{fig:naive_cp3d}
}
\subfloat[]{
\centering
\includegraphics[width=0.3\textwidth]{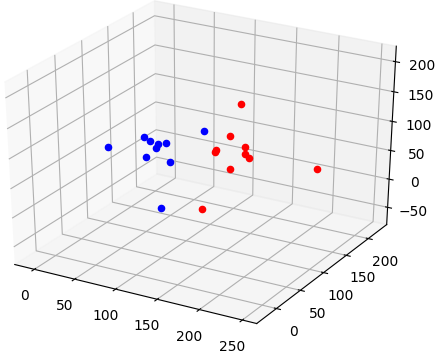}
\label{fig:paspd_bp3d}
}
\subfloat[]{
\centering
\includegraphics[width=0.3\textwidth]{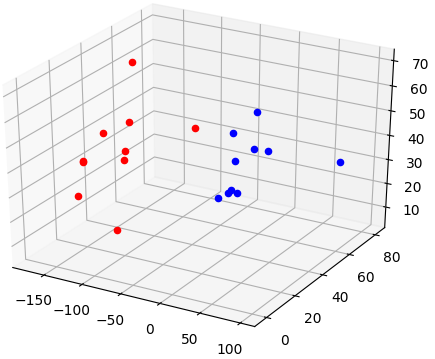}
\label{fig:paspd_cp3d}
}
\caption[Two simple graphs that guide the design of a proper FastMap distance function for block modeling.]{Shows two simple graphs that guide the design of a proper FastMap distance function for block modeling. (a) shows a fully-connected bipartite graph with the red and blue vertices indicating the two partitions. (b) shows a core-periphery graph with the red vertices indicating the core and the blue vertices indicating the periphery. All pairs of red vertices are connected by edges (not all shown to avoid clutter). (c) and (d) show the FastMap Euclidean embeddings of the graphs in (a) and (b), respectively, using the shortest-path distance function. This naive FastMap distance function fails for block modeling. Red and blue points correspond to red and blue vertices of the graphs, respectively. Many vertices are mapped to the same point. (e) and (f) show the FastMap Euclidean embeddings of the graphs in (a) and (b), respectively, when using the PASPD function. This FastMap distance function is appropriate for block modeling.}
\label{fig:paspd}
\end{figure}

In this section, we describe FMBM, our novel algorithm for block modeling based on FastMap~\cite{lfkk19}. As mentioned before, FMBM works in two phases. In the first phase, FMBM uses FastMap to efficiently embed vertices in a $\kappa$-dimensional Euclidean space, preserving the probabilistically-amplified shortest-path distances between them. In the second phase, FMBM identifies the required blocks in the resulting Euclidean space using GMM clustering.

To facilitate the description of FMBM, we first examine what happens when FastMap is used naively in the first phase, that is, when it is used to embed the vertices of a given undirected unweighted graph in a $\kappa$-dimensional Euclidean space for preserving the pairwise shortest-path distances. This naive attempt fails even in relatively simple cases. For example, Figures~\ref{fig:bipartite}-\ref{fig:naive_cp3d} show that it fails on a bipartite graph and a core-periphery graph. This is so because preserving the pairwise shortest-path distances in Euclidean space does not necessarily help GMM clustering to identify the two blocks (partitions). In fact, in a bipartite graph, the closest neighbors of a vertex are in the other partition.

\subsection{Probabilistically-Amplified Shortest-Path Distances}

From the foregoing discussion, it is clear that the shortest-path distance between two vertices $v_i$ and $v_j$ is not a viable distance function for block modeling. Therefore, in this subsection, we create a new distance function $D(v_i, v_j)$ for pairs of vertices based on the following intuitive guidelines: (a) The smaller the shortest-path distance between $v_i$ and $v_j$, the smaller the distance $D(v_i, v_j)$ should be; (b) The more paths exist between $v_i$ and $v_j$, the smaller the distance $D(v_i, v_j)$ should be; and (c) The complement graph\footnote{The complement graph $\bar{G}$ has the same vertices as the original graph $G$ but represents every edge in $G$ as a non-edge and every non-edge in $G$ as an edge.} $\bar{G}$ of the given graph $G$ should yield the same distance function as $G$: The distance function should be independent of the arbitrary choice of representing a relationship between two vertices as either an edge or a non-edge. Intuitively, these guidelines capture an effective ``resistance'' between vertices and facilitate the subsequent embedding to represent relative ``potentials'' of vertices in Euclidean space. The effectiveness of these guidelines is validated through test cases in this section and comprehensive experiments in the next section.

In adherence with these guidelines, we design the novel PASPD function $D_P(v_i, v_j)$ as follows:
\begin{equation}
D_P(v_i, v_j) = \sum_{\mathbb{G} \in G_{set}}d_{\mathbb{G}}(v_i, v_j).
\end{equation}
Here, $d_{\mathbb{G}}(v_i, v_j)$ represents the shortest-path distance between $v_i$ and $v_j$ in an undirected graph $\mathbb{G}$. $G_{set}$ represents a collection of undirected graphs, each derived from either the given graph $G$ or its complement $\bar{G}$. In particular, each graph in $G_{set}$ is an edge-induced subgraph of either $G$ or $\bar{G}$.\footnote{An edge-induced subgraph of $G$ has the same vertices as $G$ but a subset of its edges.} The edge-induced subgraphs are created by probabilistically dropping edges from $G$ or $\bar{G}$.

Intuitively, the use of shortest-path distances on multiple graphs that are probabilistically derived from the same input graph $G$ accounts for $D_P(\cdot, \cdot)$. Indeed, the smaller $d_G(v_i, v_j)$, the smaller $D_P(v_i, v_j)$ also is. Similarly, the more paths between $v_i$ and $v_j$ in $G$, the more likely it is for such paths to survive in its edge-induced subgraphs, and the smaller $D_P(v_i, v_j)$ consequently is. Moreover, since the subgraphs in $G_{set}$ are derived from both $G$ and $\bar{G}$, $D_P(\cdot, \cdot)$ satisfies all the intuitive guidelines mentioned above. From an efficiency perspective, the use of multiple graphs does not create much overhead if the number of graphs does not depend on the size of $G$. However, $\bar{G}$ can have significantly more edges than $G$ if $G$ is sparse. In such cases, if $G$ has $n$ vertices and $m < \binom{n}{2} / 2$ edges, $\bar{G}$ itself is probabilistically derived from $G$ by randomly retaining only $m$ out of the $\binom{n}{2} - m$ edges that it would otherwise have. This keeps the size of $\bar{G}$ upper-bounded by the size of the input.

Although more details on FMBM are presented in the next subsection, the benefits of using the PASPD function are visually apparent in Figures~\ref{fig:paspd_bp3d} and~\ref{fig:paspd_cp3d}. In both cases, the red and blue vertices are mapped to linearly-separable red and blue points, respectively, in Euclidean space. Its benefits can also be seen in Figure~\ref{fig:airport_3d}, where the core red vertices are mapped to a core set of red points and the peripheral blue vertices are mapped to a peripheral set of blue points, respectively, in Euclidean space. In this case, although the red and blue points are not linearly separable, GMM clustering~\cite{m12} in the second phase of FMBM is capable of separating them using two overlapping but different Gaussian distributions.

\subsection{Main Algorithm}

\begin{algorithm}[t!]
\caption{{\sc SS-PASPD}: An algorithm for computing the single-source probabilistically-amplified shortest-path distances.}
\label{alg:PASPD}
\textbf{Input}: $G = (V, E)$ and $v_s \in V$\\
\textbf{Parameters}: $L$ and $F$\\
\textbf{Output}: $d_{si}$ for each $v_i \in V$
\begin{algorithmic}[1]
\STATE Let $\bar{G} = (V, \bar{E})$ be the complement graph of $G$.
\STATE $G_{set} \leftarrow \{\}$ and $T_{set} \leftarrow \{\}$.
\FOR{$l = 1, 2 \ldots L$}\label{alg1:Lloop_start}
\STATE $\mathbb{G} \leftarrow G$ and $\bar{\mathbb{G}} \leftarrow \bar{G}$.
\IF{$|\bar{E}| > |E|$}\label{alg1:drop_start}
\STATE Drop $|\bar{E}| - |E|$ randomly chosen edges from $\bar{\mathbb{G}}$.
\ENDIF\label{alg1:drop_end}
\STATE $G_{set} \leftarrow G_{set}\cup\{\mathbb{G}\}$ and $f \leftarrow |E|/F$.\label{alg1:repeat_start}
\WHILE{$\mathbb{G}$ has edges}
\STATE Drop $f$ randomly chosen edges from $\mathbb{G}$ to obtain $\hat{G}$.
\STATE $G_{set} \leftarrow G_{set}\cup\{\hat{G}\}$.
\STATE $\mathbb{G} \leftarrow \hat{G}$.
\ENDWHILE\label{alg1:repeat_end}
\STATE Repeat lines~\ref{alg1:repeat_start}-\ref{alg1:repeat_end} for $\bar{\mathbb{G}}$.\label{line:repeat_for_G_bar}
\ENDFOR\label{alg1:Lloop_end}
\FOR{$G_i \in G_{set}$}\label{alg1:sum_start}
\STATE $T_i \leftarrow \mbox{SS-ShortestPathDistance}(G_i, v_s)$.\label{alg1:ss-spd}
\STATE $T_{set} \leftarrow T_{set}\cup\{T_i\}$.
\ENDFOR
\FOR{each $v_j \in V$}
\STATE $d_{sj} \leftarrow \sum_{T_i \in T_{set}} T_i(v_j)$.\label{alg1:ss-spd_set}
\ENDFOR
\STATE \textbf{return} $d_{si}$ for each $v_i \in V$.\label{alg1:sum_end}
\end{algorithmic}
\end{algorithm}

\begin{algorithm}[t!]
\caption{{\sc FMBM}: A FastMap-based block modeling algorithm.}
\label{alg:FMBM}
\textbf{Input}: $G = (V, E)$ and $k$\\
\textbf{Parameters}: $L$, $F$, $T$, $\kappa$, and $\epsilon$\\
\textbf{Output}: $c_i$ for each $v_i \in V$
\begin{algorithmic}[1]
\STATE $obj_{min} \leftarrow \infty$ and $\textbf{C}_{best} \leftarrow \emptyset$.
\FOR{$t = 1, 2 \ldots T$}
\STATE $\textbf{P} \leftarrow \mbox{FastMap}(G, \kappa, \epsilon)$.\label{alg2:mc_start}
\STATE $\textbf{C} \leftarrow \mbox{GMM}(\textbf{P}, k)$.\label{alg2:gmm}
\STATE $obj \leftarrow \mbox{GetObjectiveValue}(G, \textbf{C})$.\label{alg2:objective}
\IF {$obj \leq obj_{min}$}
\STATE $obj_{min} \leftarrow obj$.
\STATE $\textbf{C}_{best} \leftarrow \textbf{C}$.
\ENDIF\label{alg2:mc_end}
\ENDFOR
\STATE \textbf{return} $c_i$ for each $v_i \in V$ according to $\textbf{C}_{best}$.\label{alg2:return}
\end{algorithmic}
\end{algorithm}

Algorithm~\ref{alg:PASPD} shows the pseudocode for computing the PASPD function $D_P(\cdot, \cdot)$ parameterized by $L$ and $F$. Like the shortest-path distance function, it, too, can be computed efficiently (in one shot) for all pairs $(v_s, v_i)$, for a specified source $v_s$ and all $v_i \in V$. On Lines~\ref{alg1:Lloop_start}-\ref{alg1:Lloop_end}, the algorithm populates $G_{set}$ with $L$ lineages of $F$ nested edge-induced subgraphs of $G$ and $\bar{G}$. On Lines~\ref{alg1:drop_start}-\ref{alg1:drop_end}, the algorithm constructs the complement graph $\bar{G}$ but probabilistically retains at most $|E|$ of its edges. On Lines~\ref{alg1:sum_start}-\ref{alg1:sum_end}, it uses the single-source shortest-path distance function to compute and return the sum of the shortest-path distances from $v_s$ to $v_i$ in all $\mathbb{G} \in G_{set}$, for all $v_i \in V$. If $v_s$ and $v_i$ are disconnected in any graph $\mathbb{G} \in G_{set}$, $d_{\mathbb{G}}(v_s, v_i)$ is technically equal to $\infty$. However, for practical reasons in such cases, $d_{\mathbb{G}}(v_s, v_i)$ is set to twice the maximum shortest-path distance from $v_s$ to any other vertex connected to it in $\mathbb{G}$. $T_i$ on Line~\ref{alg1:ss-spd} refers to the array of shortest-path distances from $v_s$ in $G_i$. $T_i(v_j)$ on Line~\ref{alg1:ss-spd_set} is the array element that corresponds to vertex $v_j$.

Algorithm~\ref{alg:FMBM} shows the pseudocode for FMBM. On Line~\ref{alg2:mc_start}, it essentially implements FastMap as described in Algorithm~\ref{alg:fastmap} (from Chapter~\ref{ch:fastmap}) but calls $\mbox{SS-PASPD}(\cdot, \cdot)$ in Algorithm~\ref{alg:PASPD} instead of the regular single-source shortest-path distance function. $L$ and $F$ are simply passed to Algorithm~\ref{alg:PASPD} in the function call $\mbox{SS-PASPD}(\cdot, \cdot)$. Because Algorithm~\ref{alg:FMBM} employs randomization, it qualifies as a Monte Carlo algorithm. It implements an outer loop to boost the performance of FMBM using $T$ independent trials. On Lines~\ref{alg2:mc_start}-\ref{alg2:mc_end}, each trial invokes the GMM clustering algorithm and evaluates the results on the objective function in Equation~\ref{eqn:nullmodel-bm},\footnote{$\textbf{M}$ can be computed from $\textbf{A}$ and $\textbf{C}$ in $O(|E| + k^2)$ time while evaluating the objective function in Equation~\ref{eqn:nullmodel-bm}.} keeping record of the best value. The results of the best trial, that is, the block assignment $c_i$ for each $v_i \in V$, are returned on Line~\ref{alg2:return}.\footnote{The domain of each $c_i$ is $\{1, 2 \ldots k\}$. Block $\mathcal{B}_h$ refers to the collection of all vertices $v_i \in V$ such that $c_i = h$.}

A formal time complexity analysis of FMBM is evasive since Line~\ref{alg2:gmm} of Algorithm~\ref{alg:FMBM} calls the GMM clustering procedure, which has no defined time complexity. Therefore, we only claim to be able to reformulate the block modeling problem on graphs to its Euclidean version in $O(LF\kappa(|E| + |V|\log |V|))$ time in each of the $T$ iterations. Here, the factor $LF$ comes from the cardinality of $G_{set}$ in Algorithm~\ref{alg:PASPD}, and the factor $\kappa(|E| + |V|\log |V|)$ comes from the complexity of FastMap that uses $\mbox{SS-PASPD}(\cdot, \cdot)$ on Line~\ref{alg2:mc_start} of Algorithm~\ref{alg:FMBM}. The time complexity of $\mbox{GetObjectiveValue}(\cdot, \cdot)$ on Line~\ref{alg2:objective} is technically $O(|V|^2k + |V|k^2)$, where $k$ is the user-specified number of blocks, also passed to the GMM clustering algorithm. This time complexity comes from the matrix multiplication $\textbf{C}\textbf{M}\textbf{C}^\top$ in Equation~\ref{eqn:nullmodel-bm}. The factor $|V|^2$ in this matrix multiplication, and more generally in Equation~\ref{eqn:nullmodel-bm}, can be reduced to $O(|E|)$ by evaluating $|E|$ entries corresponding to edges and $\min(|E|,\binom{|V|}{2} - |E|)$ randomly chosen entries corresponding to non-edges in the matrix expression $(\textbf{A} - \textbf{C}\textbf{M}\textbf{C}^\top) \circ (\textbf{A} - \textbf{R})$. The matrix multiplication $\textbf{C}\textbf{M}$ takes $O(|V|k^2)$ time and results in a $|V| \times k$ matrix. $|E| + \min(|E|,\binom{|V|}{2} - |E|)$ entries in the multiplication of this matrix with $\textbf{C}^\top$ can be computed in $O(|E|k)$ time. Overall, therefore, the reformulation to Euclidean space can be done in near-linear time, that is, linear in $|V|$ and $|E|$, after ignoring logarithmic factors.

\section{Experimental Results}

In this section, we present empirical results on the comparative performances of FMBM and three other state-of-the-art solvers for block modeling: Graph-Tool, DANMF, and CPLNS. We also compared against two other solvers for block modeling: FactorBlock~\cite{clklbr13} and ASBlock~\cite{rscrbld20}. However, they are not competitive with the other solvers; and we exclude them from Tables~\ref{tab:benchmark},~\ref{tab:benchmark_complement},~\ref{tab:sparse},~\ref{tab:noise_dense}, and~\ref{tab:random} to save column space. Graph-Tool~\cite{p14} uses an agglomerative multi-level Markov Chain Monte Carlo algorithm and has been largely ignored in the computer science literature on block modeling; DANMF~\cite{ycz18} uses deep autoencoders; and CPLNS~\cite{mdns21} uses constraint programming with large neighborhood search.

\begin{table}[t!]
\centering
\scalebox{0.65}{
\begin{tabular}{|l|r|rrr|rrr|rrr|rrr|}
\cline{1-14}
\multirow{2}{*}{Test Case} &\multicolumn{1}{c}{\multirow{2}{*}{Size ($|V|$, $|E|$)}} &\multicolumn{3}{|c|}{FMBM} &\multicolumn{3}{|c|}{Graph-Tool} &\multicolumn{3}{|c|}{DANMF} &\multicolumn{3}{|c|}{CPLNS}\\
\cline{3-14}
&&Objective &NMI &Time &Objective &NMI &Time &Objective &NMI &Time &Objective &NMI &Time\\
\cline{1-14}
adjnoun &(112, 425) &616.86 &0.0025 &6.11 &612.98 &\bf0.2978 &\bf0.04 &636.75 &0.0083 &1.62 &\bf591.76 &0.0154 &1.51\\
baboons &(14, 23) &11.97 &0.0158 &0.54 &\bf11.49 &\bf0.2244 &\bf0.00 &15.49 &0.1341 &0.97 &12.81 &0.0172 &0.87\\
football &(115, 613) &665.97 &0.5608 &9.22 &\bf343.32 &\bf0.9150 &\bf0.03 &863.91 &0.2574 &1.55 &558.94 &0.6991 &83.33\\
karate &(34, 78) &74.66 &\bf0.6127 &1.47 &\bf64.67 &0.2512 &\bf0.00 &81.94 &0.1672 &0.77 &75.43 &0.2228 &1.06\\
polblogs &(1490, 16715) &98788.53 &0.0098 &239.33 &99014.21 &\bf0.4668 &\bf2.14 &101195.89 &0.0465 &404.02 &\bf95859.73 &0.0543 &506.29\\
polbooks &(105, 441) &522.33 &0.5329 &6.20 &\bf496.02 &\bf0.5462 &\bf0.02 &590.20 &0.3177 &1.98 &531.48 &0.2073 &2.09\\
\cline{1-14}
\end{tabular}
}
\caption[Results of competing algorithms for block modeling on real-world single-view undirected graphs.]{Shows the comparative results on real-world single-view undirected graphs.}
\label{tab:benchmark}
\end{table}

\begin{table}[t!]
\centering
\scalebox{0.65}{
\begin{tabular}{|l|r|rrr|rrr|rrr|rrr|}
\cline{1-14}
\multirow{2}{*}{Test Case} &\multicolumn{1}{c}{\multirow{2}{*}{Size ($|V|$, $|E|$)}} &\multicolumn{3}{|c|}{FMBM} &\multicolumn{3}{|c|}{Graph-Tool} &\multicolumn{3}{|c|}{DANMF} &\multicolumn{3}{|c|}{CPLNS}\\
\cline{3-14}
&&Objective &NMI &Time &Objective &NMI &Time &Objective &NMI &Time &Objective &NMI &Time\\
\cline{1-14}
adjnoun &(112, 5791) &611.04 &0.0048 &25.34 &636.34 &0.0168 &\bf0.41 &641.40 &0.0000 &8.34 &\bf591.54 &\bf0.0169 &1.85\\
baboons &(14, 68) &\bf12.64 &\bf0.0547 &0.62 &12.86 &0.0416 &\bf0.01 &15.46 &0.0500 &1.23 &13.35 &0.0316 &0.86\\
football &(115, 5944) &595.52 &0.5899 &27.53 &\bf344.38 &\bf0.9111 &\bf0.17 &815.71 &0.2229 &9.56 &525.54 &0.7040 &82.11\\
karate &(34, 483) &\bf72.73 &\bf0.7625 &2.46 &77.31 &0.2065 &\bf0.02 &84.23 &0.0914 &1.78 &75.00 &0.2439 &1.04\\
polblogs &(1490, 1094951) &26896.52 &0.0153 &3155.33 &26048.04 &0.0454 &\bf49.90 &- &- &$>1$ hour &\bf25871.42 &\bf0.0541 &470.48\\
polbooks &(105, 5019) &\bf509.88 &\bf0.5409 &21.55 &606.96 &0.0867 &\bf0.13 &631.77 &0.0141 &8.09 &531.65 &0.2056 &2.15\\
\cline{1-14}
\end{tabular}
}
\caption[Results of competing algorithms for block modeling on the complement graphs of the graphs in Table~\ref{tab:benchmark}.]{Shows the comparative results on the complement graphs of the graphs in Table~\ref{tab:benchmark}.}
\label{tab:benchmark_complement}
\end{table}

\begin{table}[t!]
\centering
\scalebox{0.65}{
\begin{tabular}{|l|r|rrr|rrr|rrr|rrr|}
\cline{1-14}
\multirow{2}{*}{Test Case} &\multicolumn{1}{c}{\multirow{2}{*}{Size ($|E|$)}} &\multicolumn{3}{|c|}{FMBM} &\multicolumn{3}{|c|}{Graph-Tool} &\multicolumn{3}{|c|}{DANMF} &\multicolumn{3}{|c|}{CPLNS}\\
\cline{3-14}
&&Objective &NMI &Time &Objective &NMI &Time &Objective &NMI &Time &Objective &NMI &Time\\
\cline{1-14}
V0400b04 &6722 &9355.91 &0.1622 &83.22 &\bf8385.61 &\bf0.6565 &\bf1.49 &9465.48 &0.0111 &10.88 &9400.46 &0.0534 &39.74\\
V0800b04 &14723 &24201.35 &0.1775 &199.14 &\bf22848.79 &\bf0.6599 &\bf3.99 &24387.24 &0.0019 &62.50 &24303.43 &0.0394 &195.08\\
V1600b04 &25103 &45849.52 &0.2357 &391.92 &\bf44598.02 &\bf0.9667 &\bf7.04 &46379.74 &0.0043 &753.62 &46292.59 &0.0206 &1018.54\\
V3200b04 &70973 &134217.82 &0.0348 &1376.41 &\bf131751.34 &\bf0.6654 &\bf108.08 &- &- &$>1$ hour &- &- &$>1$ hour\\
V0400b10 &3246 &5461.99 &0.1217 &40.9 &\bf4728.69 &\bf0.8542 &\bf0.63 &5489.8 &0.0509 &7.25 &5364.07 &0.1289 &101.01\\
V0800b10 &7499 &13623.69 &0.0425 &108.69 &\bf12612.44 &\bf0.9596 &\bf2.01 &13636.29 &0.0156 &47.49 &13485.93 &0.0734 &423.03\\
V1600b10 &15118 &28782.35 &0.0691 &272.65 &\bf27828.70 &\bf0.8556 &\bf4.82 &28829.58 &0.0117 &537.27 &28682.31 &0.0384 &2019.76\\
V3200b10 &36170 &70292.36 &0.0369 &782.44 &\bf68653.51 &\bf0.9173 &\bf17.62 &70315.35 &0.0074 &3272.65 &- &- &$>1$ hour\\
V0400b20 &2297 &4048.03 &0.1639 &29.66 &\bf3632.92 &\bf0.6256 &\bf0.54 &4064.77 &0.1859 &8.03 &- &- &$>1$ hour\\
V0800b20 &5049 &9451.30 &0.0848 &72.92 &\bf8960.16 &\bf0.5857 &\bf1.90 &9460.80 &0.0828 &62.32 &- &- &$>1$ hour\\
V1600b20 &11575 &22305.01 &0.0457 &251.06 &\bf21591.83 &\bf0.6718 &\bf3.91 &22315.63 &0.0445 &444.49 &- &- &$>1$ hour\\
V3200b20 &24639 &48321.14 &0.0212 &579.90 &\bf47650.73 &\bf0.6067 &\bf12.89 &- &- &$>1$ hour &- &- &$>1$ hour\\
\cline{1-14}
\end{tabular}
}
\caption[Results of competing algorithms for block modeling on sparse single-view undirected graphs using Generative Model 1.]{Shows the comparative results on sparse single-view undirected graphs using Generative Model 1.}
\label{tab:sparse}
\end{table}

\begin{table}[t!]
\centering
\scalebox{0.65}{
\begin{tabular}{|l|r|rrr|rrr|rrr|rrr|}
\cline{1-14}
\multirow{2}{*}{Test Case} &\multicolumn{1}{c}{\multirow{2}{*}{Size ($|E|$)}} &\multicolumn{3}{|c|}{FMBM} &\multicolumn{3}{|c|}{Graph-Tool} &\multicolumn{3}{|c|}{DANMF} &\multicolumn{3}{|c|}{CPLNS}\\
\cline{3-14}
&&Objective &NMI &Time &Objective &NMI &Time &Objective &NMI &Time &Objective &NMI &Time\\
\cline{1-14}
V0100b06 &4125 &815.58 &0.1308 &20.06 &818.83 &0.1291 &\bf0.46 &829.20 &0.0829 &5.85 &\bf750.28 &\bf0.2727 &3.50\\
V0300b06 &41771 &\bf4702.60 &\bf0.2061 &176.65 &4819.43 &0.0311 &\bf1.93 &4828.49 &0.0149 &143.10 &4746.32 &0.0629 &18.14\\
V0500b06 &118536 &10473.00 &\bf0.0537 &513.31 &10489.79 &0.0193 &\bf4.86 &10497.63 &0.0053 &658.45 &\bf10419.98 &0.0411 &41.13\\
V0100b08 &4212 &782.91 &0.2182 &19.23 &761.68 &0.3024 &\bf0.33 &807.05 &0.1532 &6.02 &\bf712.92 &\bf0.3331 &4.59\\
V0300b08 &42078 &4468.67 &\bf0.0944 &175.55 &4487.12 &0.0443 &\bf1.48 &4498.23 &0.0151 &144.96 &\bf4397.28 &0.0895 &19.95\\
V0500b08 &120166 &8188.55 &\bf0.1503 &505.41 &8278.07 &0.0267 &\bf4.94 &8290.82 &0.0056 &680.16 &\bf8175.52 &0.0582 &59.04\\
V0100b10 &4268 &731.61 &0.3446 &19.22 &720.38 &0.3290 &\bf0.32 &779.34 &0.1570 &6.41 &\bf662.42 &\bf0.4100 &6.93\\
V0300b10 &42385 &4069.42 &\bf0.1652 &171.93 &4126.01 &0.0569 &\bf1.47 &4141.94 &0.0292 &146.23 &\bf4009.88 &0.1160 &29.28\\
V0500b10 &120366 &7931.97 &\bf0.1174 &516.16 &7985.47 &0.0347 &\bf4.03 &8000.60 &0.0000 &616.74 &\bf7872.77 &0.0761 &60.22\\
\cline{1-14}
\end{tabular}
}
\caption[Results of competing algorithms for block modeling on dense single-view undirected graphs using Generative Model 1.]{Shows the comparative results on dense single-view undirected graphs using Generative Model 1.}
\label{tab:noise_dense}
\end{table}

\begin{table}[t!]
\centering
\scalebox{0.65}{
\begin{tabular}{|l|r|rrr|rrr|rrr|rrr|}
\cline{1-14}
\multirow{2}{*}{Test Case} &\multicolumn{1}{c}{\multirow{2}{*}{Size ($|E|$)}} &\multicolumn{3}{|c|}{FMBM} &\multicolumn{3}{|c|}{Graph-Tool} &\multicolumn{3}{|c|}{DANMF} &\multicolumn{3}{|c|}{CPLNS}\\
\cline{3-14}
&&Objective &NMI &Time &Objective &NMI &Time &Objective &NMI &Time &Objective &NMI &Time\\
\cline{1-14}
V0400b04 &7176 &9843.84 &0.0327 &75.09 &\bf9444.48 &\bf0.8214 &\bf1.77 &9855.97 &0.0116 &12.23 &9786.67 &0.0246 &34.55\\
V0800b04 &16780 &27040.21 &0.0087 &192.97 &26982.65 &\bf0.0698 &\bf5.67 &27053.24 &0.0008 &69.68 &\bf26945.04 &0.0087 &243.28\\
V1600b04 &28183 &51531.41 &\bf0.0146 &386.32 &51556.38 &0.0043 &\bf8.04 &51559.96 &0.0025 &684.58 &\bf51467.00 &0.0073 &1098.12\\
V3200b04 &72182 &136376.53 &0.0087 &1196.30 &\bf135967.24 &\bf0.5287 &\bf32.92 &- &- &$>1$ hour &- &- &$>1$ hour\\
V0400b10 &7069 &9729.49 &0.0639 &73.76 &\bf9349.76 &\bf0.4827 &\bf1.52 &9753.04 &0.0625 &10.69 &9585.60 &0.0837 &127.68\\
V0800b10 &15613 &25528.33 &0.0385 &181.56 &\bf25022.25 &\bf0.4893 &\bf4.05 &25553.52 &0.0274 &86.58 &25353.08 &0.0418 &494.10\\
V1600b10 &32294 &58279.85 &0.0195 &428.40 &58244.14 &\bf0.0305 &\bf13.92 &58305.57 &0.0157 &687.84 &\bf58103.92 &0.0224 &2459.38\\
V3200b10 &79173 &\bf148741.84 &\bf0.0082 &1307.21 &148745.65 &0.0065 &\bf32.88 &- &- &$>1$ hour &- &- &$>1$ hour\\
V0400b20 &6829 &9453.81 &0.1425 &72.78 &\bf9139.14 &\bf0.2039 &\bf1.38 &9506.35 &0.1790 &9.98 &- &- &$>1$ hour\\
V0800b20 &15106 &24810.19 &0.0784 &176.60 &\bf24521.35 &\bf0.1251 &\bf3.49 &24866.54 &0.0918 &62.93 &- &- &$>1$ hour\\
V1600b20 &30462 &55265.42 &0.0436 &420.49 &\bf55207.39 &0.0366 &\bf9.38 &55310.70 &\bf0.0440 &606.39 &- &- &$>1$ hour\\
V3200b20 &67675 &128267.45 &\bf0.0232 &1199.32 &\bf128234.10 &0.0168 &\bf40.91 &- &- &$>1$ hour &- &- &$>1$ hour\\
\cline{1-14}
\end{tabular}
}
\caption[Results of competing algorithms for block modeling on single-view undirected graphs using Generative Model 2.]{Shows the comparative results on single-view undirected graphs using Generative Model 2.}
\label{tab:random}
\end{table}

We used the following hyperparameter values for FMBM: $L = 4$, $F = 10$, $T = 10$, $\kappa = 4$, and $\epsilon = 10^{-4}$. These values are only important as ballpark estimates. We observed that the performance of FMBM often stays stable within broad ranges of hyperparameter values, imparting robustness to FMBM. Moreover, only a few different hyperparameter settings had to be examined to determine the best one. The value of $k$, that is, the number of blocks, was given as input for all the solvers in the experiments.\footnote{Although Graph-Tool does not require a user-specified value of $k$, it has a tendency to produce trivial solutions with $k = 1$, resulting in $0$ NMI values when the value of $k$ is not explicitly specified.} We used three metrics for comparison: the value of the objective function stated in Equation~\ref{eqn:nullmodel-bm}, the Normalized Mutual Information (NMI) value with respect to the ground truth, and the running time in seconds. Unlike other methods, FMBM is an anytime algorithm since it uses multiple trials. Each trial takes roughly $(1/T)$'th, that is, one-tenth, of the time reported for FMBM in the experimental results. For each method and test case, we averaged the results over $10$ runs. All experiments were conducted on a laptop with a 3.1 GHz Quad-Core Intel Core i7 processor and 16 GB LPDDR3 memory. Our implementation of FMBM was done in Python3 with NetworkX~\cite{hss08}.

Although the underlying theory of FMBM can be generalized to directed edge-weighted graphs~\cite{gckk20} and to multi-view graphs, the current version of FMBM is operational only on singe-view undirected unweighted graphs, sufficient to illustrate the power of FastMap embeddings. Therefore, only such test cases are borrowed from other commonly used datasets~\cite{n06,rscrbld20}. However, we also created new synthetic test cases to be able to do a more comprehensive analysis.\footnote{\url{https://github.com/leon-angli/Synthetic-Block-Modeling-Dataset}}

The synthetic test cases were generated according to two similar stochastic block models~\cite{a17} as follows. In Generative Model 1, given a user-specified number of vertices $|V|$ and a user-specified number of blocks $k$, we first assign each vertex to a block chosen uniformly at random to obtain the membership matrix $\textbf{C}$, representing the ground truth. The image matrix $\textbf{M}$ is drafted using certain ``block structural characteristics'' designed for that instance with a parameter $p$. Each entry $\textbf{M}_{ij}$ is set to either $p$ or $10p$ according to a rule explained below. If $\textbf{M}_{ij}$ is set to $p$ ($10p$), the two blocks $\mathcal{B}_i$ and $\mathcal{B}_j$ are weakly (strongly) connected to each other with respect to $p$. The adjacency matrix $\textbf{A}$, representing the entire graph, is constructed from $\textbf{C}$ and $\textbf{M}$ by connecting any two vertices $v_s \in \mathcal{B}_i$ and $v_t \in \mathcal{B}_j$ with probability $\textbf{M}_{ij}$. In Generative Model 2, each entry $\textbf{M}_{ij}$ is set to $cp$, where $c$ is an integer chosen uniformly at random from the interval $[1, 10]$.

Tables~\ref{tab:benchmark},~\ref{tab:benchmark_complement},~\ref{tab:sparse},~\ref{tab:noise_dense}, and~\ref{tab:random} show the comparative performances of FMBM, Graph-Tool, DANMF, and CPLNS.\footnote{DANMF did not assign any block membership to a few vertices in some synthetic test cases. We assign Block $\mathcal{B}_1$ by default to such vertices.} Table~\ref{tab:benchmark} contains commonly used real-world test cases from~\cite{n06} and~\cite{rscrbld20}. Here, FMBM outperforms DANMF and CPLNS with respect to the value of the objective function on 3 out of 6 instances, despite the fact that it uses the expression in Equation~\ref{eqn:nullmodel-bm} only for evaluation on Line~\ref{alg2:objective} of Algorithm~\ref{alg:FMBM}. Graph-Tool performs well on all the instances. Table~\ref{tab:benchmark_complement} shows the comparative performances on the complement graphs of the graphs in Table~\ref{tab:benchmark}. This is done to test the robustness of the solvers against encoding the same relationships between vertices as either edges or non-edges. While the value of the objective function and the running time are expected to change, the NMI value is expected to be stable. We observe that FMBM and CPLNS are the only solvers that convincingly pass this test. Moreover, FMBM outperforms the other solvers on more instances than in Table~\ref{tab:benchmark}. Tables~\ref{tab:benchmark} and~\ref{tab:benchmark_complement} do not test scalability since $|V|$ is small in these test cases.

Table~\ref{tab:sparse} contains synthetic sparse test cases from Generative Model 1, named ``V$n$b$k$'', where $n$ indicates the number of vertices and $k$ indicates the number of blocks. These test cases have the following block structural characteristics. Each block is strongly connected to two other randomly chosen blocks and weakly connected to the remaining ones (including itself). We set $p = (\ln |V|) / |V|$, making $|E| = O(|V|\log |V|)$ in expectation. After generating $\textbf{A}$, we also add some noise to it by flipping each of its entries independently with probability $0.05 / |V|$. FMBM outperforms DANMF and CPLNS with respect to both the value of the objective function and the NMI value on 8 out of 12 instances. We also begin to see FMBM's advantages in scalability. However, Graph-Tool outperforms all other methods by a significant margin on all the instances. Table~\ref{tab:noise_dense} contains synthetic dense test cases from Generative Model 1 constructed by setting $p = (\ln |V|) / |V|$, modifying each entry $\textbf{M}_{ij}$ to $1 - \textbf{M}_{ij}$, and adding noise, as before. We observe that the performance of Graph-Tool is poor on such dense graphs. FMBM outperforms DANMF and CPLNS with respect to the NMI value on 6 out of 9 instances. Although CPLNS produces marginally better values of the objective function, its performance on large sparse graphs in Table~\ref{tab:sparse} is bad.

Table~\ref{tab:random} contains synthetic test cases from Generative Model 2 constructed by setting $p = (\ln |V|) / |V|$. FMBM outperforms DANMF and CPLNS with respect to the value of the objective function on 6 out of 12 instances. It also outperforms DANMF and CPLNS with respect to the NMI value on a different set of 6 instances. Graph-Tool performs comparatively well on all the instances but occasionally produces low NMI values.

\subsection{Visualization}

\begin{figure}[t!]
\centering
\subfloat[standard graph visualization of blocks in an instance with $1,600$ vertices and $4,353$ edges]{
\begin{minipage}[b]{0.35\textwidth}
\centering
\includegraphics[width=0.8\textwidth]{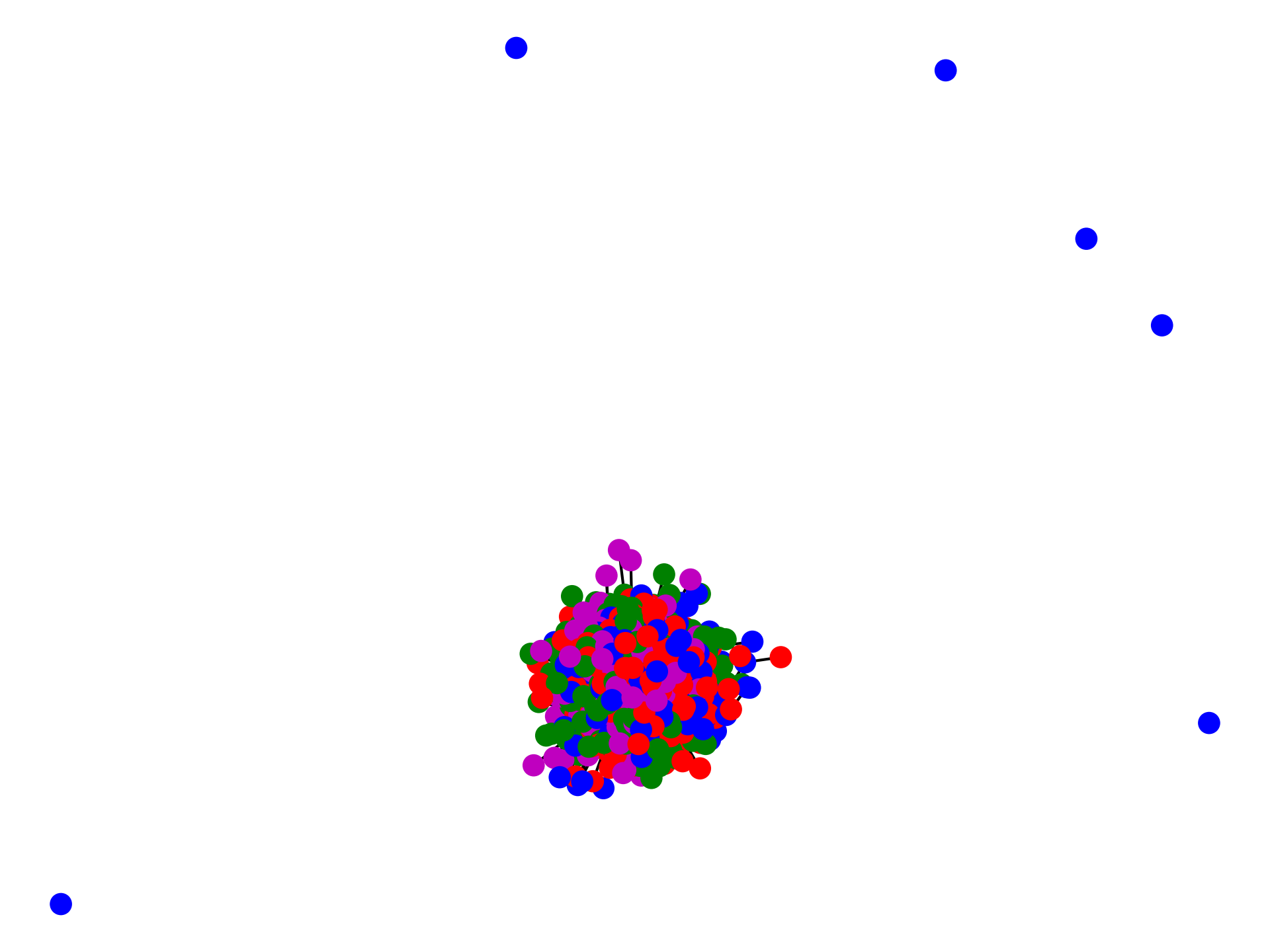}
\label{fig:gt_visual}
\end{minipage}
}
\hspace{0.05\textwidth}
\subfloat[FMBM visualization of blocks in Euclidean space for the instance from~\ref{fig:gt_visual}]{
\begin{minipage}[b]{0.35\textwidth}
\centering
\includegraphics[width=0.8\textwidth]{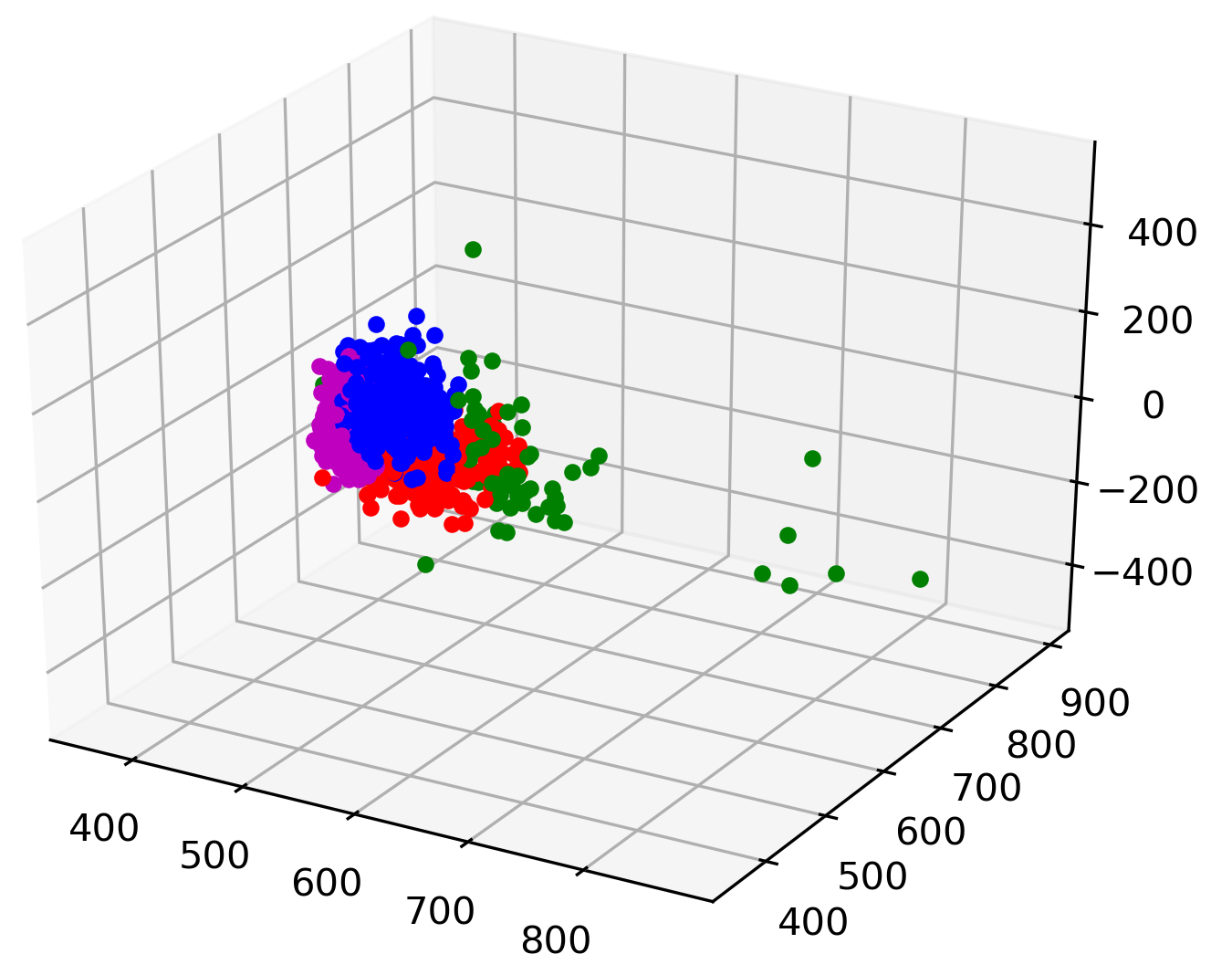}
\label{fig:fm_visual}
\end{minipage}
}

\subfloat[standard graph visualization of blocks in an instance with $1,600$ vertices and $25,103$ edges]{
\begin{minipage}[b]{0.35\textwidth}
\centering
\includegraphics[width=0.8\textwidth]{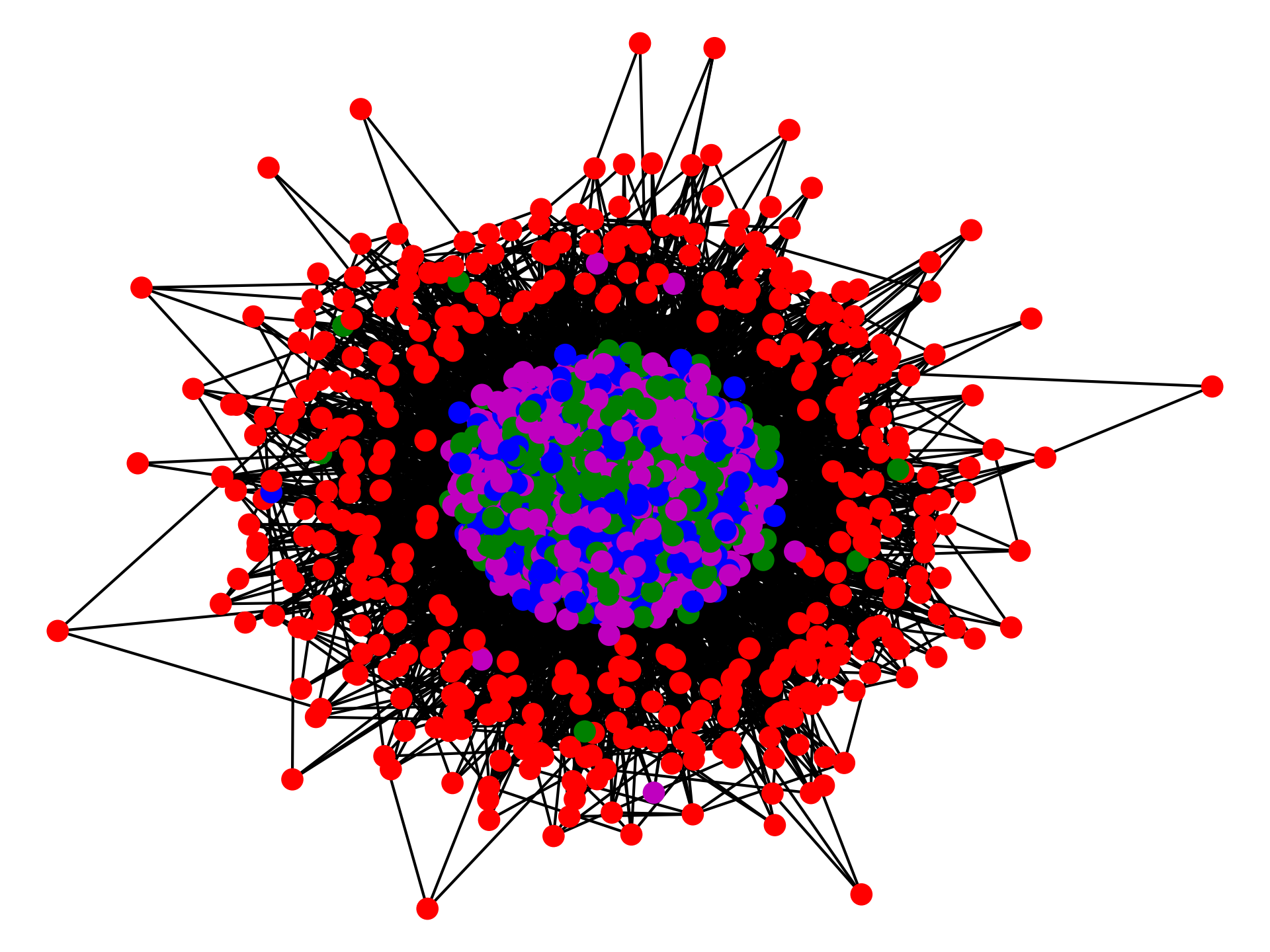}
\label{fig:gt_visual_sparse}
\end{minipage}
}
\hspace{0.05\textwidth}
\subfloat[FMBM visualization of blocks in Euclidean space for the instance from~\ref{fig:gt_visual_sparse}]{
\begin{minipage}[b]{0.35\textwidth}
\centering
\includegraphics[width=0.8\textwidth]{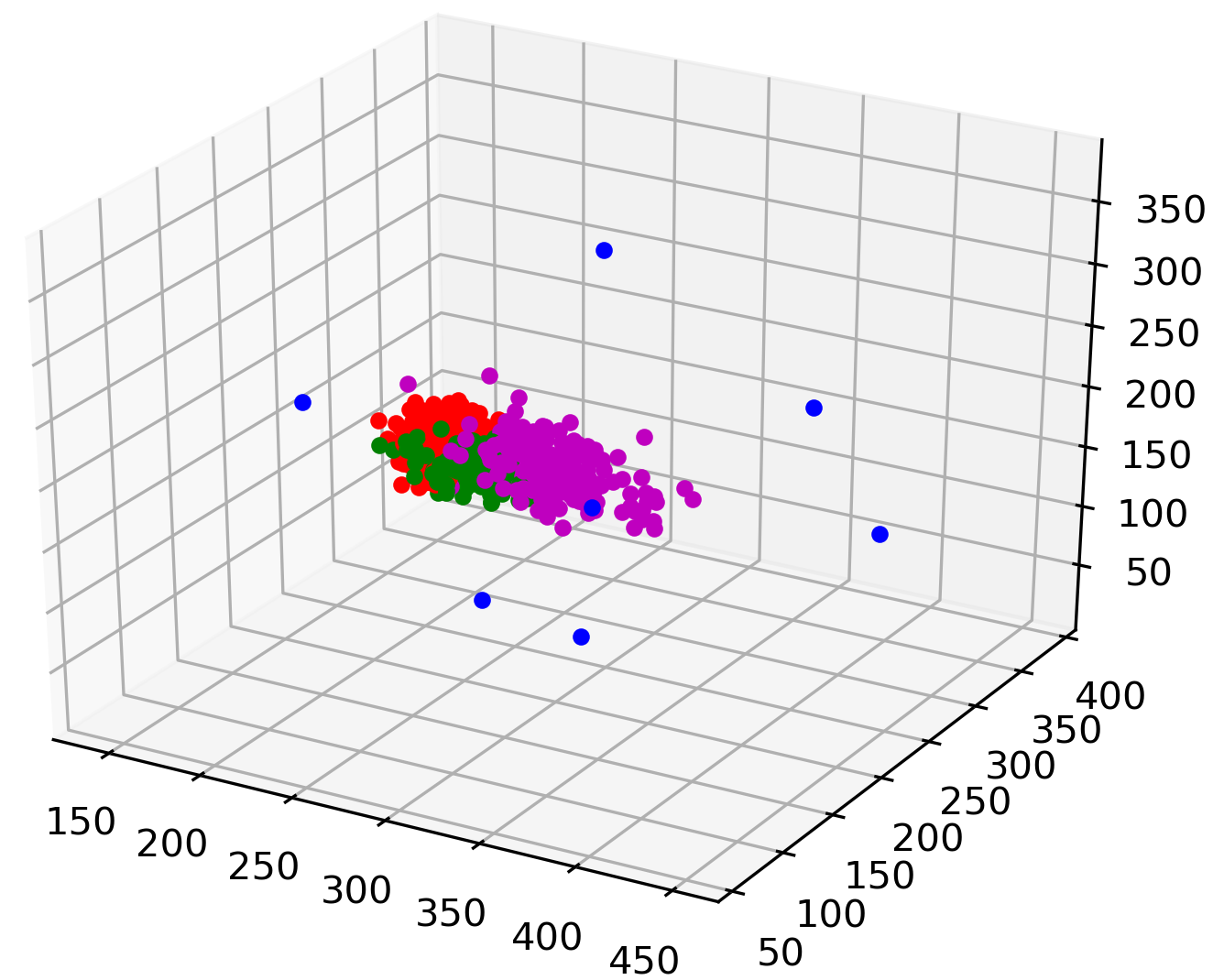}
\label{fig:fm_visual_sparse}
\end{minipage}
}
\caption[A comparison of the visualization produced by FMBM against that of a competing method.]{Compares the visualization produced by FMBM against that of a competing method. (a) and (c) show visualizations of two different instances with four blocks obtained by a standard graph visualization procedure in NetworkX used with Graph-Tool. (b) and (d) show visualizations of the same two instances obtained in the Euclidean embedding by FMBM. Four different colors are used to indicate the four different blocks. The FMBM visualization is more helpful for gauging the spread of blocks, both individually and relative to each other.}
\label{fig:visualization}
\end{figure}

In addition to identifying blocks, their visualization is important for uncovering trends, patterns, and outliers in large graphs. A good visualization aids human intuition for gauging the spread\footnote{The spread here refers to how a block extends from its center to its periphery.} of blocks, both individually and relative to each other. In market analysis, for example, a representative element can be chosen from each block with proper visualization. Figure~\ref{fig:visualization} shows that FMBM provides a much more perspicuous visualization compared to a standard graph visualization procedure in NetworkX\footnote{\url{https://networkx.org/documentation/stable/reference/generated/networkx.drawing.nx_pylab.draw.html}} used with Graph-Tool, even though Graph-Tool shows good overall performance in Tables~\ref{tab:benchmark},~\ref{tab:sparse}, and~\ref{tab:random}. This is so because FMBM solves the block modeling problem in Euclidean space, while other approaches use abstract methods that are harder to visualize.

\section{Conclusions}

In this chapter, we proposed FMBM, a FastMap-based algorithm for block modeling. In the first phase, FMBM adapts FastMap to embed a given undirected unweighted graph into a Euclidean space in near-linear time such that the pairwise Euclidean distances between vertices approximate a probabilistically-amplified graph-based distance function between them. In doing so, it avoids having to directly work on the given graphs and instead reformulates the graph block modeling problem to a Euclidean version. In the second phase, FMBM uses GMM clustering for identifying clusters (blocks) in the resulting Euclidean space. Empirically, FMBM outperforms other state-of-the-art methods like FactorBlock, Graph-Tool, DANMF, and CPLNS on many benchmark and synthetic test cases. FMBM also enables a perspicuous visualization of the blocks in the graphs, not provided by other methods.

\begin{subappendices}

\section{Table of Notations}

\begin{table}[h]
\centering
\begin{tabular}{|l|p{0.75\linewidth}|}
\cline{1-2}
Notation &Description\\
\cline{1-2}
$G = (V, E)$ &An undirected unweighted graph with vertices $V = \{v_1, v_2 \ldots v_n\}$ and edges $E = \{e_1, e_2 \ldots e_m\} \subseteq V \times V$.\\
\cline{1-2}
$\textbf{A}$ &The adjacency matrix representation of $G$, where $\textbf{A}_{ij} = 1$ iff $(v_i, v_j) \in E$.\\
\cline{1-2}
$k$ &The user-specified number of blocks.\\
\cline{1-2}
$\textbf{C}$ &The membership matrix for block modeling on $G$, where $\textbf{C}_{ij} = 0$ and $\textbf{C}_{ij} = 1$ represent vertex $v_i$ being absent from and being present in partition $j$, respectively.\\
\cline{1-2}
$\textbf{M}$ &The image matrix, where $\textbf{M}_{ij}$ represents the likelihood of an edge between a vertex in partition $i$ and a vertex in partition $j$.\\
\cline{1-2}
$\| \cdot \|_F$ &The Frobenius norm.\\
\cline{1-2}
$\circ$ &The element-wise multiplication for two matrices.\\
\cline{1-2}
$\textbf{R}$ &A matrix $\in [0, 1]^{n \times n}$, where $\textbf{R}_{ij} = \frac{m}{n^2}$.\\
\cline{1-2}
$\kappa$ &The user-specified number of dimensions of the FastMap embedding.\\
\cline{1-2}
$\bar{G}$ &The complement graph of $G$.\\
\cline{1-2}
$D_P(\cdot, \cdot)$ &The PASPD function.\\
\cline{1-2}
$d_G(v_i, v_j)$ &The shortest-path distance between $v_i$ and $v_j$ in $G$.\\
\cline{1-2}
$G_{set}$ &A collection of undirected graphs, each derived from either the given graph $G$ or its complement $\bar{G}$.\\
\cline{1-2}
$L$ &The number of lineages of $G$ and $\bar{G}$ in $G_{set}$.\\
\cline{1-2}
$F$ &The number of nested edge-induced subgraphs in each lineage.\\
\cline{1-2}
$\epsilon$ &The threshold parameter in FastMap that is used to detect large values of $\kappa$ that have diminishing returns on the accuracy of approximating the pairwise distances between the vertices.\\
\cline{1-2}
$T$ &The number of independent trials used in FMBM.\\
\cline{1-2}
$\mathcal{B}_h$ &The block that refers to the collection of all vertices with membership $h$.\\
\cline{1-2}
\end{tabular}
\caption[Notations used in Chapter~\ref{ch:block_modeling}.]{Describes the notations used in Chapter~\ref{ch:block_modeling}.}
\label{tab:block_modeling_notations}
\end{table}

\end{subappendices}

\chapter{FastMap for Graph Convex Hull Computations}
\label{ch:convex_hull}
Given an undirected edge-weighted graph $G$ and a subset of vertices $S$ in it, the graph convex hull $CH^G_S$ of $S$ in $G$ is the set of vertices obtained by the process of initializing $CH^G_S$ to $S$ and iteratively adding until convergence all vertices on all shortest paths between all pairs of vertices in $CH^G_S$ of one iteration to constitute $CH^G_S$ of the next iteration. Computing the graph convex hull has applications in shortest-path computations, active learning, and in identifying geodesic cores in social networks, among others. Unfortunately, computing it exactly is prohibitively expensive on large graphs. In this chapter, we present a FastMap-based algorithm for efficiently computing approximate graph convex hulls. Using FastMap's ability to facilitate geometric interpretations, our approach invokes the power of well-studied algorithms in Computational Geometry that efficiently compute the convex hull of a set of points in Euclidean space. Through experimental studies, we show that our approach not only is several orders of magnitude faster than the exact brute-force algorithm but also outperforms the state-of-the-art approximation algorithm, both in terms of generality and the quality of the solutions produced.

\section{Introduction}

In Computational Geometry, the convex hull of a finite set of points in Euclidean space is defined as the smallest convex polygon in that space that contains all of them. The problem of computing the convex hull of a given finite set of points is a cornerstone problem with numerous applications: In discrete Geometry, several results rely on convex hulls~\cite{s17}. In Mathematics, convex hulls are used to study polynomials~\cite{p04} and matrix eigenvalues~\cite{j76}. In Statistics, they are used to define risk sets~\cite{h71}. They also play a key role in polyhedral combinatorics~\cite{k92}.

While convex hulls have been traditionally studied in geometric spaces, they can also be defined on graphs. In particular, given an undirected edge-weighted graph $G = (V, E, w)$, where $V$ is the set of vertices, $E$ is the set of edges, and for any edge $e \in E$, $w(e)$ is the non-negative weight on it, and a subset of vertices $S \subseteq V$, the~\emph{graph convex hull} of $S$ in $G$ is the smallest set of vertices $CH^G_S$ that contains $S$ and all vertices that appear on any shortest path between any pair of vertices in $CH^G_S$. Procedurally, $CH^G_S$ can be obtained by the process of initializing it to $S$ and iteratively adding until convergence all vertices on all shortest paths between all pairs of vertices in $CH^G_S$ of one iteration to constitute $CH^G_S$ of the next iteration.

\begin{figure}[t!]
\centering
\subfloat[geometric environment]{
\begin{minipage}[b]{0.38\textwidth}
\centering
\includegraphics[width=\columnwidth]{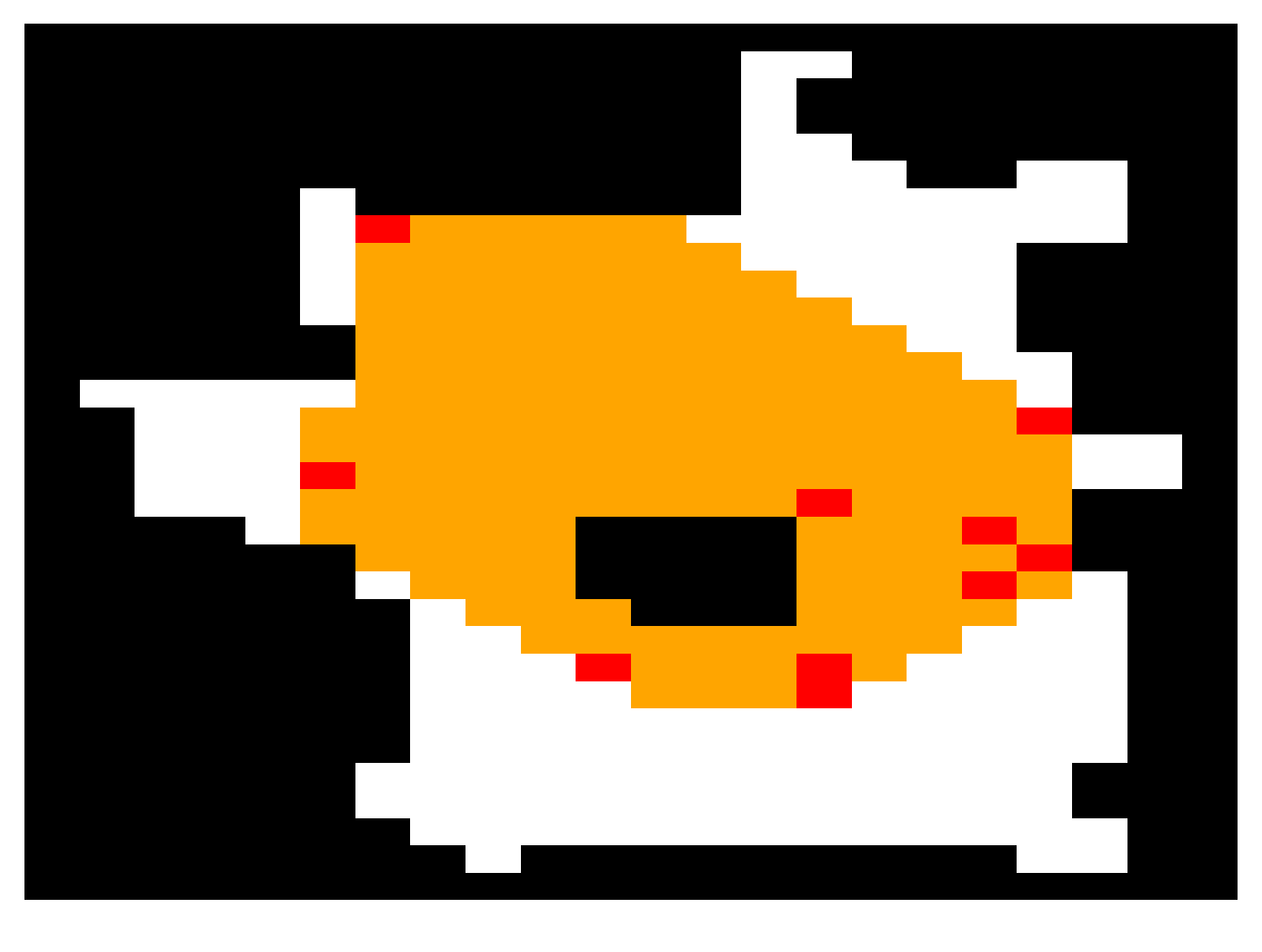}
\label{fig:orz106d}
\end{minipage}
}
\subfloat[graph environment]{
\begin{minipage}[b]{0.38\textwidth}
\centering
\includegraphics[width=\columnwidth]{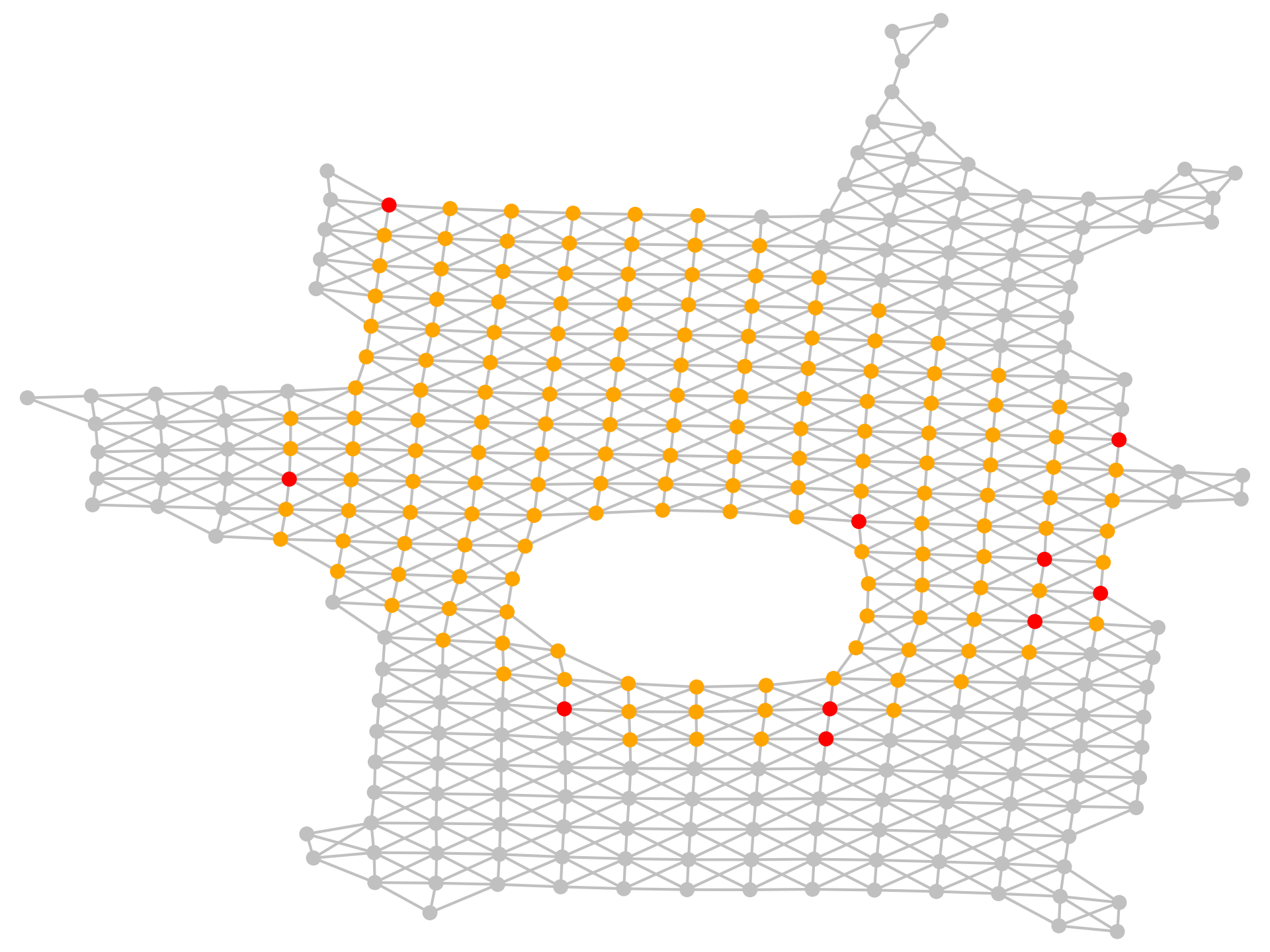}
\label{fig:orz106d_G}
\end{minipage}
}
\caption[An illustration of a graph convex hull.]{Illustrates a graph convex hull. (a) shows a geometric environment with obstacles (black regions) that is discretized as a grid-world. (b) shows a graph representation $G$ of the environment in (a), with vertices representing the top-left corners of the free cells (non-black regions) and edges, weighted by their Euclidean lengths, connecting pairs of vertices on the boundary of the same free cell. In both (a) and (b), the red dots indicate $S$ and the union of the red and orange dots indicates $CH^G_S$.}
\label{fig:graph_convex_hull}
\end{figure}

\begin{figure}[t!]
\centering
\subfloat[geometric convex hulls]{
\begin{minipage}[b]{0.38\textwidth}
\centering
\includegraphics[width=\columnwidth]{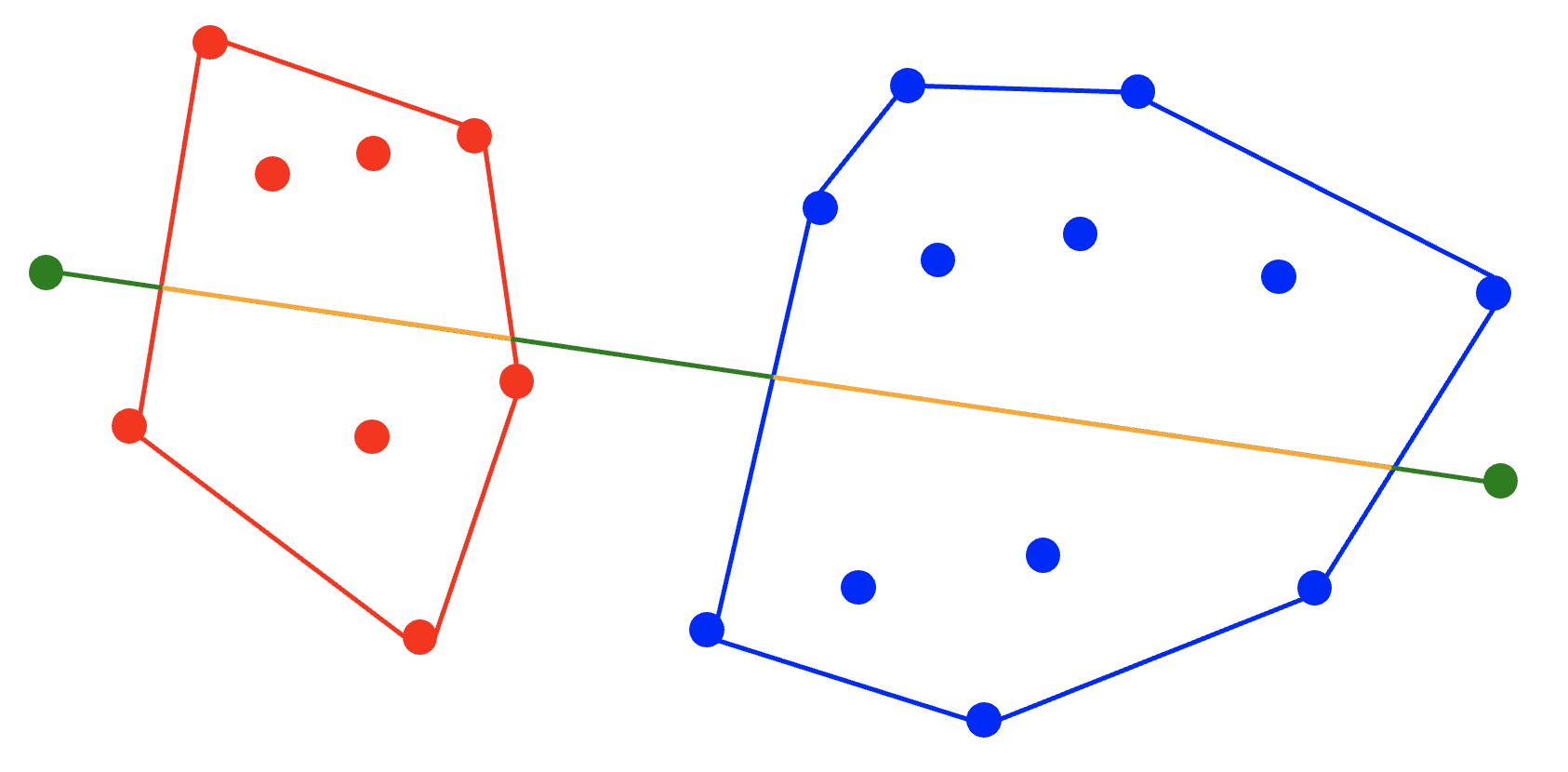}
\label{fig:convex_hulls_path}
\end{minipage}
}
\subfloat[graph convex hulls]{
\begin{minipage}[b]{0.38\textwidth}
\centering
\includegraphics[width=\columnwidth]{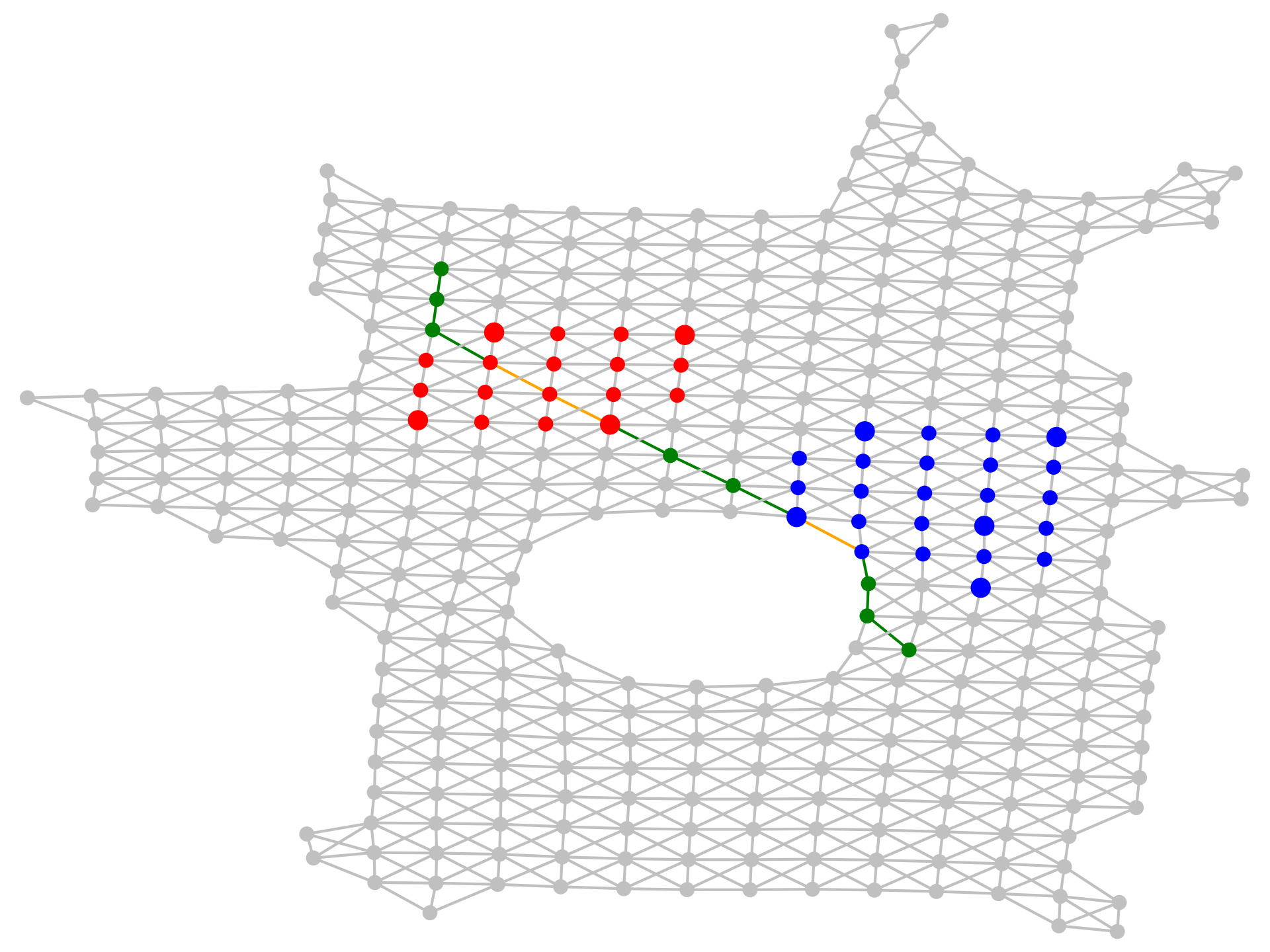}
\label{fig:orz106d_path}
\end{minipage}
}
\caption[An important property of graph convex hulls analogous to geometric convex hulls.]{Illustrates an important property of graph convex hulls analogous to geometric convex hulls. (a) shows that the straight line (shown in green) joining two points in Euclidean space intersects a geometric convex hull (one shown in red and one shown in blue) in exactly one continuum (shown in orange). Here, the red and blue dots indicate the set of points for which the red and blue geometric convex hulls are computed, respectively. (b) shows that any shortest path (shown in green) between two vertices intersects a graph convex hull (one shown in red and one shown in blue) in exactly one continuum (shown in orange). Here, the larger red and blue dots indicate the set of vertices for which the red and blue graph convex hulls are computed, respectively.}
\label{fig:convex_hulls_path_analogy}
\end{figure}

Modulo discretization, graphs are capable of representing complex manifolds and geometric spaces. Hence, graph convex hulls ``generalize'' geometric convex hulls. For example, Figure~\ref{fig:graph_convex_hull} shows a graph convex hull computed on a graph that represents a $2$-dimensional Euclidean space with obstacles. Graph convex hulls have many important properties and applications that are analogous to those of geometric convex hulls. Figure~\ref{fig:convex_hulls_path_analogy} shows one such important analogous property: Every shortest path between any two query vertices on the graph intersects a graph convex hull in a single continuum. Moreover, graph convex hulls have many other applications in active learning~\cite{tg21} and identifying geodesic cores in social networks~\cite{shw22}, among others.

While geometric convex hulls can be computed efficiently, little is known about efficient algorithms for computing graph convex hulls. In particular, it is well known that geometric convex hulls can be computed with the following complexities: Given input points $S$, Qhull~\cite{bdh96} can compute the geometric convex hull with corners $C$ in $O(|S| \log |C|)$ time in $2$-dimensional and $3$-dimensional Euclidean spaces and in $O(|S|f(|C|)/|C|)$ time in higher-dimensional Euclidean spaces. Here, the function $f(|C|)$ returns the maximum number of faces of a convex polytope with $|C|$ corners and is given by the expression $f(|C|) = O(|C|^{\lfloor \kappa/2 \rfloor} / \lfloor \kappa/2 \rfloor!)$ for a $\kappa$-dimensional Euclidean space. In contrast, computing graph convex hulls may not be that efficient. In fact, it is folklore that computing graph convex hulls on a general graph $G = (V, E, w)$ takes at least $O(|V||E|)$ time~\cite{p13}. Hence, brute-force approaches for computing graph convex hulls exactly are prohibitively expensive on large graphs.

In this chapter, we present a novel algorithm for efficiently computing approximate graph convex hulls based on FastMap. Since FastMap facilitates geometric interpretations of graph-theoretic problems, our proposed approach utilizes the efficient algorithms mentioned above for computing the geometric convex hull of a set of points in Euclidean space, particularly in $2$-dimensional and $3$-dimensional Euclidean spaces.

Although our FastMap-based transformation of the graph convex hull problem to the geometric convex hull problem is promising, it does not guarantee exactness. Thus, as a further contribution in this chapter, we advance this approach using an iterative refinement procedure. This procedure significantly improves the recall without compromising the precision. Hence, our iterative FastMap-based algorithm has much better Jaccard scores compared to the naive FastMap-based algorithm. It also runs several orders of magnitude faster than the exact brute-force algorithm. Moreover, we compare our iterative FastMap-based algorithm against the state-of-the-art approximation algorithm and demonstrate two advantages over it. First, our approach is applicable to undirected edge-weighted graphs while the competing approach is applicable only to undirected unweighted graphs. Second, even on undirected unweighted graphs, our iterative FastMap-based algorithm experimentally produces higher-quality solutions and is faster on large graphs.

\section{FastMap-Based Algorithms for Graph Convex Hull Computations}

In this section, we first introduce a naive version of our FastMap-based algorithm. We then improve it to an iterative version using certain geometric intuitions, which, in turn, are also enabled by FastMap. This iterative version of our FastMap-based algorithm is the final product we use in our experimental comparisons against the state-of-the-art competing approach.

The naive version of our FastMap-based algorithm computes an approximation of the graph convex hull as follows: (1) It embeds the vertices of the given graph $G = (V, E, w)$ in a Euclidean space with $\kappa$ dimensions, typically for $\kappa = 2$, $3$, or $4$; (2) It computes the geometric convex hull of the points corresponding to the vertices in $S$; and (3) It reports all the vertices that map to the interior\footnote{The interior includes the boundaries and the corners of the geometric convex hull.} of this geometric convex hull as the required approximation of the graph convex hull. This algorithm is very efficient, especially if $\kappa = 2$ or $3$: Step (1) runs in $O(|E| + |V| \log |V|)$ time; Step (2) runs in $O(|S| \log |C|)$ time; and Step (3) runs in $O(|V||C|)$ time.\footnote{To check if a given point is inside a convex polytope, we have to check its relationship to each of the convex polytope's $f(|C|)$ faces. $f(|C|) = O(|C|)$ for $\kappa = 2$ or $3$.} In Steps (2) and (3), $|C|$ is upper-bounded by $|S|$ since the computation is done in Euclidean space.

The foregoing algorithm is not guaranteed to return an under-approximation or an over-approximation of the vertices in the graph convex hull. Hence, we introduce the measures of precision, recall, and Jaccard score. Here, the precision refers to the fraction of reported vertices that belong to the ground-truth graph convex hull. The recall refers to the fraction of vertices in the ground-truth graph convex hull that are reported. The Jaccard score refers to the ratio of the number of reported vertices that are in the ground-truth graph convex hull to the number of vertices that are either reported or in the ground-truth graph convex hull. Empirically, we observe that even this naive version of our FastMap-based algorithm generally yields very high precision values on a wide variety of graphs. However, there is a leeway for improving its recall and, consequently, its Jaccard score. Towards this end, we design an iterative version of our FastMap-based algorithm drawing intuitions from the geometric convex hull.

The iterative version of our FastMap-based algorithm computes an approximation of the graph convex hull as follows: In the first iteration, it follows Steps (1) and (2) mentioned above. However, it does not terminate by merely identifying the vertices mapped to the interior of the geometric convex hull. Instead, it identifies the vertices mapped to the corners of the geometric convex hull and computes all shortest paths between all pairs of them directly on the input graph $G$. Vertices on any of these shortest paths that are not already in $S$ are considered in addition to $S$ for the second iteration of computing the geometric convex hull. The algorithm may not terminate by identifying the vertices mapped to the interior of the new geometric convex hull produced in the second iteration, either. In such a case, it identifies the vertices mapped to the corners of the new geometric convex hull and aims to compute all shortest paths between all pairs of them directly on the input graph $G$. In doing so, it avoids any redundant computations for pairs of vertices with cached results from the first iteration\footnote{a previous iteration, in general}. The algorithm continues this process until convergence, that is, no new vertices are added to $S$ for the next iteration. Upon convergence, it reports all the vertices mapped to the interior of the geometric convex hull from the last iteration as the required approximation of the graph convex hull. Moreover, the algorithm is guaranteed to converge since: (a) The set of vertices inducted into the graph convex hull in any iteration subsumes that of the previous iteration; and (b) $G$ has a finite number of vertices. Before convergence, the algorithm can also be terminated after a user-specified number of iterations.

Overall, our iterative FastMap-based algorithm hybridizes the exact brute-force algorithm and the naive FastMap-based algorithm. While the exact brute-force algorithm performs all its convex hull-related computations directly on the input graph $G$, the naive FastMap-based algorithm performs all its convex hull-related computations on the Euclidean embedding of $G$. The iterative FastMap-based algorithm hybridizes them and performs its convex hull-related computations partly on $G$ and partly on its Euclidean embedding, interleaving them intelligently so that the shortest paths on $G$ are computed only between pairs of vertices mapped to the corners of the geometric convex hull obtained in the previous iteration. On the one hand, it is significantly more efficient than the exact brute-force algorithm that computes all shortest paths between all pairs of vertices in every iteration until convergence. On the other hand, it is more informed than the naive FastMap-based algorithm, which may occasionally miss qualifying vertices\textemdash and, consequently, their substantial downstream effects\textemdash when they are placed even marginally outside the geometric convex hull in the Euclidean embedding of $G$.

\begin{figure}[!t]
\centering
\setcounter{subfigure}{0}
\subfloat[]{
\begin{minipage}[b]{0.3\textwidth}
\centering
\includegraphics[width=\columnwidth]{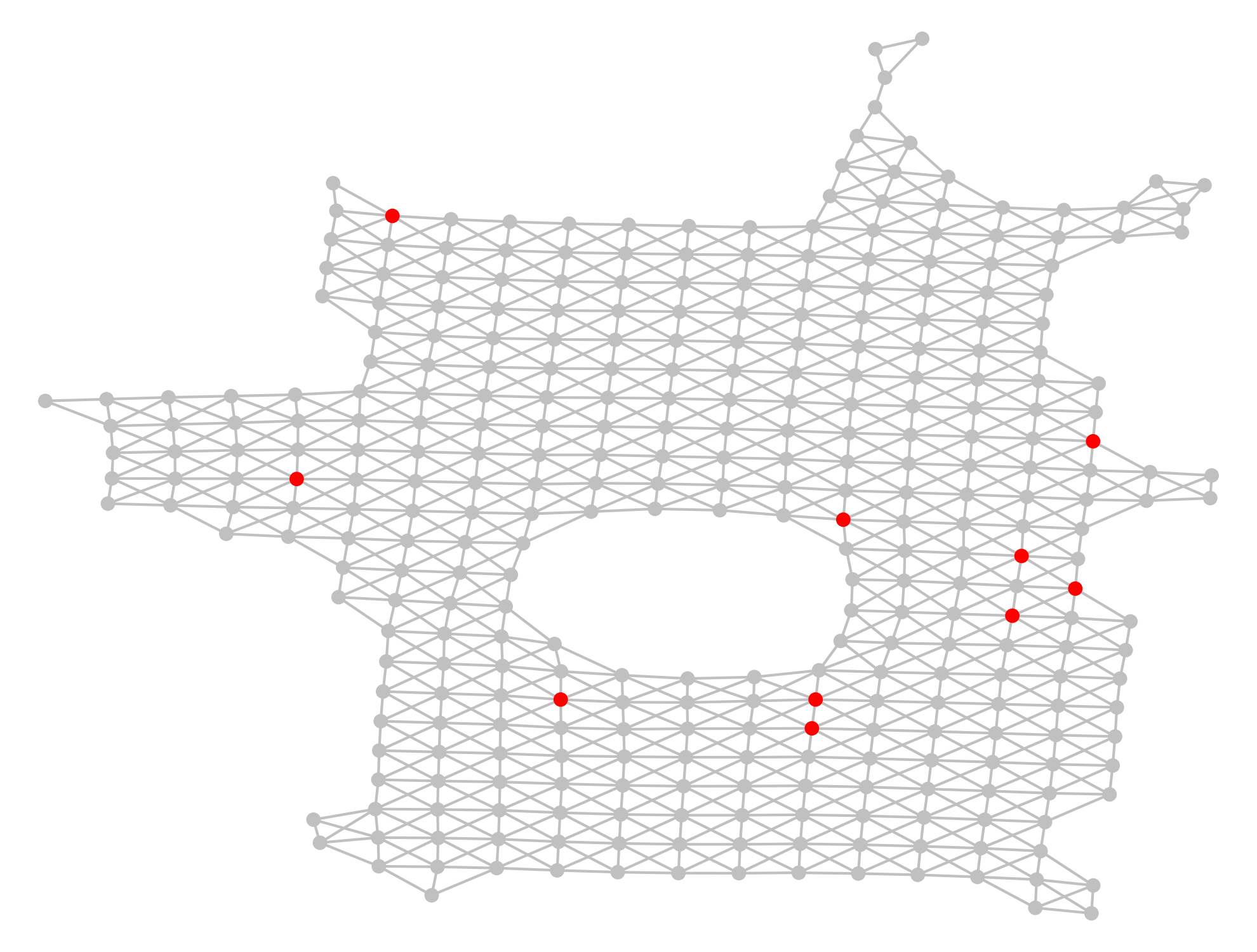}
\label{fig:iter1_1}
\end{minipage}
}
\setcounter{subfigure}{4}
\subfloat[]{
\begin{minipage}[b]{0.3\textwidth}
\centering
\includegraphics[width=\columnwidth]{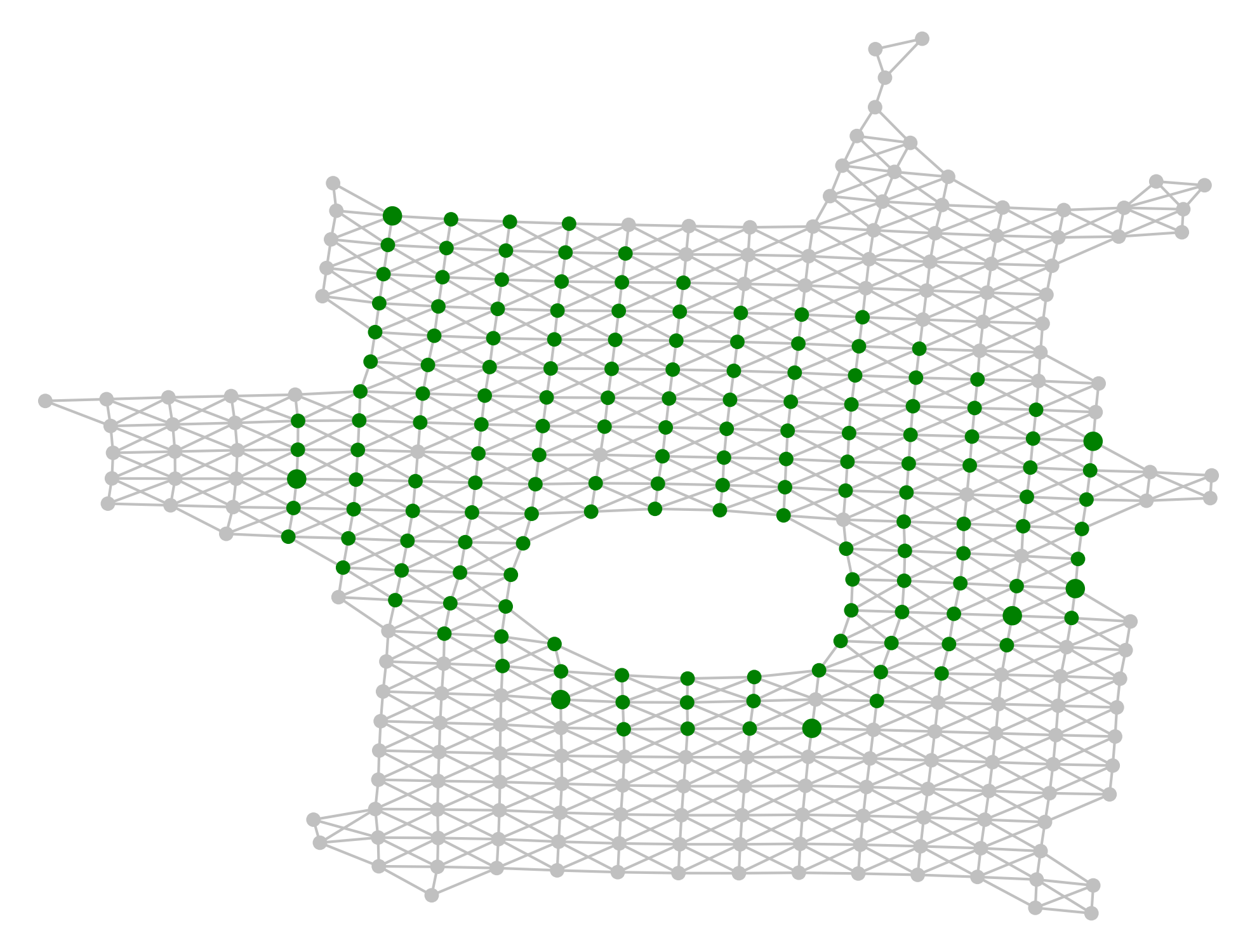}
\label{fig:iter2_1}
\end{minipage}
}
\setcounter{subfigure}{8}
\subfloat[]{
\begin{minipage}[b]{0.3\textwidth}
\centering
\includegraphics[width=\columnwidth]{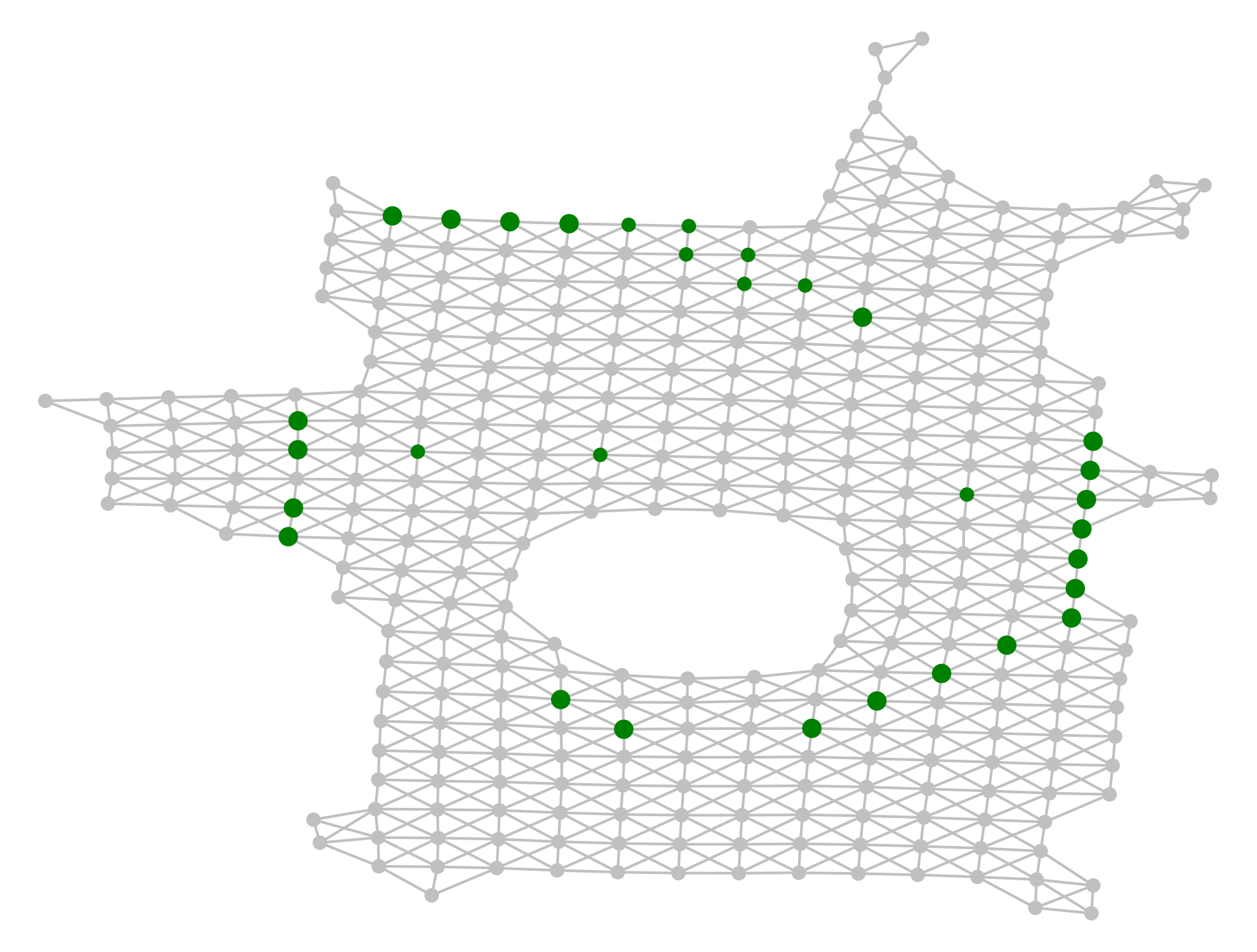}
\label{fig:iter3_1}
\end{minipage}
}
\\
\setcounter{subfigure}{1}
\subfloat[]{
\begin{minipage}[b]{0.3\textwidth}
\centering
\includegraphics[width=\columnwidth]{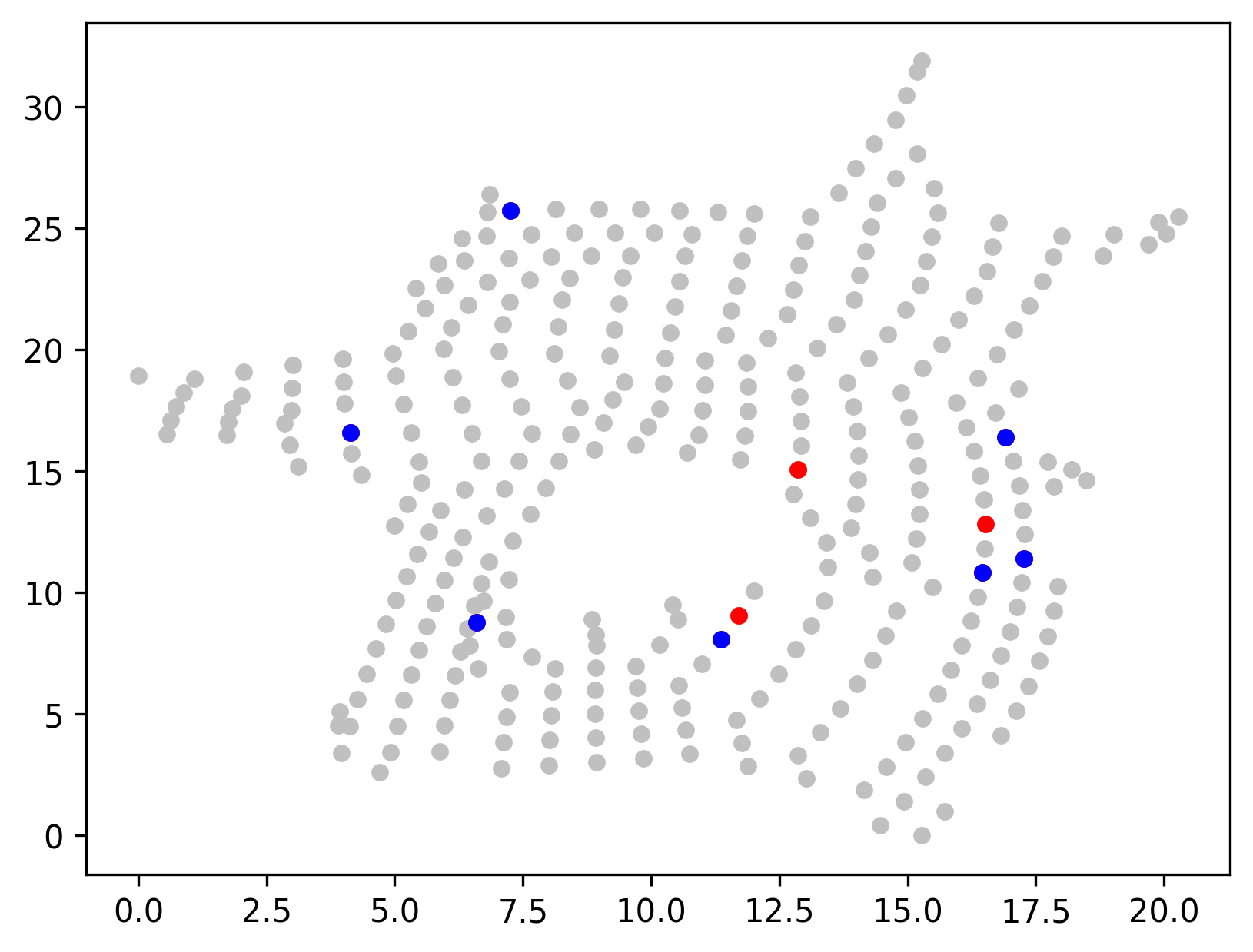}
\label{fig:iter1_2}
\end{minipage}
}
\setcounter{subfigure}{5}
\subfloat[]{
\begin{minipage}[b]{0.3\textwidth}
\centering
\includegraphics[width=\columnwidth]{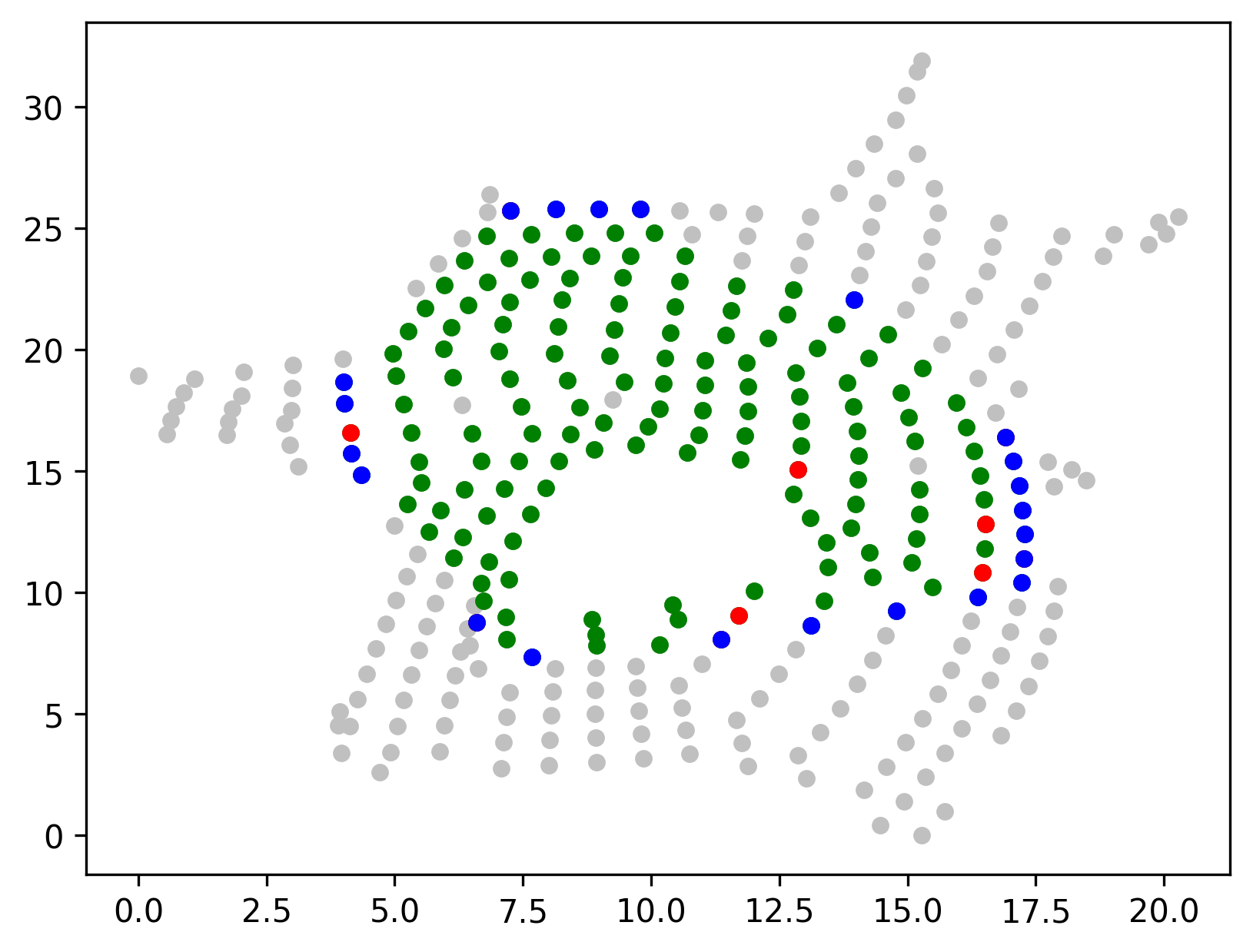}
\label{fig:iter2_2}
\end{minipage}
}
\setcounter{subfigure}{9}
\subfloat[]{
\begin{minipage}[b]{0.3\textwidth}
\centering
\includegraphics[width=\columnwidth]{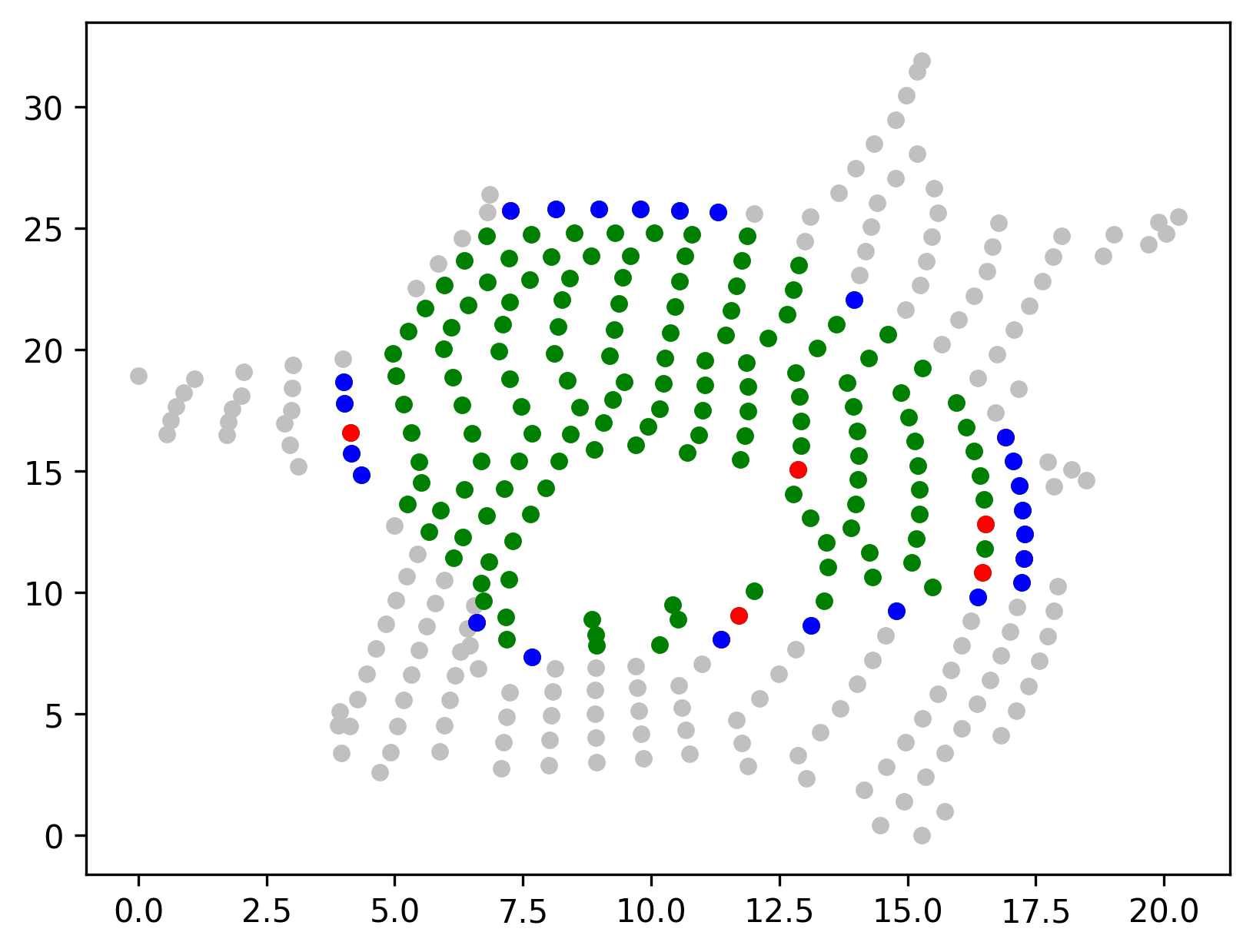}
\label{fig:iter3_2}
\end{minipage}
}
\\
\setcounter{subfigure}{2}
\subfloat[]{
\begin{minipage}[b]{0.3\textwidth}
\centering
\includegraphics[width=\columnwidth]{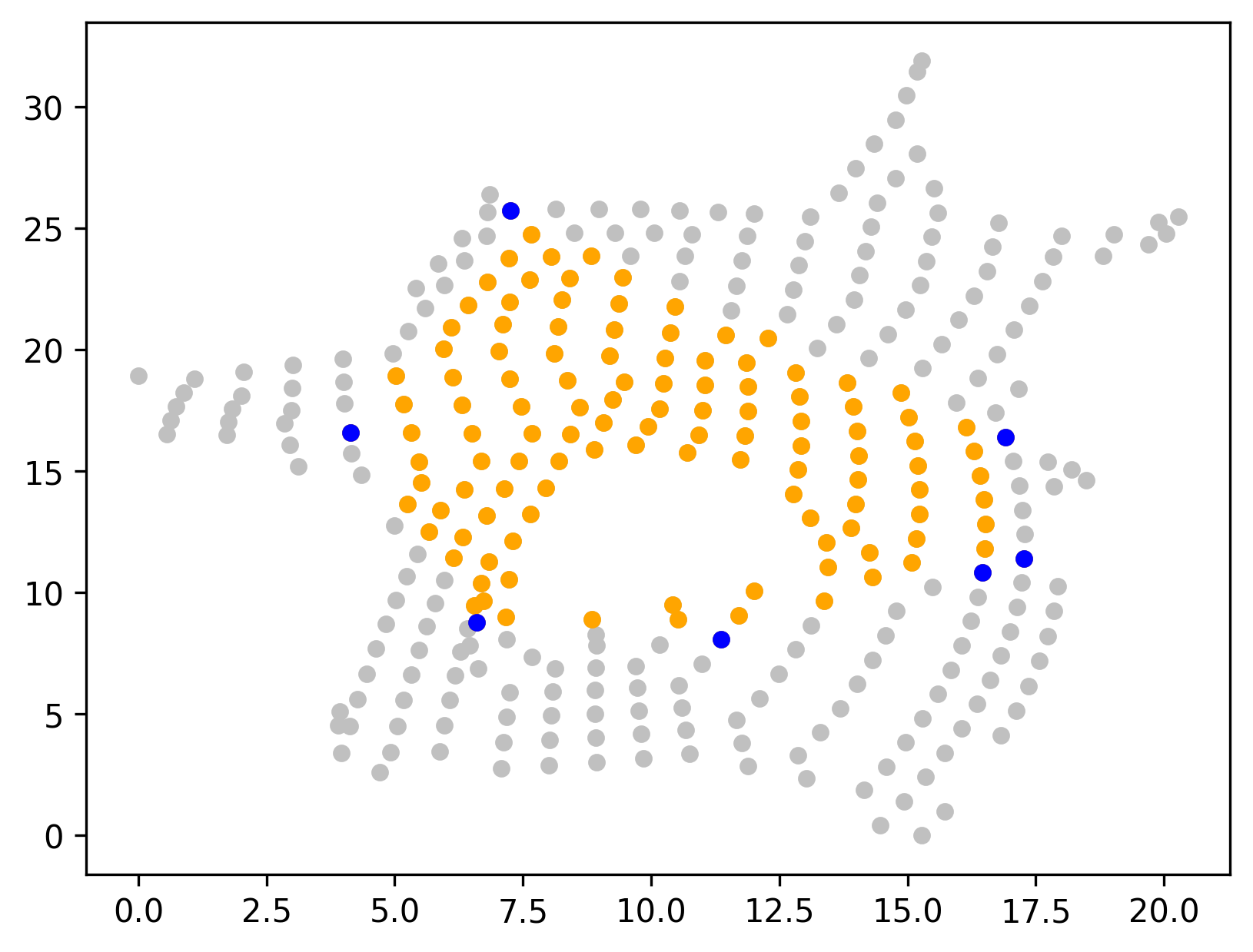}
\label{fig:iter1_3}
\end{minipage}
}
\setcounter{subfigure}{6}
\subfloat[]{
\begin{minipage}[b]{0.3\textwidth}
\centering
\includegraphics[width=\columnwidth]{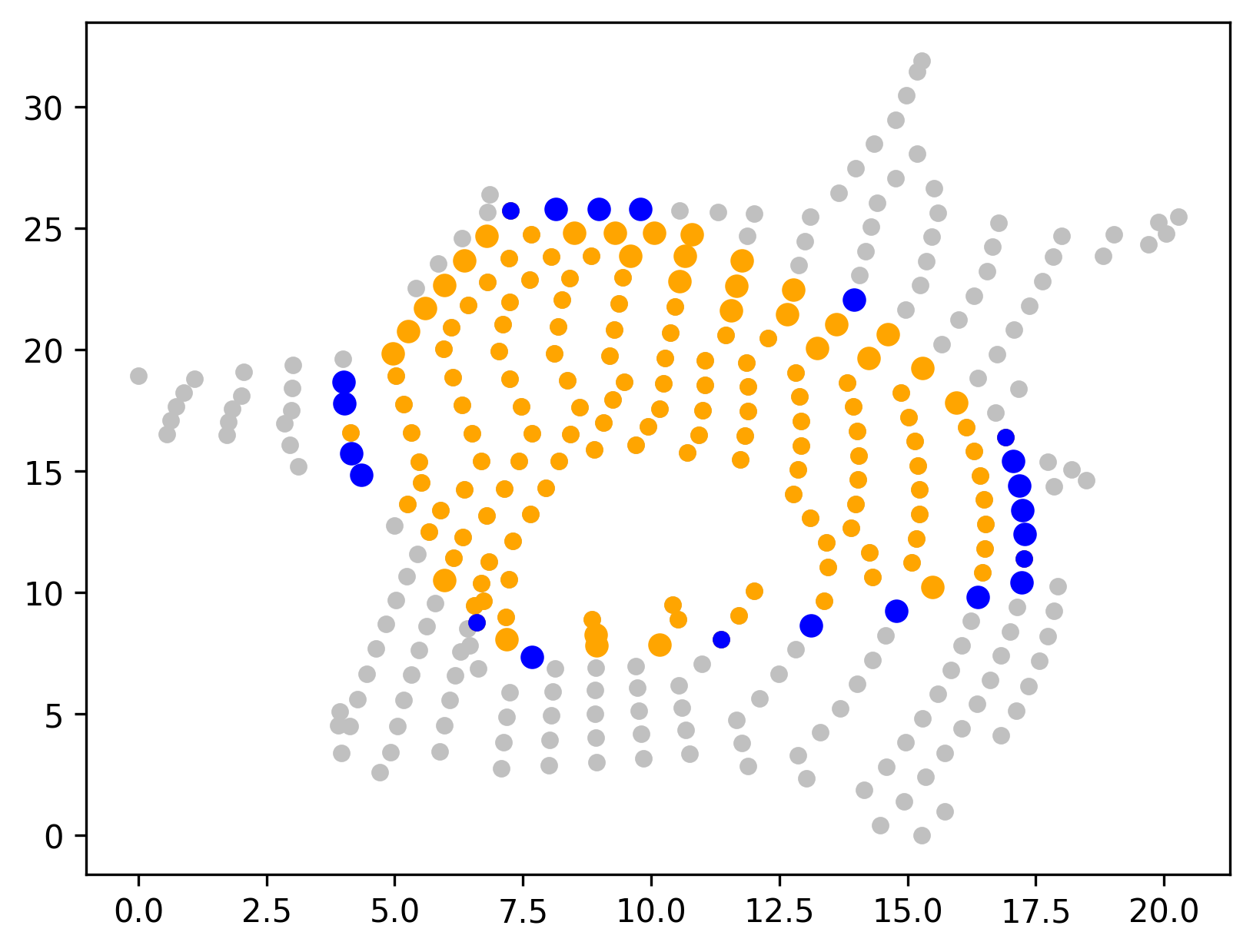}
\label{fig:iter2_3}
\end{minipage}
}
\setcounter{subfigure}{10}
\subfloat[]{
\begin{minipage}[b]{0.3\textwidth}
\centering
\includegraphics[width=\columnwidth]{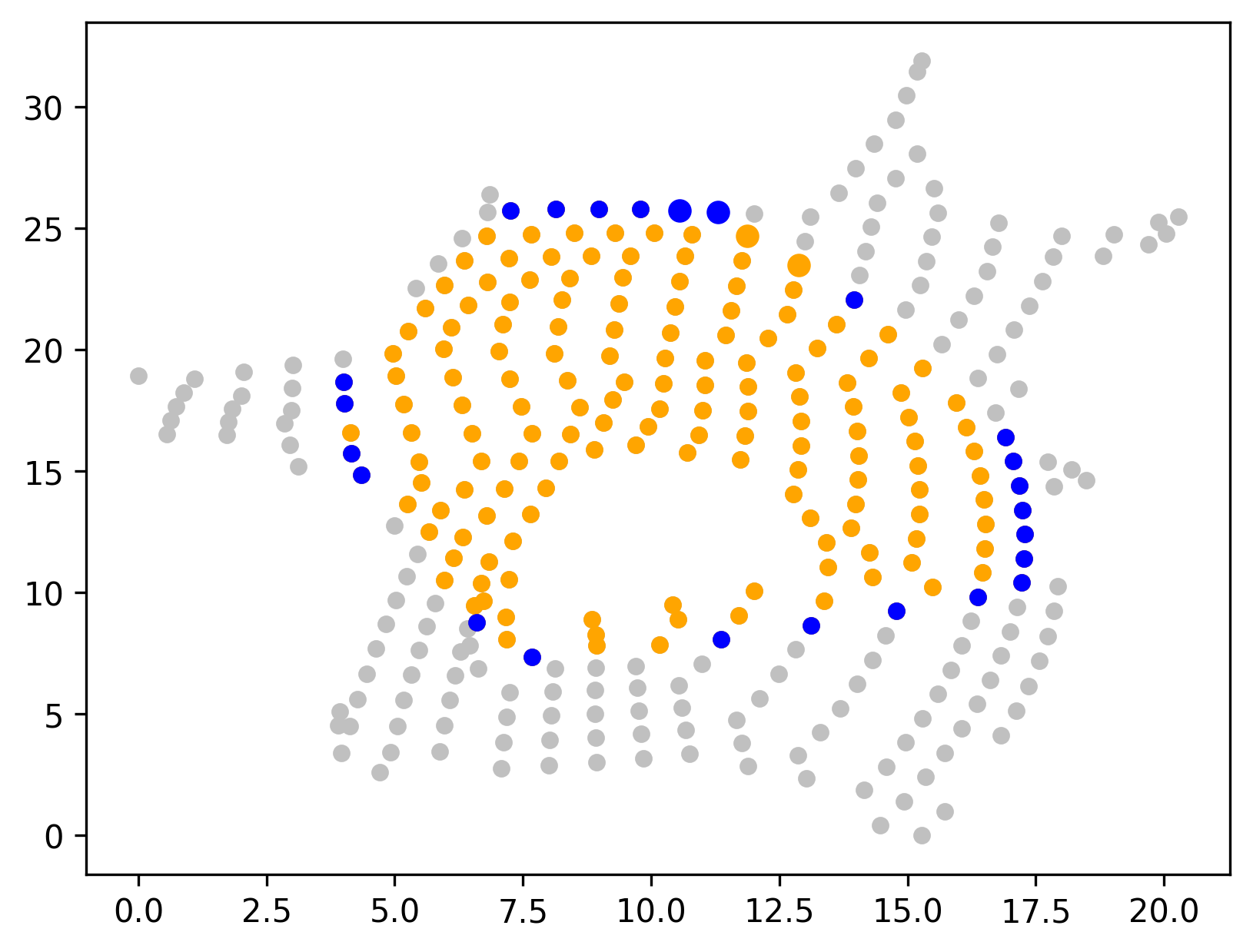}
\label{fig:iter3_3}
\end{minipage}
}
\\
\setcounter{subfigure}{3}
\subfloat[]{
\begin{minipage}[b]{0.3\textwidth}
\centering
\includegraphics[width=\columnwidth]{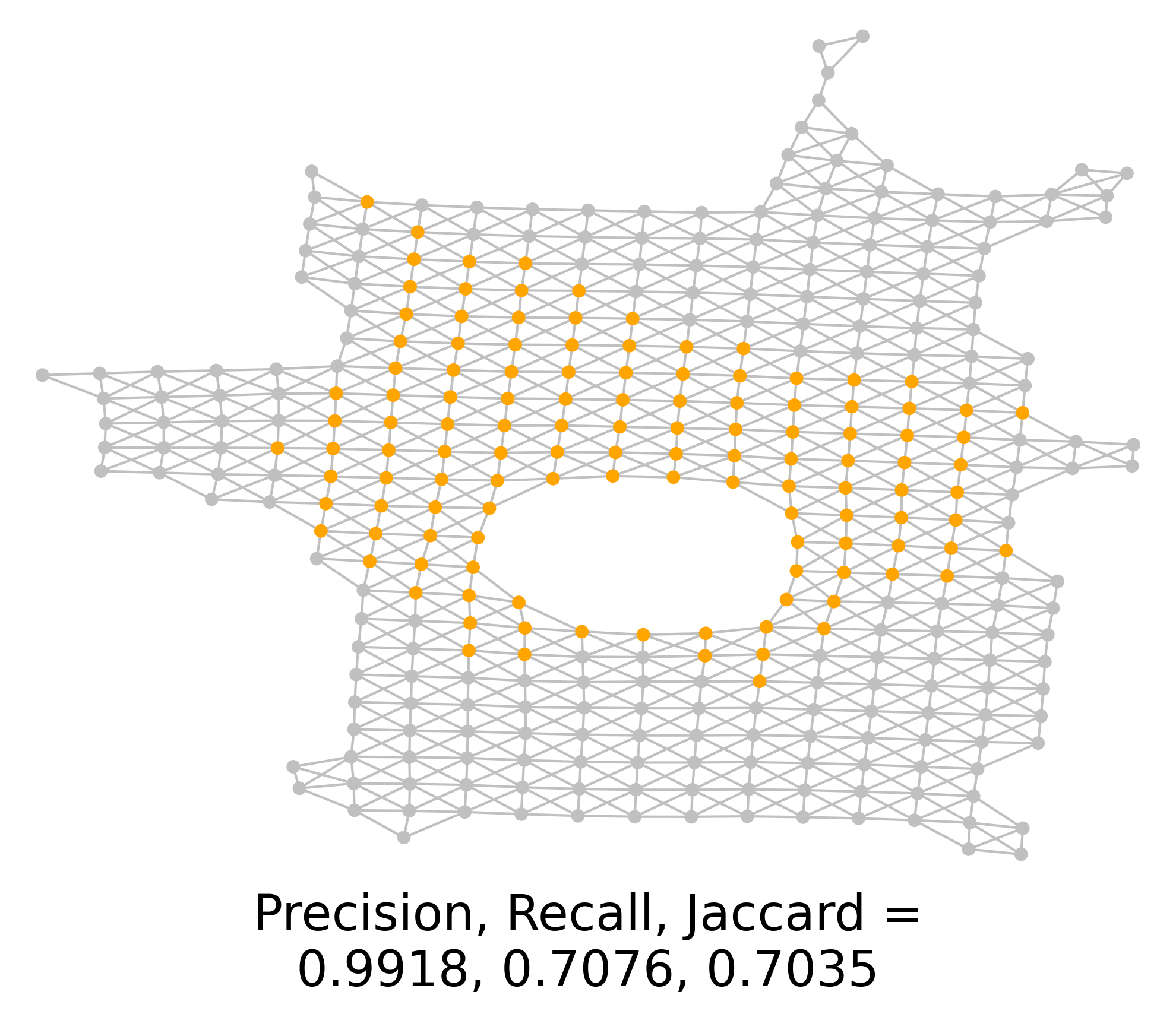}
\label{fig:iter1_4}
\end{minipage}
}
\setcounter{subfigure}{7}
\subfloat[]{
\begin{minipage}[b]{0.3\textwidth}
\centering
\includegraphics[width=\columnwidth]{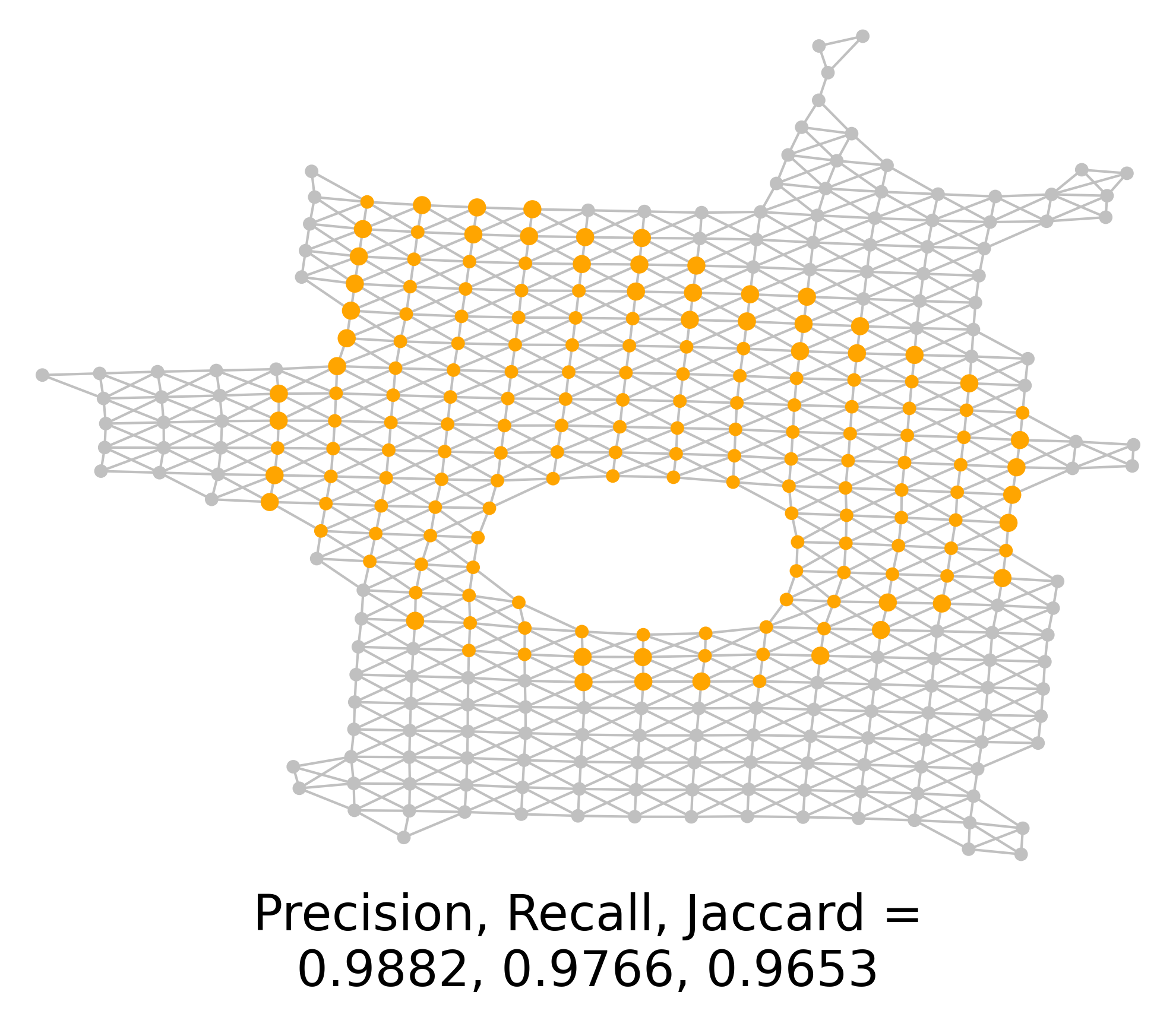}
\label{fig:iter2_4}
\end{minipage}
}
\setcounter{subfigure}{11}
\subfloat[]{
\begin{minipage}[b]{0.3\textwidth}
\centering
\includegraphics[width=\columnwidth]{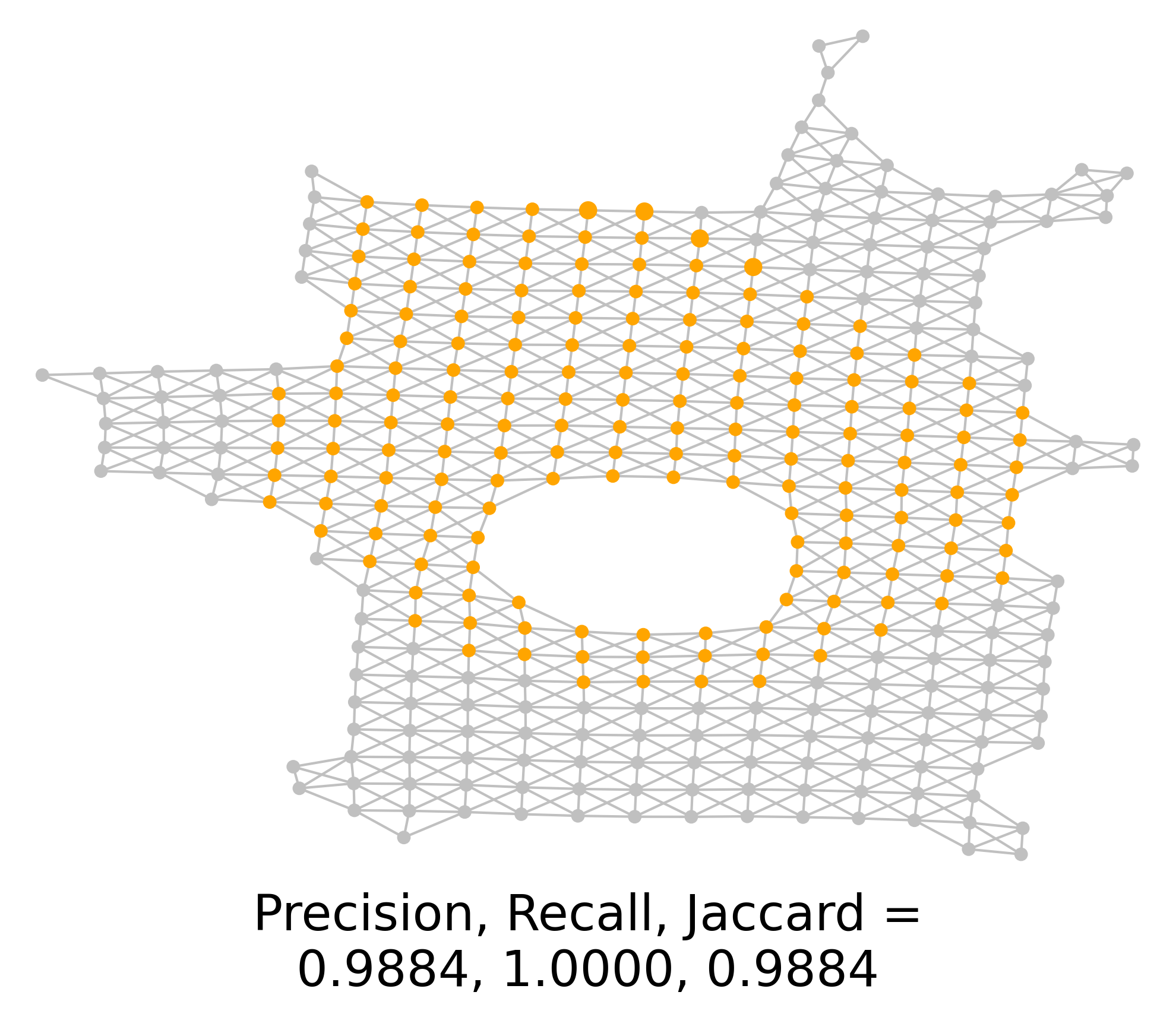}
\label{fig:iter3_4}
\end{minipage}
}
\caption[An illustration of the iterative FastMap-based algorithm for computing graph convex hulls.]{Shows the behavior of our iterative FastMap-based algorithm on the running example from Figure~\ref{fig:graph_convex_hull}. The individual panels are explained in the main text of the chapter. The precision, recall, and Jaccard score are reported after each iteration.}
\label{fig:graph_convex_iterations}
\end{figure}

Figure~\ref{fig:graph_convex_iterations} shows the step-wise working of our iterative FastMap-based algorithm on an example graph. Figure~\ref{fig:iter1_1} shows the input graph $G$ with the set $S$ indicated in red. Figure~\ref{fig:iter1_2} indicates the geometric convex hull of the red points from Figure~\ref{fig:iter1_1} in the FastMap embedding of $G$. The corners of the geometric convex hull are shown in the overriding color blue. Figure~\ref{fig:iter1_3} shows the interior points of the geometric convex hull from Figure~\ref{fig:iter1_2}. The blue points are carried over from Figure~\ref{fig:iter1_2} and the orange points indicate the rest of the interior points. Figure~\ref{fig:iter1_4} shows the interior (blue and orange) points from Figure~\ref{fig:iter1_3} identified on $G$ in orange. This set of orange vertices is the algorithm's approximation of $CH^G_S$ after the first iteration.

Figure~\ref{fig:iter2_1} marks the blue points from Figure~\ref{fig:iter1_3} as larger green vertices. It also shows all vertices that appear on any of the shortest paths between any pair of these larger green vertices as regular green vertices. However, this set of regular green vertices excludes the vertices in $S$, since we compute only the incremental update to the set of vertices that are deemed to be in $CH^G_S$. Figure~\ref{fig:iter2_2} shows the green vertices from Figure~\ref{fig:iter2_1} and the red vertices from Figure~\ref{fig:iter1_1} as the vertices deemed to be in $CH^G_S$ at this stage. The corners of the geometric convex hull of the points corresponding to these vertices are shown in the overriding color blue. Figure~\ref{fig:iter2_3} is similar to Figure~\ref{fig:iter1_3} and carries over the blue points from Figure~\ref{fig:iter2_2}. However, the new internal points identified in this iteration, compared to the previous one, are shown as larger points. Figure~\ref{fig:iter2_4} is similar to Figure~\ref{fig:iter1_4} and is derived from Figure~\ref{fig:iter2_3}. The new set of orange vertices is the algorithm's approximation of $CH^G_S$ after the second iteration, where the new orange vertices are larger.

Figure~\ref{fig:iter3_1} marks the blue points from Figure~\ref{fig:iter2_3} as larger green vertices. It also shows all vertices that appear on any of the shortest paths between any pair of these larger green vertices as regular green vertices. However, this set of regular green vertices excludes the vertices in $S$ (red vertices from Figure~\ref{fig:iter1_1}) and the green vertices computed in the previous iterations (green vertices from Figure~\ref{fig:iter2_1}), since we compute only the incremental update to the set of vertices that are deemed to be in $CH^G_S$. Figure~\ref{fig:iter3_2} shows the green vertices from Figures~\ref{fig:iter3_1} and~\ref{fig:iter2_1} and the red vertices from Figure~\ref{fig:iter1_1} as the vertices deemed to be in $CH^G_S$ at this stage. The corners of the geometric convex hull of the points corresponding to these vertices are shown in the overriding color blue. Figure~\ref{fig:iter3_3} is similar to Figure~\ref{fig:iter2_3} and carries over the blue points from Figure~\ref{fig:iter3_2}. However, the new internal points identified in this iteration, compared to the previous one, are shown as larger points. Figure~\ref{fig:iter3_4} is similar to Figure~\ref{fig:iter2_4} and is derived from Figure~\ref{fig:iter3_3}. The new set of orange vertices is the algorithm's approximation of $CH^G_S$ after the third iteration, where the new orange vertices are larger. At this stage, the algorithm converges; and, in general, would continue either until convergence or for a user-specified number of iterations.

\begin{algorithm}[!tb]
\caption{{\sc FMGCH} (FastMap-Based Graph Convex Hull): A FastMap-based algorithm for computing graph convex hulls.}
\label{alg:FMGCH}
\textbf{Input}: $G = (V, E, w)$ and $S \subseteq V$\\
\textbf{Parameter}: $\kappa$ and $\epsilon$\\ 
\textbf{Output}: $\overline{CH}^G_S$
\begin{algorithmic}[1]
\STATE $P \leftarrow \mbox{FastMap}(G, \kappa, \epsilon)$.\label{line:fastmap}
\STATE $\tilde{S} \leftarrow S$.
\STATE $P_{\tilde{S}} \leftarrow \{p_i \in P: v_i \in \tilde{S}\}$.\label{line:set_P_S}
\STATE $CH_{\tilde{S}} \leftarrow \mbox{ConvexHull}(P_{\tilde{S}})$.\label{line:set_CH_S}
\STATE $CH'_{\tilde{S}} \leftarrow \{\}$.\label{line:init_CH'_S}
\STATE $Dicts \leftarrow dictionary()$.\label{line:init_Dicts}
\WHILE{$CH_{\tilde{S}} \neq CH'_{\tilde{S}}$}\label{line:iter_start}
\STATE $CH'_{\tilde{S}} \leftarrow CH_{\tilde{S}}$.\label{line:set_CH'_S}
\STATE $pairs \leftarrow \{(p_i, p_j): p_i, p_j \in CH'_{\tilde{S}}, i < j, \mbox{ and } (p_i, p_j) \mbox{ is not cached}\}$.\label{line:get_pairs}
\STATE $\tilde{S}' \leftarrow \tilde{S}$.
\FOR{$(p_i, p_j) \in pairs$}
\IF{$v_i \in Dicts$}\label{line:dict_start}
\STATE $SPD_{v_i} \leftarrow Dicts[v_i]$.
\ELSE
\STATE $SPD_{v_i} \leftarrow \mbox{ShortestPathDictionary}(G, v_i)$.
\STATE $Dicts[v_i] \leftarrow SPD_{v_i}$.
\ENDIF\label{line:dict_end}
\STATE $S_{\Delta} \leftarrow \mbox{VerticesOnAllShortestPaths}(SPD_{v_i}, v_j)$.\label{line:vertices_of_paths}
\STATE ${\tilde{S}} \leftarrow {\tilde{S}} \cup S_{\Delta}$.\label{line:update_S}
\ENDFOR
\IF{${\tilde{S}} = {\tilde{S}}'$}\label{line:check_S_start}
\STATE \textbf{break}
\ENDIF\label{line:check_S_end}
\STATE $P_{\tilde{S}} \leftarrow \{p_i \in P: v_i \in \tilde{S}\}$.\label{line:update_P_S}
\STATE $CH_{\tilde{S}} \leftarrow \mbox{ConvexHull}(P_{\tilde{S}})$.\label{line:update_CH_S}
\ENDWHILE\label{line:iter_end}
\STATE $\overline{CH}^G_S \leftarrow \{v_i: p_i \in \mbox{PointsWithinHull}(CH_{\tilde{S}}, P)$\}.\label{line:get_CHGS}
\STATE \textbf{return} $\overline{CH}^G_S$.\label{line:return_CHGS}
\end{algorithmic}
\end{algorithm}

Algorithm~\ref{alg:FMGCH} shows the pseudocode for our iterative FastMap-based algorithm (FMGCH). It takes as input the graph $G = (V, E, w)$ and a set of vertices $S \subseteq V$, for which the graph convex hull needs to be computed. The input parameters $\kappa$ and $\epsilon$ are pertinent to the FastMap embedding, as in Algorithm~\ref{alg:fastmap} (from Chapter~\ref{ch:fastmap}). The output $\overline{CH}^G_S$ is the required graph convex hull or an approximation of it. The algorithm initializes and maintains $\tilde{S}$ to represent the set of vertices deemed to be in $\overline{CH}^G_S$. In addition, it initializes and maintains $CH_{\tilde{S}}$ to represent the corners of the geometric convex hull of $\tilde{S}$ in the FastMap embedding of $G$.

On Line~\ref{line:fastmap}, the algorithm calls Algorithm~\ref{alg:fastmap} and creates a $\kappa$-dimensional Euclidean embedding of $G$, with vertex $v_i \in V$ mapped to point $p_i \in \mathbb{R}^{\kappa}$. On Lines~\ref{line:set_P_S} and~\ref{line:set_CH_S}, the algorithm identifies the points corresponding to the specified vertices in $S$ and computes their geometric convex hull. Here, the function $\mbox{ConvexHull}(\cdot)$ returns only the corners of the geometric convex hull. The algorithm then performs some initializations on Lines~\ref{line:init_CH'_S} and~\ref{line:init_Dicts} and begins the iterative process on Lines~\ref{line:iter_start}-\ref{line:iter_end} until convergence is detected on Lines~\ref{line:iter_start} or~\ref{line:check_S_start}. In each iteration, the old value of $CH_{\tilde{S}}$, that is, $CH'_{\tilde{S}}$, is first replaced by $CH_{\tilde{S}}$ on Line~\ref{line:set_CH'_S}. Subsequently, $CH_{\tilde{S}}$ is updated on Lines~\ref{line:get_pairs}-\ref{line:update_CH_S}. This update starts on Line~\ref{line:get_pairs} by identifying the necessary pairs of vertices between which all vertices on all shortest paths need to be computed, while avoiding any redundant computations with respect to the previous iterations. On Lines~\ref{line:dict_start}-\ref{line:dict_end}, the algorithm then computes the shortest-path dictionary rooted at $v_i$, for each necessary pair $(p_i, p_j)$ with $i < j$, if this dictionary is not already cached in the `$Dicts$' data structure. The function $\mbox{ShortestPathDictionary}(\cdot, \cdot)$ returns a list of predecessors of each vertex that lead to the root vertex $v_i$ along a shortest path.\footnote{In Python3, this can be realized using the `dijkstra\_predecessor\_and\_distance()' function of NetworkX~\cite{hss08}.} On Lines~\ref{line:vertices_of_paths} and~\ref{line:update_S}, the algorithm calls the function $\mbox{VerticesOnAllShortestPaths}(\cdot, \cdot)$ to gather all vertices that appear on any of the shortest paths from $v_i$ to $v_j$ and adds them incrementally to $\tilde{S}$. On Lines~\ref{line:check_S_start}-\ref{line:check_S_end}, the algorithm checks for convergence and breaks the iterative loop if necessary.\footnote{In such a case, the convergence condition on Line~\ref{line:iter_start} is also satisfied. However, the work on Lines~\ref{line:update_P_S} and~\ref{line:update_CH_S} can be avoided.} On Lines~\ref{line:update_P_S} and~\ref{line:update_CH_S}, it updates the geometric convex hull in preparation for the next iteration. Upon termination of the iterative loop, on Lines~\ref{line:get_CHGS} and~\ref{line:return_CHGS}, the algorithm computes and returns the entire interior of the geometric convex hull from the last iteration. The function $\mbox{PointsWithinHull}(\cdot, \cdot)$ determines which of the specified points belong to the interior of a geometric convex hull specified by its corners. It does so by computing the faces of the geometric convex hull from its corners and querying each point with respect to these faces.

As analyzed before, the running time complexity of the naive FastMap-based algorithm is superior to the folklore results for the exact computation of graph convex hulls. Although it is much harder to analyze the running time complexity of our iterative FastMap-based algorithm, we provide an analysis here under certain realistic assumptions. Let $CH^G_S$ be the ground truth and $\kappa = 2$ or $3$.\footnote{For higher values of $\kappa$, the analysis has to explicitly factor in the number of faces of $\kappa$-dimensional convex polytopes.} We assume that the algorithm runs for $\tau$ iterations, for a small constant $\tau$, has a high precision in all iterations, with $|\overline{CH}^G_S| \lessapprox |CH^G_S|$, and that $CH_{\tilde{S}}$ has at most $\bar{c}$ corners in all iterations, with $\bar{c} \ll |CH^G_S|$. These assumptions have been observed to be true in extensive experimental studies. Hence, in each iteration, the algorithm computes the geometric convex hull in $O(|\tilde{S}| \log \bar{c})$ time, in which $|\tilde{S}|$ is upper-bounded by $|CH^G_S|$. In addition, in each iteration, the algorithm computes all vertices on the shortest paths between all pairs of vertices corresponding to the points in $CH_{\tilde{S}}$. It does so in two phases. First, it computes the shortest-path dictionaries rooted at each of the vertices corresponding to the points in $CH_{\tilde{S}}$ in $O(\bar{c}(|E| + |V|\log |V|))$ time. Second, it post-processes each such dictionary with respect to each of the destination vertices corresponding to the points in $CH_{\tilde{S}}$ in time linear in the size of the dictionary. This takes $O(\bar{c}^2(|E| + |V|))$ time. Finally, the algorithm computes the interior of the geometric convex hull in $O(\bar{c}|V|)$ time. Therefore, the overall time complexity is $O(\tau (\log \bar{c}|CH^G_S| + \bar{c}(|V|\log |V|) + \bar{c}^2(|E| + |V|)))$. This complexity is still better than the folklore results for the exact computation of graph convex hulls. Moreover, the real benefits of the iterative FastMap-based algorithm become evident in the experimental studies presented in the next section.

\section{An Efficient Implementation of the Exact Brute-Force Algorithm}
\label{sec:exact_algorithm}

\begin{algorithm}[!tb]
\caption{{\sc ExactGraphConvexHull}: An exact algorithm with implementation-level optimizations for computing graph convex hulls.}
\label{alg:GCH}
\textbf{Input}: $G = (V, E, w)$ and $S \subseteq V$\\
\textbf{Output}: $CH^G_S$
\begin{algorithmic}[1]
\STATE $CH^G_S \leftarrow S$.\label{line:init_start}
\STATE $V_{unexpanded} \leftarrow S$.
\STATE $V_{expanded} \leftarrow \{\}$.
\STATE $Pairs \leftarrow \{\}$.\label{line:init_end}
\WHILE{$V_{unexpanded} \neq \{\}$}\label{line:outer_loop_start}
\STATE Randomly select $v_s$ from $V_{unexpanded}$.\label{line:select_vs}
\STATE $SPD_{v_s} \leftarrow \mbox{ShortestPathDictionary}(G, v_s)$.
\STATE $V_{new} \leftarrow \{\}$.\label{line:init_vnew}
\FOR{$v_t \in CH^G_S$}\label{line:inner_loop_start}
\IF{$v_s = v_t$ or $(v_s, v_t) \in Pairs$}\label{line:skip_start}
\STATE Continue.
\ENDIF\label{line:skip_end}
\STATE $V_{s, t} \leftarrow \mbox{VerticesOnAllShortestPaths}(SPD_{v_s}, v_t)$.\label{line:all_vertices}
\STATE $Pairs_{s, t} \leftarrow \mbox{PairsOnAllShortestPaths}(SPD_{v_s}, v_t)$.\label{line:all_pairs}
\STATE $Pairs \leftarrow Pairs \cup Pairs_{s, t}$.\label{line:update_pairs}
\FOR{$v \in V_{s, t}$}\label{line:check_vst_start}
\IF{$v \notin V_{expanded}$ and $v \notin V_{unexpanded}$}
\STATE Add $v$ to $V_{unexpanded}$.\label{line:update_vunexpanded}
\STATE Add $v$ to $V_{new}$.\label{line:update_vnew}
\ENDIF
\ENDFOR\label{line:check_vst_end}
\ENDFOR\label{line:inner_loop_end}
\STATE $CH^G_S \leftarrow CH^G_S \cup V_{new}$.\label{line:add_vnew}
\STATE Add $v_s$ to $V_{expanded}$.\label{line:update_vexpanded}
\STATE Remove $v_s$ from $V_{unexpanded}$.\label{line:decrease_vunexpanded}
\ENDWHILE\label{line:outer_loop_end}
\STATE \textbf{return} $CH^G_S$.\label{line:return_graph_convex_hull}
\end{algorithmic}
\end{algorithm}

In this section, we present an exact brute-force algorithm for computing graph convex hulls. This procedure not only serves as a baseline competing method but also produces the ground truth. Although it is a brute-force algorithm, it incorporates several implementation-level optimizations derived from the use of dictionaries and caching. Some of these optimizations are also used in FMGCH and, in fact, may further be useful for other algorithms developed in the future.

Algorithm~\ref{alg:GCH} presents the pseudocode of the brute-force procedure. It takes as input the graph $G = (V, E, w)$ and a set of vertices $S \subseteq V$, for which the graph convex hull needs to be computed. The output $CH^G_S$ is the required graph convex hull. The algorithm uses shortest-path dictionaries as a critical data structure: A shortest-path dictionary rooted at the vertex $v$ is denoted by $SPD_v$. It is similar to the shortest-path tree rooted at $v$ but represents all shortest paths from $v$ to all other vertices. It does this by maintaining a list of predecessors of each vertex that lead to the root vertex $v$ along a shortest path. The algorithm also maintains a set of vertices $V_{unexpanded}$, rooted at each of which it intends to compute a shortest-path dictionary. The set $V_{expanded}$ gathers the vertices for which this task has been accomplished. Moreover, the algorithm uses $Pairs$ to maintain a set of pairs of vertices between which all shortest paths have been generated.

On Lines~\ref{line:init_start}-\ref{line:init_end}, Algorithm~\ref{alg:GCH} initializes $CH^G_S$, $V_{unexpanded}$, $V_{expanded}$, and $Pairs$. On Lines~\ref{line:outer_loop_start}-\ref{line:outer_loop_end}, the algorithm loops until convergence while computing a new shortest-path dictionary in each iteration. The new shortest-path dictionary provides information on some vertices that provably belong to $CH^G_S$: These vertices are gathered in the set $V_{new}$; and all other data structures are appropriately updated. On Lines~\ref{line:select_vs}-\ref{line:init_vnew}, the algorithm picks a random vertex $v_s$ from $V_{unexpanded}$, calls the function $\mbox{ShortestPathDictionary}(\cdot, \cdot)$ to compute $SPD_{v_s}$, the shortest-path dictionary rooted at it, and initializes $V_{new}$. On Lines~\ref{line:inner_loop_start}-\ref{line:inner_loop_end}, the algorithm considers every possible vertex $v_t$ that is currently in $CH^G_S$ and pairs it with $v_s$. On Lines~\ref{line:skip_start}-\ref{line:skip_end}, the algorithm skips the pair $(v_s, v_t)$ if $v_s = v_t$ or $(v_s, v_t)$ has been processed before and, hence, has been recorded in $Pairs$. Otherwise, on Line~\ref{line:all_vertices}, the algorithm calls the function $\mbox{VerticesOnAllShortestPaths}(\cdot, \cdot)$ to gather into $V_{s, t}$ all vertices that appear on any of the shortest paths from $v_s$ to $v_t$.

Moreover, on Line~\ref{line:all_pairs}, the algorithm calls the function $\mbox{PairsOnAllShortestPaths}(\cdot, \cdot)$ to gather all pairs of vertices that both appear on any of the shortest paths from $v_s$ to $v_t$. This is achieved by: (1) Creating a directed acyclic graph $G_{DAG}$ on all vertices that are reachable via predecessor (parent) relationships from $v_t$ in $SPD_{v_s}$; (2) Repeating until there are no more vertices, the process of identifying a vertex $v_i$ with no predecessors, computing all of its descendants via breadth-first search, recording the pair $(v_i, v_j)$ for each descendant $v_j$, and removing $v_i$ with all its edges from $G_{DAG}$; and (3) Returning all pairs of vertices recorded in the previous step as the output. On Line~\ref{line:update_pairs}, the algorithm updates $Pairs$ accordingly: This is used in future iterations on Line~\ref{line:skip_start} to avoid redundant work.

On Lines~\ref{line:check_vst_start}-\ref{line:check_vst_end}, the algorithm considers each vertex $v$ in $V_{s, t}$. If $v$ has not yet been considered for computing a shortest-path dictionary rooted at it, that is, if $v$ does not appear in $V_{unexpanded}$ or $V_{expanded}$, it is added to $V_{unexpanded}$ for future consideration. It is also added to $V_{new}$. On Line~\ref{line:add_vnew}, the algorithm adds $V_{new}$ to $CH^G_S$, the correctness of which is proved by the following arguments. From Lines~\ref{line:update_vunexpanded},~\ref{line:update_vnew}, and~\ref{line:add_vnew}, it is easy to infer that any vertex added to $V_{unexpanded}$ is also added to $V_{new}$ and $CH^G_S$. Hence, $v_s$ chosen on Line~\ref{line:select_vs} belongs to $CH^G_S$. Moreover, from Line~\ref{line:inner_loop_start}, $v_t$ also belongs to $CH^G_S$. Since all vertices in $V_{s, t}$ and $V_{new}$ appear on a shortest path from $v_s$ to $v_t$, Line~\ref{line:add_vnew} correctly adds $V_{new}$ to $CH^G_S$ in accordance with the definition of the graph convex hull. On Lines~\ref{line:update_vexpanded} and~\ref{line:decrease_vunexpanded}, the algorithm updates $V_{expanded}$ and $V_{unexpanded}$, respectively. On Line~\ref{line:return_graph_convex_hull}, it returns $CH^G_S$ upon convergence.

\section{Experimental Results}

In this section, we present tabular experimental results that compare FMGCH, that is, Algorithm~\ref{alg:FMGCH}, against the state-of-the-art algorithm for computing graph convex hulls, which is encapsulated within GCoreApproximation (GCA)~\cite{shw22}\footnote{\url{https://github.com/fseiffarth/GCoreApproximation}}. However, GCA is also an approximation algorithm. Hence, to produce the ground truth (GT), we invoked Algorithm~\ref{alg:GCH} presented in Section~\ref{sec:exact_algorithm}. We do not include our naive FastMap-based algorithm in the tabular results to avoid clutter. However, as expected, it is more efficient than FMGCH, our iterative FastMap-based algorithm, but produces a lower recall and Jaccard score. We implemented FMGCH in Python3. For computing the geometric convex hull of a collection of points in Euclidean space, we used the `Qhull' library~\cite{bdh96}. We invoked GCA using a simple Python wrapper function. We conducted all experiments on a laptop with an Apple M2 Max chip and 96 GB memory.

\begin{sidewaystable}[]
\centering
\scalebox{0.62}{
\begin{tabular}{|l|r|r|rrrr|rrrrr|rrrrr|rrrrr|}
\cline{1-22}
\multirow{3}{*}{Instance} &\multirow{3}{*}{Size ($|V|$, $|E|$)} &\multirow{3}{*}{GT (s)} &\multicolumn{4}{c|}{\multirow{2}{*}{GCA}} &\multicolumn{15}{c|}{FMGCH}\\
\cline{8-22}
&&&&&&&\multicolumn{5}{c|}{$\kappa=2$} &\multicolumn{5}{c|}{$\kappa=3$} &\multicolumn{5}{c|}{$\kappa=4$}\\
\cline{4-22}
&&&time (s) &jacc &prec &recall &pre (s) &query (s) &jacc &prec &recall &pre (s) &query (s) &jacc &prec &recall &pre (s) &query (s) &jacc &prec &recall\\
\cline{1-22}
miles1000 &(128, 3216) &0.4754 &0.0016 &0.9922 &1.0000 &0.9922 &0.0091 &0.0146 &1.0000 &1.0000 &1.0000 &0.0172 &0.0394 &1.0000 &1.0000 &1.0000 &0.0227 &0.0657 &1.0000 &1.0000 &1.0000\\
myciel7 &(191, 2360) &0.5030 &0.0009 &1.0000 &1.0000 &1.0000 &0.0074 &0.0049 &1.0000 &1.0000 &1.0000 &0.0099 &0.0151 &1.0000 &1.0000 &1.0000 &0.0131 &0.0300 &1.0000 &1.0000 &1.0000\\
queen16\_16 &(256, 6320) &1.1260 &0.0034 &1.0000 &1.0000 &1.0000 &0.0154 &0.0264 &1.0000 &1.0000 &1.0000 &0.0346 &0.1073 &1.0000 &1.0000 &1.0000 &0.0392 &0.1241 &1.0000 &1.0000 &1.0000\\
le450\_25d &(450, 17425) &5.5958 &0.0240 &1.0000 &1.0000 &1.0000 &0.0647 &0.0601 &0.9133 &1.0000 &0.9133 &0.1187 &0.2570 &0.9889 &1.0000 &0.9889 &0.0936 &0.4376 &0.9978 &1.0000 &0.9978\\
\cline{1-22}
orz601d &(1890, 3473) &48.2478 &0.0307 &0.7801 &0.8664 &0.8867 &0.0156 &0.0169 &0.6456 &0.6456 &1.0000 &0.0278 &0.0862 &0.7724 &0.7731 &0.9988 &0.0373 &0.2785 &0.9115 &0.9125 &0.9988\\
lak106d &(1909, 3589) &54.9086 &0.0431 &0.7510 &0.8528 &0.8628 &0.0181 &0.0410 &0.8722 &0.8736 &0.9982 &0.0319 &0.1404 &0.9104 &0.9111 &0.9991 &0.0401 &0.2190 &0.9568 &0.9576 &0.9991\\
hrt001d &(3705, 6914) &1342.2423 &0.1715 &0.6319 &0.7551 &0.7948 &0.0276 &0.0714 &0.6375 &0.6453 &0.9815 &0.0516 &0.2105 &0.7344 &0.7344 &1.0000 &0.0595 &0.6886 &0.9316 &0.9354 &0.9956\\
orz000d &(4057, 7744) &1936.5250 &0.2134 &0.9439 &0.9711 &0.9711 &0.0315 &0.0780 &0.9683 &0.9683 &1.0000 &0.0944 &0.2351 &0.9873 &0.9884 &0.9988 &0.0774 &0.9722 &0.9891 &0.9949 &0.9941\\
\cline{1-22}
ca-GrQc &(4158, 13428) &26.1400 &0.2220 &0.1945 &0.3254 &0.3259 &0.0548 &0.1805 &0.3307 &0.3330 &0.9792 &0.1027 &0.5365 &0.3389 &0.3403 &0.9881 &0.0959 &1.1852 &0.3554 &0.3566 &0.9903\\
ca-HepTh &(8638, 24827) &219.1975 &1.5639 &0.2704 &0.4258 &0.4255 &0.0975 &0.2371 &0.4219 &0.4329 &0.9432 &0.2595 &1.2079 &0.4299 &0.4313 &0.9928 &0.2274 &3.5873 &0.4372 &0.4378 &0.9967\\
wiki-Vote &(7066, 100736) &862.9993 &2.7505 &0.5034 &0.6698 &0.6696 &0.2689 &0.4216 &0.6173 &0.6613 &0.9027 &0.4657 &2.1628 &0.6408 &0.6550 &0.9673 &0.5820 &5.9597 &0.6696 &0.6854 &0.9666\\
ca-HepPh &(11204, 117649) &799.8486 &3.5258 &0.3324 &0.4989 &0.4990 &0.3205 &1.1677 &0.4352 &0.4389 &0.9810 &0.6013 &3.8834 &0.4379 &0.4404 &0.9872 &0.6396 &13.0594 &0.4446 &0.4465 &0.9907\\
\cline{1-22}
wm04000 &(4000, 20109) &207.3266 &1.2004 &0.9995 &0.9997 &0.9997 &0.0651 &0.1952 &0.9945 &0.9997 &0.9947 &0.1505 &0.8467 &0.9975 &1.0000 &0.9975 &0.1726 &2.7267 &0.9992 &1.0000 &0.9992\\
wm08000 &(8000, 40014) &1061.7808 &5.3560 &0.9998 &0.9999 &0.9999 &0.2337 &0.5732 &0.9972 &0.9999 &0.9974 &0.3414 &2.3460 &0.9986 &0.9999 &0.9987 &0.3716 &5.6938 &0.9987 &1.0000 &0.9987\\
wm10000 &(10000, 49980) &1755.0927 &8.7739 &0.9994 &0.9997 &0.9997 &0.2436 &0.5254 &0.9991 &0.9998 &0.9993 &0.3281 &2.0855 &0.9983 &0.9997 &0.9986 &0.3831 &7.6367 &0.9987 &0.9997 &0.9990\\
wm12000 &(12000, 59853) &2603.6225 &12.9154 &0.9991 &0.9996 &0.9995 &0.2663 &0.9198 &0.9929 &0.9997 &0.9932 &0.4050 &3.1991 &0.9992 &0.9997 &0.9994 &0.4999 &9.7890 &0.9988 &0.9999 &0.9989\\
\toprule
\bottomrule
miles1000 &(128, 3216) &0.2592 &-&-&-&- &0.0068 &0.0056 &0.9444 &1.0000 &0.9444 &0.0130 &0.0269 &0.9683 &1.0000 &0.9683 &0.0174 &0.0696 &0.9921 &1.0000 &0.9921\\
myciel7 &(191, 2360) &0.2875 &-&-&-&- &0.0075 &0.0099 &0.8066 &0.8957 &0.8902 &0.0096 &0.0282 &0.8268 &0.9080 &0.9024 &0.0164 &0.1033 &0.9266 &0.9266 &1.0000\\
queen16\_16 &(256, 6320) &0.9702 &-&-&-&- &0.0127 &0.0515 &0.9922 &1.0000 &0.9922 &0.0258 &0.1423 &0.9922 &1.0000 &0.9922 &0.0370 &0.2167 &0.9961 &1.0000 &0.9961\\
le450\_25d &(450, 17425) &4.6087 &-&-&-&- &0.0580 &0.1069 &0.9376 &1.0000 &0.9376 &0.0875 &0.4325 &0.9866 &1.0000 &0.9866 &0.0978 &0.6410 &0.9889 &1.0000 &0.9889\\
\cline{1-22}
orz601d &(1890, 6746) &22.4946 &-&-&-&- &0.0255 &0.0340 &0.5961 &0.5961 &1.0000 &0.0373 &0.0930 &0.7192 &0.7192 &1.0000 &0.0516 &0.3527 &0.8906 &0.8916 &0.9987\\
lak106d &(1909, 7029) &16.6713 &-&-&-&- &0.0268 &0.0929 &0.7819 &0.7819 &1.0000 &0.0379 &0.1915 &0.8666 &0.8675 &0.9988 &0.0571 &0.7088 &0.9330 &0.9362 &0.9964\\
hrt001d &(3705, 13498) &318.5500 &-&-&-&- &0.0586 &0.0769 &0.8019 &0.8109 &0.9864 &0.0801 &0.4049 &0.8297 &0.8821 &0.9332 &0.1104 &1.5441 &0.9596 &0.9629 &0.9965\\
orz000d &(4057, 15208) &439.4339 &-&-&-&- &0.0560 &0.1739 &0.6183 &0.6183 &1.0000 &0.1362 &0.7398 &0.9270 &0.9275 &0.9995 &0.1106 &2.1894 &0.9439 &0.9439 &1.0000\\
\cline{1-22}
ca-GrQc &(4158, 13428) &27.2310 &-&-&-&- &0.0560 &0.0823 &0.3304 &0.3491 &0.8604 &0.0707 &0.5626 &0.3424 &0.3454 &0.9748 &0.1090 &1.6527 &0.3586 &0.3605 &0.9859\\
ca-HepTh &(8638, 24827) &219.0672 &-&-&-&- &0.0961 &0.2631 &0.4261 &0.4339 &0.9599 &0.1570 &1.0497 &0.4288 &0.4341 &0.9723 &0.1943 &3.2234 &0.4352 &0.4379 &0.9859\\
wiki-Vote &(7066, 100736) &867.2150 &-&-&-&- &0.2691 &0.2840 &0.5799 &0.6571 &0.8316 &0.4012 &1.8959 &0.6610 &0.6735 &0.9727 &0.6367 &5.1408 &0.6642 &0.6849 &0.9566\\
ca-HepPh &(11204, 117649) &805.3969 &-&-&-&- &0.3188 &0.4995 &0.4281 &0.4569 &0.8717 &0.4761 &4.9446 &0.4385 &0.4414 &0.9853 &0.6435 &10.3627 &0.4446 &0.4466 &0.9901\\
\cline{1-22}
wm04000 &(4000, 20109) &139.1687 &-&-&-&- &0.1148 &0.2276 &0.8095 &0.9981 &0.8107 &0.1256 &1.2747 &0.9815 &0.9980 &0.9835 &0.2004 &3.2818 &0.9900 &0.9980 &0.9920\\
wm08000 &(8000, 40014) &628.2476 &-&-&-&- &0.1793 &0.4662 &0.8419 &0.9988 &0.8427 &0.3319 &3.6184 &0.9895 &0.9987 &0.9907 &0.3983 &7.6990 &0.9901 &0.9989 &0.9912\\
wm10000 &(10000, 49980) &1021.6094 &-&-&-&- &0.2362 &0.9538 &0.9012 &0.9979 &0.9029 &0.4013 &4.3646 &0.9922 &0.9979 &0.9943 &0.5599 &11.0225 &0.9936 &0.9980 &0.9956\\
wm12000 &(12000, 59853) &1544.6484 &-&-&-&- &0.4505 &0.9459 &0.8570 &0.9986 &0.8580 &0.5633 &6.6372 &0.9795 &0.9984 &0.9811 &0.6056 &14.4132 &0.9935 &0.9984 &0.9951\\
\cline{1-22}
\end{tabular}
}
\caption[Performance results of FMGCH with respect to graph convex hull computations.]{Shows the performance results of FMGCH, GCA, and the GT procedure. FMGCH is shown with different values of $\kappa$. The columns `prec', `recall', and `jacc' indicate the precision, recall, and Jaccard score, respectively. `GT (s)', `time (s)' under `GCA', and `pre (s)' and `query (s)' under `FMGCH' represent the running time of GT, the running time of GCA, the precomputation time of FMGCH, and the query time of FMGCH, respectively, in seconds. The top half of the rows are unweighted graphs and the bottom half are their weighted counterparts. Within each half, the rows are divided into the categories: DIMACS, movingAI, SNAP, and Waxman.}
\label{tab:convex_hull}
\end{sidewaystable}

While FMGCH is applicable to both unweighted and (edge-)weighted graphs, GCA is applicable to only unweighted graphs~\cite{shw22}. That is, FMGCH already has the advantage of being a more general algorithm compared to GCA. Hence, we perform two kinds of experiments. First, we compare FMGCH against GCA on unweighted graphs. Second, we study the performance of FMGCH on weighted graphs. In both cases, the GT procedure is included as the baseline.

We used four datasets in our experiments from which both unweighted and weighted graphs can be derived: the DIMACS, movingAI, SNAP, and the Waxman graphs. The DIMACS graphs\footnote{\url{https://mat.tepper.cmu.edu/COLOR/instances.html}} are a standard benchmark collection of unweighted graphs. They can be converted to weighted graphs by assigning an integer weight chosen uniformly at random from the interval $[1, 10]$ to each edge. The movingAI graphs model grid-worlds with obstacles~\cite{s12}. They can be used as unweighted graphs if the grid-worlds are assumed to be four-connected (with only horizontal and vertical connections) or as weighted graphs if the grid-worlds are assumed to be eight-connected (with additional diagonal connections). The SNAP dataset refers to the Stanford Large Network Dataset Collection~\cite{lk14}, some graphs from which were chosen as undirected unweighted graphs for experimentation in~\cite{shw22}. We use the same graphs for a fair comparison with GCA. Moreover, these graphs can be converted to weighted graphs by assigning an integer weight chosen uniformly at random from the interval $[1, 10]$ to each edge. Waxman graphs are used to generate realistic communication networks~\cite{w88}. Here, we generated Waxman graphs using NetworkX~\cite{hss08} with parameter values $\alpha = 100/|V|$ and $\beta = 0.1$, within a rectangular domain of $100 \times 100$, and with the weight on each edge set to the Euclidean distance between its endpoints. These graphs are naturally weighted but can be made unweighted by ignoring the weights.

Table~\ref{tab:convex_hull} shows the performance results of FMGCH, GCA, and the GT procedure. FMGCH is shown with three subdivisions, for $\kappa = 2$, $3$, and $4$. In each case, the table reports the precomputation time and the query time. The precomputation time is the time required by the FastMap component of FMGCH to generate the $\kappa$-dimensional Euclidean embedding. This precomputation is done only once per graph and can serve the purpose of answering many queries on the same graph with different input $S$. Only representative results are shown on selected graphs from each dataset. Queries were formulated on a given graph by randomly choosing $10$ vertices to constitute the input set $S$. In most cases, FMGCH required $\leq 10$ iterations for convergence.

In the first set of experiments (top half of Table~\ref{tab:convex_hull}), it is easy to observe that both FMGCH and GCA are orders of magnitude faster than GT, with the efficiency gaps being more pronounced on large graphs. In fact, FMGCH is also faster than GCA on large graphs. On the DIMACS dataset, both FMGCH and GCA produce very high-quality solutions. On the movingAI dataset, GCA does not produce high-quality solutions. While the same is true for FMGCH with $\kappa = 2$, the quality increases for $\kappa = 3$ and increases further for $\kappa = 4$. In fact, FMGCH with $\kappa = 4$ produces very high-quality solutions. On the SNAP dataset, GCA produces low-quality solutions and FMGCH produces better-quality solutions, particularly on recall. However, this quality deterioration is not related to the larger sizes of the SNAP graphs. In fact, on the Waxman graphs, which are also large, both FMGCH and GCA produce very high-quality solutions.

In the second set of experiments (bottom half of Table~\ref{tab:convex_hull}), GCA is not applicable at all. In contrast, FMGCH is fully applicable and can be evaluated using GT. Even here, it is easy to observe that FMGCH is orders of magnitude faster than GT. The qualities of the solutions that it produces follow the same patterns as in the unweighted case. On some weighted graphs, FMGCH and/or GT may run faster than on their unweighted counterparts. This happens because the number of shortest paths between a pair of vertices is usually more on unweighted graphs and, consequently, computing all of them is more expensive.

\section{Conclusions}

The graph convex hull problem is an important graph-theoretic problem that is analogous to the geometric convex hull problem. The two problems also share many important analogous properties and real-world applications. Yet, while the geometric convex hull problem is very well studied, the graph convex hull problem has not received much attention thus far. Moreover, while geometric convex hulls can be computed very efficiently in low-dimensional Euclidean spaces, folklore results for algorithms that compute graph convex hulls exactly make them prohibitively expensive on large graphs. In this chapter, we presented a FastMap-based algorithm for efficiently computing approximate graph convex hulls. Our FastMap-based algorithm utilizes FastMap's ability to facilitate geometric interpretations. While the naive version of our algorithm uses a single shot of such a geometric interpretation, the iterative version of our algorithm repeatedly interleaves the graph and geometric interpretations to reinforce one with the other. This iterative version was encapsulated in our solver, FMGCH, and experimentally compared against the state-of-the-art solver, GCA. On a variety of graphs, we showed that FMGCH not only runs several orders of magnitude faster than a highly-optimized exact algorithm but also outperforms GCA, both in terms of generality and the quality of the solutions produced. It is also faster than GCA on large graphs.

\begin{subappendices}

\section{Table of Notations}

\begin{table}[h]
\centering
\begin{tabular}{|l|p{0.75\linewidth}|}
\cline{1-2}
Notation &Description\\
\cline{1-2}
$G = (V, E, w)$ &An undirected edge-weighted graph, where $V$ is the set of vertices, $E$ is the set of edges, and for any edge $e \in E$, $w(e)$ is the non-negative weight on it.\\
\cline{1-2}
$S$ &A subset of the vertices $V$.\\
\cline{1-2}
$CH^G_S$ &The graph convex hull of $S$ in $G$.\\
\cline{1-2}
$C$ &The corners of a geometric convex hull.\\
\cline{1-2}
$f(|C|)$ &The function that returns the maximum number of faces of a convex polytope with $|C|$ corners.\\
\cline{1-2}
$\kappa$ &The user-specified number of dimensions of the FastMap embedding.\\
\cline{1-2}
$\epsilon$ &The threshold parameter in FastMap that is used to detect large values of $\kappa$ that have diminishing returns on the accuracy of approximating the pairwise distances between the vertices.\\
\cline{1-2}
$\overline{CH}^G_S$ &The required graph convex hull of $S$ in $G$ or an approximation of it.\\
\cline{1-2}
$CH_S$ &The corners of the geometric convex hull of a collection of points corresponding to $S$.\\
\cline{1-2}
\end{tabular}
\caption[Notations used in Chapter~\ref{ch:convex_hull}.]{Describes the notations used in Chapter~\ref{ch:convex_hull}.}
\label{tab:convex_hull_notations}
\end{table}

\end{subappendices}

\chapter{FastMapSVM: Combining FastMap and Support Vector Machines}
\label{ch:fastmapsvm}
NNs and related DL methods are currently at the leading edge of technologies used for classifying complex objects. However, they generally demand large amounts of data and time for model training; and their learned models can sometimes be difficult to interpret. In this chapter, we introduce FastMapSVM as an interpretable lightweight alternative ML framework for classifying complex objects. FastMapSVM combines the complementary strengths of FastMap and SVMs while extending the applicability of SVMs to domains with complex objects. It is particularly useful when it is easier to measure the dissimilarity between pairs of objects in the domain via a well-defined distance function on them than it is to identify and reason about complex characteristic features of individual objects. As a case study, we demonstrate the success of FastMapSVM in the Earthquake Science domain, where the objects are seismograms that need to be classified as earthquake signals or noise signals. We show that FastMapSVM outperforms other state-of-the-art methods for classifying seismograms when training data or time is limited. FastMapSVM also provides an insightful visualization of seismogram clustering behavior and the earthquake classification boundaries. We expect it to be viable for classification tasks in many other real-world domains as well.

\section{Introduction}

Various ML and DL methods, such as NNs, are popularly used for classifying complex objects. For example, a Convolutional NN (CNN) is used for classifying Sunyaev-Zel'dovich galaxy clusters~\cite{lhawtcn21}, a densely connected CNN is used for classifying images~\cite{hlvw17}, and a deep NN is used for differentiating the chest X-rays of Covid-19 patients from other cases~\cite{entfhmwpn20,saskk22}. However, they generally demand large amounts of data and time for model training; and their learned models can sometimes be difficult to interpret.

In this chapter, we introduce FastMapSVM\textemdash an interpretable ML framework for classifying complex objects\textemdash as a lightweight alternative to NNs for classification tasks in which training data or time is limited and a suitable distance function can be defined. While most ML algorithms learn to identify~\emph{characteristic features} of~\emph{individual} objects in a class, FastMapSVM leverages a domain-specific~\emph{distance function} on~\emph{pairs} of objects. It does this by combining the strengths of FastMap and SVMs. In its first phase, FastMapSVM invokes FastMap for mapping complex objects to points in a Euclidean space while preserving pairwise domain-specific distances between them. In its second phase, it invokes SVMs and kernel methods for learning to classify the points in this Euclidean space.

Our FastMapSVM framework is conceptually similar to the SupFM-SVM method~\cite{bka09}, the application of which to complex objects was anticipated by the original authors. Here, we present the first such application to complex objects by using FastMapSVM to classify seismograms. We compare the performance characteristics of FastMapSVM against state-of-the-art NN alternatives in the Earthquake Science domain on a commonly used benchmark dataset. We show that FastMapSVM outperforms these NN alternatives when training data or time is limited. FastMapSVM also provides an insightful visualization of seismogram clustering behavior and the earthquake classification boundaries. We further demonstrate that FastMapSVM can be easily deployed for different real-world classification tasks in the seismogram domain.

Backed by our results in the Earthquake Science domain, we propound FastMapSVM as a potentially advantageous alternative to NNs in other real-world domains as well when training data or time is limited and a suitable distance function can be defined.

\section{FastMapSVM}

\begin{figure}[t!]
\centering
\includegraphics[width=0.9\columnwidth]{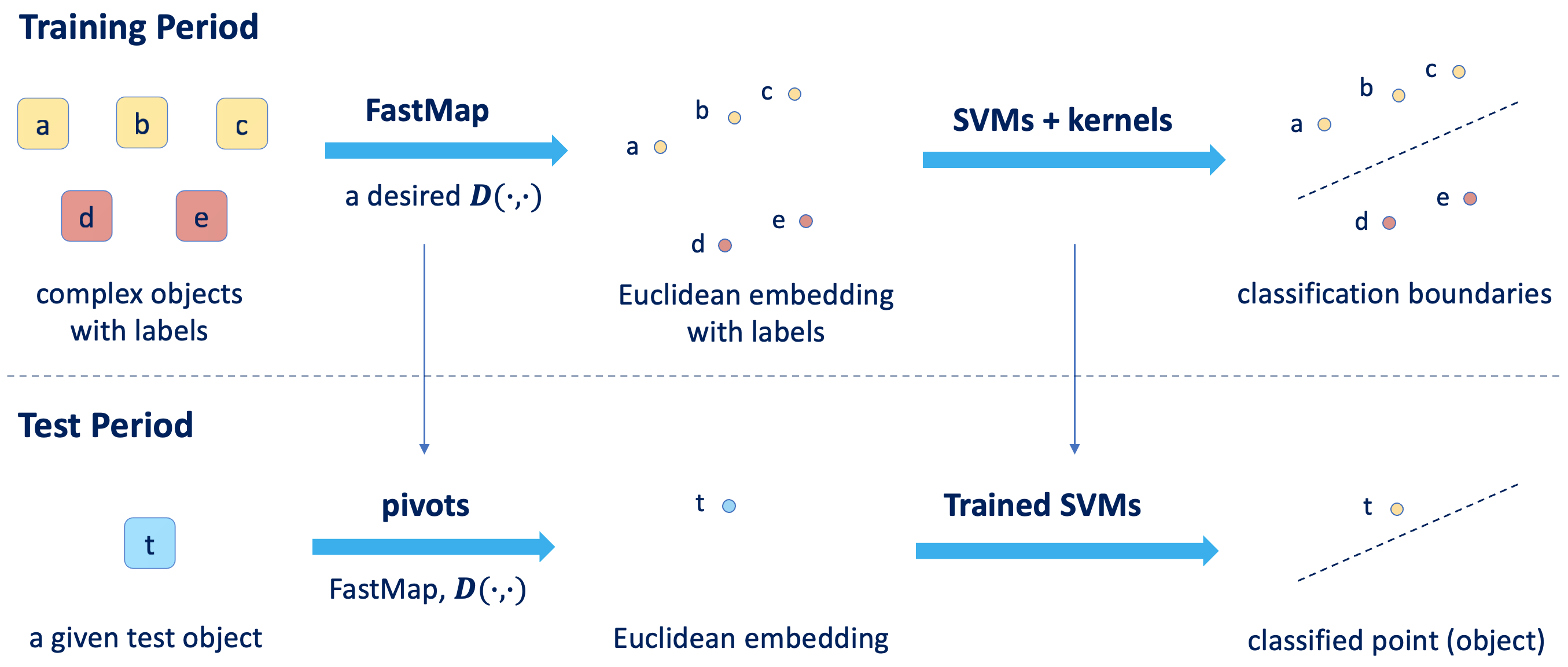}
\caption[An illustration of the overall architecture of FastMapSVM.]{Illustrates the overall architecture of FastMapSVM. The top half illustrates the training period of FastMapSVM. During this period, it ingests training instances in the form of complex objects with their associated labels. It first invokes FastMap with a desired domain-specific distance function for efficiently mapping the complex objects to points in a Euclidean space; it also stores explicit references to the pivots during this process. It then invokes and trains SVMs with kernel methods for learning to classify the resulting points in this Euclidean space. The bottom half illustrates the test period of FastMapSVM. During this period, it takes a given test object as input. It first uses the same distance function to compute the distances between the test object and the stored pivots generated in the training period. Using these distances, it maps the test object to a point in the same Euclidean space generated in the training period. It then invokes the trained SVMs to classify the point\textemdash and, consequently, the test object\textemdash in the Euclidean space.}
\label{fig:fastmapsvm}
\end{figure}

In this section, we propose FastMapSVM as a novel supervised ML framework that elegantly combines the strengths of FastMap and SVMs. FastMapSVM gets as input training instances in the form of complex objects with their associated labels. During training, it operates in two phases. In the first phase, it invokes FastMap for efficiently mapping the complex objects to points in a Euclidean space while preserving pairwise domain-specific distances between them. In the second phase, it invokes SVMs and kernel methods for learning to classify the points in this Euclidean space. Later, for classifying a test object, FastMapSVM once again operates in two phases. In the first phase, it calls the same distance function to compute the distances between the given test object and the pivots generated during the training period. These distances allow it to map the test object to a point in the same Euclidean space in which the training instances reside. In the second phase, FastMapSVM classifies this point using the trained SVMs. Figure~\ref{fig:fastmapsvm} illustrates the overall architecture of FastMapSVM during the training and test periods. Because of its methodology, it has several advantages over other ML frameworks.

First, FastMapSVM leverages domain-specific knowledge via a distance function. There are many real-world domains in which feature selection for~\emph{individual} objects is hard. While a domain expert can occasionally identify and incorporate domain-specific features of the objects to be classified, doing so becomes increasingly hard with increasing complexity of the objects. Therefore, many existing ML algorithms for classification find it hard to leverage domain knowledge when used off the shelf. However, in many real-world domains with complex objects, a distance function on~\emph{pairs} of objects is well defined and easy to compute. In such domains, FastMapSVM is more easily applicable than other ML algorithms that focus on the features of individual objects. FastMapSVM also enables domain experts to incorporate their domain knowledge via a distance function instead of relying entirely on complex ML models to infer the underlying structure in the data. As mentioned before, examples of such real-world objects include audio signals, seismograms, DNA sequences, ECGs, and MRIs. While these objects are complex and may have many subtle features that are hard to recognize, there exists a well-defined distance function on pairs of objects that is easy to compute. For instance, individual DNA sequences have many complex and subtle features but, as discussed in Chapter~\ref{ch:fastmap}, the edit distance~\cite{ry98} between two DNA sequences is well defined and easy to compute. Similarly, as discussed in Chapter~\ref{ch:fastmap}, the Minkowski distance~\cite{lkd07} is well defined for images and the cosine similarity~\cite{rka12} is well defined for text documents.

Second, FastMapSVM facilitates interpretability, explainability, and visualization. Many existing ML algorithms produce results that are hard to interpret or explain. For example, in NNs, a large number of interactions between neurons with nonlinearities makes a meaningful interpretation or explanation of the results very hard. In fact, the very complexity of the objects in the domain can hinder interpretability and explainability. FastMapSVM mitigates these challenges and thereby supports interpretability and explainability. While the objects themselves may be complex, FastMapSVM embeds them in a Euclidean space by considering only the distance function defined on pairs of objects. In effect, it simplifies the description of the objects by assigning Euclidean coordinates to them. Moreover, since the distance function is itself user-defined and encapsulates domain knowledge, FastMapSVM naturally facilitates interpretability and explainability. In fact, it even provides a perspicuous visualization of the objects and the classification boundaries between them. This aids human interpretation of the data and results. It also enables a human-in-the-loop framework for refining the processes of learning and decision making. As a hallmark, FastMapSVM produces the visualization very efficiently since it invests only linear time in generating the Euclidean embedding.

Third, FastMapSVM uses significantly smaller amounts of data and time for model training compared to other ML algorithms. While NNs and other ML algorithms store abstract representations of the training data in their model parameters, FastMapSVM stores explicit references to the pivots. While making predictions, objects in the test instances are compared directly to the pivots using the user-defined distance function. Therefore, FastMapSVM obviates the need to learn a complex transformation of the input data and thus significantly reduces the amount of data and time required for model training. Moreover, given $N$ training instances, that is, $N$ objects and their classification labels, FastMapSVM leverages $O(N^2)$ pieces of information via the distance function that is defined on every pair of objects. In contrast, ML algorithms that focus on individual objects leverage only $O(N)$ pieces of information.

Fourth, FastMapSVM extends the applicability of SVMs and kernel methods to complex objects. Generally speaking, SVMs are particularly good for classification tasks. When combined with kernel methods, they recognize and represent complex nonlinear classification boundaries very elegantly~\cite{s03}. Moreover, soft-margin SVMs with kernel methods~\cite{pc13} can be used to recognize both outliers and inherent nonlinearities in the data. While the SVM machinery is very effective, it requires the objects in the classification task to be represented as points in a Euclidean space. Often, it is very difficult to represent complex objects as precise geometric points without introducing inaccuracy or losing domain-specific representational features. In such cases, deep NNs have gained more popularity compared to SVMs for the reason that it is unwieldy for SVMs to represent all the features of complex objects in Euclidean space. However, FastMapSVM revives the SVM approach by leveraging a distance function and creating a low-dimensional Euclidean embedding of the complex objects.

\section{Case Study: FastMapSVM for Classifying Seismograms}

In this section, we illustrate the advantages of FastMapSVM in the context of classifying seismograms. This is a particularly illustrative domain because seismograms are complex objects with subtle features indicating diverse energy sources such as earthquakes, ocean-Earth interactions, atmospheric phenomena, and human-related activities. There are two fundamental, perennial questions in seismology: (a) Does a given seismogram record an earthquake?~and (b) Which type of wave motion, that is,~\emph{compressional} (P-wave) versus~\emph{shear strain} (S-wave), is predominant in an earthquake seismogram? In Earthquake Science, answering these questions is referred to as ``detecting earthquakes'' and ``identifying phases'', respectively. The development of efficient, reliable, and automated solution procedures that can be easily adapted to new environments is essential for modern research and engineering applications in this field, such as in building Earthquake Early Warning Systems. Moreover, an ML framework that imposes only modest demands on the training data aids the analysis of signal classes, such as ``icequakes'', stick-slip events at the base of landslides, and nuisance signals recorded during temporary seismometer deployments, for which large training datasets are unavailable.

Towards this end, we have shown that FastMapSVM is indeed a viable ML framework~\cite{wslkn23}. Through experiments, we have also demonstrated that it (a) outperforms state-of-the-art NNs for classifying seismograms when training data or time is limited, (b) can be rapidly deployed for different real-world classification tasks, and (c) is robust against noisy perturbations to the input signals. However, for the purposes of this chapter, we avoid a deep dive into the details of the problem domain: Instead, we focus only on a few exemplary results that illustrate the benefits of FastMapSVM. As mentioned before, in-depth details of the application of FastMapSVM in the Earthquake Science domain can be found in~\cite{wslkn23}.

\subsection{Distance Function on Seismograms}

\begin{figure}[t!]
\centering
\includegraphics[width=0.8\textwidth]{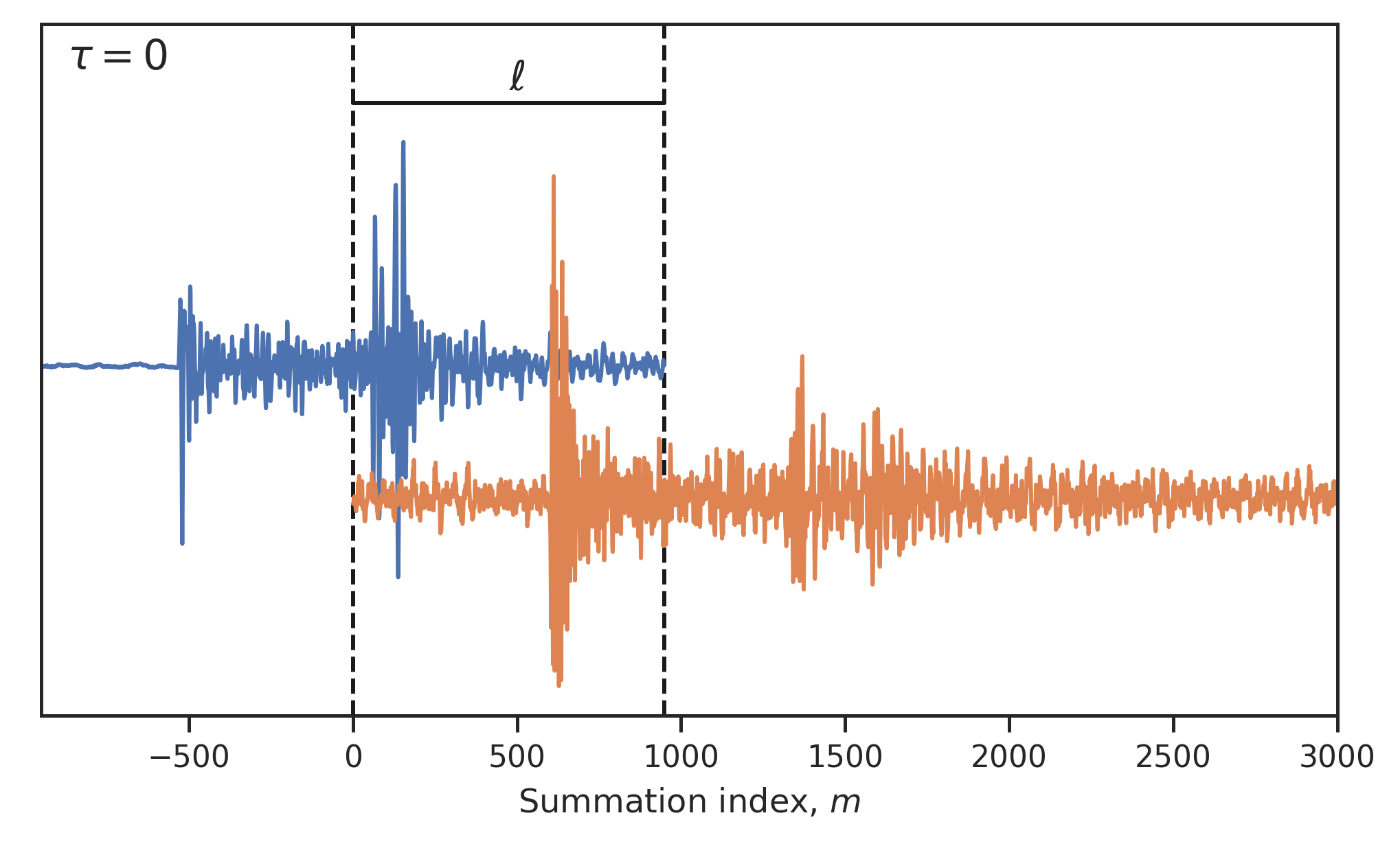}
\caption[An illustration of a distance function on seismograms.]{Illustrates some quantities in Equation~\ref{eqn:cross_correlation_operator}. The longer waveform in orange is aligned with a shorter waveform in blue to examine the normalized cross-correlation with respect to $\tau = 0$. The quantity $\ell$ measures about half the length of the shorter waveform.}
\label{fig:xcorr_demo}
\end{figure}

The FastMap component of FastMapSVM requires the distance function to be symmetric, yield non-negative values for all pairs of seismograms, and yield $0$ for identical seismograms. In the subsection, we describe one such appropriate distance function on seismograms that can be used in the Earthquake Science domain.

In Earthquake Science, the normalized cross-correlation operator, denoted here by $\star$, is popularly used to measure similarity between two waveforms. For two $0$-mean, single-component seismograms $O_i$ and $O_j$ with lengths $n_i$ and $n_j$, respectively, and starting with index $0$, the normalized cross-correlation is defined with respect to a lag $\tau$ as follows:
\begin{align}
(O_i \star O_j)[\tau] & \triangleq \frac{1}{\sigma_i \sigma_j} \sum_{m = 0}^{n_i - 1} O_i[m]\widehat{O}_j[m + \ell - \tau],
\label{eqn:cross_correlation_operator}
\end{align}
\noindent in which, without loss of generality, we assume that $n_i \ge n_j$. $\sigma_i$ and $\sigma_j$ are the standard deviations of $O_i$ and $O_j$, respectively. Moreover, $\ell$ and $\widehat{O}_j$ are defined as follows:
\begin{equation}
\ell \triangleq \frac{n_j - n_j \Mod{2}}{2} - (n_i \Mod{2})(1 - n_j \Mod{2})
\label{eqn:ell}
\end{equation}
\noindent and
\begin{equation}
\widehat{O}_j[m] \triangleq
\begin{cases}
O_j[m] & \text{if}~0 \le m < n_j\\
0 & \text{otherwise}
\end{cases}.
\end{equation}
\noindent The quantity $\ell$ in Equation~\ref{eqn:ell} is defined as a subtraction. The first term is approximately half of $n_j$. The second term is $0$ or $1$ depending on whether $n_i$ and $n_j$ are odd or even. Therefore, $\ell$ measures about half the length of the shorter waveform and ensures at least $50\%$ overlap between the two waveforms for computing the normalized cross-correlation at any $\tau$. Figure~\ref{fig:xcorr_demo} shows a schematic illustration of some quantities in Equation~\ref{eqn:cross_correlation_operator}.

Equipped with this knowledge, we first define the following distance function that is appropriate for waveforms with a single component:
\begin{align}
\mathcal{D}(O_i, O_j) & \triangleq 1 - \max_{0 \leq \tau \leq n_i - 1}\left|(O_i \star O_j)[\tau]\right|.
\end{align}
\noindent Based on this, we define the following distance function that is appropriate for waveforms with $L$ components:
\begin{align}
\mathcal{D}(O_i, O_j) & \triangleq 1 - \frac{1}{L} \max_{0 \leq \tau \leq n_i - 1}\left|\sum_{l = 1}^L (O^l_i \star O^l_j)[\tau]\right|.
\label{eqn:correlation_distance}
\end{align}
\noindent Here, each component $O^l_i$ of $O_i$, or $O^l_j$ of $O_j$, is a channel representing a $1$-dimensional data stream. A channel is associated with a single standalone sensor or a single sensor in a multi-sensor array. $L$ is typically equal to $3$, since three-component (3C) seismograms are popularly used in earthquake datasets. The same value of $L$ is also extensively used in matched filters in Earthquake Science~\cite{gr06,sbi07,seh16,sfkzyzm19}.

We can also use a variety of other distance functions on seismograms. In fact, three other distance functions, the Wasserstein distance, a maxflow-based distance, and the Minkowski distance, are studied in~\cite{slwk23}. We can even potentially derive new distance functions from the Jensen-Shannon~\cite{mppp97} or the Kullback-Leibler~\cite{vh14} measures of divergence. Furthermore, we can also encapsulate more domain-specific knowledge in the distance functions, if required.

\subsection{Earthquake Datasets}

We assess the performance of FastMapSVM using seismograms from two datasets. All seismograms used in this section record ground velocity at a sampling rate of~\SI{100}{\per\second} and are bandpass filtered between~\SI{1}{\hertz} and~\SI{20}{\hertz} before analysis using a $0$-phase Butterworth filter with four poles; we refer to this frequency band as our passband.

\subsubsection{The Stanford Earthquake Dataset}

The Stanford Earthquake Dataset (STEAD)~\cite{mszb19} is a benchmark dataset with more than $1.2 \times 10^6$ carefully curated 3C~\SI{60}{\second} seismograms for training and testing algorithms in Earthquake Science. We select a balanced subset of $65,536$ seismograms from the STEAD comprising $32,768$ earthquake seismograms and $32,768$ noise seismograms. Earthquake seismograms record ground motions induced by a nearby earthquake; whereas noise seismograms record no known earthquake-related ground motions. We randomly select earthquake seismograms using a selection probability that is inversely proportional to the kernel density estimate of the $5$-dimensional joint distribution over (a) the epicentral distance, (b) the event magnitude, (c) the event depth, (d) the time interval between the P- and S-wave arrivals, and (e) the signal-to-noise ratio (SNR). This scheme is designed to yield a broad distribution of seismograms. All earthquake seismograms are recorded by a seismometer within~\SI{100}{\kilo\meter} of the epicenter, have a hypocentral depth of less than~\SI{30}{\kilo\meter}, and have P- and S-wave arrival times manually identified by a trained analyst. Distributions in this subset that are skewed towards shallow earthquakes with low magnitude and low SNR reflect the distribution of natural earthquakes and their recordings. Noise seismograms are randomly selected to maximize the geographic diversity of recording locations.

From this base dataset of $65,536$ seismograms, we draw a simple random sample (SRS) of $16,384$ earthquake seismograms and an equal-sized SRS of noise seismograms for model training. The $32,768$ remaining seismograms make up the test dataset. The test seismograms are trimmed to~\SI{30}{\second} per seismogram, including an amount of time uniformly distributed between~\SI{4}{\second} and~\SI{15}{\second} preceding the P-wave arrival for earthquake seismograms. We note that both the training dataset and the test dataset have balanced numbers of earthquake seismograms and noise seismograms. We recursively draw SRSs from the training dataset to create multiple smaller balanced training datasets, each of which is half the size of the set from which it is drawn. Thus, we have nested balanced training datasets with sample sizes $2^{15}, 2^{14} \ldots 2^6$.

\subsubsection{Ridgecrest Dataset}

The Ridgecrest dataset comprises data recorded by station CI.CLC of the Southern California Seismic Network on July 5th, 2019, the first day of the aftershock sequence following the 2019 Ridgecrest, CA, earthquake pair and on December 5th, 2019, five months after the mainshocks. We use the earthquake catalog published by the Southern California Earthquake Data Center to extract $512$ 3C~\SI{8}{\second} seismograms, $256$ of which record both P- and S-wave phase arrivals from a nearby\textemdash that is, with epicentral distance between~\SI{4.5}{\kilo\meter} and~\SI{27.6}{\kilo\meter}\textemdash aftershock and the remaining $256$ of which record only noise. All $512$ of these seismograms are recorded on July 5th, 2019. Earthquake magnitudes represented in the Ridgecrest dataset range between $0.5$ and $4.0$, earthquake depths range between~\SI{900}{\meter} above sea level and~\SI{9.75}{\kilo\meter} below sea level, and SNRs range between~\SI{-8}{\deci\bel} and~\SI{73}{\deci\bel}. The maximum peak ground acceleration recorded in the Ridgecrest dataset is~\SI{0.197}{\meter\per\second^2}.

\subsection{Experimental Results}

We now present experimental results on the STEAD and the Ridgecrest dataset. Whereas the analysis on the STEAD demonstrates FastMapSVM's performance on a benchmark, the analysis on the Ridgecrest dataset provides an example of a more realistic use case of FastMapSVM: After handpicking a small number of earthquake and noise signals\textemdash a task that even a novice analyst can perform in a few hours\textemdash continually arriving seismic data can be automatically scanned for additional earthquake signals. Hence, even when earthquake signals are difficult to discern by the human eye, FastMapSVM can often reliably detect them.

\subsubsection{Results on the Stanford Earthquake Dataset}

\begin{figure}[t!]
\centering
\includegraphics[width=\textwidth]{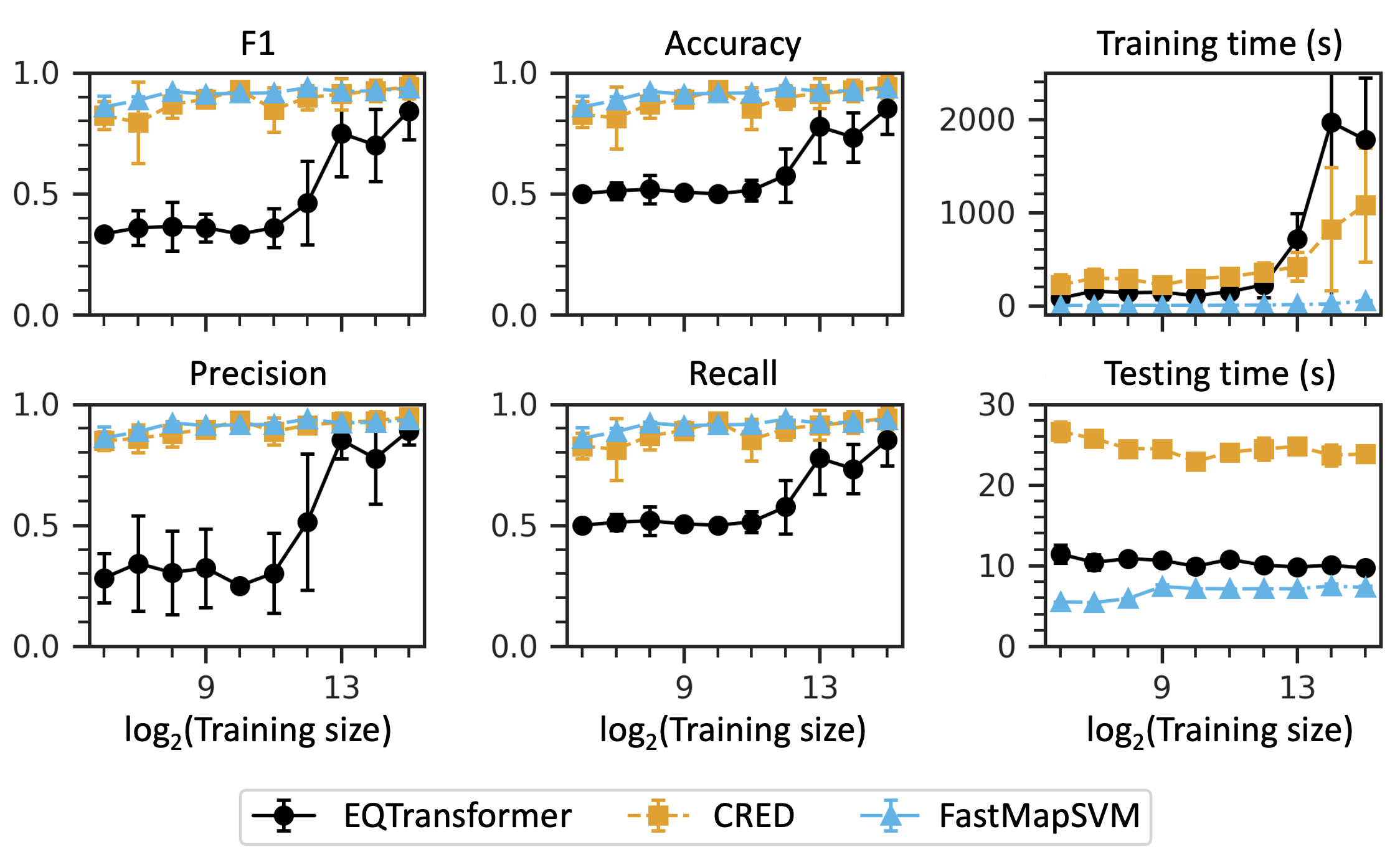}
\caption[The performance statistics of FastMapSVM, EQTransformer, and CRED on the Stanford Earthquake Dataset with varying training data size.]{Shows the performance statistics of FastMapSVM, EQTransformer, and CRED on the STEAD with varying training data size. Error bars represent the standard deviation of the measurements. (The recall is identical to the accuracy in our case.)}
\label{fig:STEAD_performance}
\end{figure}

The EQTransformer model~\cite{mezcb20} is a DL model trained for simultaneously detecting earthquakes and identifying phase arrivals. It is arguably the most accurate publicly available model for this pair of tasks. The authors of EQTransformer report perfect precision and recall scores for detecting earthquakes in $10\%$ of the STEAD waveforms after training its roughly $3.72 \times 10^5$ model parameters with $85\%$ of the STEAD waveforms; $5\%$ of the STEAD waveforms were reserved for model validation.\footnote{We note that the authors of EQTransformer used a version of the STEAD with $1 \times 10^6$ and $3 \times 10^5$ earthquake and noise waveforms, respectively, which differs slightly from the newer version of the STEAD that we use.} The CRED model~\cite{mzsb19} is another DL model trained for detecting earthquakes, which scored perfect precision and $0.96$ recall using the same training and test data as EQTransformer~\cite{mezcb20}. However, the CRED model does not identify phase arrivals. We choose these two DL models for comparison because EQTransformer is popularly used~\cite{jffl21,jzwpm22} and represents the state of the art in general practice, CRED is designed specifically for detecting earthquakes, and the pretrained models are readily available through the SeisBench package~\cite{wmtrlbdghjmss22}.

To compare the performance of FastMapSVM against EQTransformer and CRED, we train multiple instances of each model using varying amounts of training data and test them on the same test data: the $32,768$ test seismograms selected from the STEAD, as described above. All models are trained and tested using an NVIDIA RTX A6000 GPU. We train and test each model multiple times for each training data size to estimate the statistics for the following performance measures: F1 score, accuracy, precision, recall, training time, and testing time. For FastMapSVM, we set the number of trials to $20$ for each training data size; but, for EQTransformer and CRED, we limit the number of trials to $10$ for each training data size because training them is prohibitively time-consuming for large training data sizes. The training data for FastMapSVM are trimmed to~\SI{30}{\second} per seismogram, including~\SI{4}{\second} of data preceding the P-wave arrival for earthquake seismograms. The FastMapSVM model uses a $4$-dimensional Euclidean embedding. We set the probability threshold for the decision boundary to $0.5$. The performance scores are averaged over both labels. Figure~\ref{fig:STEAD_performance} shows the statistics of the various performance measures.

FastMapSVM consistently outperforms EQTransformer and CRED using less training time for all training data sizes. FastMapSVM training times are $1$-$3$ orders of magnitude smaller than those for EQTransformer and CRED. The respective performances of EQTransformer and CRED approach that of FastMapSVM as the training data size increases; however, they do so at the cost of rapidly increasing training times. The testing times of EQTransformer and CRED are $1.33$-$2.08$ and $3.17$-$4.82$ times the testing time of FastMapSVM, respectively. FastMapSVM also exhibits more stable performance, that is, less variance between trials, than the other two models because the final performance of these models is sensitive to the initial random values of their model parameters.

The performance of FastMapSVM can be further improved by increasing the dimensionality of the Euclidean embedding, as demonstrated in~\cite{wslkn23}.

\subsubsection{Results on the Ridgecrest Dataset}

\begin{figure}[t!]
\centering
\includegraphics[width=\textwidth]{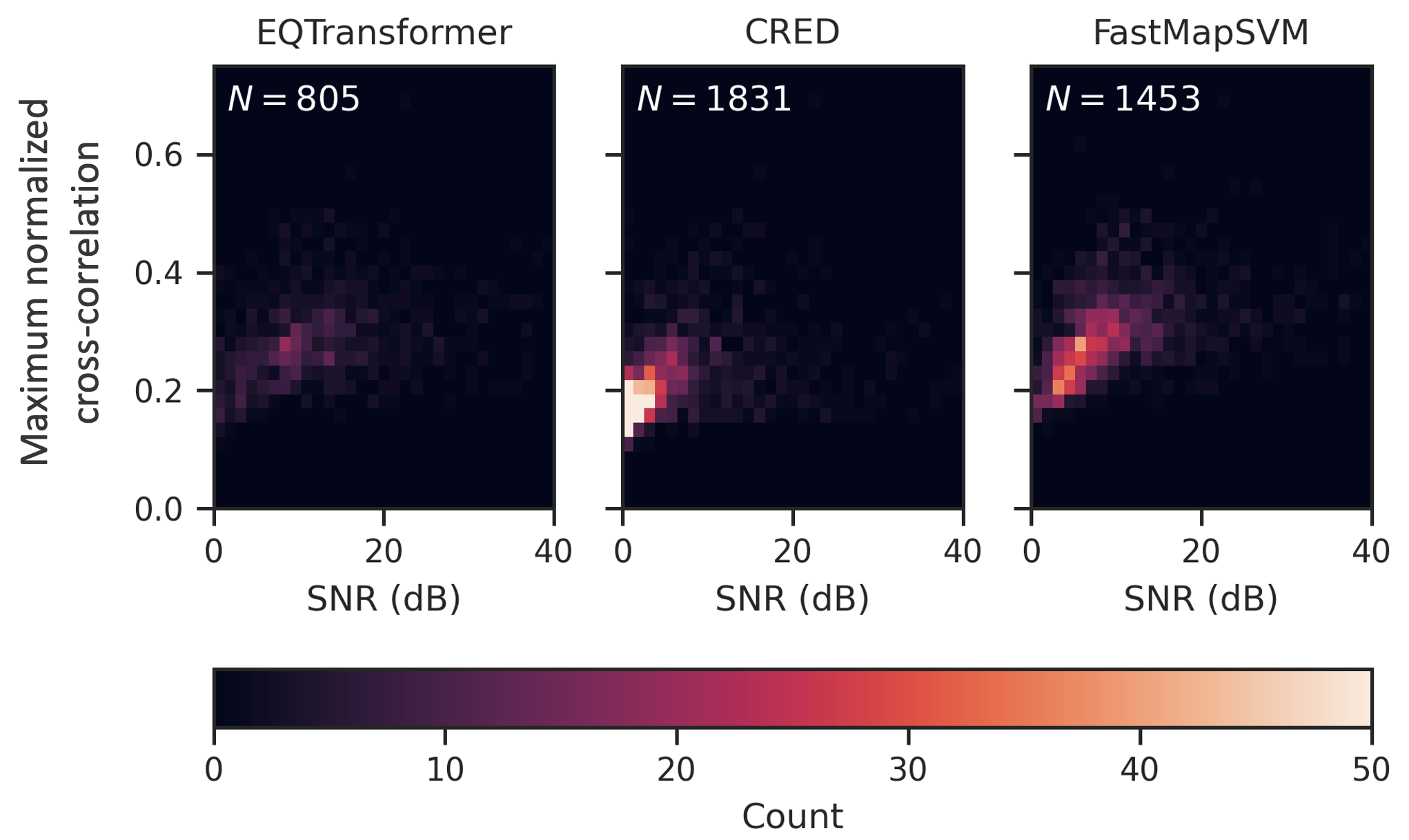}
\caption[A comparison of the automatic scanning results produced by EQTransformer, CRED, and FastMapSVM on the Ridgecrest dataset.]{Shows a comparison of the automatic scanning results produced by EQTransformer, CRED, and FastMapSVM on the Ridgecrest dataset. The panels show the joint distribution of the maximum SNR and the maximum normalized cross-correlation coefficient for detections registered by EQTransformer, CRED, and FastMapSVM in an automatic scan of~\SI{24}{\hour} of data.}
\label{fig:blind_scan}
\end{figure}

We demonstrate a use case-inspired application of FastMapSVM using a $32$-dimensional model trained on $256$ earthquake seismograms and $256$ noise seismograms from the Ridgecrest dataset. We apply the model to~\SI{8}{\second} windows extracted from a~\SI{24}{\hour} continuous 3C seismogram with $25\%$ overlap and register detections for windows with detection probability $> 0.95$. The test seismogram was recorded by station CI.CLC between 00:00:00 and 23:59:59 (UTC) on December 5th, 2019. We also apply the pretrained EQTransformer~\cite{mezcb20} and CRED~\cite{mzsb19} models on the same data.

For each detection, we compute two quantities: (1) the maximum SNR and (2) the maximum normalized cross-correlation coefficient measured against the $256$ earthquake seismograms used to train FastMapSVM. We define the SNR as follows:
\begin{equation}
\text{SNR} = 10 \log_{10} \left( \frac{P_{signal}}{P_{noise}} \right).
\label{eqn:snr}
\end{equation}
\noindent Here, $P_{signal}$ and $P_{noise}$ represent the average power of the signal and noise, respectively, which are measured in~\SI{1}{\second} and~\SI{10}{\second} sliding windows, respectively.

Figure~\ref{fig:blind_scan} shows a comparison of the automatic scanning results produced by EQTransformer, CRED, and FastMapSVM. CRED registers the largest number of detections ($1,831$), EQTransformer registers the fewest ($805$), and FastMapSVM registers an intermediate number ($1,453$). Although CRED registers the largest number of detections, many of them correspond to signals with very low SNR ($< 2.5$) and low normalized cross-correlation coefficients ($< 0.2$). This implies that many of its detections are likely false positives. Indeed, visual inspection of an SRS of these detections confirms this. FastMapSVM also registers a significant number of detections with relatively low SNR ($< 5$); however, these detections are generally associated with higher normalized cross-correlation coefficients ($> 0.2$) and marginally higher SNR ($> 2.5$).

The majority of detections registered by FastMapSVM are associated with low to moderate SNR between $2.5$ and $10$ and normalized cross-correlation coefficients between $0.2$ and $0.4$. This is an expected consequence of the Gutenberg-Richter statistics that describe earthquake magnitude-frequency distributions. However, perhaps surprisingly, FastMapSVM also registers a greater number of detections with high SNR ($> 10$) than both EQTransformer and CRED. Visual inspection confirms that FastMapSVM seldom misses a high-SNR event detected by EQTransformer or CRED, whereas EQTransformer and CRED do occasionally miss high-SNR events detected by FastMapSVM.

Results presented in Figure~\ref{fig:blind_scan} suggest that (1) EQTransformer has the lowest detection and false detection rates, (2) CRED has the highest detection and false detection rates, (3) FastMapSVM has relatively high detection and low false detection rates, and (4) FastMapSVM detects high-SNR events with greater fidelity than EQTransformer and CRED.

\section{Conclusions}

In this chapter, we introduced FastMapSVM as an interpretable ML framework that combines the complementary strengths of FastMap and SVMs. We posited that it is an advantageous, lightweight alternative to existing methods, such as NNs, for classifying complex objects when training data or time is limited. FastMapSVM offers several advantages. First, it enables domain experts to incorporate their domain knowledge using a distance function. This avoids relying on complex ML models to infer the underlying structure in the data entirely. Second, because the distance function encapsulates domain knowledge, FastMapSVM naturally facilitates interpretability and explainability. In fact, it even provides a perspicuous visualization of the objects and the classification boundaries between them. Third, FastMapSVM uses significantly smaller amounts of data and time for model training compared to other ML algorithms. Fourth, it extends the applicability of SVMs and kernel methods to domains with complex objects.

We demonstrated the efficiency and effectiveness of FastMapSVM in the context of classifying seismograms. On the STEAD, we showed that FastMapSVM performs comparably to state-of-the-art NN models in terms of the precision, recall/accuracy, and the F1 score. It also uses significantly smaller amounts of data and time for model training compared to other methods. Moreover, it can also be faster at testing time. On the Ridgecrest dataset, we demonstrated its ability to reliably detect new ``microearthquakes'' that are otherwise difficult to detect even by the human eye.

\chapter{FastMapSVM in the Constraint Satisfaction Problem Domain}
\label{ch:constraint}
Recognizing the satisfiability of CSPs is NP-hard. Although several ML approaches have attempted this task by casting it as a binary classification problem, they have had only limited success for a variety of challenging reasons. First, the NP-hardness of the task does not make it amenable to straightforward approaches. Second, CSPs come in various forms and sizes while many ML algorithms impose the same form and size on their training and test instances. Third, the representation of a CSP instance is not unique since the variables and their domain values are unordered. In this chapter, we demonstrate the success of the FastMapSVM framework\textemdash proposed in the previous chapter\textemdash on the task of predicting the satisfiability of CSP instances. Since FastMapSVM leverages a distance function on pairs of objects in the problem domain, we define a novel distance function between two CSP instances using maxflow computations. This distance function is well defined for CSPs of different sizes. It is also invariant to the ordering on the variables and their domain values. Therefore, our framework has broader applicability compared to other approaches. We discuss various representational and combinatorial advantages of FastMapSVM. Through experiments, we also show that it outperforms other state-of-the-art ML approaches.

\section{Introduction}

Constraints constitute a very natural and general means for formulating regularities in the real world. A fundamental combinatorial structure used for reasoning with constraints is that of the CSP. The CSP formally models a set of variables, their corresponding domains, and a collection of constraints between subsets of the variables. Each constraint restricts the set of allowed combinations of values of the participating variables. A solution of a given CSP instance is an assignment of values to all the variables from their respective domains such that all the constraints are satisfied. Technologies for efficiently solving CSPs bear immediate and important implications on how fast we can solve computational problems that arise in several other areas of research, including Computer Vision, spatial and temporal reasoning, model-based diagnosis, planning and scheduling, and language understanding.

Unfortunately, solving CSPs is NP-hard since they generalize the Satisfiability (SAT) problem. Although many technologies have been developed for solving CSPs in practice~\cite{d03}, they do not sufficiently harness the power of ML techniques. While there have been a lot of attempts to apply ML techniques to CSPs, none of these attempts have yielded spectacular results: They do not consistently produce high-quality outcomes. Some ML approaches used in the CSP domain include the application of SVMs~\cite{ahs10}, linear regression~\cite{xhhl08}, decision tree learning~\cite{gm04,gjkmmnp10}, clustering~\cite{pt07,kmst10}, $k$-nearest neighbors~\cite{ohhno08}, and others~\cite{k16}. However, most of these approaches have had limited success for a variety of challenging reasons.

First, from a complexity theory perspective, the NP-hardness of the task does not make it amenable to straightforward ML approaches. For example, since an NN is essentially a continuous differentiable form of a circuit, it is not straightforward to make NNs effective in the CSP domain. Second, CSPs come in various forms and sizes while many ML algorithms use an architectural framework that imposes the same form and size on their training and test instances. For example, an NN may have a fixed input layer that it uses for the training and test instances alike. Third, the representation of a CSP instance is not unique since the variables and their domain values are unordered. This poses a significant combinatorial challenge for ML algorithms since they have to learn the permutation invariance with respect to orderings on the variables and their domain values. For example, an NN may pose the overhead of having to be trained on all permutations of the same CSP instance to become effective.

In this chapter, we consider the problem of predicting the satisfiability of CSP instances using ML. In ML terminology, this is essentially a binary classification problem defined on CSPs with the two possible classification labels `satisfiable' and `unsatisfiable'. This classification problem is a cornerstone task for addressing the combinatorics of CSPs. It also serves as a stepping stone for the task of solving CSPs. In fact, any ML framework expected to be viable for solving CSPs should likely first demonstrate its success on solving the aforementioned classification problem on CSPs.

We propose to solve the above classification problem on binary CSPs\footnote{Binary CSPs have at most two variables per constraint but are allowed to have non-Boolean variables. Binary CSPs are representationally as powerful as general CSPs.} using FastMapSVM. In applying FastMapSVM to the CSP domain, we define a novel distance function between two CSP instances. This distance function uses maxflow computations and is well defined for CSP instances of different sizes. It is also invariant to the ordering on the variables and their domain values in the CSP instances. Therefore, FastMapSVM has broader applicability compared to other ML approaches in the CSP domain. Moreover, since it encapsulates the intelligence of FastMap, SVMs, kernel methods, and maxflow computations, it is able to significantly outperform competing ML approaches. It is also able to outperform procedures that invest polynomial time in establishing local consistency\textemdash such as arc-consistency\textemdash to discover unsatisfiable CSP instances. This demonstrates that a trained FastMapSVM model acquires an intelligence beyond that of polynomial-time procedures.\footnote{This is an important hallmark of an ML algorithm. In~\cite{xkk18}, a deep NN model is presented to recognize the satisfiability of CSP instances with Boolean variables and binary constraints. However, this class of CSP instances is equivalent to 2-SAT and can be solved in polynomial time, diminishing the advantages of an ML framework over polynomial-time reasoning.} We discuss various other representational and combinatorial advantages of FastMapSVM and, through experiments, we also demonstrate its superior performance.

\section{Preliminaries and Definitions}

\begin{figure}[t!]
\centering
\begin{minipage}[b]{0.45\textwidth}
\centering
\includegraphics[width=0.6\textwidth]{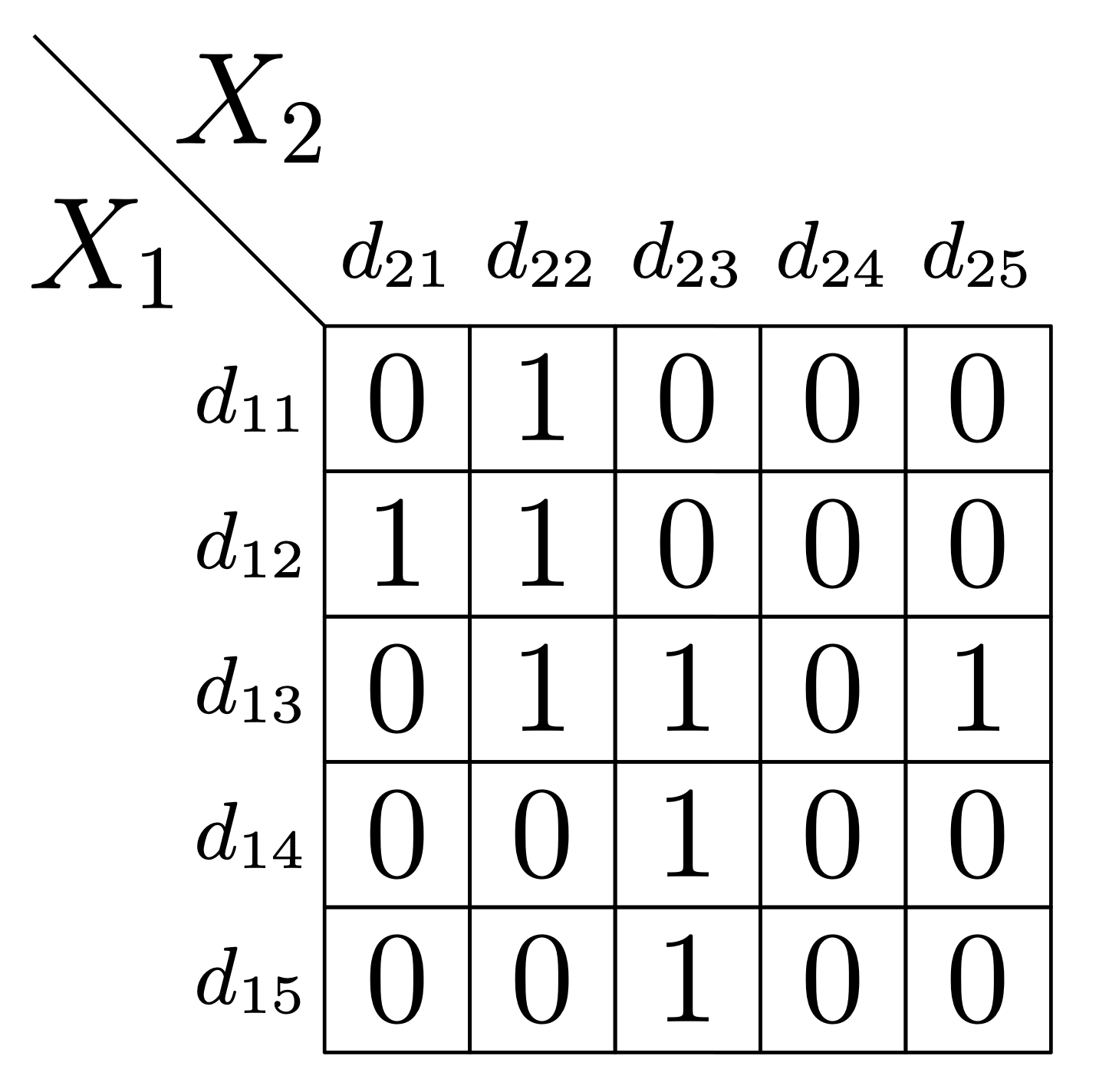}
\end{minipage}
\hspace{0.02\textwidth}
\begin{minipage}[b]{0.45\textwidth}
\centering
\includegraphics[width=0.8\textwidth]{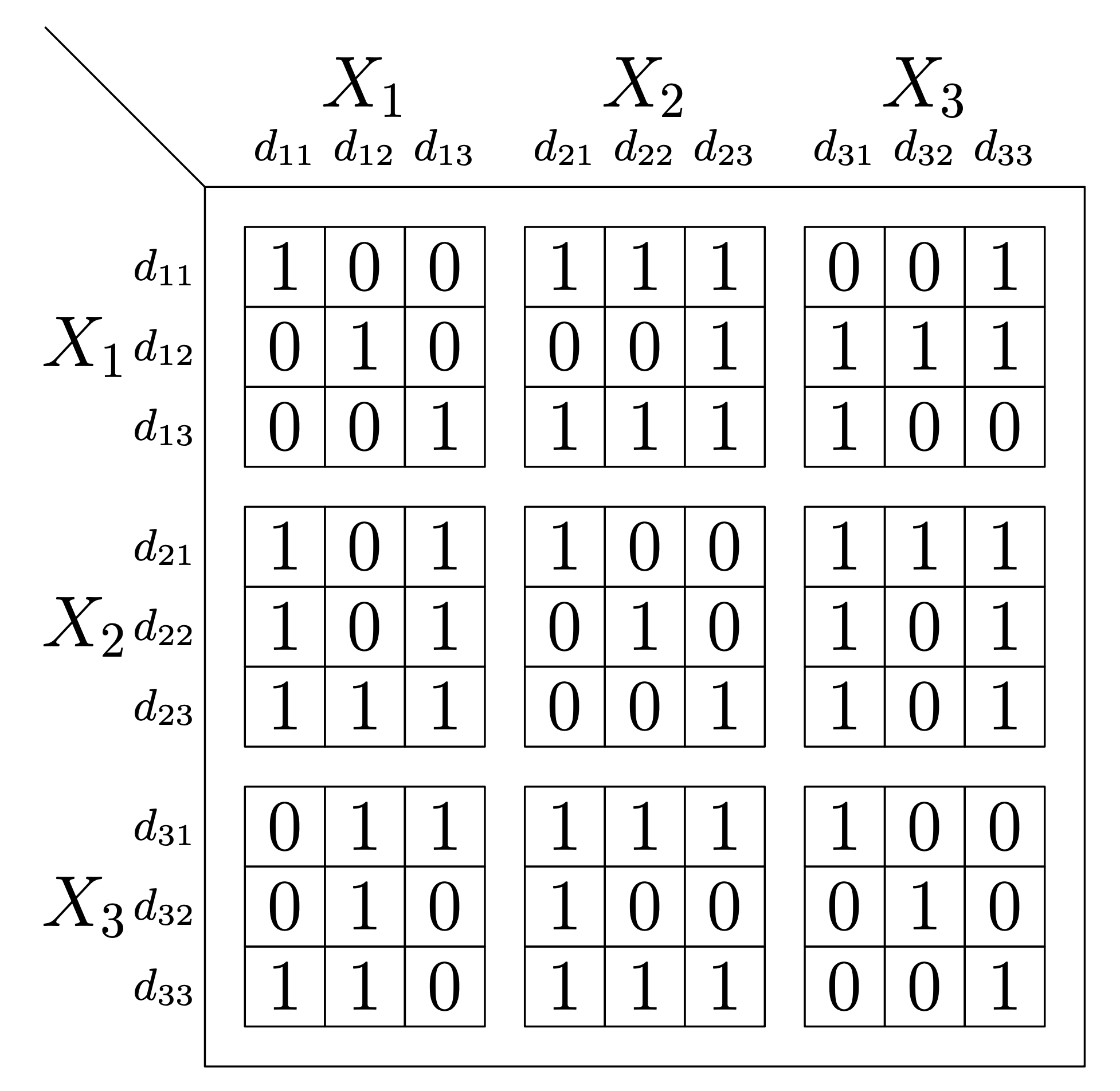}
\end{minipage}
\caption[The $(0, 1)$-matrix representation of a constraint and a Constraint Satisfaction Problem.]{Shows the $(0, 1)$-matrix representation of a constraint and a CSP instance. The left panel shows the $(0, 1)$-matrix representation of a single constraint on $X_1$ and $X_2$. The right panel shows the $(0, 1)$-matrix representation of an entire binary CSP instance.}
\label{fig:csp_rep}
\end{figure}

A CSP instance is defined by a triplet $\langle \mathcal X, \mathcal D, \mathcal C \rangle$, where $\mathcal X = \{X_1, X_2 \ldots X_N\}$ is a set of variables and $\mathcal C = \{C_1, C_2 \ldots C_M\}$ is a set of constraints on subsets of them. Each variable $X_i$ is associated with a discrete-valued domain $D_i \in \mathcal D$, and each constraint $C_i$ is a pair $\langle S_i, R_i \rangle$ defined on a subset of variables $S_i \subseteq \mathcal X$, called the~\emph{scope} of $C_i$. $R_i \subseteq D_{S_i}$ ($D_{S_i} = \times_{X_j \in S_i}D_j$) denotes all compatible tuples of $D_{S_i}$ allowed by the constraint. The absence of a constraint on a certain subset of the variables is equivalent to a constraint on the same subset of the variables that allows all combinations of values to them. A~\emph{solution} of a CSP instance is an assignment of values to all the variables from their respective domains such that all the constraints are satisfied. A binary CSP instance has at most two variables per constraint. Binary CSPs are representationally as powerful as general CSPs~\cite{d92}.

A binary CSP is~\emph{arc-consistent} if and only if for all variables $X_i$ and $X_j$, and for every instantiation of $X_i$, there exists an instantiation of $X_j$ such that the direct constraint between them is satisfied. Similarly, a binary CSP is~\emph{path-consistent} if and only if for all variables $X_i$, $X_j$ and $X_k$, and for every instantiation of $X_i$ and $X_j$ that satisfies the direct constraint between them, there exists an instantiation of $X_k$ such that the direct constraints between $X_i$ and $X_k$ and between $X_j$ and $X_k$ are also satisfied.

For a given binary CSP instance, we can build a matrix representation for it using a simple mechanism. First, we assume that the domain values of each variable are ordered in some way. (We can simply use the order in which the domain values of each variable are specified.) Under such an ordering, we can represent each binary constraint as a $2$-dimensional matrix with all its entries set to either $1$ or $0$ based on whether the corresponding combination of values to the participating variables is allowed or not by that constraint. The left panel of Figure~\ref{fig:csp_rep} shows the $(0, 1)$-matrix representation of a binary constraint between two variables $X_1$ and $X_2$ with domain sizes of $5$ each. The combination of values $(X_1 \leftarrow d_{12}, X_2 \leftarrow d_{21})$ is an allowed combination, and the corresponding entry in the matrix is therefore set to $1$. However, the combination of values $(X_1 \leftarrow d_{14}, X_2 \leftarrow d_{22})$ is a disallowed combination, and the corresponding entry is therefore set to $0$. In general, $d_{ip}$ denotes the $p$-th domain value of $X_i$ assuming an index ordering on the domain values of $X_i$.

The $(0, 1)$-matrix representation of an entire binary CSP instance can be constructed simply by stacking up the $(0, 1)$-matrix representations of the individual constraints into a bigger ``block'' matrix. The right panel of Figure~\ref{fig:csp_rep} illustrates how a binary CSP instance on three variables $X_1$, $X_2$ and $X_3$ can be represented as a ``mega-matrix'' with three sets of rows and three sets of columns. Each block-entry inside this mega-matrix is the $(0, 1)$-matrix representation of the direct constraint between the corresponding row and column variables. In essence, therefore, the $(0, 1)$-matrix representation of an entire binary CSP instance has $\sum_{i=1}^N |D_i|$ rows and $\sum_{i=1}^N |D_i|$ columns.

\section{Distance Function on Constraint Satisfaction Problems}

\begin{figure}[t!]
\centering
\includegraphics[width=\linewidth]{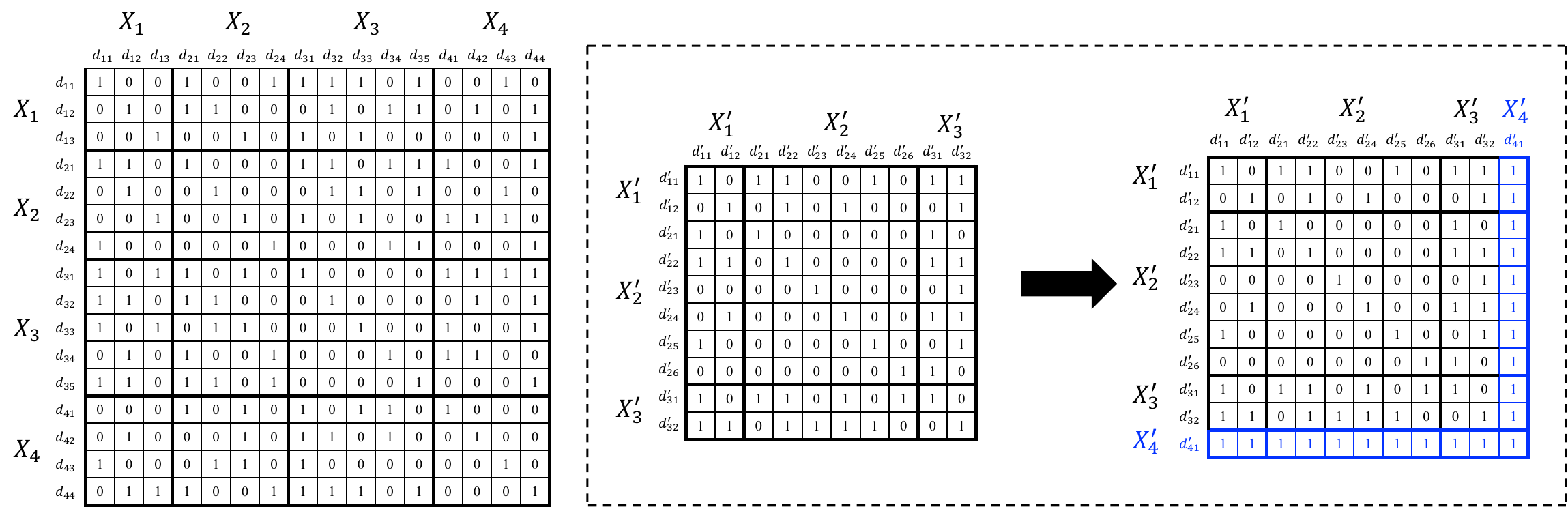}
\includegraphics[width=\linewidth]{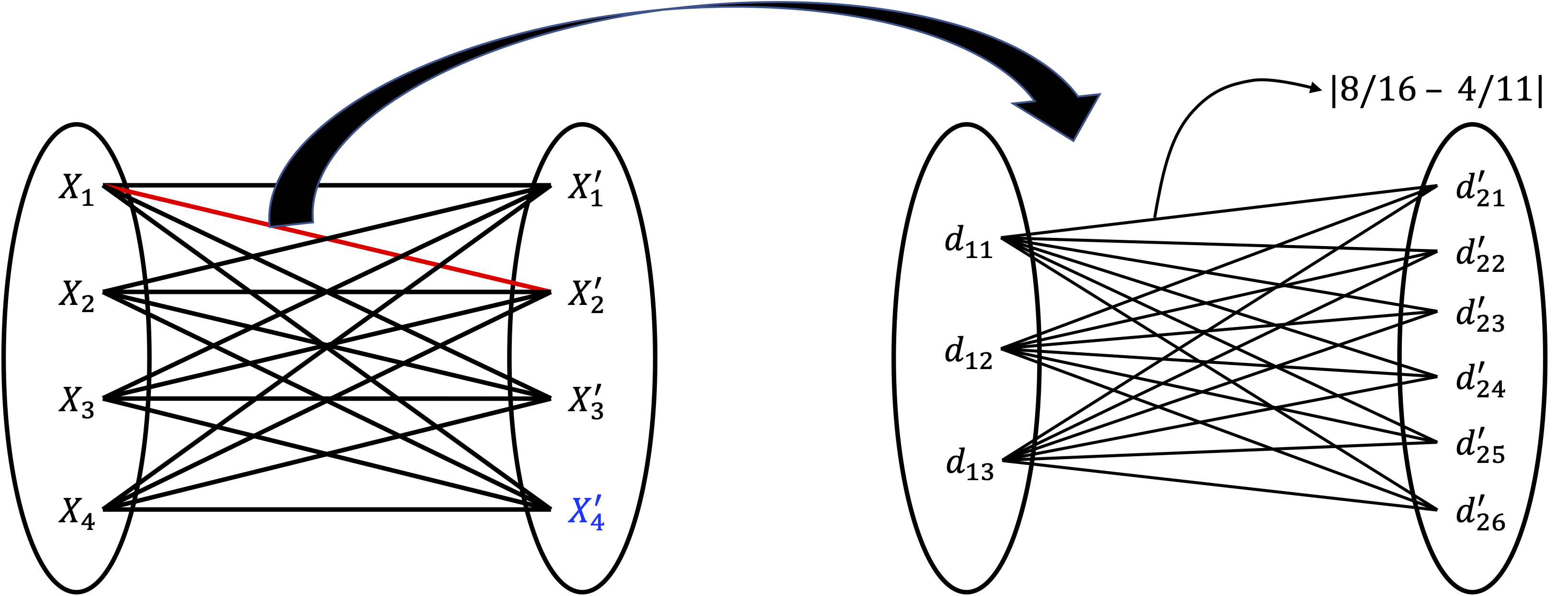}
\caption[An illustration of our novel distance function between two binary Constraint Satisfaction Problems.]{Illustrates the distance function between two binary CSP instances. The top panel shows two CSP instances with variables $\{X_1, X_2, X_3, X_4\}$ (left) and $\{X'_1, X'_2, X'_3\}$ (middle), respectively. A dummy variable $X'_4$ with a singleton domain is added to the CSP instance with fewer variables (right). The bottom panel (left) shows how a `maximum matching of minimum cost' problem is posed on a complete bipartite graph with the variables of the two CSP instances in each partition. The cost annotating the edge between $X_i$ and $X'_j$ is itself derived from a `maximum matching of minimum cost' problem posed on the domain values of $X_i$ and $X'_j$. The bottom panel (right) shows this `maximum matching of minimum cost' problem for the variables $X_1$ and $X'_2$. It is posed on a complete bipartite graph with the domain values $\{d_{11}, d_{12}, d_{13}\}$ and $\{d'_{21}, d'_{22}, d'_{23}, d'_{24}, d'_{25}, d'_{26}\}$ constituting the two partitions. The cost annotating the edge between $d_{11}$ and $d'_{21}$ is the absolute value of the difference between the average compatibility of $d_{11}$ and that of $d'_{21}$.}
\label{fig:distance_function}
\end{figure}

In this section, we describe a novel distance function on binary CSPs. As explained in Chapter~\ref{ch:fastmapsvm}, the FastMap component of FastMapSVM requires the distance function to be symmetric, yield non-negative values for all pairs of CSP instances, and yield $0$ for identical CSP instances. Satisfying these requirements, our distance function is based on maxflow computations and is illustrated in Figure~\ref{fig:distance_function}. It is well defined for CSP instances $\mathcal{I}_1$ and $\mathcal{I}_2$ that may have different sizes. The maxflow computations are utilized in: (a) a single high-level `maximum matching of minimum cost' problem posed on the variables of $\mathcal{I}_1$ and $\mathcal{I}_2$; and (b) multiple low-level `maximum matching of minimum cost' problems posed on the domain values of pairs of variables, one from each of $\mathcal{I}_1$ and $\mathcal{I}_2$.

The high-level `maximum matching of minimum cost' problem is posed on a complete bipartite graph, in which the two partitions of the bipartite graph correspond to the variables of $\mathcal{I}_1$ and $\mathcal{I}_2$, respectively. If the number of variables in $\mathcal{I}_1$ does not match the number of variables in $\mathcal{I}_2$, dummy variables are added to the CSP instance with fewer variables. The top panel of Figure~\ref{fig:distance_function} illustrates this for $\mathcal{I}_1$ and $\mathcal{I}_2$ with variables $\{X_1, X_2, X_3, X_4\}$ and $\{X'_1, X'_2, X'_3\}$, respectively. A dummy variable $X'_4$ is added to $\mathcal{I}_2$. The dummy variable has a single domain value that is designed to be consistent with all domain values of all other variables, since this does not change the CSP instance.

The distance between $\mathcal{I}_1$ and $\mathcal{I}_2$ is defined to be the cost of the `maximum matching of minimum cost' on the high-level bipartite graph. This bipartite graph has an edge between every $X_i$ in $\mathcal{I}_1$ and every $X'_j$ in $\mathcal{I}_2$. The cost annotating an edge between $X_i$ and $X'_j$ is itself set to be the cost of the `maximum matching of minimum cost' posed at the low level on the domain values of $X_i$ and $X'_j$. The bottom-left panel of Figure~\ref{fig:distance_function} shows the high-level bipartite graph and highlights an edge between $X_1$ and $X'_2$ for explanation of the low-level `maximum matching of minimum cost'.

The low-level `maximum matching of minimum cost' problem posed on the domain values of $X_i$ and $X'_j$ also uses a complete bipartite graph. The two partitions consist of the domain values of $X_i$ and $X'_j$, respectively. The cost annotating the edge between $d_{ip}$ and $d'_{jq}$ is the absolute value of the difference between the average compatibility of $d_{ip}$ and the average compatibility of $d'_{jq}$. The bottom-right panel of Figure~\ref{fig:distance_function} shows the low-level `maximum matching of minimum cost' problem posed on the domain values of $X_1$ and $X'_2$. The domains of $X_1$ and $X'_2$ are $\{d_{11}, d_{12}, d_{13}\}$ and $\{d'_{21}, d'_{22}, d'_{23}, d'_{24}, d'_{25}, d'_{26}\}$, respectively. Consider the edge between $d_{11}$ and $d'_{21}$. The average compatibility of $d_{11}$ is the fraction of `$1$'s in the column `$d_{11}$' in the $(0, 1)$-matrix representation of $\mathcal{I}_1$. This fraction is equal to $8/16$. The average compatibility of $d'_{21}$ is the fraction of `$1$'s in the column `$d'_{21}$' in the $(0, 1)$-matrix representation of $\mathcal{I}_2$ after adding the dummy variable $X'_4$. This fraction is equal to $4/11$. Therefore, the cost annotating the edge between $d_{11}$ and $d'_{21}$ is equal to $|8/16 - 4/11|$.

The `maximum matching of minimum cost' problems in the high level and the low level are posed on bipartite graphs. Since the two partitions of any bipartite graph can be viewed interchangeably without affecting the `maximum matching of minimum cost', the overall distance function is symmetric. Moreover, since the cost annotating any edge in the high level or the low level is non-negative, the distance function always yields non-negative values. Similarly, it is also easy to observe that the distance function yields $0$ for two identical CSP instances. These properties satisfy all the conditions on the distance function required by the FastMap component of FastMapSVM. In addition, the high-level bipartite graph is invariant to the orderings on the elements within each partition. That is, it is invariant to the orderings on the variables of $\mathcal{I}_1$ and $\mathcal{I}_2$. For a similar reason, all low-level bipartite graphs are also invariant to the orderings on the domain values of the participating variables. Therefore, the overall distance function has the additional property of being invariant to variable-orderings as well as domain value-orderings.

The above properties of the distance function allow us to bypass data augmentation methods typically required for training other ML models. Data augmentation refers to transforming data without changing their labels, known as label-preserving transformations. For example, to generate more training data serving object recognition tasks in Computer Vision applications, an image can be augmented by translating it or reflecting it horizontally without changing its label~\cite{ksh17}. In the context of CSPs, a CSP instance is typically augmented by changing the ordering on its variables or the ordering on the domain values of individual variables. However, doing so generates an exponential number of CSP training instances within the same equivalence class. This drawback of traditional ML algorithms of having to learn equivalence classes is now intelligently addressed within the framework of FastMapSVM by utilizing a distance function that is invariant to both variable-orderings and domain value-orderings.

We note that the above distance function could have been defined in many other ways. For example, we could have introduced dummy domain values in the low-level `maximum matching of minimum cost' problems to equalize the domain sizes of the participating variables. We could have also chosen not to use dummy variables in the high-level `maximum matching of minimum cost' problem. In addition, we could have defined the costs annotating the edges of the bipartite graphs using many other characteristics of the CSP instances. These variations of the distance function are not of fundamental importance to this chapter. Instead, in this chapter, we focus on the advantages of the FastMapSVM framework as a whole. The study of more refined distance functions is delegated to future work.

\section{Experimental Results}

In this section, we describe the comparative performance of FastMapSVM against other state-of-the-art ML approaches for predicting CSP satisfiability.

\subsection{Experimental Setup}

\begin{figure}[t!]
\centering
\includegraphics[width=0.9\textwidth]{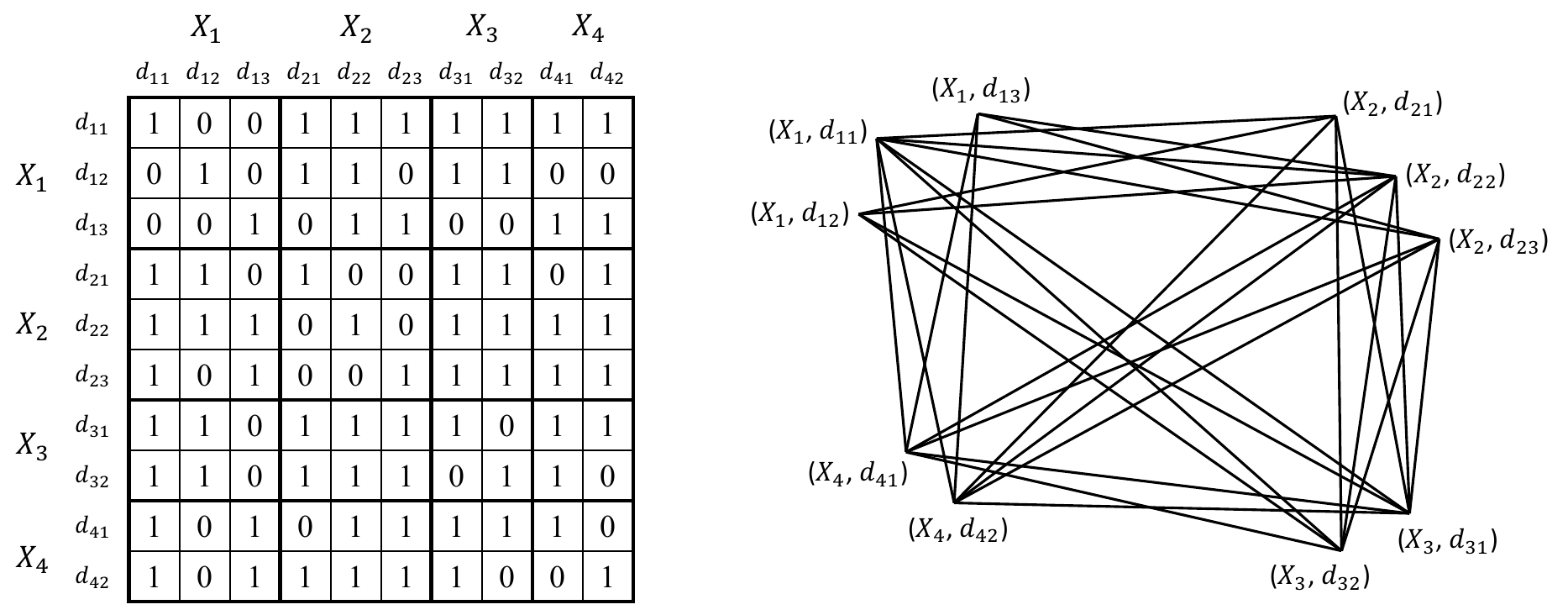}
\caption[The graphical representation of a binary Constraint Satisfaction Problem obtained from its $(0, 1)$-matrix representation.]{Shows the graphical representation of a binary CSP instance obtained from its $(0, 1)$-matrix representation. The left panel shows the $(0, 1)$-matrix representation of a binary CSP instance. The right panel shows its graphical representation. The vertices represent domain values and are clustered into four groups, corresponding to the four variables $\{X_1, X_2, X_3, X_4\}$.}
\label{fig:2graph}
\end{figure}

We evaluate FastMap against three competing approaches. The first is a state-of-the-art deep graph convolutional neural network (DGCNN)~\cite{zcnc18}. The second is a state-of-the-art graph isomorphism network (GIN)~\cite{xhlj18}. Both these networks ingest a CSP instance in the form of a graph, as shown in Figure~\ref{fig:2graph}. In the graphical representation of a binary CSP instance, a vertex represents a domain value and is tagged with the name of the variable that it belongs to. Information in these tags is utilized by DGCNN and GIN. An edge between two vertices $v_1$ and $v_2$ with tags $X_i$ and $X_j$, respectively, represents the compatible combination $(X_i \leftarrow v_1, X_j \leftarrow v_2)$ allowed by the direct constraint between $X_i$ and $X_j$.\footnote{The graphical representation of a binary CSP is obtained by parsing its $(0, 1)$-matrix representation. Thus, we correctly represent the compatible tuples of domain values between every pair of variables, even if there does not exist a direct constraint between those variables.} DGCNN and GIN do not require the CSP training and test instances to be of the same size.

The third is a polynomial-time algorithm based on establishing arc-consistency. This algorithm first establishes arc-consistency and then checks whether any variable's domain is annihilated. If so, it declares the CSP instance to be `unsatisfiable'. Otherwise, it declares the CSP instance to be `satisfiable'. This algorithm is used in our evaluation to demonstrate that FastMapSVM's capabilities go beyond that of a polynomial-time algorithm.\footnote{This is done to avoid the pitfalls of~\cite{xkk18}, as mentioned before.} Of course, a polynomial-time algorithm based on establishing path-consistency also could have been used. But we excluded this algorithm since arc-consistency already provides the required proof of concept and establishing path-consistency is prohibitively expensive.

We implemented FastMapSVM and arc-consistency in Python3 and ran them on a laptop with an Apple M2 chip and 16 GB memory. We ran DGCNN and GIN on a Linux system with an Intel(R) Xeon(R) Silver 4216 CPU at 2.10 GHz. The different platforms are inconsequential to the comparative performances of these algorithms with respect to effectiveness. For each dataset, we trained DGCNN and GIN for $100$ epochs with a learning rate of $0.0001$ and a minibatch size of $100$ to obtain representative results.

\subsection{Instance Generation}

We generate the binary CSP instances for both training and testing using the Model A method in~\cite{sd96,xkk18}. We generate a CSP instance by first picking the number of variables $N$ uniformly at random to be an integer within the range $[1, 100]$. Then, we pick the domain size of each variable independently and uniformly at random to be an integer within the range $[1, 10]$. We use a probability parameter $P_1$ to independently determine the existence of a direct constraint between every pair of distinct variables. That is, for every pair of distinct variables $X_i$ and $X_j$, we introduce a direct constraint between them with probability $P_1$. Moreover, we use a probability parameter $P_2$ to determine the compatible tuples of a direct constraint. For a pair of variables $X_i$ and $X_j$ with a direct constraint between them, each tuple $(X_i \leftarrow d_{ip}, X_j \leftarrow d_{jq})$ is independently deemed to be compatible with probability $1 - P_2$. We set $P_1 = 1$ and $P_2 = 0.4$ to obtain representative results for all approaches.

Model A has a tendency to produce mostly unsatisfiable CSP instances with increasing $N$~\cite{sd96,xkk18}. Therefore, we use a ``hidden solution'' method to generate satisfiable CSP instances whenever required. In this method, a set of hidden solutions of the CSP instance are chosen a priori.\footnote{The CSP instance can have other solutions as well.} A hidden solution $(X_1 \leftarrow d_{1p_1}, X_2 \leftarrow d_{2p_2} \ldots X_N \leftarrow d_{Np_N})$ is utilized as follows: While generating the direct constraints using Model A, a direct constraint between variables $X_i$ and $X_j$ reserves the tuple $(X_i \leftarrow d_{ip_i}, X_j \leftarrow d_{jp_j})$ as being compatible before the other tuples are set using the probability parameter $P_2$. Therefore, $(X_1 \leftarrow d_{1p_1}, X_2 \leftarrow d_{2p_2} \ldots X_N \leftarrow d_{Np_N})$ satisfies all the direct constraints and, consequently, qualifies as a solution. Similarly, multiple hidden solutions can be utilized with the following modification in the generation procedure: A direct constraint between variables $X_i$ and $X_j$ reserves multiple tuples as being compatible. For generating satisfiable CSP instances, we pick the number of hidden solutions uniformly at random to be an integer within the range $[1, 10]$. We pick a hidden solution itself by assigning a domain value chosen independently and uniformly at random for each variable from its domain.

We generate three datasets: Dataset-1, Dataset-2, and Dataset-3. For each dataset, we generate $1,000$ training instances and $1,000$ test instances. Each training and test set has an equal number of satisfiable and unsatisfiable instances.

In Dataset-1, we generate the instances using Model A. Since Model A frequently generates unsatisfiable instances, we use a complete CSP solver to identify and collect such instances. We generate the satisfiable instances using the hidden solution method, as described above.

In Dataset-2, we generate the satisfiable instances as in Dataset-1. However, we design and generate the unsatisfiable instances to be more challenging. We do this by hiding two complementary pseudo-solutions $(X_1 \leftarrow d_{1p_1}, X_2 \leftarrow d_{2p_2} \ldots X_N \leftarrow d_{Np_N})$ and $(X_1 \leftarrow d_{1q_1}, X_2 \leftarrow d_{2q_2} \ldots X_N \leftarrow d_{Nq_N})$. We identify a pair of distinct variables $X_i$ and $X_j$ such that $d_{ip_i} \neq d_{iq_i}$ and $d_{jp_j} \neq d_{jq_j}$. All direct constraints between distinct variables $X_s$ and $X_t$ such that $\{X_s, X_t\} \neq \{X_i, X_j\}$ are generated as before by reserving the tuples $(X_s \leftarrow d_{sp_s}, X_t \leftarrow d_{tp_t})$ and $(X_s \leftarrow d_{sq_s}, X_t \leftarrow d_{tq_t})$ as being compatible. However, the direct constraint between $X_i$ and $X_j$ reserves the tuples $(X_i \leftarrow d_{ip_i}, X_j \leftarrow d_{jq_j})$ and $(X_i \leftarrow d_{iq_i}, X_j \leftarrow d_{jp_j})$ as being compatible and reserves the tuples $(X_i \leftarrow d_{ip_i}, X_j \leftarrow d_{jp_j})$ and $(X_i \leftarrow d_{iq_i}, X_j \leftarrow d_{jq_j})$ as being not compatible. We finally use a complete CSP solver to verify that such a CSP instance is indeed unsatisfiable.\footnote{This procedure frequently generates unsatisfiable instances, as required. However, satisfiable instances that are generated occasionally are filtered out by the CSP solver.}

In Dataset-3, we generate the satisfiable instances as in Dataset-1. However, we design and generate the unsatisfiable instances differently from in Dataset-2. We do this by first hiding two complementary pseudo-solutions $(X_1 \leftarrow d_{1p_1}, X_2 \leftarrow d_{2p_2} \ldots X_N \leftarrow d_{Np_N})$ and $(X_1 \leftarrow d_{1q_1}, X_2 \leftarrow d_{2q_2} \ldots X_N \leftarrow d_{Nq_N})$, as in Dataset-2. However, we gather all variables $X_{r_1}, X_{r_2} \ldots X_{r_{\bar{M}}}$ for which the two pseudo-solutions have different assignments of domain values, that is, $d_{r_mp_{r_m}} \neq d_{r_mq_{r_m}}$, for all $1 \leq m \leq \bar{M}$. For any two distinct variables $X_i$ and $X_j$ in $\{X_{r_1}, X_{r_2} \ldots X_{r_{\bar{M}}}\}$, we reserve the tuples $(X_i \leftarrow d_{ip_i}, X_j \leftarrow d_{jq_j})$ and $(X_i \leftarrow d_{iq_i}, X_j \leftarrow d_{jp_j})$ as being not compatible. Finally, we pick two distinct variables $X_s$ and $X_t$ from $\{X_{r_1}, X_{r_2} \ldots X_{r_{\bar{M}}}\}$ and reserve the tuples $(X_s \leftarrow d_{sp_s}, X_t \leftarrow d_{tq_t})$ and $(X_s \leftarrow d_{sq_s}, X_t \leftarrow d_{tp_t})$ as being compatible and reserve the tuples $(X_s \leftarrow d_{sp_s}, X_t \leftarrow d_{tp_t})$ and $(X_s \leftarrow d_{sq_s}, X_t \leftarrow d_{tq_t})$ as being not compatible. As before, we use a complete CSP solver to verify that such a CSP instance is indeed unsatisfiable.

\subsection{Results}

We show three sets of results pertaining to FastMapSVM. First, we show the $2$-dimensional and the $3$-dimensional embeddings that FastMapSVM produces to aid visualization. Second, we show the behavior of FastMapSVM with respect to the hyperparameter $\kappa$, that is, the number of dimensions and with respect to the size of the training data. Third, we show the comparative performance of FastMapSVM against DGCNN, GIN, and arc-consistency.

\begin{figure}[!t]
\centering
\subfloat[Dataset-1 CSP test instances embedded in a $2$-dimensional Euclidean space by FastMapSVM]{
\begin{minipage}[b]{0.45\textwidth}
\centering
\includegraphics[width=0.9\textwidth]{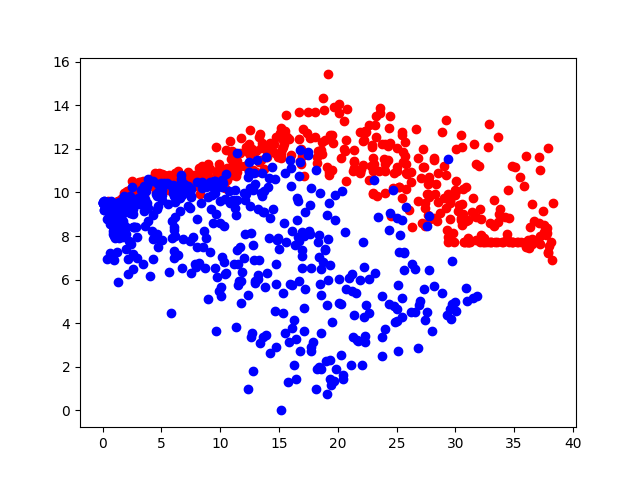}
\label{fig:ds1_2d}
\end{minipage}
}
\hspace{0.05\textwidth}
\subfloat[Dataset-1 CSP test instances embedded in a $3$-dimensional Euclidean space by FastMapSVM]{
\begin{minipage}[b]{0.45\textwidth}
\centering
\includegraphics[width=0.9\textwidth]{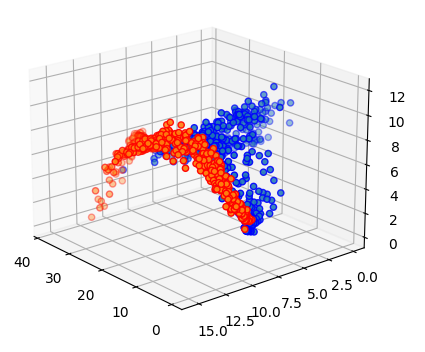}
\label{fig:ds1_3d}
\end{minipage}
}
\\
\subfloat[Dataset-2 CSP test instances embedded in a $2$-dimensional Euclidean space by FastMapSVM]{
\begin{minipage}[b]{0.45\textwidth}
\centering
\includegraphics[width=0.9\textwidth]{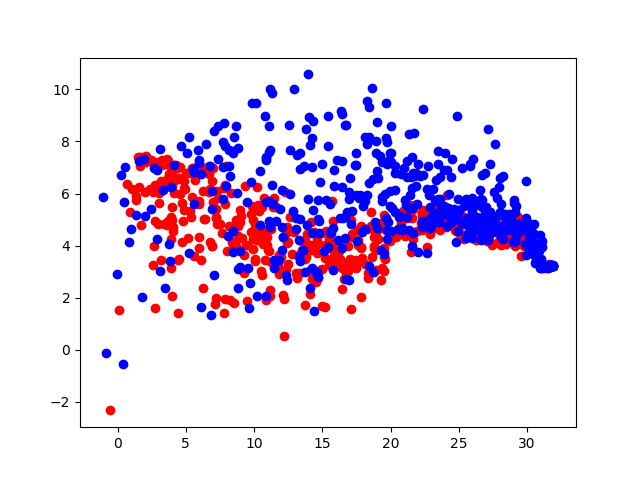}
\label{fig:ds2_2d}
\end{minipage}
}
\hspace{0.05\textwidth}
\subfloat[Dataset-2 CSP test instances embedded in a $3$-dimensional Euclidean space by FastMapSVM]{
\begin{minipage}[b]{0.45\textwidth}
\centering
\includegraphics[width=0.9\textwidth]{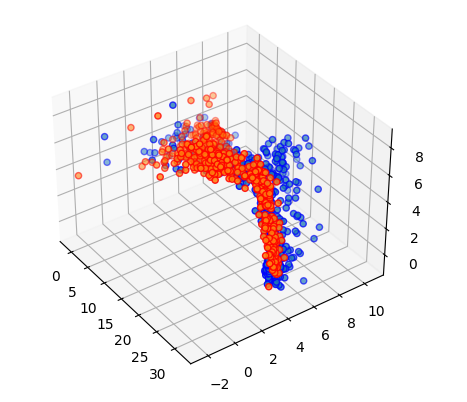}
\label{fig:ds2_3d}
\end{minipage}
}
\\
\subfloat[Dataset-3 CSP test instances embedded in a $2$-dimensional Euclidean space by FastMapSVM]{
\begin{minipage}[b]{0.45\textwidth}
\centering
\includegraphics[width=0.9\textwidth]{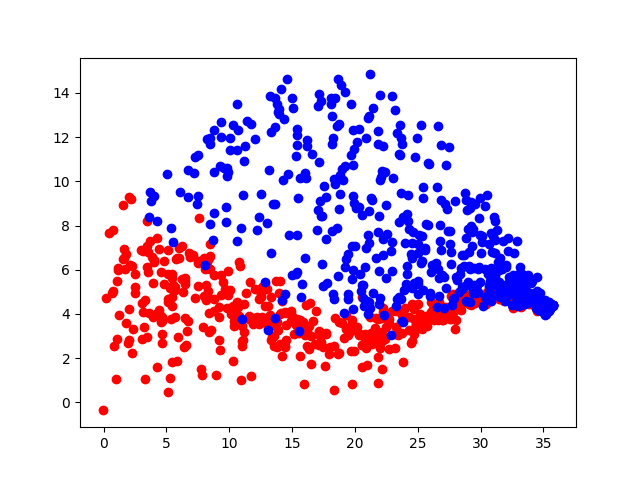}
\label{fig:ds3_2d}
\end{minipage}
}
\hspace{0.05\textwidth}
\subfloat[Dataset-3 CSP test instances embedded in a $3$-dimensional Euclidean space by FastMapSVM]{
\begin{minipage}[b]{0.45\textwidth}
\centering
\includegraphics[width=0.9\textwidth]{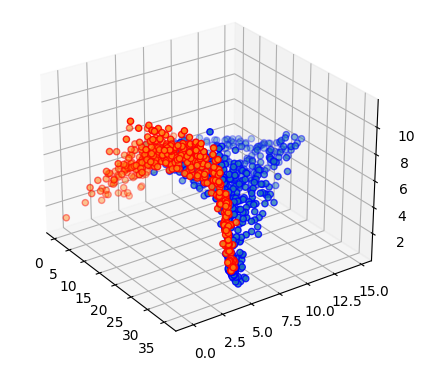}
\label{fig:ds3_3d}
\end{minipage}
}
\caption[The $2$-dimensional and $3$-dimensional Euclidean embeddings produced by FastMapSVM for classifying Constraint Satisfaction Problems.]{Shows the $2$-dimensional and $3$-dimensional Euclidean embeddings produced by FastMapSVM for classifying CSP instances. Mostly, there is a clear separation of satisfiable instances (blue) and unsatisfiable instances (red).}
\label{fig:csp_visualization}
\end{figure}

Figure~\ref{fig:csp_visualization} shows a perspicuous visualization of the CSP test instances for all three datasets. This visualization capability is unique to FastMapSVM. We note that while the accuracy, recall, precision, and the F1 score of FastMapSVM typically increase with increasing $\kappa$, $\kappa = 2$ and $\kappa = 3$ are the only two values that support visualization. Still, in most cases, Figure~\ref{fig:csp_visualization} shows a clear separation between the satisfiable and unsatisfiable instances. Moreover, the separation is clearer in the $3$-dimensional embeddings compared to their $2$-dimensional counterparts.

\begin{figure}[t!]
\centering
\subfloat[influence of the number of dimensions on the performance of FastMapSVM]{
\begin{minipage}{0.8\textwidth}
\centering
\includegraphics[width=1.0\textwidth]{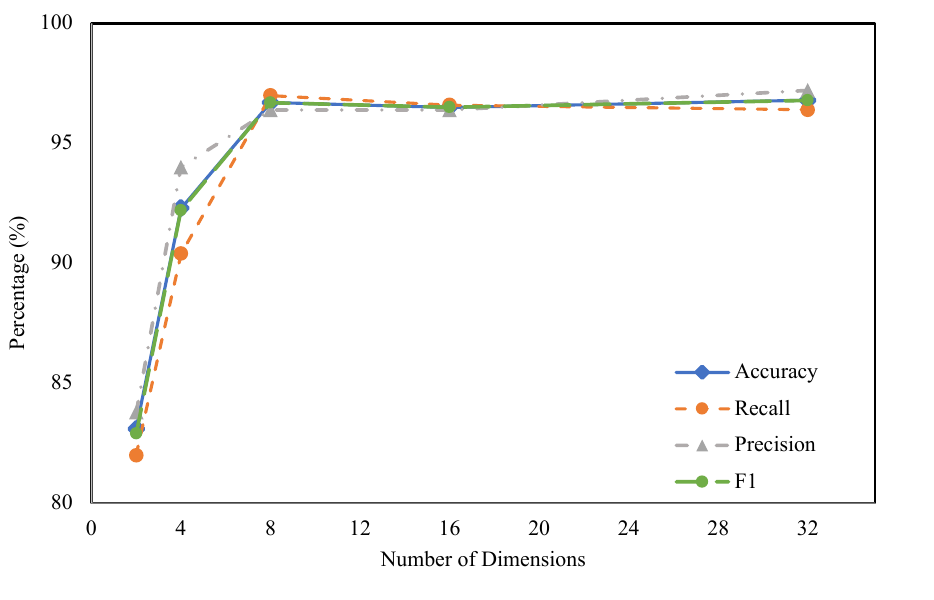}
\label{fig:dimension}
\end{minipage}
}
\\
\subfloat[influence of the size of the training data on the performance of FastMapSVM]{
\begin{minipage}{0.8\textwidth}
\centering
\includegraphics[width=1.0\textwidth]{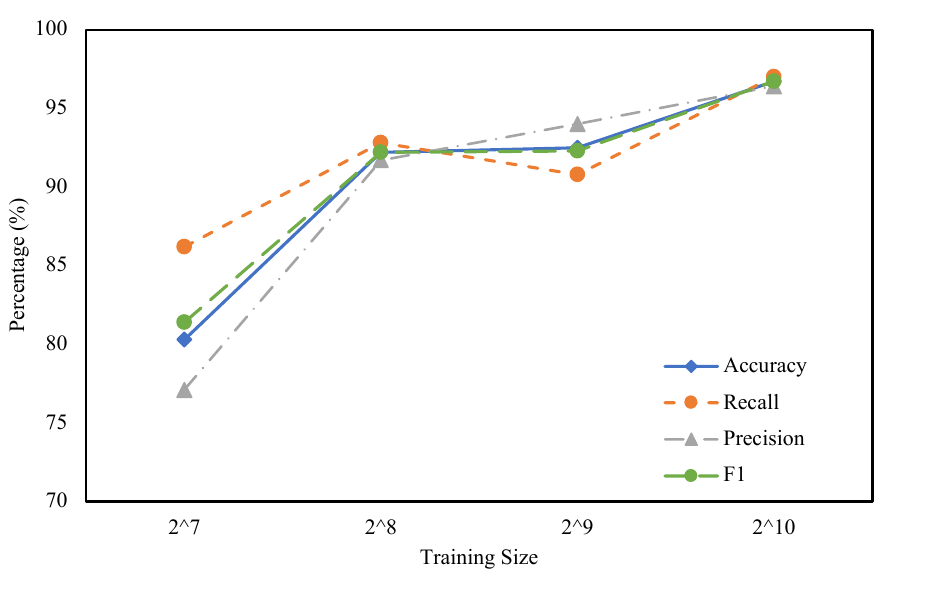}
\label{fig:size}
\end{minipage}
}
\caption[The behavior of FastMapSVM in the Constraint Satisfaction Problem domain with respect to the number of dimensions and with respect to the size of the training data.]{Shows the behavior of FastMapSVM in the CSP domain with respect to the number of dimensions and with respect to the size of the training data. The performance metrics include the accuracy, recall, precision, and the F1 score.}
\label{fig:dimension_size}
\end{figure}

Figure~\ref{fig:dimension} shows the behavior of FastMapSVM with respect to the number of dimensions $\kappa$ on Dataset-1. Its behavior on the other datasets is similar. The performance metrics, that is, the accuracy, recall, precision, and the F1 score, improve with increasing $\kappa$. This is intuitively expected since the distances between the CSP instances can be embedded with lower distortion in higher dimensions. However, Figure~\ref{fig:dimension} also shows that a point of diminishing returns is attained rather quickly at around $\kappa = 8$. This shows that $\kappa = 8$, $9$, or $10$ is good enough for the CSP domain. Finally, Figure~\ref{fig:dimension} also shows that the improvements in the performance metrics are significant between $\kappa = 2$ and $\kappa = 8$.

Figure~\ref{fig:size} shows the behavior of FastMapSVM with respect to the size of the training data on Dataset-1. Its behavior on the other datasets is similar. The performance metrics improve with increasing size of the training data. Figure~\ref{fig:size} also shows that the improvements in the performance metrics are significant between $128$ and $256$ training data instances. Further improvements are gradual between $256$ and $1,000$ training data instances. This shows that FastMapSVM has the capability to achieve good performance from relatively small amounts of training data and training time.

\begin{table}[t!]
\footnotesize
\centering
\scalebox{1.0}{
\begin{tabular}{|l|l|r|r|r|r|}
\hline
Dataset &Model &Accuracy &Recall &Precision &F1\\
\hline
\multirow{6}{*}{Dataset-1} &FastMapSVM &96.7\% &97.0\% &96.4\% &96.7\%\\
&AC &\bf99.3\% &\bf100.0\% &96.7\% &\bf98.3\%\\
&DGCNN (unlabeled) &94.2\% &92.2\% &96.0\% &94.1\%\\
&DGCNN (labeled) &82.3\% &82.0\% &82.5\% &82.2\%\\
&GIN (unlabeled) &84.1\% &67.0\% &\bf99.4\% &80.0\%\\
&GIN (labeled) &56.4\% &52.6\% &56.9\% &54.7\%\\
\hline
\multirow{6}{*}{Dataset-2} &FastMapSVM &\bf82.9\% &72.8\% &\bf91.2\% &\bf81.0\%\\
&AC &50.1\% &\bf100.0\% &50.1\% &66.8\%\\
&DGCNN (unlabeled) &73.4\% &61.4\% &80.8\% &69.8\%\\
&DGCNN (labeled) &53.9\% &51.2\% &54.1\% &52.6\%\\
&GIN (unlabeled) &71.9\% &49.0\% &90.4\% &63.6\%\\
&GIN (labeled) &54.4\% &51.6\% &54.7\% &53.1\%\\
\hline
\multirow{6}{*}{Dataset-3} &FastMapSVM &\bf95.4\% &94.4\% &96.3\% &\bf95.3\%\\
&AC &50.0\% &\bf100.0\% &50.0\% &66.7\%\\
&DGCNN (unlabeled) &90.3\% &86.6\% &93.5\% &89.9\%\\
&DGCNN (labeled) &74.7\% &71.8\% &76.2\% &73.9\%\\
&GIN (unlabeled) &78.4\% &53.6\% &\bf98.5\% &69.4\%\\
&GIN (labeled) &57.4\% &53.8\% &58.0\% &55.8\%\\
\hline
\end{tabular}
}
\caption[The accuracy, recall, precision, and the F1 score of FastMapSVM and its competing methods on three datasets of Constraint Satisfaction Problems.]{Shows all performance metrics of all competing methods on all datasets. `AC' represents arc-consistency.}
\label{tab:comparison}
\end{table}

Table~\ref{tab:comparison} shows a comparison of all the competing methods on all three datasets with respect to all of the performance metrics. It uses $\kappa = 8$ for FastMapSVM. It also shows two versions of DGCNN and GIN: the `labeled' version and the `unlabeled' version.

We recollect that in the graphical representation of a binary CSP instance, a vertex represents a domain value and is tagged with the name of the variable that it belongs to. Information in these tags is available to be utilized by DGCNN and GIN. The labeled versions of DGCNN and GIN utilize this information while the unlabeled versions ignore this information. Table~\ref{tab:comparison} shows that the unlabeled versions perform better than their labeled counterparts. While this is a little surprising, it is likely that the unlabeled versions perform permutation reasoning on the tags (names of variables) much more efficiently.

Table~\ref{tab:comparison} shows that FastMapSVM generally outperforms all other competing methods by a significant margin. Even on a particular dataset where it is not the top performer with respect to a particular performance metric, it is a close second. Overall, Dataset-1 seems to be the easiest for all methods and Dataset-2 seems to be the hardest for all methods.

In comparison to arc-consistency, FastMapSVM is significantly better on Dataset-2 and Dataset-3. On these datasets, arc-consistency declares all test instances as being `satisfiable', leading to a perfect recall score but very poor accuracy, precision, and F1 scores. On the one hand, this shows that arc-consistency is ineffective in recognizing unsatisfiable CSP instances. On the other hand, it also shows that CSP instances generated as in Dataset-1 are insufficient to conclusively evaluate competing ML methods. In contrast, FastMapSVM performs well on all three datasets.

In comparison to DGCNN and GIN, FastMapSVM is significantly better on all three datasets. On the accuracy, recall, and F1 scores, FastMapSVM is better than DGCNN, which in turn is better than GIN. GIN generally has high precision scores but very poor recall scores. This shows that it is poor in identifying satisfiable instances but is mostly correct when it does so. FastMapSVM does not have this drawback. Moreover, on the accuracy and F1 scores, FastMapSVM outperforms the closest competitor (DGCNN) by larger margins with increasing hardness of the CSP instances, that is, in the order of Dataset-1, Dataset-3, and Dataset-2.

Even on the metric of efficiency, FastMapSVM outperforms DGCNN and GIN.\footnote{DGCNN and GIN ran on a different platform compared to FastMapSVM. However, the ballpark results are still conclusive.} For example, FastMapSVM took only~\SI{2965}{\second} for training and testing on Dataset-1 while DGCNN and GIN took~\SI{5440}{\second} and~\SI{10865}{\second}, respectively, for the same task.

\section{Conclusions}

In this chapter, we demonstrated the success of FastMapSVM on the task of predicting CSP satisfiability. FastMapSVM overcomes the hurdles faced by other ML approaches in the CSP domain. It leverages a distance function on CSPs that is defined via maxflow computations. FastMapSVM is applicable to CSP training and test instances of different sizes and is invariant to both variable-orderings and domain value-orderings. This allows it to bypass the onus of having to learn equivalence classes of CSP instances and, therefore, requires significantly smaller amounts of data and time for model training compared to other ML algorithms. FastMapSVM also encapsulates the intelligence of FastMap, SVMs, kernel methods, and maxflow computations, accounting for its superior empirical performance, even over state-of-the-art graph neural networks. Moreover, it facilitates a perspicuous visualization of the CSP instances, their distribution, and the classification boundaries between them. Overall, the FastMapSVM framework for CSPs has broader applicability and various representational and combinatorial advantages compared to other ML approaches.

\begin{subappendices}

\section{Table of Notations}

\begin{table}[h]
\centering
\begin{tabular}{|l|p{0.75\linewidth}|}
\cline{1-2}
Notation &Description\\
\cline{1-2}
$\langle \mathcal X, \mathcal D, \mathcal C \rangle$ &A CSP instance, where $\mathcal X = \{X_1, X_2 \ldots X_N\}$ is a set of variables and $\mathcal C = \{C_1, C_2 \ldots C_M\}$ is a set of constraints on subsets of them. Each variable $X_i$ is associated with a discrete-valued domain $D_i \in \mathcal D$.\\
\cline{1-2}
$d_{ip}$ &The $p$-th domain value of variable $X_i$ assuming an index ordering on the domain values of $X_i$.\\
\cline{1-2}
$\mathcal{I}$ &A binary CSP instance.\\
\cline{1-2}
$\kappa$ &The user-specified number of dimensions of the FastMap embedding.\\
\cline{1-2}
\end{tabular}
\caption[Notations used in Chapter~\ref{ch:constraint}.]{Describes the notations used in Chapter~\ref{ch:constraint}.}
\label{tab:constraint_notations}
\end{table}

\end{subappendices}

\chapter{Conclusions and Future Work}
\label{ch:conclusions}
In this chapter, we present our conclusions on the foregoing chapters of this dissertation and the directions of our future work that are intended to go beyond this dissertation. The `Summary' section reiterates our contributions and validates the overall hypothesis of this dissertation: The ability of FastMap to efficiently embed complex objects or graphs in a Euclidean space harnesses many powerful algorithmic techniques developed in AI, ML, Computational Geometry, Mathematics, Operations Research, and Theoretical Computer Science towards efficiently and effectively solving important combinatorial problems that arise in various real-world problem domains. The `Future Work' section lists some major directions in which further relevant research can be conducted. The `Concluding Remarks' section presents some final words on this dissertation.

\section{Summary}

In this section, we summarize our contributions from each of the previous chapters for the benefit of the reader. These are as follows:

\begin{itemize}
\item In Chapter~\ref{ch:fastmap}, we revisited the popular Data-Mining version of FastMap and a graph version of it: also conveniently referred to as FastMap. FastMap's ability to embed the vertices of a graph as points in a Euclidean space is complemented by the ability of LSH to map any point of interest in the Euclidean space back on the graph. Hence, we proposed our FastMap+LSH framework as an important way to harness the intelligence of algorithms that work in Euclidean space towards solving combinatorial problems on graphs. Such algorithms include clustering algorithms from ML, algorithms from Computational Geometry, and analytical techniques, among others.
\item In Chapter~\ref{ch:facility_location}, we studied four representative FLPs: the MAM, VKM, WVKM, and the CVKM problems. We used FastMap to reformulate FLPs defined on a graph to FLPs defined in a Euclidean space without obstacles. Subsequently, we used standard clustering algorithms to solve the problems in the resulting Euclidean space and LSH to interpret the solutions back on the original graph. We showed that our FastMap+LSH approach produces high-quality solutions with orders-of-magnitude speedup over state-of-the-art competing algorithms.
\item In Chapter~\ref{ch:centrality}, we generalized various measures of centrality on explicit graphs to corresponding measures of projected centrality on implicit graphs. We used our FastMap+LSH framework to compute the top-$K$ pertinent vertices with the highest projected centrality values. We designed different distance functions to be used by FastMap for different measures of projected centrality and invoked various procedures for computing analytical solutions in the FastMap embedding. We experimentally demonstrated that our FastMap+LSH framework is both efficient and effective for many popular measures of centrality and their generalizations to projected centrality. In addition, and unlike other methods, our FastMap framework is not tied to a specific measure of projected centrality.
\item In Chapter~\ref{ch:block_modeling}, we proposed FMBM, a FastMap-based algorithm for block modeling. In the first phase, FMBM adapts FastMap to embed a given undirected unweighted graph into a Euclidean space such that the pairwise Euclidean distances between vertices approximate a probabilistically-amplified graph-based distance function between them. In the second phase, FMBM uses GMM clustering for identifying clusters (blocks) in the resulting Euclidean space. We showed that FMBM empirically outperforms other state-of-the-art methods like FactorBlock, Graph-Tool, DANMF, and CPLNS on many benchmark and synthetic test cases. FMBM also enables a perspicuous visualization of the blocks in the graphs, not provided by other methods.
\item In Chapter~\ref{ch:convex_hull}, we presented a FastMap-based algorithm for efficiently computing approximate graph convex hulls by utilizing FastMap's ability to facilitate geometric interpretations of graphs. While the naive version of our algorithm uses a single shot of such a geometric interpretation, the iterative version of our algorithm repeatedly interleaves the graph and geometric interpretations to reinforce one with the other. This iterative version was encapsulated in our solver, FMGCH, and experimentally compared against the state-of-the-art solver, GCA. On a variety of graphs, we showed that FMGCH not only runs several orders of magnitude faster than a highly-optimized exact algorithm but also outperforms GCA, both in terms of generality and the quality of the solutions produced. It is also faster than GCA on large graphs.
\item In Chapter~\ref{ch:fastmapsvm}, we introduced FastMapSVM as an interpretable ML framework that combines the complementary strengths of FastMap and SVMs. We posited that it is an advantageous, lightweight alternative to existing methods, such as NNs, for classifying complex objects when training data or time is limited. FastMapSVM offers several advantages. First, it enables domain experts to incorporate their domain knowledge using a distance function. This avoids relying on complex ML models to infer the underlying structure in the data entirely. Second, because the distance function encapsulates domain knowledge, FastMapSVM naturally facilitates interpretability and explainability. In fact, it even provides a perspicuous visualization of the objects and the classification boundaries between them. Third, FastMapSVM uses significantly smaller amounts of data and time for model training compared to other ML algorithms. Fourth, it extends the applicability of SVMs and kernel methods to domains with complex objects. We demonstrated the efficiency and effectiveness of FastMapSVM in the context of classifying seismograms using significantly smaller amounts of data and time for model training compared to other methods. We also demonstrated its ability to reliably detect new microearthquakes that are otherwise difficult to detect even by the human eye.
\item In Chapter~\ref{ch:constraint}, we demonstrated the success of FastMapSVM on the task of predicting CSP satisfiability by leveraging a distance function on CSPs that is defined via maxflow computations. In this context, our FastMapSVM framework encapsulates the intelligence of FastMap, SVMs, kernel methods, and maxflow computations, accounting for its superior empirical performance, even over state-of-the-art graph neural networks. Moreover, it is applicable to CSP training and test instances of different sizes and is invariant to both variable-orderings and domain value-orderings. This allows it to bypass the onus of having to learn equivalence classes of CSP instances and, therefore, requires significantly smaller amounts of data and time for model training compared to other ML algorithms.
\end{itemize}

\section{Future Work}

In this section, we describe some directions of future work. First, we describe a few directions that are relevant to each of the previous chapters. Second, we describe four broader directions that are relevant to the dissertation as a whole. The chapter-wise future directions are as follows:

\begin{itemize}
\item In future work relevant to Chapter~\ref{ch:facility_location}, we will use our FastMap framework to solve many other kinds of FLPs. We will also consider FLPs that arise in the real world. Moreover, we will try to use our techniques for supply chain optimization in manufacturing and distribution management.
\item In future work relevant to Chapter~\ref{ch:centrality}, we will apply our FastMap framework to various other measures of projected centrality not discussed in that chapter. Such measures may include the betweenness centrality, Katz centrality, and the page rank centrality.
\item In future work relevant to Chapter~\ref{ch:block_modeling}, we will generalize FMBM to work on directed graphs and multi-view graphs. We will also apply FMBM and its generalizations to real-world graphs from various domains, including social and biological networks.
\item In future work relevant to Chapter~\ref{ch:convex_hull}, we will enhance FMGCH with other geometric intuitions derived from the FastMap embedding of the graphs. We will also explore the importance of the graph convex hull problem in relation to other graph-theoretic problems. This can be done by following the analogy of the importance of the geometric convex hull problem in Computational Geometry.
\item In future work relevant to Chapter~\ref{ch:fastmapsvm}, we will apply FastMapSVM in Earthquake Science to analyze and learn from data obtained during temporary deployments of large-$N$ nodal arrays and distributed acoustic sensing. The efficiency and effectiveness of FastMapSVM also make it suitable for real-time deployment in dynamic environments in applications such as Earthquake Early Warning Systems. In Computational Astrophysics, we anticipate the use of FastMapSVM for identifying galaxy clusters based on cosmological observations.
\item In future work relevant to Chapter~\ref{ch:constraint}, we will enhance our current distance function on CSPs with local consistency algorithms. We will also apply FastMapSVM to optimization variants of CSPs and generally facilitate the integration of constraint reasoning and ML algorithms.
\end{itemize}

\subsection{Downsampling Graphs}

\begin{figure}[!t]
\centering
\subfloat[input graph with $168$ vertices and $546$ edges]{
\begin{minipage}[b]{0.45\textwidth}
\centering
\includegraphics[width=\columnwidth]{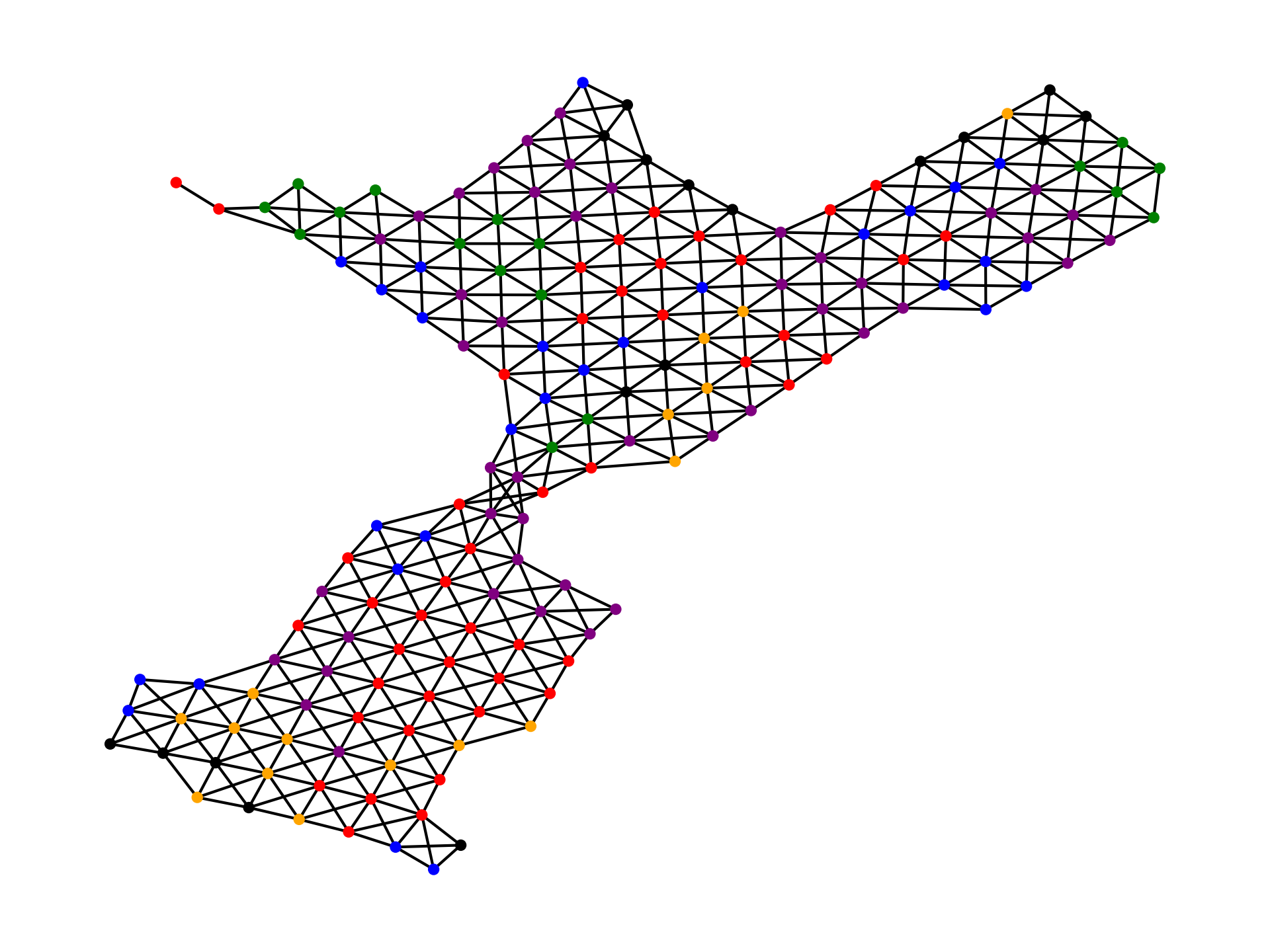}
\label{fig:lak110d}
\end{minipage}
}
\subfloat[FastMap embedding of the input graph]{
\begin{minipage}[b]{0.45\textwidth}
\centering
\includegraphics[width=\columnwidth]{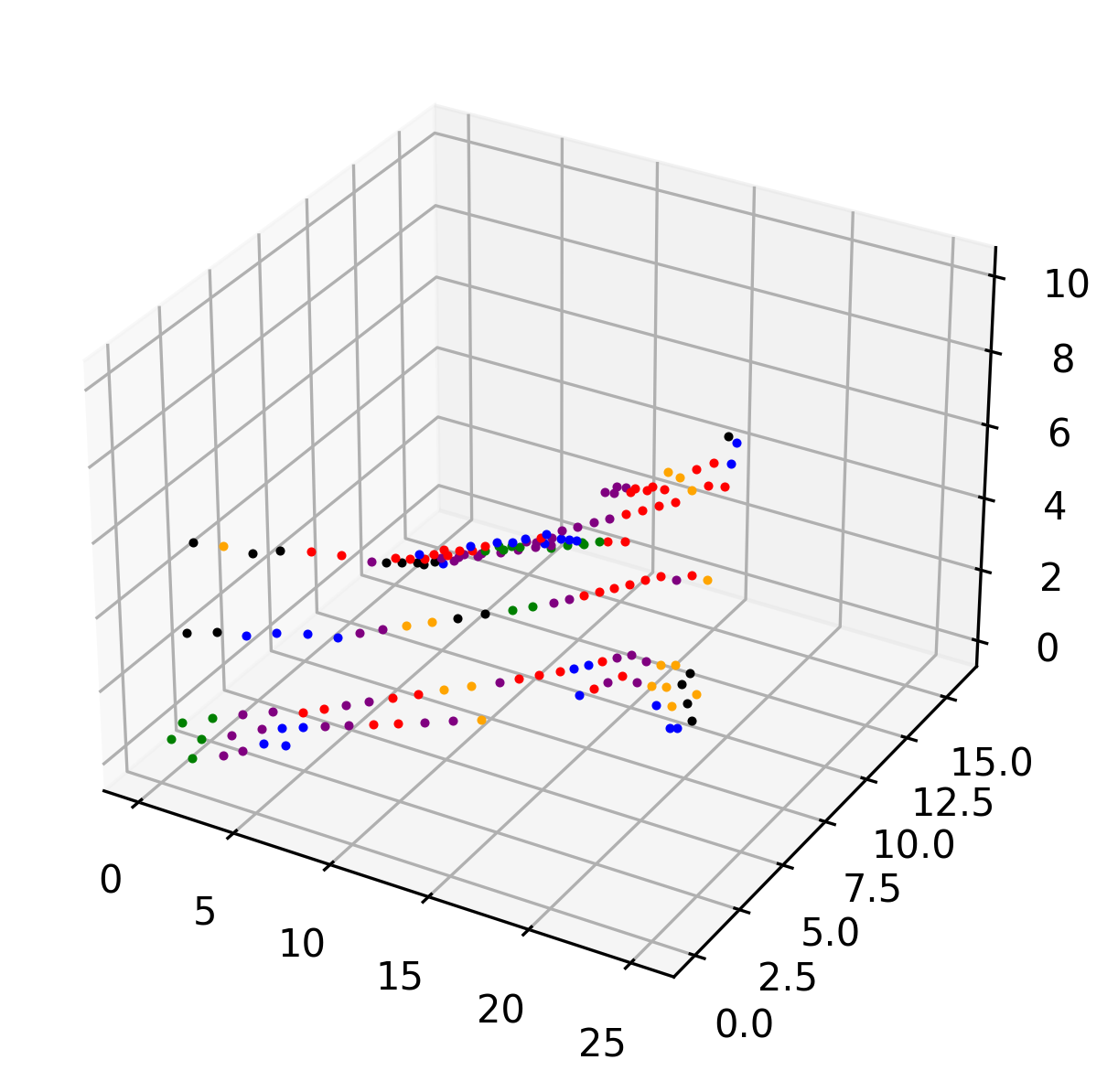}
\label{fig:lak110d_P}
\end{minipage}
}
\\
\subfloat[downsampled FastMap embedding]{
\begin{minipage}[b]{0.45\textwidth}
\centering
\includegraphics[width=\columnwidth]{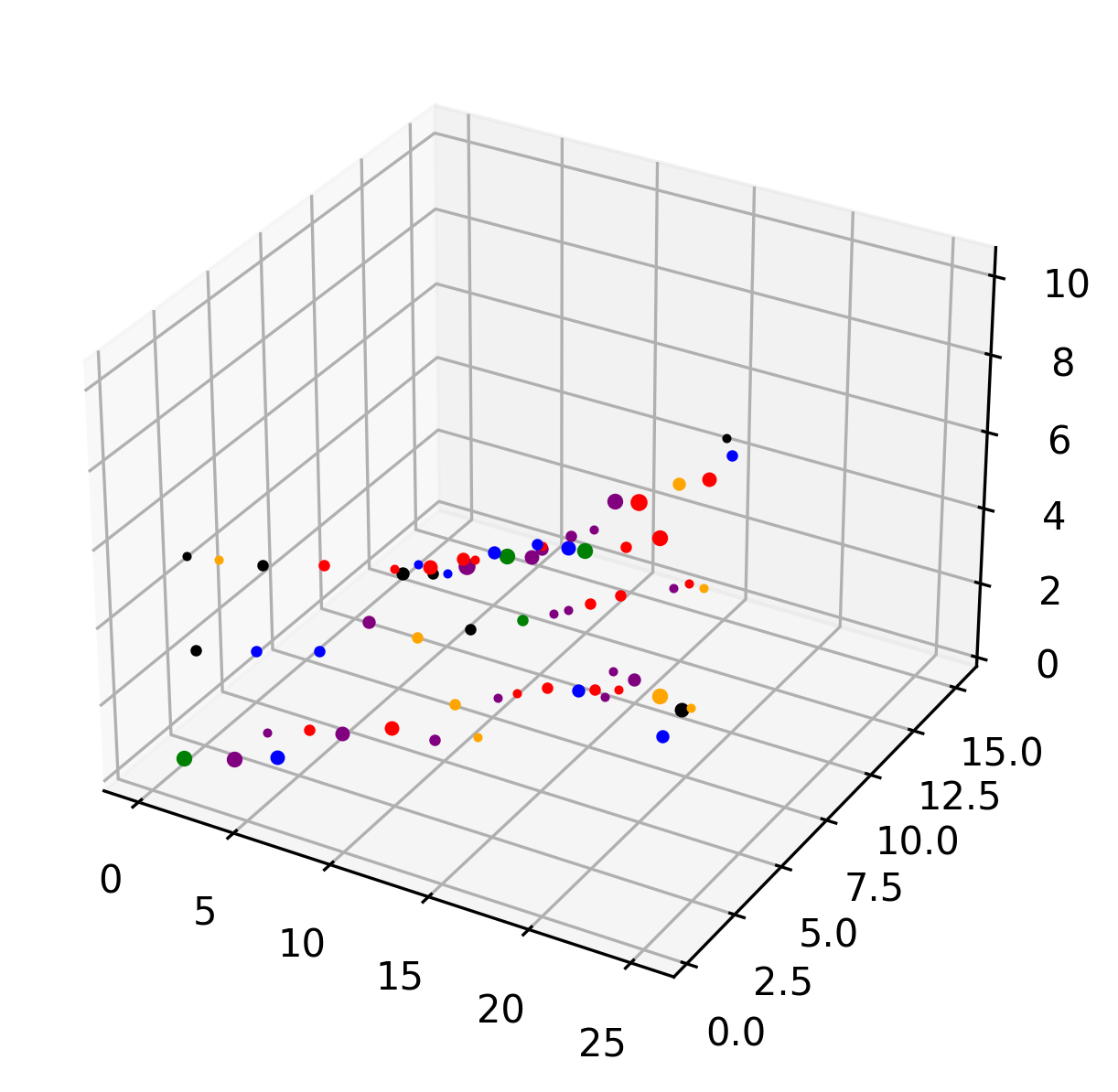}
\label{fig:lak110d_downP}
\end{minipage}
}
\subfloat[downsampled graph with $68$ vertices and $185$ edges]{
\begin{minipage}[b]{0.45\textwidth}
\centering
\includegraphics[width=\columnwidth]{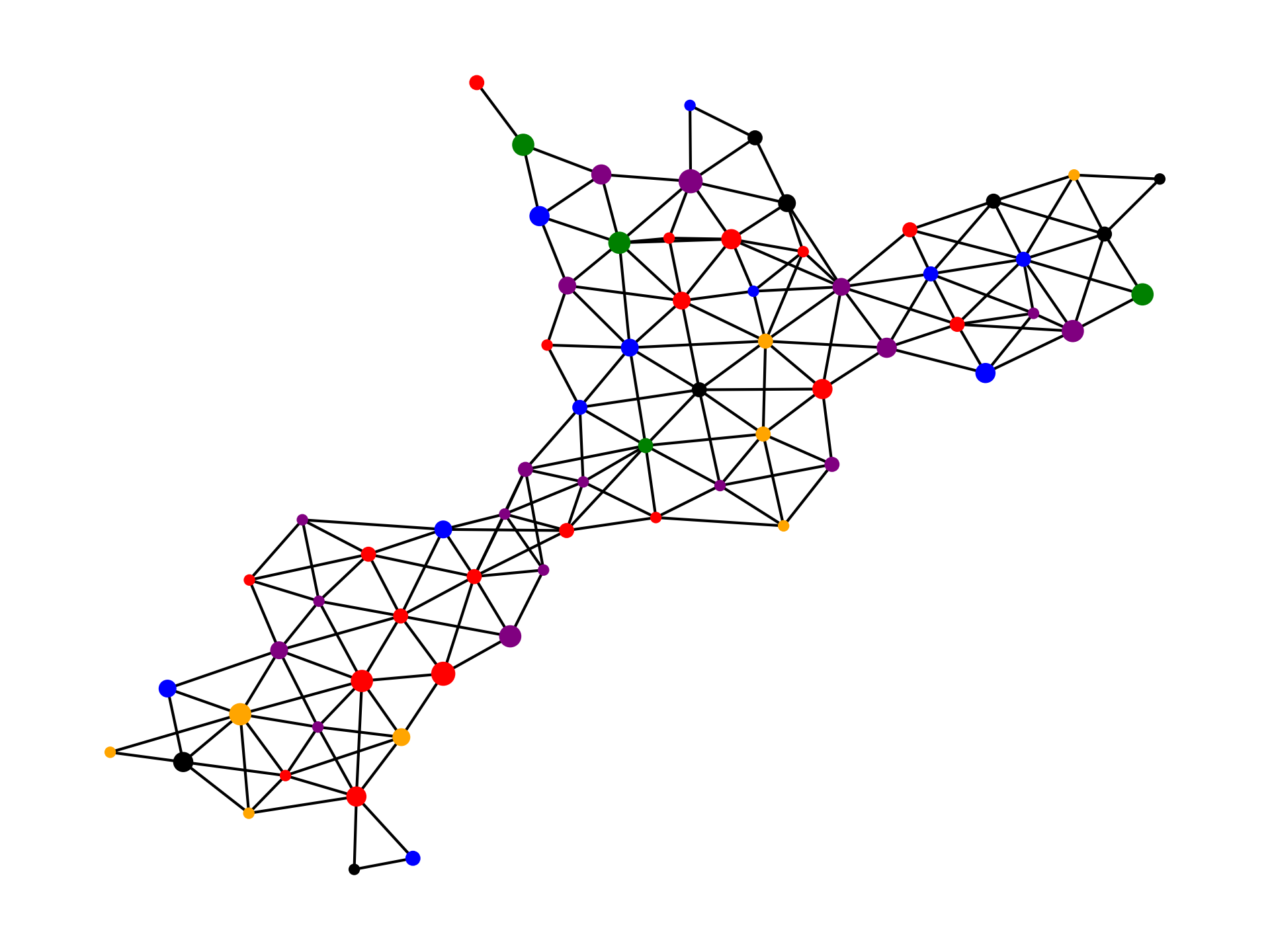}
\label{fig:lak110d_down}
\end{minipage}
}
\caption[An illustration of downsampling graphs via FastMap.]{Illustrates the process of downsampling graphs via FastMap. (a) shows the input graph. (b) shows the $3$-dimensional FastMap embedding of the input graph. (c) shows the downsampled points obtained from (b) by using a point cloud downsampling procedure on it with a voxel size of $2$. (d) shows a graph reconstructed from (a) and (c). End to end, (d) represents a downsampled version of the input graph from (a). We color each vertex of the downsampled graph using a randomly chosen color from the set red, orange, green, blue, purple, and black. The same colors are traced back from (d) to (c), from (c) to (b), and from (b) to (a), for each vertex. The colors are used to aid a visual mapping of various regions on the input graph to their corresponding regions on the downsampled graph. In (c) and (d), the radius of a dot is proportional to the number of vertices of the input graph that are mapped to it. In (d), the length of an edge represents the minimum length over all edges of the input graph that are mapped to it.}
\label{fig:downsampling_graph}
\end{figure}

Downsampling generally refers to the idea of reducing high-resolution spatial and/or temporal data to a lower resolution that is determined by storage, transmission bandwidth, or other computational restrictions of the problem domain. As a general concept, it is extensively used in many fields: in the analysis of time series data for summarization~\cite{s13}, in image processing for size reduction of images or videos~\cite{ld06}, and in acoustic and signal processing for approximation and compression~\cite{gss99}, among many others.

Downsampling can also be defined on graphs. Given an undirected edge-weighted graph $G = (V, E, w)$, where $w(e)$ is the non-negative weight on edge $e \in E$, the downsampling task is to create an undirected edge-weighted graph $G^d = (V^d, E^d, w^d)$, where $w^d(e^d)$ is the non-negative weight on edge $e^d \in E^d$, and a vertex mapping function $f^d: V \rightarrow V^d$, such that: (a) $|V^d|$ is within a user-specified range $\leq |V|$; (b) the edge $(v_s^d, v_t^d) \in E^d$ summarizes all edges $(v_i, v_j) \in E$, where $f^d(v_i) = v_s^d$ and $f^d(v_j) = v_t^d$; and (c) $G^d$ retains all the fundamental ``characteristics'' of $G$. The characteristics of $G$ that need to be retained in $G^d$ depend on the problem domain and may include graph-theoretic measures such as the diameter of the graph, the pairwise distances between its vertices, its spectral properties, and the behavior of certain kinds of algorithms on the graph, among many others.

Although some work has been done in~\cite{gd07,no11,nd14} for downsampling graphs, we propose a FastMap-based framework for doing so as one major direction of our future work. Figure~\ref{fig:downsampling_graph} illustrates our proposed framework. The main idea is to first embed the given graph $G$ in a $3$-dimensional Euclidean space using FastMap and then view it as a point cloud for downsampling.

A~\emph{point cloud} is a collection of data points in a $3$-dimensional Euclidean space that often represent point locations on the surface of an object or a landscape obtained by optical scanning or photogrammetry~\cite{wkwl02}. Downsampling of point clouds while retaining their fundamental shape characteristics is well studied in Computer Vision~\cite{zpk18}. Such procedures require a user-specified parameter referred to as the voxel size: All points that are within the same voxel are aggregated to a single point in the downsampling procedure.

Such a procedure applied on the FastMap embedding of $G$ would readily yield the downsampled points, that is, the vertices $V^d$ of the downsampled graph $G^d$. However, the edges $E^d$ of $G^d$ should be constructed in cognizance of the edges $E$ of $G$. There are many ways to do this. One way is to iterate over each edge $(v_i, v_j) \in E$ and incorporate it into the aggregate edge $(f^d(v_i), f^d(v_j)) \in E^d$. In doing so, $w^d((v_s^d, v_t^d))$ can be defined in many ways. For example, $w^d((v_s^d, v_t^d))$ can be defined to be the sum, average, or minimum over all $w(v_i, v_j)$ such that $f^d(v_i) = v_s^d$ and $f^d(v_j) = v_t^d$.

We will use our FastMap-based downsampling algorithm on graphs for various applications. First, if $w^d((v_s^d, v_t^d))$ is defined using the `minimum' operator, $G^d$ can be used to boost shortest-path computations. In particular, the all-pairs shortest-path distances on $G^d$ can be precomputed and used as admissible and consistent heuristic values to boost $A^*$ search at query time. Second, if $w^d((v_s^d, v_t^d))$ is defined using the `average' operator, $G^d$ can potentially be used to significantly speed up the computation of the centrality values of the vertices of $G$ albeit with a little distortion. Third, our ability to downsample graphs can also overcome some of the most important limitations of ML algorithms on them. Many existing ML and DL frameworks, including deep NNs, require the training and test graph instances to be of the same size (in terms of the number of vertices). This serious limitation can be overcome by using our downsampling method as a normalization procedure for the training and test graph instances. Fourth, our downsampling method can be used to create an appropriate testbed of smaller graphs on which various algorithms can be evaluated before their deployment on larger graphs. For instance, different network slicing algorithms can first be comparatively evaluated on a small substrate network with $50$-$60$ compute nodes; and the best algorithm can then be chosen for deployment on the real-world substrate network, which may be several orders of magnitude larger.

\subsection{FastMap Enhancements}

In another major direction of our future work, we will consider various enhancements to the algorithmic core of FastMap.

One such enhancement has been presented in~\cite{mas23} for the $L_1$ version of FastMap. The $L_1$ version of FastMap serves as a preprocessing technique to boost $A^*$ search for shortest-path computations~\cite{cujakk18}. In this context, it differs slightly from Algorithm~\ref{alg:fastmap} (from Chapter~\ref{ch:fastmap}) to ensure the admissibility and the consistency of the Euclidean distances that approximate the shortest-path distances on the graph. The enhancement to this $L_1$ version of FastMap comes from the incorporation of differential heuristics in the last iteration~\cite{mas23}. In future work, we will adapt the same enhancement to Algorithm~\ref{alg:fastmap} and evaluate it for better accuracy of the Euclidean embedding. If such an enhancement is indeed beneficial, we will likely be able to improve our performance metrics on all the problems discussed in the previous chapters.

A second enhancement has been presented in~\cite{gckk20} for directed graphs. Since Euclidean distances are inherently symmetric, Euclidean embeddings cannot be used as such for directed graphs. Hence, FastMap-D~\cite{gckk20} generalizes FastMap to directed graphs by using a potential field to capture the asymmetry in the pairwise distances between their vertices. It uses a self-supervised ML module and learns a potential function, by which it defines the potential field. Our future work will be motivated by FastMap-D: We will consider generalizing our FastMap-based algorithms presented in the previous chapters to directed graphs.

A third possible enhancement is impelled by the task of having to embed graphs in manifolds. It is well known that many kinds of real-world graphs can be embedded in nonlinear manifolds with better accuracy than in Euclidean spaces. For example, hyperbolic spaces are more suitable for embedding social networks compared to Euclidean spaces~\cite{vs14}. In future work, we will generalize FastMap to generate manifold embeddings of graphs. A promising way to do this is to examine every step of the FastMap algorithm and accomplish it using only dot products. If this can be done successfully, kernel functions popularly used with SVMs, can also be used with FastMap to implicitly create nonlinear transformations required for complex manifold embeddings.

\subsection{Mirroring and Solving Computational Geometry Problems on Graphs}

Although the field of Computational Geometry is in many ways as old as Geometry itself, it has evolved rapidly in the past thirty years. These rapid strides have come from advancements in the design and analysis of algorithms and their interplay with Complexity Theory, Geometry, and other mathematical techniques. There are many cornerstone problems in Computational Geometry, the studies of which have been motivated by their frequent occurrence in various real-world problem domains. Hence, Computational Geometry has many important problem formulations as well as techniques to offer to other areas of research. Indeed, such problems and techniques have already found plenty of relevance in AI and ML~\cite{mpc23}, Robotics~\cite{hks17}, Graphics~\cite{se02}, Computer-Aided Design~\cite{f85}, and Statistics~\cite{rh15}, among others.

In the third major direction of our future work, we will mirror and solve Computational Geometry problems on graphs. First, we will formulate the cornerstone problems from Computational Geometry in graph-theoretic terms. A good example of this is the graph convex hull problem that was already discussed in Chapter~\ref{ch:convex_hull}. We will consider the graph-theoretic counterparts of other Computational Geometry problems such as the generation of Voronoi diagrams, coresets\footnote{In Computational Geometry, a coreset is a smaller set of points that approximates the shape of a larger set of points.}, triangulations, meshes, and other tessellations. Second, we will study their relevance in real-world problem domains across different areas of research. For example, graph Voronoi diagrams may have applications in identifying the closest points of interest on a traffic network or the closest influencers on a social network; and graph coresets may be intimately related to downsampling graphs. Third, we will study the complexity of solving them directly on the input graph and develop baseline methods for them. Fourth, we will develop FastMap-based algorithms for solving them and compare our algorithms to the baseline methods and other state-of-the-art procedures. Finally, we will develop various connections between the graph counterparts of these problems by following the analogy of how they relate to each other in Computational Geometry.

\subsection{FastMapSVM: Further Applications and Enhancements}

\begin{figure}[t!]
\centering
\includegraphics[width=\columnwidth]{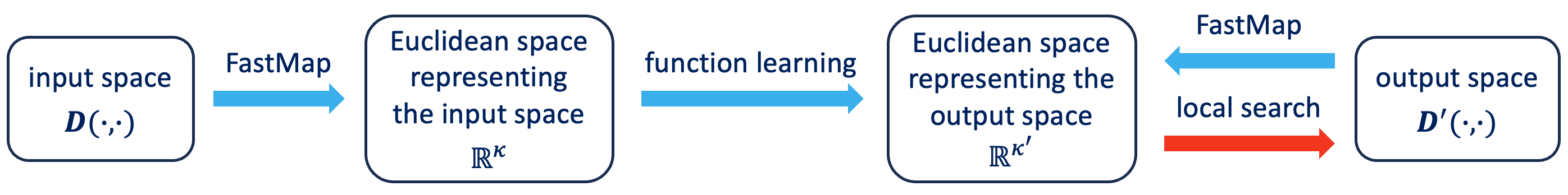}
\caption[A possible FastMapSVM enhancement that not only maps the input space to a Euclidean space but also maps the output space to a Euclidean space.]{Illustrates a possible FastMapSVM enhancement that not only maps the input space to a Euclidean space but also maps the output space to a Euclidean space. The complex objects in the input space are mapped to a Euclidean space $\mathbb{R}^{\kappa}$ using the distance function $D(\cdot, \cdot)$. Similarly, the complex objects in the output space are mapped to a Euclidean space $\mathbb{R}^{\kappa'}$ using the distance function $D'(\cdot, \cdot)$. The overall framework reduces the task of learning the mapping from the complex objects in the input space to the complex objects in the output space to the simpler task of learning a function that maps $\mathbb{R}^{\kappa}$ to $\mathbb{R}^{\kappa'}$. For a test instance, the point in $\mathbb{R}^{\kappa'}$ identified by the learned function can be transformed to a complex object in the required output space via a local search procedure that is guided by $D'(\cdot, \cdot)$.}
\label{fig:fastmapsvm_future}
\end{figure}

FastMapSVM is nascent: It allows for the use of very sophisticated distance functions and has a very fertile ground for its general applicability. Hence, in the fourth major direction of our future work, we will not only apply FastMapSVM in many other problem domains but will also enhance it to be able to produce complex outputs when required.

In the field of Optimization, we will apply FastMapSVM on Weighted CSPs since they can represent a wide range of optimization problems~\cite{k08}. A FastMapSVM-based classifier on Weighted CSPs can yield an algorithm selector: a meta-level procedure that can be used to choose from a portfolio of algorithms a specific algorithm that is best suited to solve a given instance of the problem. Similarly, a FastMapSVM-based regressor on Weighted CSPs can yield guidance for branch-and-bound search. In the more fundamental sciences, such as in Biochemistry, FastMapSVM can play a key role in facilitating ML techniques for tasks such as structure prediction and automated drug discovery. In such problem domains, a primary reason for its potential advantage over other methods is its ability to utilize ``chemical'' distance functions that have been developed by experts in those sciences~\cite{gkta10}.

While classifiers output discrete classification labels and regressors output real-valued numbers, there are many problem domains in which the output is also required to be a complex object. A good example of such a domain is Multi-Agent Path Finding (MAPF)~\cite{ssfkmwlackbb19}. In the MAPF problem, we are given a team of agents and an undirected graph that models their shared environment. Each agent has to move from a distinct start vertex to a distinct goal vertex while avoiding collisions with the other agents. Solving the MAPF problem optimally for minimum cost or minimum makespan is NP-hard~\cite{mtskk16}. In this problem domain, the input space has complex objects in the form of graphs and the concomitant agents' start and goal vertices. Moreover, the output space also has complex objects in the form of entire plans for the coordinating team of agents.

In the previous chapters, while we used FastMapSVM to simplify the representation of the complex objects in the input space, the output space was restricted to be a set of classification labels. However, in future work, we will enhance FastMapSVM on the output side as well to make it effective in MAPF and other domains where the output space also has complex objects. That is, we will not only use FastMap to simplify the representation of the complex objects in the input space but will also use it to simplify the representation of the complex objects in the output space. While the input space can be simplified to the Euclidean space $\mathbb{R}^{\kappa}$ using a distance function $D(\cdot, \cdot)$, the output space can be simplified to a different Euclidean space $\mathbb{R}^{\kappa'}$ using a different distance function $D'(\cdot, \cdot)$. Figure~\ref{fig:fastmapsvm_future} illustrates this possible enhancement.

The above generalization of FastMapSVM has an important potential benefit: It reduces the task of learning the mapping from the complex objects in the input space to the complex objects in the output space to the simpler task of learning a function that maps $\mathbb{R}^{\kappa}$ to $\mathbb{R}^{\kappa'}$. Hence, it may require significantly smaller amounts of data and time for model training compared to other ML and DL methods. For a test instance in this framework, the learned function identifies a point in $\mathbb{R}^{\kappa'}$ that still needs to be transformed to a complex object in the required output space. Towards this end, we can use a local search procedure that explores the subspace of the output space that maps to a point close to the identified point in $\mathbb{R}^{\kappa'}$. Local search operators in the output space can be used to morph a complex object to ``nearby'' but valid complex objects in the same space, neighborhoods in which can be defined using the distance function $D'(\cdot, \cdot)$. This FastMapSVM framework can also potentially serve as a ``generative'' AI framework because of its ability to produce complex objects.

\section{Concluding Remarks}

In this dissertation, we showed several advantages of FastMap and successfully leveraged them towards many real-world applications. However, the big question on whether FastMap is the holy grail for solving large-scale combinatorial problems on graphs, or on complex objects in other domains, still remains open. We hope that future research will enable us to conclusively answer this big question. At the same time, it is also evident that FastMap's ability to efficiently generate simplified representations of the vertices of a graph, or complex objects in other domains, enables many powerful downstream algorithms developed in diverse research communities such as AI, ML, Computational Geometry, Mathematics, Operations Research, and Theoretical Computer Science. Hence, we envision that FastMap can facilitate and harness the confluence of these algorithms and find future applications in many other problem domains that are not necessarily discussed here.

\clearpage
\chapter*{Bibliography}
\addcontentsline{toc}{chapter}{Bibliography}

\begin{singlespace}
\patchcmd{\bibsetup}{\interlinepenalty=5000}{\interlinepenalty=10000}{}{}
\setlength\bibitemsep{0.9\baselineskip}
\raggedright
\printbibliography[
heading = none
]

@inproceedings{ycz18,
title={Deep autoencoder-like nonnegative matrix factorization for community detection},
author={Ye, Fanghua and Chen, Chuan and Zheng, Zibin},
booktitle={Proceedings of the 27th ACM International Conference on Information and Knowledge Management},
year={2018}
}

@article{n06,
title={Finding community structure in networks using the eigenvectors of matrices},
author={Newman, Mark E.~J.},
journal={Physical Review E},
year={2006}
}

@inproceedings{rscrbld20,
title={Improving single and multi-view blockmodelling by algebraic simplification},
author={Ramteke, Rishabh and Stuckey, Peter and Chan, Jeffrey and Ramamohanarao, Kotagiri and Bailey, James and Leckie, Christopher and Demirovi{\'c}, Emir},
booktitle={Proceedings of the International Joint Conference on Neural Networks},
year={2020}
}

@techreport{hss08,
title={Exploring network structure, dynamics, and function using {NetworkX}},
author={Hagberg, Aric and Swart, Pieter J. and Schult, Daniel A.},
year={2008},
institution={Los Alamos National Lab, Los Alamos, NM (United States)}
}

@inproceedings{cujakk18,
title={The {FastMap} algorithm for shortest path computations},
author={Cohen, Liron and Uras, Tansel and Jahangiri, Shiva and Arunasalam, Aliyah and Koenig, Sven and Kumar, T.~K.~Satish},
booktitle={Proceedings of the 27th International Joint Conference on Artificial Intelligence},
year={2018}
}

@inproceedings{lfkk19,
title={Using {FastMap} to solve graph problems in a {Euclidean} space},
author={Li, Jiaoyang and Felner, Ariel and Koenig, Sven and Kumar, T.~K.~Satish},
booktitle={Proceedings of the 29th International Conference on Automated Planning and Scheduling},
year={2019}
}

@inproceedings{fl95,
title={{FastMap}: A fast algorithm for indexing, data-mining and visualization of traditional and multimedia datasets},
author={Faloutsos, Christos and Lin, King-Ip},
booktitle={Proceedings of the 1995 ACM SIGMOD International Conference on Management of Data},
year={1995}
}

@manual{d14,
title={USAir97},
author={Davis, Tim},
url={https://www.cise.ufl.edu/research/sparse/matrices/Pajek/USAir97},
year={2014}
}

@article{a16,
title={Dynamic range in the {C.}~elegans brain network},
author={Antonopoulos, Chris G.},
journal={Chaos: An Interdisciplinary Journal of Nonlinear Science},
year={2016}
}

@article{a17,
title={Community detection and stochastic block models: Recent developments},
author={Abbe, Emmanuel},
journal={Journal of Machine Learning Research},
year={2017}
}

@book{m12,
title={Machine learning: A probabilistic perspective},
author={Murphy, Kevin P.},
year={2012},
publisher={The MIT Press}
}

@article{ft87,
title={Fibonacci heaps and their uses in improved network optimization algorithms},
author={Fredman, Michael L. and Tarjan, Robert Endre},
journal={Journal of the ACM},
year={1987}
}

@article{lgl13,
title={Improved network community structure improves function prediction},
author={Lee, Juyong and Gross, Steven P. and Lee, Jooyoung},
journal={Scientific Reports},
year={2013}
}

@inproceedings{clklbr13,
title={Discovering latent blockmodels in sparse and noisy graphs using non-negative matrix factorisation},
author={Chan, Jeffrey and Liu, Wei and Kan, Andrey and Leckie, Christopher and Bailey, James and Ramamohanarao, Kotagiri},
booktitle={Proceedings of the 22nd ACM International Conference on Information \& Knowledge Management},
year={2013}
}

@inproceedings{lhwy15,
title={Understanding community effects on information diffusion},
author={Lin, Shuyang and Hu, Qingbo and Wang, Guan and Yu, Philip S.},
booktitle={Proceedings of the 19th Pacific-Asia Conference on Knowledge Discovery and Data Mining},
year={2015}
}

@article{gn02,
title={Community structure in social and biological networks},
author={Girvan, Michelle and Newman, Mark E.~J.},
journal={National Academy of Sciences},
year={2002}
}

@article{p14,
title={Efficient {Monte Carlo} and greedy heuristic for the inference of stochastic block models},
author={Peixoto, Tiago P.},
journal={Physical Review E},
year={2014}
}

@article{mdns21,
title={Generic constraint-based block modeling using constraint programming},
author={Mattenet, Alex and Davidson, Ian and Nijssen, Siegfried and Schaus, Pierre},
journal={Journal of Artificial Intelligence Research},
year={2021}
}

@inproceedings{gckk20,
title={Embedding directed graphs in potential fields using {FastMap-D}},
author={Gopalakrishnan, Sriram and Cohen, Liron and Koenig, Sven and Kumar, T.~K.~Satish},
booktitle={Proceedings of the 13th International Symposium on Combinatorial Search},
year={2020}
}

@article{d92,
title={Constraint networks},
author={Dechter, Rina},
journal={Artificial Intelligence},
year={1992}
}

@book{fh09,
title={Facility location: Concepts, models, algorithms and case studies},
author={Farahani, Reza Zanjirani and Hekmatfar, Masoud},
year={2009},
publisher={Springer Science \& Business Media}
}

@article{a91,
title={Voronoi diagrams--A survey of a fundamental geometric data structure},
author={Aurenhammer, Franz},
journal={ACM Computing Surveys},
year={1991}
}

@article{ry98,
title={Learning string-edit distance},
author={Ristad, Eric Sven and Yianilos, Peter N.},
journal={IEEE Transactions on Pattern Analysis and Machine Intelligence},
year={1998}
}

@article{lkd07,
title={Computation of {Minkowski} measures on {2D} and {3D} binary images},
author={Legland, David and Ki{\^e}u, Ki{\^e}n and Devaux, Marie-Fran{\c{c}}oise},
journal={Image Analysis \& Stereology},
year={2007}
}

@inproceedings{rka12,
title={Semantic cosine similarity},
author={Rahutomo, Faisal and Kitasuka, Teruaki and Aritsugi, Masayoshi},
booktitle={Proceedings of the 7th International Student Conference on Advanced Science and Technology},
year={2012}
}

@article{f62,
title={Algorithm 97: Shortest path},
author={Floyd, Robert W.},
journal={Communications of the ACM},
year={1962}
}

@article{gx98,
title={$K$-center and {$K$}-median problems in graded distances},
author={Guo-Hui, Lin and Xue, Guoliang},
journal={Theoretical Computer Science},
year={1998}
}

@inproceedings{kr87,
title={Clustering by means of medoids},
author={Kaufman, L. and Rousseeuw, Peter J.},
booktitle={Proceedings of the Statistical Data Analysis Based on the L1 Norm Conference, Neuchatel, Switzerland},
year={1987}
}

@article{sr21,
title={Fast and eager $k$-medoids clustering: $O(k)$ runtime improvement of the {PAM}, {CLARA}, and {CLARANS} algorithms},
author={Schubert, Erich and Rousseeuw, Peter J.},
journal={Information Systems},
year={2021}
}

@article{psc05,
title={The {Perron-Frobenius} theorem: Some of its applications},
author={Pillai, S.~Unnikrishna and Suel, Torsten and Cha, Seunghun},
journal={IEEE Signal Processing Magazine},
year={2005}
}

@techreport{pbmw99,
title={The {PageRank} citation ranking: Bringing order to the web},
author={Page, Lawrence and Brin, Sergey and Motwani, Rajeev and Winograd, Terry},
year={1999},
institution={Stanford InfoLab}
}

@article{k53,
title={A new status index derived from sociometric analysis},
author={Katz, Leo},
journal={Psychometrika},
year={1953}
}

@article{ew06,
title={Fast approximation of centrality},
author={Eppstein, David and Wang, Joseph},
journal={Graph Algorithms and Applications},
year={2006}
}

@inproceedings{y14,
title={Almost linear-time algorithms for adaptive betweenness centrality using hypergraph sketches},
author={Yoshida, Yuichi},
booktitle={Proceedings of the 20th ACM SIGKDD International Conference on Knowledge Discovery and Data Mining},
year={2014}
}

@inproceedings{gss08,
title={Better approximation of betweenness centrality},
author={Geisberger, Robert and Sanders, Peter and Schultes, Dominik},
booktitle={Proceedings of the 10th Workshop on Algorithm Engineering and Experiments},
year={2008}
}

@inproceedings{bms15,
title={Approximating betweenness centrality in large evolving networks},
author={Bergamini, Elisabetta and Meyerhenke, Henning and Staudt, Christian L.},
booktitle={Proceedings of the 17th Workshop on Algorithm Engineering and Experiments},
year={2015}
}

@inproceedings{bm15,
title={Fully-dynamic approximation of betweenness centrality},
author={Bergamini, Elisabetta and Meyerhenke, Henning},
booktitle={Proceedings of Algorithms-ESA 2015: The 23rd Annual European Symposium},
year={2015}
}

@inproceedings{bdr16,
title={Centrality measures on big graphs: Exact, approximated, and distributed algorithms},
author={Bonchi, Francesco and De Francisci Morales, Gianmarco and Riondato, Matteo},
booktitle={Proceedings of the 25th International Conference Companion on World Wide Web},
year={2016}
}

@inproceedings{diim04,
title={Locality-sensitive hashing scheme based on p-stable distributions},
author={Datar, Mayur and Immorlica, Nicole and Indyk, Piotr and Mirrokni, Vahab S.},
booktitle={Proceedings of the 20th Annual Symposium on Computational Geometry},
year={2004}
}

@article{f79,
title={Centrality in social networks conceptual clarification},
author={Freeman, Linton C.},
journal={Social Networks},
year={1979}
}

@article{bv14,
title={Axioms for centrality},
author={Boldi, Paolo and Vigna, Sebastiano},
journal={Internet Mathematics},
year={2014}
}

@inproceedings{bf05,
title={Centrality measures based on current flow},
author={Brandes, Ulrik and Fleischer, Daniel},
booktitle={Proceedings of the 22nd Annual Symposium on Theoretical Aspects of Computer Science},
year={2005}
}

@article{sz89,
title={Rethinking centrality: Methods and examples},
author={Stephenson, Karen and Zelen, Marvin},
journal={Social Networks},
year={1989}
}

@article{b87,
title={Power and centrality: A family of measures},
author={Bonacich, Phillip},
journal={American Journal of Sociology},
year={1987}
}

@article{b01,
title={A faster algorithm for betweenness centrality},
author={Brandes, Ulrik},
journal={Journal of Mathematical Sociology},
year={2001}
}

@article{b08,
title={On variants of shortest-path betweenness centrality and their generic computation},
author={Brandes, Ulrik},
journal={Social Networks},
year={2008}
}

@article{f77,
title={A set of measures of centrality based on betweenness},
author={Freeman, Linton C.},
journal={Sociometry},
year={1977}
}

@inproceedings{wwlhl13,
title={A theoretical analysis of {NDCG} type ranking measures},
author={Wang, Yining and Wang, Liwei and Li, Yuanzhi and He, Di and Liu, Tie-Yan},
booktitle={Proceedings of the 26th Annual Conference on Learning Theory},
year={2013}
}

@article{r91,
title={{TSPLIB}--A traveling salesman problem library},
author={Reinelt, Gerhard},
journal={ORSA Journal on Computing},
year={1991}
}

@article{s12,
title={Benchmarks for grid-based pathfinding},
author={Sturtevant, Nathan},
journal={Transactions on Computational Intelligence and AI in Games},
year={2012}
}

@article{nw99,
title={Renormalization group analysis of the small-world network model},
author={Newman, Mark E.~J. and Watts, Duncan J.},
journal={Physics Letters A},
year={1999}
}

@article{r87,
title={Silhouettes: A graphical aid to the interpretation and validation of cluster analysis},
author={Rousseeuw, Peter J.},
journal={Journal of Computational and Applied Mathematics},
year={1987}
}

@book{l10,
title={Random walk and the heat equation},
author={Lawler, Gregory F.},
year={2010},
publisher={American Mathematical Soc.}
}

@inproceedings{sfbsb09,
title={Memory-based heuristics for explicit state spaces},
author={Sturtevant, Nathan and Felner, Ariel and Barrer, Max and Schaeffer, Jonathan and Burch, Neil},
booktitle={Proceedings of the 21st International Joint Conference on Artificial Intelligence},
year={2009}
}

@inproceedings{cdpw14,
title={Computing classic closeness centrality, at scale},
author={Cohen, Edith and Delling, Daniel and Pajor, Thomas and Werneck, Renato F.},
booktitle={Proceedings of the 2nd ACM Conference on Online Social Networks},
year={2014}
}

@article{bbcmm19,
title={Computing top-$k$ closeness centrality faster in unweighted graphs},
author={Bergamini, Elisabetta and Borassi, Michele and Crescenzi, Pierluigi and Marino, Andrea and Meyerhenke, Henning},
journal={ACM Transactions on Knowledge Discovery from Data},
year={2019}
}

@book{d03,
title={Constraint processing},
author={Dechter, Rina},
year={2003},
publisher={Morgan Kaufmann}
}

@inproceedings{ahs10,
title={Continuous search in constraint programming},
author={Arbelaez, Alejandro and Hamadi, Youssef and Sebag, Michele},
booktitle={Proceedings of the 22nd IEEE International Conference on Tools with Artificial Intelligence},
year={2010}
}

@article{xhhl08,
title={{SATzilla}: Portfolio-based algorithm selection for {SAT}},
author={Xu, Lin and Hutter, Frank and Hoos, Holger H. and Leyton-Brown, Kevin},
journal={Journal of Artificial Intelligence Research},
year={2008}
}

@inproceedings{gjkmmnp10,
title={Learning when to use lazy learning in constraint solving},
author={Gent, Ian Philip and Jefferson, Christopher Anthony and Kotthoff, Lars and Miguel, Ian James and Moore, Neil Charles Armour and Nightingale, Peter and Petrie, Karen},
booktitle={Proceedings of the 19th European Conference on Artificial Intelligence},
year={2010}
}

@inproceedings{kmst10,
title={{ISAC}--Instance-specific algorithm configuration},
author={Kadioglu, Serdar and Malitsky, Yuri and Sellmann, Meinolf and Tierney, Kevin},
booktitle={Proceedings of the 19th European Conference on Artificial Intelligence},
year={2010}
}

@inproceedings{pt07,
title={A multi-engine solver for quantified {Boolean} formulas},
author={Pulina, Luca and Tacchella, Armando},
booktitle={Proceedings of the 13th International Conference on Principles and Practice of Constraint Programming},
year={2007}
}

@inproceedings{ohhno08,
title={Using case-based reasoning in an algorithm portfolio for constraint solving},
author={O'Mahony, Eoin and Hebrard, Emmanuel and Holland, Alan and Nugent, Conor and O'Sullivan, Barry},
booktitle={Proceedings of the 19th Irish Conference on Artificial Intelligence and Cognitive Science},
year={2008}
}

@article{k16,
title={Algorithm selection for combinatorial search problems: A survey},
author={Kotthoff, Lars},
journal={Data Mining and Constraint Programming: Foundations of a Cross-Disciplinary Approach},
year={2016}
}

@inproceedings{gm04,
title={Learning techniques for automatic algorithm portfolio selection},
author={Guerri, Alessio and Milano, Michela},
booktitle={Proceedings of the 16th European Conference on Artificial Intelligence},
year={2004}
}

@inproceedings{xkk18,
title={Towards effective deep learning for constraint satisfaction problems},
author={Xu, Hong and Koenig, Sven and Kumar, T.~K.~Satish},
booktitle={Proceedings of the 24th International Conference on Principles and Practice of Constraint Programming},
year={2018}
}

@article{s03,
title={Advanced support vector machines and kernel methods},
author={S{\'a}nchez A., V.~David},
journal={Neurocomputing},
year={2003}
}

@inproceedings{pc13,
title={{SVM} kernel functions for classification},
author={Patle, Arti and Chouhan, Deepak Singh},
booktitle={Proceedings of the International Conference on Advances in Technology and Engineering},
year={2013}
}

@article{ksh17,
title={{ImageNet} classification with deep convolutional neural networks},
author={Krizhevsky, Alex and Sutskever, Ilya and Hinton, Geoffrey E.},
journal={Communications of the ACM},
year={2017}
}

@inproceedings{zcnc18,
title={An end-to-end deep learning architecture for graph classification},
author={Zhang, Muhan and Cui, Zhicheng and Neumann, Marion and Chen, Yixin},
booktitle={Proceedings of the 32nd AAAI Conference on Artificial Intelligence},
year={2018}
}

@article{xhlj18,
title={How powerful are graph neural networks?},
author={Xu, Keyulu and Hu, Weihua and Leskovec, Jure and Jegelka, Stefanie},
journal={arXiv preprint arXiv:1810.00826},
year={2018}
}

@article{sd96,
title={Locating the phase transition in binary constraint satisfaction problems},
author={Smith, Barbara M. and Dyer, Martin E.},
journal={Artificial Intelligence},
year={1996}
}

@inproceedings{bka09,
title={Sparse kernel feature analysis using {FastMap} and its variants},
author={Ban, Tao and Kadobayashi, Youki and Abe, Shigeo},
booktitle={Proceedings of the International Joint Conference on Neural Networks},
year={2009}
}

@inproceedings{hlvw17,
title={Densely connected convolutional networks},
author={Huang, Gao and Liu, Zhuang and Van Der Maaten, Laurens and Weinberger, Kilian Q.},
booktitle={Proceedings of the IEEE Conference on Computer Vision and Pattern Recognition},
year={2017}
}

@article{entfhmwpn20,
title={The performance of deep neural networks in differentiating chest {X-rays} of {COVID-19} patients from other bacterial and viral pneumonias},
author={Elgendi, Mohamed and Nasir, Muhammad Umer and Tang, Qunfeng and Fletcher, Richard Ribon and Howard, Newton and Menon, Carlo and Ward, Rabab and Parker, William and Nicolaou, Savvas},
journal={Frontiers in Medicine},
year={2020}
}

@article{gr06,
title={The detection of low magnitude seismic events using array-based waveform correlation},
author={Gibbons, Steven J. and Ringdal, Frode},
journal={Geophysical Journal International},
year={2006},
}

@article{mszb19,
title={{STanford EArthquake} dataset ({STEAD}): A global data set of seismic signals for {AI}},
author={Mousavi, S.~Mostafa and Sheng, Yixiao and Zhu, Weiqiang and Beroza, Gregory C.},
journal={IEEE Access},
year={2019}
}

@article{mzsb19,
title={{CRED}: A deep residual network of convolutional and recurrent units for earthquake signal detection},
author={Mousavi, S.~Mostafa and Zhu, Weiqiang and Sheng, Yixiao and Beroza, Gregory C.},
journal={Scientific Reports},
year={2019}
}

@article{mezcb20,
title={Earthquake transformer--An attentive deep-learning model for simultaneous earthquake detection and phase picking},
author={Mousavi, S.~Mostafa and Ellsworth, William L. and Zhu, Weiqiang and Chuang, Lindsay Y. and Beroza, Gregory C.},
journal={Nature Communications},
year={2020}
}

@article{sbi07,
title={Non-volcanic tremor and low-frequency earthquake swarms},
author={Shelly, David R. and Beroza, Gregory C. and Ide, Satoshi},
journal={Nature},
year={2007}
}

@article{seh16,
title={Fluid-faulting evolution in high definition: Connecting fault structure and frequency-magnitude variations during the 2014 {Long Valley Caldera}, {California}, earthquake swarm},
author={Shelly, David R. and Ellsworth, William L. and Hill, David P.},
journal={Journal of Geophysical Research: Solid Earth},
year={2016}
}

@article{wmtrlbdghjmss22,
title={SeisBench--A toolbox for machine learning in seismology},
author={Woollam, Jack and M{\"u}nchmeyer, Jannes and Tilmann, Frederik and Rietbrock, Andreas and Lange, Dietrich and Bornstein, Thomas and Diehl, Tobias and Giunchi, Carlo and Haslinger, Florian and Jozinovi{\'c}, Dario and Michelini, Alberto and Saul, Joachim and Soto, Hugo},
journal={Seismological Society of America},
year={2022}
}

@article{jzwpm22,
title={A detailed earthquake catalog for {Banda Arc}--{Australian} plate collision zone using machine-learning phase picker and an automated workflow},
author={Jiang, Chengxin and Zhang, Ping and White, Malcolm C.~A. and Pickle, Robert and Miller, Meghan S.},
journal={The Seismic Record},
year={2022}
}

@article{jffl21,
title={Comparison of the earthquake detection abilities of {PhaseNet} and {EQTransformer} with the {Yangbi} and {Maduo} earthquakes},
author={Jiang, Ce and Fang, Lihua and Fan, Liping and Li, Boren},
journal={Earthquake Science},
year={2021},
}

@article{aflssk23,
title={Conflict-tolerant and conflict-free multi-agent meeting},
author={Atzmon, Dor and Felner, Ariel and Li, Jiaoyang and Shperberg, Shahaf and Sturtevant, Nathan and Koenig, Sven},
journal={Artificial Intelligence},
year={2023}
}

@article{ailrs15,
title={Practical and optimal {LSH} for angular distance},
author={Andoni, Alexandr and Indyk, Piotr and Laarhoven, Thijs and Razenshteyn, Ilya and Schmidt, Ludwig},
journal={Advances in Neural Information Processing Systems},
year={2015}
}

@techreport{bbd00,
title={Constrained k-means clustering},
author={Bradley, Paul S. and Bennett, Kristin P. and Demiriz, Ayhan},
year={2000},
institution={Microsoft Research, Redmond}
}

@article{fmb05,
title={On the continuous {Fermat-Weber} problem},
author={Fekete, S{\'a}ndor P. and Mitchell, Joseph S.~B. and Beurer, Karin},
journal={Operations Research},
year={2005}
}

@article{dm85,
title={Geometrical properties of the {Fermat-Weber} problem},
author={Durier, Roland and Michelot, Christian},
journal={European Journal of Operational Research},
year={1985}
}

@misc{g23,
title={{Gurobi Optimizer Reference Manual}},
author={{Gurobi Optimization, LLC}},
url={https://www.gurobi.com},
year={2023}
}

@article{m18,
title={Cluster: Cluster analysis basics and extensions},
author={Maechler, Martin},
journal={R Package Version 2.0.7--1},
year={2018}
}

@article{hgoa16,
title={Optimal any-angle pathfinding in practice},
author={Harabor, Daniel Damir and Grastien, Alban and {\"O}z, Dindar and Aksakalli, Vural},
journal={Journal of Artificial Intelligence Research},
year={2016}
}

@article{od98,
title={Strategic facility location: A review},
author={Owen, Susan Hesse and Daskin, Mark S.},
journal={European Journal of Operational Research},
year={1998}
}

@article{lhawtcn21,
title={{DeepSZ}: Identification of {Sunyaev--Zel'dovich} galaxy clusters using deep learning},
author={Lin, Zhen and Huang, Nicholas and Avestruz, Camille and Wu, W.~L.~Kimmy and Trivedi, Shubhendu and Caldeira, Jo{\~a}o and Nord, Brian},
journal={Monthly Notices of the Royal Astronomical Society},
year={2021},
}

@article{sfkzyzm19,
title={Super‐efficient cross‐correlation ({SEC-C}): A fast matched filtering code suitable for desktop computers},
author={Senobari, Nader Shakibay and Funning, Gareth J. and Keogh, Eamonn and Zhu, Yan and Yeh, Chin‐Chia Michael and Zimmerman, Zachary and Mueen, Abdullah},
journal={Seismological Research Letters},
year={2019}
}

@article{saskk22,
title={{AI-driven} quantification of ground glass opacities in lungs of {COVID-19} patients using {3D} computed tomography imaging},
author={Saha, Monjoy and Amin, Sagar B. and Sharma, Ashish and Kumar, T.~K.~Satish and Kalia, Rajiv K.},
journal={PLoS One},
year={2022}
}

@article{ffrha19,
title={{OR} models in urban service facility location: A critical review of applications and future developments},
author={Farahani, Reza Zanjirani and Fallah, Samira and Ruiz, Rub{\'e}n and Hosseini, Sara and Asgari, Nasrin},
journal={European Journal of Operational Research},
year={2019}
}

@book{mc03,
title={Theories of communication networks},
author={Monge, Peter R. and Contractor, Noshir S.},
year={2003},
publisher={Oxford University Press, USA}
}

@inproceedings{mas23,
title={Analyzing and improving the use of the {FastMap} embedding in pathfinding tasks},
author={Mashayekhi, Reza and Atzmon, Dor and Sturtevant, Nathan},
booktitle={Proceedings of the 37th AAAI Conference on Artificial Intelligence},
year={2023}
}

@article{pvgmtgbpwdvpcbpd11,
title={Scikit-learn: Machine learning in {Python}},
author={Pedregosa, F. and Varoquaux, G. and Gramfort, A. and Michel, V. and Thirion, B. and Grisel, O. and Blondel, M. and Prettenhofer, P. and Weiss, R. and Dubourg, V. and Vanderplas, J. and Passos, A. and Cournapeau, D. and Brucher, M. and Perrot, M. and Duchesnay, E.},
journal={Journal of Machine Learning Research},
year={2011}
}

@article{w88,
title={Routing of multipoint connections},
author={Waxman, Bernard M.},
journal={IEEE Journal on Selected Areas in Communications},
year={1988}
}

@article{wslkn23,
title={Classifying seismograms using the {FastMap} algorithm and support-vector machines},
author={White, Malcolm C.~A. and Sharma, Kushal and Li, Ang and Kumar, T.~K.~Satish and Nakata, Nori},
journal={Communications Engineering},
year={2023}
}

@inproceedings{slwk23,
title={A study of distance functions in {FastMapSVM} for classifying seismograms},
author={Sharma, Kushal and Li, Ang and White, Malcolm C.~A. and Kumar, T.~K.~Satish},
booktitle={Proceedings of the 22nd International Conference on Machine Learning and Applications},
year={2023}
}

@article{mppp97,
title={The {Jensen-Shannon} divergence},
author={Men{\'e}ndez, M.~L. and Pardo, J.~A. and Pardo, L. and Pardo, M.~C.},
journal={Journal of the Franklin Institute},
year={1997}
}

@article{vh14,
title={R{\'e}nyi divergence and {Kullback-Leibler} divergence},
author={Van Erven, Tim and Harremos, Peter},
journal={IEEE Transactions on Information Theory},
year={2014}
}

@article{bdh96,
title={The quickhull algorithm for convex hulls},
author={Barber, C.~Bradford and Dobkin, David P. and Huhdanpaa, Hannu},
journal={ACM Transactions on Mathematical Software},
year={1996}
}

@article{shw22,
title={A fast heuristic for computing geodesic cores in large networks},
author={Seiffarth, Florian and Horv{\'a}th, Tam{\'a}s and Wrobel, Stefan},
journal={arXiv preprint arXiv:2206.07350},
year={2022}
}

@misc{lk14,
title={{SNAP Datasets: Stanford Large Network Dataset Collection}},
author={Leskovec, Jure and Krevl, Andrej},
url={http://snap.stanford.edu/data},
year={2014}
}

@book{p13,
title={Geodesic convexity in graphs},
author={Pelayo, Ignacio M.},
year={2013},
publisher={Springer}
}

@article{tg21,
title={Active learning of convex halfspaces on graphs},
author={Thiessen, Maximilian and G{\"a}rtner, Thomas},
journal={Advances in Neural Information Processing Systems},
year={2021}
}

@incollection{s17,
title={Convex hull computations},
author={Seidel, Raimund},
booktitle={Handbook of Discrete and Computational Geometry},
year={2017},
publisher={Chapman and Hall/CRC}
}

@book{p04,
title={Polynomials},
author={Prasolov, Victor V.},
year={2004},
publisher={Springer Science \& Business Media}
}

@article{j76,
title={Normality and the numerical range},
author={Johnson, Charles R.},
journal={Linear Algebra and its Applications},
year={1976}
}

@incollection{h71,
title={Mathematical models for statistical decision theory},
author={Harris, Bernard},
booktitle={Optimizing Methods in Statistics},
year={1971},
publisher={Elsevier}
}

@article{k92,
title={Bicriteria network optimization problems},
author={Katoh, Naoki},
journal={IEICE Transactions on Fundamentals of Electronics, Communications and Computer Sciences},
year={1992}
}

@inproceedings{tlkrkk22,
title={The {FastMap} pipeline for facility location problems},
author={Thakoor, Omkar and Li, Ang and Koenig, Sven and Ravi, Srivatsan and Kline, Erik and Kumar, T.~K.~Satish},
booktitle={Proceedings of the 24th International Conference on Principles and Practice of Multi-Agent Systems},
year={2022}
}

@article{mlh03,
title={Solving the p-{Center} problem with {Tabu Search} and {Variable Neighborhood Search}},
author={Mladenovi{\'c}, Nenad and Labb{\'e}, Martine and Hansen, Pierre},
journal={Networks: An International Journal},
year={2003}
}

@phdthesis{s13,
title={Downsampling time series for visual representation},
author={Steinarsson, Sveinn},
year={2013}
}

@article{ld06,
title={Adaptive downsampling to improve image compression at low bit rates},
author={Lin, Weisi and Dong, Li},
journal={IEEE Transactions on Image Processing},
year={2006}
}

@inproceedings{gss99,
title={Multiresolution signal processing for meshes},
author={Guskov, Igor and Sweldens, Wim and Schr{\"o}der, Peter},
booktitle={Proceedings of the 26th Annual Conference on Computer Graphics and Interactive Techniques},
year={1999}
}

@inproceedings{no11,
title={Downsampling graphs using spectral theory},
author={Narang, Sunil K. and Ortega, Antonio},
booktitle={Proceedings of the 36th IEEE International Conference on Acoustics, Speech and Signal Processing},
year={2011}
}

@article{nd14,
title={Downsampling of signals on graphs via maximum spanning trees},
author={Nguyen, Ha Q. and Do, Minh N.},
journal={IEEE Transactions on Signal Processing},
year={2014}
}

@article{gd07,
title={Spectral coarse graining of complex networks},
author={Gfeller, David and De Los Rios, Paolo},
journal={Physical Review Letters},
year={2007}
}

@article{wkwl02,
title={A new segmentation method for point cloud data},
author={Woo, H. and Kang, E. and Wang, Semyung and Lee, Kwan H.},
journal={International Journal of Machine Tools and Manufacture},
year={2002}
}

@article{zpk18,
title={{Open3D}: A modern library for {3D} data processing},
author={Zhou, Qian-Yi and Park, Jaesik and Koltun, Vladlen},
journal={arXiv preprint arXiv:1801.09847},
year={2018}
}

@inproceedings{vs14,
title={Metric embedding, hyperbolic space, and social networks},
author={Verbeek, Kevin and Suri, Subhash},
booktitle={Proceedings of the 30th Annual Symposium on Computational Geometry},
year={2014}
}

@article{mpc23,
title={Challenges and opportunities in machine learning for geometry},
author={Magdalena-Benedicto, Rafael and P{\'e}rez-D{\'i}az, Sonia and Costa-Roig, Adri{\`a}},
journal={Mathematics},
year={2023}
}

@incollection{hks17,
title={Robotics},
author={Halperin, Dan and Kavraki, Lydia E. and Solovey, Kiril},
booktitle={Handbook of Discrete and Computational Geometry},
year={2017},
publisher={CRC Press}
}

@book{se02,
title={Geometric tools for computer graphics},
author={Schneider, Philip and Eberly, David H.},
year={2002},
publisher={Elsevier}
}

@book{f85,
title={Computational geometry and computer-aided design},
author={Fay, T.~H.},
year={1985},
publisher={National Aeronautics and Space Administration, Scientific and Technical Information Branch}
}

@article{rh15,
title={Statistical depth meets computational geometry: A short survey},
author={Rousseeuw, Peter J. and Hubert, Mia},
journal={arXiv preprint arXiv:1508.03828},
year={2015}
}

@inproceedings{k08,
title={A framework for hybrid tractability results in {Boolean} weighted constraint satisfaction problems},
author={Kumar, T.~K.~Satish},
booktitle={Proceedings of the 14th International Conference on Principles and Practice of Constraint Programming},
year={2008}
}

@article{gkta10,
title={Spatial chemical distance based on atomic property fields},
author={Grigoryan, A.~V. and Kufareva, Irina and Totrov, Maxim and Abagyan, R.~A.},
journal={Journal of Computer-Aided Molecular Design},
year={2010}
}

@inproceedings{ssfkmwlackbb19,
title={Multi-agent pathfinding: Definitions, variants, and benchmarks},
author={Stern, Roni and Sturtevant, Nathan and Felner, Ariel and Koenig, Sven and Ma, Hang and Walker, Thayne and Li, Jiaoyang and Atzmon, Dor and Cohen, Liron and Kumar, T.~K.~Satish and Boyarski, Eli and Bart{\'a}k, Roman},
booktitle={Proceedings of the 12th International Symposium on Combinatorial Search},
year={2019}
}

@inproceedings{mtskk16,
title={Multi-agent path finding with payload transfers and the package-exchange robot-routing problem},
author={Ma, Hang and Tovey, Craig and Sharon, Guni and Kumar, T.~K.~Satish and Koenig, Sven},
booktitle={Proceedings of the 30th AAAI Conference on Artificial Intelligence},
year={2016}
}
\end{singlespace}

\clearpage
\appendix
\section*{Appendices}
\addcontentsline{toc}{chapter}{Appendices}
\renewcommand{\thesection}{\Alph{section}}
\counterwithin{table}{section}

\section{Table of Abbreviations}

\begin{table}[h]
\centering
\begin{tabular}{|l|l|}
\cline{1-2}
Abbreviation &Expansion\\
\cline{1-2}
3C &three-component\\
AI &Artificial Intelligence\\
CNN &Convolutional Neural Network\\
CSPs &Constraint Satisfaction Problems\\
CVKM &Capacitated Vertex $K$-Median\\
DGCNN &deep graph convolutional neural network\\
DL &Deep Learning\\
DNA &deoxyribonucleic acid\\
ECGs &electrocardiograms\\
FMBM &FastMap-Based Block Modeling\\
FMGCH &FastMap-Based Graph Convex Hull\\
GIN &graph isomorphism network\\
GMM &Gaussian Mixture Model\\
ILP &Integer Linear Programming\\
LSH &Locality Sensitive Hashing\\
MAM &Multi-Agent Meeting\\
MAPF &Multi-Agent Path Finding\\
ML &Machine Learning\\
MRIs &magnetic resonance images\\
nDCG &normalized Discounted Cumulative Gain\\
NLP &Natural Language Processing\\
NMI &Normalized Mutual Information\\
NNs &Neural Networks\\
PAM &Partition Around Medoids\\
PASPD &probabilistically-amplified shortest-path distance\\
PCA &Principal Component Analysis\\
SAT &Satisfiability\\
SNR &signal-to-noise ratio\\
SRS &simple random sample\\
STEAD &Stanford Earthquake Dataset\\
SVMs &Support Vector Machines\\
VKC &Vertex $K$-Center\\
VKM &Vertex $K$-Median\\
WVKM &Weighted Vertex $K$-Median\\
\cline{1-2}
\end{tabular}
\caption[Abbreviations used in the dissertation.]{Describes the abbreviations used in the dissertation.}
\label{tab:abbrs}
\end{table}

\end{document}